\documentclass{article}
\usepackage[english]{babel}

\usepackage[letterpaper,top=2cm,bottom=2cm,left=2cm,right=2cm,marginparwidth=1.75cm]{geometry}

\usepackage{amssymb}
\usepackage{amsmath}
\usepackage{amsthm}
\usepackage{subcaption}
\usepackage{multirow}
\usepackage{graphicx}
\usepackage{amsfonts}
\usepackage{mathtools}
\usepackage{mathrsfs}
\usepackage{placeins}
\usepackage{ulem}

\usepackage[mathscr]{euscript}
\usepackage[dvipsnames,table]{xcolor}
\usepackage[    
    bookmarks=true,
    bookmarksopen=true,
    hidelinks
]{hyperref}
\usepackage{pgfplots}
\usetikzlibrary{external}
\usepackage{authblk}
\usepackage{todonotes}
\usepackage{array}
\usepackage{comment}
\usepackage{lscape}
\usepackage{tabularx}
\usepackage{longtable} 
\usepackage{colortbl}
\usepackage{cancel}
\usepackage{siunitx}
\usepackage{nicematrix}

\usepackage{bm}
\usepackage{tabularx}
\usepackage{float}
\usepackage[font=footnotesize,labelfont=bf]{caption}
\usepackage[overload]{empheq}

\usepackage{tikz}
\usepackage{tikz-3dplot}
\usetikzlibrary{shapes.geometric,arrows.meta, decorations.pathmorphing, decorations.pathreplacing, positioning, spy, patterns, patterns.meta, shapes.misc, calligraphy, backgrounds, fit, shapes.multipart, matrix, chains, positioning,decorations.pathreplacing,arrows,calc,graphs,graphs.standard,quotes}  
\pgfdeclarelayer{foreground}
\pgfsetlayers{main,foreground}

\usepackage[ruled]{algorithm2e}

\newcommand{\norm}[1]{\left\lVert#1\right\rVert}
\newcommand{\abs}[1]{\left\vert#1\right\vert}
\newtheorem{remark}{Remark}
\let\oldremark\remark
\renewcommand{\remark}{\oldremark\normalfont}

\colorlet{neworange}{orange!85!black}
\colorlet{newgreen}{green!60!black}
\definecolor{newblue}{RGB}{30, 70, 150}
\definecolor{newred}{RGB}{180, 40, 40}

\colorlet{fluidblue}{newblue!10!white}
\definecolor{wallgrey}{RGB}{200,200,200}
\definecolor{commentgreen}{RGB}{34,139,34}
\definecolor{keywordblue}{RGB}{0,51,153}    

\makeatletter
\newcommand{\CenterComment}[1]{%
    \par 
    \begingroup
        \small\ttfamily\color{commentgreen}
        /*\cleaders\hbox{-}\hfill\ 
        \textbf{#1} 
        \ \cleaders\hbox{-}\hfill*/
    \endgroup
    \par
}

\newcolumntype{C}{>{\centering\arraybackslash}p{2.5cm}}

\graphicspath{{./img/}}

\pgfplotsset{compat=1.18}

\providecommand{\keywords}[1]
{
  \small	
  \textbf{Keywords:} #1
}

\title{A Volume of Fluid Immersed Boundary Method for Industrial Polymer Mixing}

\author[1]{Emilia Capuano}
\author[2]{Daniele Cerroni}
\author[3]{Holger Marschall}
\author[2]{Giorgio Negrini}
\author[1]{Nicola Parolini}
\author[1]{Marco Verani}

\affil[1]{MOX, Department of Mathematics, Politecnico di Milano, Piazza Leonardo Da Vinci, 32, 20133, Milano, Italy}
\affil[2]{Pirelli Tyre S.p.A., Viale Piero e Alberto Pirelli, 25, 20126, Milano, Italy}
\affil[3]{Department of Mathematics, Computational Multiphase Flow
Group, Technische Universit\"at Darmstadt, Peter-Gr\"unberg-Stra\ss e 10, 64287, Darmstadt, Hessen, Germany}

\date{\empty}

\begin{document}
\maketitle

\begin{abstract}
Free-surface numerical simulation of polymer mixing processes requires handling rotating solid boundaries of complex shapes as well as kinematic viscosity ratios between polymer melt and air up to $10^8$. We present a finite volume method that couples a Volume of Fluid (VOF) interface-capturing scheme with a non-conforming immersed boundary (IB) method, so that moving screws are represented on a fixed mesh without remeshing.
In standard segregated VOF solvers, part of the viscous stress is treated explicitly, which can lead to instabilities and severe time-step restrictions at such viscosity ratios.
To address this limitation, we integrate into the VOF-IB framework a block-coupled (BC) scheme in which the full viscous stress, including its transposed-gradient part, is treated implicitly.
The method is implemented in the finite volume library OpenFOAM.
On a three-dimensional injection-molding test case with a generalized Newtonian melt, segregated solvers either diverge or produce spurious air-phase velocities at the time steps allowed by the block-coupled solver, which instead yields a physical solution and reduces the computing time by a factor of about 2.7 with respect to the fastest segregated run that remains physical.
On the finer grid, the IB solution agrees with the conforming-mesh solution within 2\% in velocity and 0.5\% in pressure drop.
The BC-VOF-IB solver is then applied to single- and twin-screw extruders under fully and partially filled conditions, capturing the filling dynamics and the resulting pressure build-up along the screw, demonstrating the applicability of the method to industrial geometries. The current formulation is isothermal; its extension to non-isothermal flows, required for quantitative prediction of melt temperature and viscous heating, is the subject of ongoing work.
\end{abstract}


\section{Introduction}
\label{sec:intro}

Polymer mixing processes constitute a large family of industrial processes aimed at homogenizing polymeric compounds to achieve desired chemical and physical properties for applications ranging from automotive components to medical devices \cite{tadmor2013principles}.
In practice, mixing occurs in the melt state, with polymers that are heated above their melting point and blended with other polymers or additives such as fillers, pigments, plasticizers, and stabilizers \cite{tadmor2013principles}. These operations are carried out using either continuous machines, such as single-screw extruders (SSEs) and twin-screw extruders (TSEs), or batch machines, such as internal mixers. In all mixers, rotating screws or rotors subject the material to high shear and temperature, converting mechanical energy into heat through viscous dissipation, promoting diffusive and convective transport of the components \cite{rauwendaal2014TSE}.
The quality of the final product, including its mechanical, thermal, rheological and optical properties, depends critically on how well components are distributed (spatial spreading) and dispersed (break-up of agglomerates) during mixing. Poor mixing leads to inhomogeneities such as agglomerates or uneven filler distribution, resulting in defective or under-performing parts.

Computational fluid dynamics (CFD) simulations are essential for understanding, designing and optimizing polymer mixing processes, as laboratory experiments alone are often costly, time-consuming and limited in their ability to resolve internal flow patterns, temperature fields and species distributions \cite{marquez2024predicting}. Numerical models allow for the reconstruction of three-dimensional velocity, pressure, stress and temperature fields, as well as local mixing indices that are otherwise inaccessible experimentally.

However, the numerical simulation of polymer mixing processes poses several challenges. First, the non-Newtonian rheology of polymer melts requires advanced constitutive equations to capture shear-thinning and elastic effects, introducing nonlinearities and potential numerical instabilities. Second, the problem is multiphysics: temperature, viscous dissipation, chemical reactions and mass transfer are coupled and must be solved simultaneously. Third, complex three-dimensional geometries with narrow gaps and moving parts demand careful mesh generation, often relying on unstructured, moving, or non-conforming grids. Lastly, multiple fluid phases can coexist in the mixing apparatus under partial filling conditions, where the polymer melt interacts with air or another gas, leading to free‑surface dynamics that must be captured using proper multiphase models. Dedicated interface‑tracking or interface-capturing algorithms are required; they involve additional transport equations and stricter stability constraints, which increase the cost relative to single-phase simulations.

Depending on the physical regime under consideration, several numerical methods are available in the literature to perform simulations of multiphase flows, which are typically classified according to the interface topology as either \textit{dispersed flows}, where discrete bubbles, droplets, or particles are suspended in a continuous phase, or \textit{segregated flows}, where phases are separated by a well-defined interface. 
In the present context, a segregated flow regime is assumed, consisting of two immiscible incompressible fluid phases coexisting within the same spatial domain and separated by an interface. 
Segregated flows are typically addressed by interface-resolving approaches, where the position and deformation of the interface are tracked or captured as part of the solution \cite{marschall2011phdthesis, mirjalili2017}. These methods are broadly divided into interface tracking and interface capturing \cite{elgeti2015, scardovelli1999,marschall2011phdthesis}.
Interface tracking methods, such as front-tracking \cite{Tryggvason2001}, moving mesh or Arbitrary Lagrangian-Eulerian (ALE) \cite{Hirt1974}, conform the mesh to the moving interface and offer high accuracy, but become computationally demanding under large deformations \cite{Tryggvason2001, Hirt1974}. 
In contrast, interface capturing methods rely on an Eulerian description, where the interface is implicitly identified by a scalar indicator function evolving on a fixed grid \cite{rusche2002phdthesis,Sun2016,Wang2012}. Volume of Fluid (VOF) \cite{hirt1981volume}, level set \cite{Sussman1994,OsherSethian1988,parolini2005finite,dipietro2006mass} and phase-field \cite{Anderson1998} are the most prominent approaches. 

We adopt the VOF method which captures the interface by solving a transport equation for the phasic volume fraction and conserves mass inherently \cite{scardovelli1999, mirjalili2017,Maronnier1999}. It is a well-established and extensively documented approach within the OpenFOAM framework \cite{openFoamRef}, the software used throughout this study, in line with its widespread adoption in both academic and industrial studies \cite{garcia2025,scardovelli1999}.
When simulating multiphase flows interacting with moving solid bodies of complex shapes, additional numerical challenges arise, as the inherent complexities of multiphase systems must be combined with the accurate and efficient treatment of a time-dependent computational domain.

Non-conforming approaches, such as Immersed Boundary (IB) methods, offer a practical solution, allowing to solve the flow on a fixed background grid while accounting for solid boundaries through forcing terms or velocity corrections \cite{Mittal2005,ikeno2007, jasak2014design}. Only a few studies combine IB with VOF \cite{Sun2016,Patel2017,LiMa2024,horgue2014penalization} or level set \cite{Calderer2014,gutierrez2018numerical} to obtain multiphase simulations within domains with complex boundaries, and they differ mainly in how the presence of the solid boundary is imposed. 

Applications of multiphase methods to polymer processing have grown in relevance, given the prevalence of partially filled (also referred to as \textit{starve-fed}) configurations in industrial mixers and extruders \cite{tadmor2013principles, rauwendaal2014TSE}, which feature complicated geometries with rotating boundaries, as well as highly viscous non-Newtonian material rheologies, typically modeled by generalized Newtonian constitutive laws or viscoelastic models \cite{macosko1994, ostwlad1925}. 
In the context of batch processes, VOF combined with moving-mesh techniques has been used to study filling ratio effects and viscous heating in the processing of rubber and polymers \cite{Ahmed2019fill, Poudyal2019speedratio,Wang2023rubber}. Concerning continuous processes, VOF with a cut-cell approach has been applied for simulations of partially filled single-screw extruders \cite{olofsson2023cfd, Larsen2024continuous}. To our knowledge, the twin-screw kneader \cite{Sun2016} is the only application of VOF-IB to a continuous mixing geometry. The level set method has also been used to model gas injection in a co-rotating twin-screw extruder \cite{Kousemaker2024levelset}. Nonetheless. none of these studies addresses the issues arising with high viscosity contrast between polymer melt and air.

The scarcity of works simultaneously addressing free-surface flows, rotating complex geometries, and non-Newtonian rheology, particularly for continuous mixing, motivates the development of our non-conforming interface-capturing framework. 
This work builds on an existing non-conforming immersed boundary method, developed through the consolidated partnership between Pirelli Tyre S.p.A. and the MOX Laboratory at Politecnico di Milano \cite{negrini2023phdthesis, negrini2025IBM}, extending the framework by integrating a VOF approach to simulate the two-phase flow of polymer melt and air in partially filled mixing devices. The proposed methods are implemented in the in-house polymer mixing library, built on the OpenFOAM C++ CFD library \cite{openFoamRef,jasak1996phdthesis,Weller1998OpenFOAM} (OpenFOAM-10 version from The OpenFOAM Foundation), which is a  Finite Volume (FV)--based software widely used for industrial CFD applications.

Two technical challenges are addressed. First, the volume-fraction transport and the pressure–velocity coupling must be made consistent with the IB constraints, including where the interface meets the immersed surface. Second, strong viscosity contrasts between polymer melt and air destabilizes the classical segregated VOF solvers of OpenFOAM.

The main contributions of the work are: (i) a non-conforming VOF-IB solver, where the IB conditions for velocity and volume fraction are imposed through the sharp WLS reconstruction of \cite{negrini2023phdthesis,negrini2025IBM} directly in the discrete systems and the pressure equation is derived to be consistent with the IB constraint; (ii) a block-coupled (BC) scheme for the momentum equation that employs a fully implicit treatment of the viscous diffusion term, including
its transposed-gradient part, to improve stability under high viscosity contrasts. The scheme extends the block-coupled finite volume discretization of Cardiff et al. \cite{cardiff2016block},
developed for linear elasticity, to incompressible two-phase flows with viscosity jumps across the interface, and its integration into the PIMPLE algorithm yields the conforming BC-VOF solver; (iii) a BC-VOF-IB solver, which combines (i) and (ii) and imposes the IB constraint on the block-coupled momentum system; (iv) a quantitative assessment on a three-dimensional injection-molding test case, comparing the block-coupled solvers with the standard segregated solver and with a segregated variant sharing the same algorithmic structure, on conforming and IB meshes; the BC-VOF-IB solver is then applied to an SSE and a co-rotating TSE under fully and partially filled conditions.
A graphical overview of the different VOF solvers, distinguishing between segregated and block-coupled, as well as conforming and IB approaches, is presented in Figure \ref{fig:ch1-graphicalIndexSolvers}.

\begin{figure}[H]
    \centering

\begin{tikzpicture}[
    thick,
    nodes={scale=0.8}, 
    base/.style={align=center, inner sep=4pt,thick},
    box/.style={draw, thick, rectangle, rounded corners=3pt, base},
    arrow/.style={thick},
    line/.style={thick},
    node distance=0.3cm and 0.1cm 
]

    \node[font=\bfseries] (top) {VOF solvers};

    \node[box, below left=0.35cm and 0.5cm of top,fill=newblue!10] (segregated) {
        \textbf{\textcolor{newblue!70!black}{Segregated}}
    };
    
    \node[box, below right=0.35cm and 0.8cm of top,fill=newred!10] (block) {
        \textbf{\textcolor{newred!70!black}{Block--coupled}}
    };

    \node[box, below=0.35cm of segregated, xshift=-1.8cm,fill=newblue!5] (vof) {
        \textbf{VOF}\\{\footnotesize \texttt{interFoam}}
    };
    
    \node[box, below=0.35cm of segregated, xshift=1.8cm,fill=newblue!5,draw=newgreen, dashed, ultra thick] (vofib) {
        \textbf{VOF--IB}\\{\footnotesize\texttt{interIbFoam}}
    };

    \node[box, below=0.35cm of block, xshift=-1.8cm,fill=newred!5,draw=newgreen, dashed, ultra thick] (bcvof) {
        \textbf{BC--VOF}\\{\footnotesize\texttt{UCInterFoam}}
    };
    
    \node[box, below=0.35cm of block, xshift=1.8cm,fill=newred!5,draw=newgreen, dashed, ultra thick] (bcvofib) {
        \textbf{BC--VOF--IB}\\{\footnotesize\texttt{UCInterIbFoam}}
    };

    \draw[thin] (top) -- (segregated);
    \draw[thin] (top) -- (block);
    
    \draw[thin] (segregated.south) -- (vof.north);
    \draw[thin] (segregated.south) -- (vofib.north);
    
    \draw[thin] (block.south) -- (bcvof.north);
    \draw[thin] (block.south) -- (bcvofib.north);

\end{tikzpicture}

\caption{Classification of the investigated Volume of Fluid (VOF) solvers. Dashed green outlines denote the original implementations developed in the present work.}
\label{fig:ch1-graphicalIndexSolvers}
\end{figure}

The formulation of the two-phase Navier--Stokes equations is presented in Section \ref{sec:model}. Afterwards, the formulation of the VOF-IB method is discussed in Section \ref{sec:methods-VOF-IB}, by first providing an overview of the underlying non-conforming IB approach \cite{negrini2025IBM} in Section \ref{sec:IB} and, secondly, presenting the finite volume discretization of the governing equations in Section \ref{sec:VOFIB}, with the VOF-IB PIMPLE solution algorithm in Section \ref{sec:VOFIB-PIMPLE}. Section \ref{sec:methods-BC-VOF-IB} is dedicated to the derivation of the BC approach, adapted from \cite{cardiff2016block}, and its coupling with the IB constraint.
Section \ref{sec:dogbone} assesses the solvers on the injection-molding test case, and Section \ref{sec:industrial-app} presents the extruder simulations. Conclusions are drawn in Section \ref{sec:discussion}.


\section{Mathematical Model}
\label{sec:model}

We consider a time-interval $\mathcal{I} \subset \mathbb{R}^+$ and a spatial domain $\Omega \subset \mathbb{R}^d$ containing two immiscible fluid phases. Each phase occupies the volume $\Omega_{k}, \ k = 1,2$, separated by the two-phase interface $\Sigma$ and such that $\Omega = \Omega_{1} \cup \Omega_{2}$.
We denote by $\bm\Sigma= | \Sigma | \bm n_{\Sigma}$ the surface normal vector associated with the interface separating the two phases, directed from phase 2 to phase 1, with $\bm n_\Sigma = \bm n_{21} = -\bm n_{12} =  \nabla\alpha/ \lVert\nabla\alpha\rVert$ unit normal vector and $|\bm\Sigma|$ surface area of the two-phase interface (see Figure \ref{fig:ch2-multiphase-CV}).

\begin{figure}[h!]
    \centering
    \vspace{0.05cm}
    \begin{tikzpicture}[>=Stealth]
    \small
    \begin{scope}
        \clip (0,0) ellipse (3cm and 2cm);
        \fill[newblue!10!white] (0,-2) .. controls (0.5,0) .. (0,2) 
            -- (-4,2) -- (-4,-2) -- cycle;
        -- (-4,2) -- (-4,-2) -- cycle;
    \end{scope}

    \begin{scope}
        \clip (0,0) ellipse (3cm and 2cm);
        \fill[newred!10!white] (0,-2) .. controls (0.5,0) .. (0,2) -- (4,2) -- (4,-2) -- cycle;
    \end{scope}

    \node (CV) [draw, thick, ellipse, minimum width=6cm, minimum height=4cm] at (0,0) {};
    \draw[dashed,black, thick] (0, -2) .. controls (0.5, 0) .. (0, 2);

    \draw[->, thick,newblue] (0.28, 0.7) -- ++(20:0.8) node[above,xshift=0.2cm]  {\footnotesize $\bm n_{12}$};
    
    \draw[->, thick,newred] (0.28, -0.7) -- ++(160:0.8) node[above] {\footnotesize $\bm n_{21}\equiv\bm n_\Sigma$};

    \node at (-3.3, 0.7) {$\Omega$};
    \node at (0.9, 0) {\color{black}$\Sigma$\color{black}};
    \node at (-1.4, -1.4) {\color{newblue}$\Omega_{1}$\color{black}};
    \node at (1.4, -1.4) {\color{newred}$\Omega_{2}$\color{black}};
    
    \end{tikzpicture}

    \caption{Two-dimensional sketch of the domain $\Omega$.}
    \label{fig:ch2-multiphase-CV}
\end{figure}

The VOF governing equations rely on a single-field (or mixture) approach, in which a single set of incompressible Navier--Stokes equations holds throughout the entire domain $\Omega$. 

Specifically, the derivation stems from two sets of incompressible Navier--Stokes equations valid in the two phase domains, equipped with proper interfacial coupling conditions on $\Sigma$. Then, by means of a conditional averaging procedure that avoids the otherwise unfeasible local microscopic description of the two-phase interface, the two-phase Navier--Stokes equations in single-field formulation are obtained, endowed with an advection equation for the two-phase interface evolution. The rigorous derivation is based on the works proposed in \cite{hill1998phdthesis,rusche2002phdthesis,marschall2011phdthesis}. In the following we report the main definitions and the final form of the two-phase VOF equations.

The \textit{phase indicator function} $\chi(\bm x,t)$ associated with phase 1 is defined as
\begin{equation}
	\label{eq:phaseindicator}
	\chi(\bm x,t) = 
	\begin{cases}
		1 & \text{if } \bm x \in V_1, \ t \in \mathcal{I}, \\
		0 & \text{if } \bm x \in V_2, \ t \in \mathcal{I},
	\end{cases}
\end{equation} 
where $V \subset \Omega$, such that $V = V_1 \cup V_2$, is a control volume contained in $\Omega$.

The \textit{phase volume fraction} is therefore computed as the volume average of $\chi$ over each control volume $V$:
\begin{equation}
	\label{eq:phasevolumefraction}
	\alpha = \overline{\chi} = \dfrac{1}{|V|} \int_V \chi \ dV = \dfrac{|V_1|}{|V|}.
\end{equation}

This leads to the definition of the two-phase mixture density and viscosity as the weighted average of the properties of the two fluids using the phase volume fraction as the weighting factor, such that $\rho = \alpha \rho_1+ (1-\alpha) \rho_2$ and $\mu = \alpha \mu_1 + (1-\alpha) \mu_2$, respectively. The arithmetic average is the standard choice in OpenFOAM; harmonic or mixed averages, such as the one proposed in \cite{ouafa2021monolithic}, are possible alternatives, whose influence is not investigated here.

\begin{remark}
    In the context of free-surface simulations of polymeric flows, we will always consider the highly viscous polymeric material to be fluid 1 and air will be indicated as fluid 2.
    As mentioned in Section \ref{sec:intro}, we model the polymeric viscosity by means of generalized Newtonian constitutive laws that relate the viscosity $\mu_1$ to the shear rate $\dot\gamma$ \cite{urraca2017}. In this work, we consider a power-law model \cite{ostwlad1925} $\mu_1(\dot \gamma) = K \dot \gamma^{n-1}$,
    where $K \ [\unit{Pa.s}^n]$ is the \textit{consistency factor}, equal to the material viscosity when $\dot\gamma = 1 \ [\unit{s^{-1}}]$, and $n$ is the power-law exponent. Polymers usually feature a shear-thinning behavior, meaning that viscosity decreases as shear rate increases, corresponding to the case $n<1$.
\end{remark}

A modified pressure, $p_{rgh} = p - \rho \bm g \cdot \bm x$, is defined by removing the hydrostatic pressure from the total pressure $p$, aiming at improving the numerical stability and simplifying the imposition of boundary conditions \cite{rusche2002phdthesis}. The gravitational acceleration is denoted by $\bm g$, while $\bm x$ indicates the position vector.

The two-phase VOF Navier--Stokes equations read:
\begin{subequations}
	\label{eq:VOFbiphaseNSequations}
	\begin{align}[left=\empheqlbrace]
		\dfrac{\partial \rho \bm u}{\partial t} + \nabla \cdot (\rho \bm u \otimes \bm u) - \nabla \cdot \Bigl[\mu (\nabla \bm u + \nabla^{\intercal} \bm u)\Bigr] + \nabla p_{rgh} &= \bm f_{\text{VOF}} && \text{in } \Omega \times \mathcal{I}, \label{eq:VOFmomentumbalance} \\
        \nabla \cdot \bm u &= 0 && \text{in } \Omega \times \mathcal{I}, \label{eq:VOFmassbalance} \\
		\dfrac{\partial \alpha}{\partial t} + \nabla \cdot (\alpha \bm u) +  \nabla \cdot (\alpha (1 - \alpha) \bm u_c) &= 0 && \text{in } \Omega\times \mathcal{I}, \label{eq:VOFtransport} \\
        \bm u &= \bm g_D && \text{on } \Gamma^D\times \mathcal{I}, \label{eq:dirichletVOF} \\
        \bm T \bm n &= \bm g_N && \text{on } \Gamma^N\times\mathcal{I}, \label{eq:neumannVOF}\\
        \alpha &= g_\alpha && \text{on } \Gamma^{\text{in}}\times\mathcal{I}, \\
        \bm u &= \bm u_0 && \text{in } \Omega,\\
        \alpha &= \alpha_0 && \text{in } \Omega.
	\end{align}
\end{subequations}

Equation \eqref{eq:VOFmomentumbalance} is the two-phase momentum balance, where an additional source term appears at right-hand-side, whose expression is given by 
\begin{equation}
    \label{eq:VOFmomentumbalance-source}
    \bm f_{\text{VOF}} = - \nabla\rho (\bm g \cdot \bm x) + \bm f_\sigma(\alpha) = - \nabla\rho (\bm g \cdot \bm x) -\sigma \nabla \cdot \left( \dfrac{\nabla \alpha}{\|\nabla \alpha\|} \right) \nabla \alpha.
\end{equation}
The first term results from the substitution of $\nabla p$ with $\nabla p_{rgh}$, while the second term accounts for the effect of surface tension, an interfacial force that arises from the momentum jump condition at the interface. Specifically, the \textit{Continuum Surface Force} (CSF) model of Brackbill et al. \cite{BRACKBILL1992335} is employed, allowing to treat the surface tension contribution as a volume source localized at the interface, defined as
\begin{equation}
	\label{eq:CSFBrackbill}
	\bm f_\sigma 
    = \sigma \kappa \nabla \alpha
    = -\sigma \nabla \cdot \left( \dfrac{\nabla \alpha}{\|\nabla \alpha\|} \right) \nabla \alpha,
\end{equation}
with $\kappa$ denoting the \textit{interface curvature}, computed as $\kappa = \nabla\cdot\left({-\bm n_\Sigma} \right)= \nabla\cdot \left( - \frac{\nabla \alpha}{\|\nabla \alpha\|} \right)$ and interface normal defined as $\bm n_\Sigma = \frac{\nabla\alpha}{|\nabla \alpha|}$.

Equation \eqref{eq:VOFmassbalance} is the mass balance, under the assumption of incompressible fluid phases.
Equation \eqref{eq:VOFtransport} is the advection equation for the phase volume fraction $\alpha$, used to follow the free-surface evolution. An additional nonlinear transport contribution is added for numerical purposes, introducing a \textit{compression velocity} $\bm u_c$ that acts to compress the interface thickness counteracting the numerical diffusion \cite{rusche2002phdthesis,Weller2006techreport,olsson2005conservative}. Specifically, the compressive velocity is set to be in the direction of $\nabla \alpha$, namely, normal to the interface, with magnitude proportional to $\| \bm u \|$ and it only acts in the vicinity of the free-surface thanks to the product $\alpha(1 - \alpha)$. The most common definition for the compressive velocity, which is also employed in this work, is 
\begin{equation}
	\label{eq:compressionvelocityVOF}
	\bm u_c = \min \left\{ c_\alpha \|{\bm u}\|, \max(\|{\bm u}\|) \right\} \dfrac{\nabla \alpha}{\|{\nabla \alpha}\|}, 
\end{equation}
where the constant parameter is usually chosen in the range $1 \leq c_\alpha \leq 4$ \cite{marschall2011phdthesis}; in this work, $c_\alpha = 1$, the default value in OpenFOAM, is used in all simulations.

Concerning boundary conditions, the domain frontier is subdivided into a Neumann and a Dirichlet portion: $\partial \Omega = \Gamma^D \cup \Gamma^N$. In the Neumann condition \eqref{eq:neumannVOF}, $\bm T$ is the Cauchy stress tensor, which is defined as $\bm T = -p \bm I + \mu \left(\nabla \bm u + \nabla^\intercal \bm u\right)$. Additionally, the inflow boundary $\Gamma^{\text{in}} := \{ \bm x \in \partial \Omega: \bm u \cdot \bm n < 0 \}$ is identified, with $\bm n$ the outward unit normal vector to $\partial \Omega$, to impose the inflow boundary datum $g_\alpha$ for the advection equation. Specifically, $g_\alpha=1$ on the whole boundary if only fluid 1 enters and $g_\alpha = 0$ if only fluid 2 enters the domain. In contrast, when both phases are present on the inflow boundary, $g_\alpha$ will be defined as a given Heaviside function.

At solid walls, either a no-slip condition, $\bm u = \bm u_w$, where the wall velocity $\bm u_w = \bm g_D$ using the notation of \eqref{eq:dirichletVOF}, or a Navier-slip condition can be prescribed. The latter imposes a tangential velocity relative to the wall proportional to the wall shear stress
\begin{equation}
    \label{eq:navierSlipBC}
    \left\{
    \begin{aligned}
        \bm u_n &= \bm u_{w,n} & \text{on } \Gamma_w, \\
        \bm u_t + \lambda \bm \tau_w &= \bm u_{w,t}  & \text{on } \Gamma_w.
    \end{aligned}
    \right.
\end{equation}
where $\lambda$ is the slip coefficient, $\bm u_n = (\bm u \cdot \bm n_w) \cdot \bm n_w$ and $\bm u_t = (\bm I - \bm n_w \otimes \bm n_w )\bm u$, with $\bm n_w$ unit normal vector to the solid boundary $\Gamma_w$. Both conditions have been implemented in the conforming and IB frameworks; unless otherwise stated, the no-slip condition is used.


\section{A VOF Non-Conforming Immersed Boundary Solver}
\label{sec:methods-VOF-IB}

The current section presents the Volume of Fluid Immersed Boundary numerical solver, that we denote by the acronym VOF-IB, in which we combine the IB approach proposed in \cite{negrini2025IBM}, which was initially only available for single-phase flows, with the VOF method, for the numerical approximation of Equations \eqref{eq:VOFbiphaseNSequations} in a non-conforming framework.

The spatial discretization of Equations \eqref{eq:VOFbiphaseNSequations} is based on the FV method \cite{FerzigerPeric2020}, that has been widely adopted in the CFD community since the mid-1970s.

The solution domain $\Omega \in \mathbb{R}^3$ is partitioned into $N$ Control Volumes (CVs) $K_i$ such that
\[
\Omega = \bigcup_{i = 1}^N K_i, \quad K_i \cap K_j = \varnothing \quad \forall K_i, K_j. 
\]
We denote by $\mathcal{T}_h$ the polyhedral tessellation of the computational domain. We indicate with $\mathcal{F}_h$ the set of faces of $\mathcal{T}_h$, with $\mathcal{F}_i = \{ F: K_i \cap F \neq \varnothing \}$ the set of faces of cell $K_i$ and with $\mathcal{K}_i$ the set of cells $K_j, \ j \neq i$ that share a face with $K_i$. 
The set of faces $\mathcal{F}_h$ can be subdivided into $\mathcal{F}_h = \mathcal{F}_{h,I} \cup \mathcal{F}_{h,B}$, denoting with $\mathcal{F}_{h,I}$ and $\mathcal{F}_{h,B}$ the sets of internal and boundary faces, respectively.
Then, let $F_{ij} \in \mathcal{F}_{h,I}$ be the face shared by cells $K_i$ and cell $K_j$, $\bm n_{ij}$ be its unit normal vector directed from $K_i$ to $K_j$ and $\bm S_{ij} = S_{ij} \bm n_{ij}$ be the surface area vector, where $S_{ij}$ is the area of $F_{ij}$.
Let us denote by $\bm x_i$ the barycenter of cell $K_i$ and by $\bm f_{ij}$ the barycenter of face $F_{ij}$. Finally, we indicate with $\bm d_{ij} = \bm x_j - \bm x_i$ the vector connecting the centroids of cells $K_i$ and $K_j$, with $ d_{ij} = \|{\bm x_j - \bm x_i}\|$ its length, and with $\bm x_{ij}$ the intersection of vector $\bm d_{ij}$ with face $F_{ij}$.

\subsection{The Immersed Boundary Method}
\label{sec:IB}

In the present section, the IB method available in our in-house polymer mixing library, which was first proposed and implemented by Negrini \cite{negrini2023phdthesis,negrini2025IBM}, is briefly recalled.

The IB approach employs a fixed background mesh, avoiding the complications related to the generation of conforming computational grids aligned with complex wall boundaries. Triangulated STL (stereolithography) surfaces representing the solid bodies immersed into the computational grid are used to mark the domain Degrees of Freedom (DOFs) as either fluid, solid or IB cells. Subsequently, the immersed boundary conditions are enforced at IB DOFs by means of a Weighted Least Squares (WLS) interpolation.

Let $\Sigma_\text{IB}$ be a closed triangulated surface representing the immersed boundary. Then, consider $\widetilde\Omega\in \mathbb{R}^3$ such that $\Omega \subset \widetilde\Omega$ and $\Sigma_\text{IB} \cap \widetilde\Omega \neq \varnothing$. We indicate with $\overline{\Sigma}_\text{IB}$ the volume enclosed by $\Sigma_\text{IB}$ and assume that the fluid flows outside of $\overline{\Sigma}_\text{IB}$, in $\widetilde\Omega \smallsetminus \overline{\Sigma}_\text{IB}$.

Using the triangulated surface, the elements of $\mathcal{T}_h$ are subdivided into a \textit{solid set} $\mathcal{T}_S =\{ K_i \in \mathcal{T}_h : \bm x_i \in \overline{\Sigma}_\text{IB} \}$, \textit{IB set} $\mathcal{T}_\text{IB} = \{ K_i \in \mathcal{T}_h \smallsetminus \mathcal{T}_S: K_i \in \mathcal{K}_j \text{ for some } K_j \in \mathcal{T}_S \}$ and \textit{fluid set} $\mathcal{T}_F = \{ K_i \in \mathcal{T}_h \smallsetminus (\mathcal{T}_S \cup \mathcal{T}_\text{IB}) \}$. We recall that $\mathcal{K}_j$ is the set of all cells $K_i$ sharing a face with $K_j$. Thus, the IB set is composed by all cells of $\mathcal{T}_h$ that have at least one solid cell among their neighbors.

The projections of IB cell centers on surface $\Sigma_\text{IB}$ are the IB points, whose set is denoted by $\mathcal{P}_\text{IB}=\{ \bm x_{\text{IBP},i}\}_{i: \ K_i \in \mathcal{T}_\text{IB}}$. The unit normal vectors at the IB points directed inwards $\mathcal{T}_S$ are the IB normals $\bm n_\text{IB}$. 

Analogously, the solid, IB and fluid face sets are defined as $\Gamma_S = \{ F_{ij} \in \mathcal{F}: K_i \in \mathcal{T}_S \land K_j \in \mathcal{T}_S\}$, $\Gamma_{\text{IB}} = \{ F_{ij} \in \mathcal{F}: K_i \in \mathcal{T}_{\text{IB}} \ \oplus \ K_j \in \mathcal{T}_{\text{IB}}\}$ and $\Gamma_F = \{ F_{ij} \in \mathcal{F}: K_i \in \mathcal{T}_F \land K_j \in \mathcal{T}_F \}$, respectively, where $\oplus$ is the logical operator representing the exclusive or and it is used in the definition of $\Gamma_\text{IB}$ to define the IB face set in order to exclude those faces shared by two cells $K_i$ and $K_j$ that are either both belonging to $\mathcal{T}_\text{IB}$, or neither of them is. 

Moreover, we can distinguish two subsets such that $\Gamma_\text{IB} = \Gamma_{\text{IB,int}} \cup \Gamma_{\text{IB,ext}}$ defining the \textit{IB internal face set} as $\Gamma_{\text{IB,int}} = \{ F_{ij} \in \mathcal{F}: K_i \in \mathcal{T}_{\text{IB}} \land K_j \in \mathcal{T}_{F} \}$ and the \textit{IB external face set} as $\Gamma_{\text{IB,ext}} = \{ F_{ij} \in \mathcal{F}: K_i \in \mathcal{T}_{\text{IB}} \land K_j \in \mathcal{T}_S \}$.

For each IB cell $K_i \in \mathcal{T}_\text{IB}$ a stencil $\mathcal{S}_i$ is defined to construct a WLS approximation operator, selecting cells in $\mathcal{T}_\text{IB} \cup \mathcal{T}_F$ whose barycenters satisfy the \textit{spatial distance}, \textit{connectivity distance} and \textit{field of view} criteria. The reader is referred to \cite{negrini2023phdthesis,negrini2025IBM} for a detailed description of the construction of $\mathcal{S}_i$.

Figure \ref{fig:ch3-ibm-scheme} shows a schematic of the geometrical elements that have just been introduced.

\begin{figure}[ht]
\centering

\begin{tikzpicture}[
    >=Stealth,
    cell/.style={rectangle, draw=black, minimum size=1cm, inner sep=0pt},
    fluid/.style={fill=newblue!20!white},
    solid/.style={fill=newred!20!white},
    ib/.style={fill=newgreen!20!white},
    ibi/.style={fill=newgreen!40!white}
]

    \def\shift{0.5}
    
    \foreach \x in {0,1,2,3} {
        \foreach \y in {0,1,2,3} {

            \pgfmathsetmacro{\cellstyle}{
                (\x < 2 && \y < 3) ? "solid" : (
                (\x == 2 && \y < 2) ? "solid" : (
                (\x == 3 && \y < 2) ? "ib" : (
                (\x < 2 && \y == 3) ? "ib" : 
                ((\x==2 && \y ==2) ? "ibi": "fluid" ))))
            }
            
            \node[\cellstyle, cell] (c-\x-\y) at (\x, \y) {};
            \fill[black] (\x, \y) circle (1.5pt); 
            
        }
    }

    \fill[pattern=north west lines, pattern color=black!50] 
      (1-\shift,3-\shift) rectangle (2-\shift,4-\shift);

    \fill[pattern=north west lines, pattern color=black!50] 
      (2-\shift,3-\shift) rectangle (3-\shift,4-\shift);
      
    \fill[pattern=north west lines, pattern color=black!50] 
      (3-\shift,3-\shift) rectangle (4-\shift,4-\shift);

    \fill[pattern=dots, pattern color=black!50] 
      (2-\shift,2-\shift) rectangle (3-\shift,3-\shift);

    \fill[pattern=north west lines, pattern color=black!50] 
      (3-\shift,2-\shift) rectangle (4-\shift,3-\shift);
      
    \fill[pattern=north west lines, pattern color=black!70] 
      (3-\shift,1-\shift) rectangle (4-\shift,2-\shift);
        
        \draw[newred, ultra thick] (0-\shift, 3-\shift) -- (2-\shift,3-\shift) -- (2-\shift, 2-\shift) -- (3-\shift, 2-\shift) -- (3-\shift,0-\shift);

        \draw[newblue, ultra thick] (2-\shift, 4-\shift) -- (2-\shift,3-\shift) -- (3-\shift, 3-\shift) -- (3-\shift, 2-\shift) -- (4-\shift,2-\shift);

        \draw[newgreen, ultra thick] ([shift=(0:3.2)] -0.5,-0.5) arc (0:90:3.2);

    \matrix[draw=none, anchor=north west, column sep=0.3cm, row sep=0.1cm] at (4.2, 3.6) {

        \draw[fill=newblue!20!white] (0,0) rectangle (0.4,0.2); 
        \node[right] at (0.4,0.1) {\footnotesize $\mathcal{T}_F$}; & 
        \draw[newgreen, ultra thick] (0,0.1) -- (0.4,0.1); 
        \node[right] at (0.4,0.1) {\footnotesize $\Sigma_\text{IB}$}; \\

        \draw[fill=newred!20!white] (0,0) rectangle (0.4,0.2); 
        \node[right] at (0.4,0.1) {\footnotesize $\mathcal{T}_S$}; & 
        \draw[newred, ultra thick] (0,0.1) -- (0.4,0.1); 
        \node[right] at (0.4,0.1) {\footnotesize $\Gamma_\text{IB,ext}$}; \\

        \draw[fill=newgreen!20!white] (0,0) rectangle (0.4,0.2); 
        \node[right] at (0.4,0.1) {\footnotesize $\mathcal{T}_\text{IB}$}; & 
        \draw[newblue, ultra thick] (0,0.1) -- (0.4,0.1); 
        \node[right] at (0.4,0.1) {\footnotesize $\Gamma_\text{IB,int}$}; \\

        \draw[pattern=north west lines, pattern color=black!50] (0,0) rectangle (0.4,0.2); 
        \node[right] at (0.4,0.1) {\footnotesize $\mathcal{S}_\text{i}$}; & 
        \draw[neworange, thick, ->] (0,0.1) -- (0.4,0.1); 
        \node[right] at (0.4,0.1) {\footnotesize $\bm{n}_\text{IB}$}; \\

        \draw[fill=newgreen!40!white,postaction={pattern={Dots[distance=2.5pt, radius=0.5pt]}, 
        pattern color=black!70}] (0,0) rectangle (0.4,0.2);\node[right] at (0.4,0.1) {\footnotesize $K_i$}; 
        &
        \draw[fill=neworange, thick] (0.2, 0.1) circle (2pt);
        \node[right] at (0.4,0.1) {\footnotesize $\mathcal{P}_\text{IB}$}; \\
    };

    \def\origx{-0.5}
    \def\origy{-0.5}    

    \def\ax{0}
    \def\ay{3}
    \def\axx{-0.047}
    \def\ayy{2.668}
    \def\axxx{-0.160}
    \def\ayyy{1.878}
    \draw[thick,dotted] (\ax,\ay) -- (\axx,\ayy) {}; 
    \draw[neworange, ultra thick,->] (\axx,\ayy) -- (\axxx,\ayyy) {}; 
    \draw[fill=neworange,thick] (\axx, \ayy) circle (1.5pt);

    \def\ax{1}
    \def\ay{3}
    \def\axx{0.761}
    \def\ayy{2.442}
    \def\axxx{0.446}
    \def\ayyy{1.707}
    \draw[thick,dotted] (\ax,\ay) -- (\axx,\ayy) {}; 
    \draw[neworange, ultra thick,->] (\axx,\ayy) -- (\axxx,\ayyy) {}; 
    \draw[fill=neworange,thick] (\axx, \ayy) circle (1.5pt);

    \def\ax{2}
    \def\ay{2}
    \def\axx{1.763}
    \def\ayy{1.763}
    \def\axxx{1.197}
    \def\ayyy{1.197}
    \draw[thick,dotted] (\ax,\ay) -- (\axx,\ayy) {}; 
    \draw[neworange, ultra thick,->] (\axx,\ayy) -- (\axxx,\ayyy) {}; 
    \draw[fill=neworange,thick] (\axx, \ayy) circle (1.5pt);

    \def\ax{3}
    \def\ay{1}
    \def\axx{2.441}
    \def\ayy{0.760}
    \def\axxx{1.706}
    \def\ayyy{0.445}
    \draw[thick,dotted] (\ax,\ay) -- (\axx,\ayy) {}; 
    \draw[neworange, ultra thick,->] (\axx,\ayy) -- (\axxx,\ayyy) {}; 
    \draw[fill=neworange,thick] (\axx, \ayy) circle (1.5pt);

    \def\ax{3}
    \def\ay{0}
    \def\axx{2.668}
    \def\ayy{-0.047}
    \def\axxx{1.876}
    \def\ayyy{-0.160}
    \draw[thick,dotted] (\ax,\ay) -- (\axx,\ayy) {}; 
    \draw[neworange, ultra thick,->] (\axx,\ayy) -- (\axxx,\ayyy) {}; 
    \draw[fill=neworange,thick] (\axx, \ayy) circle (1.5pt);

\end{tikzpicture}

\caption{Schematic representation of solid, IB and fluid cell and face sets, IB points, IB normals and IB stencil (adapted from \cite{negrini2023phdthesis}).}
\label{fig:ch3-ibm-scheme}

\end{figure}

Without loss of generality, we construct the WLS approximant for a scalar variable $\psi$ for which either Dirichlet or Neumann conditions are imposed on $\Sigma_\text{IB}$.
Given the basis of polynomials of degree $p$ in a point $\bm x = [x,y,z]^\intercal$, $\mathbb{P}_{\bm x} = [1,x,y,z,\dots,x^p,y^p,z^p]^\intercal$, we can define a polynomial function $f_{\text{IB},i}$ that approximates the value $\psi_i$ at each $i-$th IB cell center as
\begin{equation}
    \label{eq:ch3-IBApproxF}
    f_{\text{IB},i}(\bm x) = \mathbb{P}_{\bm x}^\intercal \bm \beta^i.
\end{equation}
The coefficients $\bm\beta^i = [\beta^i_0,\dots,\beta^i_{N_\text{coeffs}}]^\intercal$ are determined by solving a WLS minimization problem on the values $\psi_j$ stored in the centroids of cells $K_j \in \mathcal{S}_i$ associated with each $K_i \in \mathcal{T}_\text{IB}$, plus the assigned IB value, namely, the linearly reconstructed IB condition from the nodes of the triangulated surface to the IB points.

\subsubsection{Non-conforming Dirichlet conditions}
\label{sec:IB-Dirichlet}

Suppose that the Dirichlet condition $\psi|_{\Sigma_\text{IB}} = g_D$ is prescribed on $\Sigma_\text{IB}$.
The optimal coefficients $\bm \beta^{i,\star}$, for each IB cell  $K_i \in \mathcal{T}_\text{IB}$, provided as solution of the WLS problem, allow to compute the vector of linear combination coefficients $\bm s^i = \mathbb{P}_{\bm x_i}^\intercal \bm \beta^{i,\star}= [s^i_\text{IB},s^i_1,\dots, s^i_n]$ that allow to define the IB interpolation operator $S_\text{IB}$, such that, for each $K_i \in \mathcal{T}_h$, the corrected variable $\widetilde{\psi}_i$ reads \cite{negrini2025IBM}
\begin{equation}
    \label{eq:ch3-IBCorrVariablei}
    \widetilde\psi_i = [S_\text{IB}(\bm g, \bm\Psi)]_i =  S^i_\text{IB}(\bm g, \bm\Psi) = 
    \begin{cases}
        s^i_\text{IB} (g_{D,\text{IB}})_i + \sum_{j=1}^n s^i_j \psi_j & \text{if } K_i \in \mathcal{T}_\text{IB}, \\
        \psi_i & \text{if } K_i \in \mathcal{T}_F,\\
        g_{S,i} & \text{if } K_i \in \mathcal{T}_S,
    \end{cases}
\end{equation}
that can be compactly written as $\widetilde{\bm\Psi} = S_\text{IB}(\bm g_D,\bm\Psi) =S_g \bm  g + S \bm \Psi$, introducing matrices $S, S_g \in \mathbb{R}^{N\times N}$ and the vector of the IB and solid data $\bm g \in \mathbb{R}^N$. Vector $\bm g_{D,\text{IB}}$ collects the IB values, namely, the vector collecting the linear reconstruction of $g_D$ from the nodes of the triangulated surface to the IB points. 

\subsubsection{Non-conforming Neumann conditions}
\label{sec:IB-Neumann}

The imposition of Neumann boundary conditions on non-conforming boundaries is handled analogously to the Dirichlet case. Suppose a Neumann boundary condition is prescribed on $\Sigma_\text{IB}$, such that $\left(\nabla \psi \cdot \bm n \right)|_{\Sigma_\text{IB}} = g_N$. We denote by $\bm g_{N,\text{IB}}$ the vector of piecewise linear reconstructions of $g_N$ from the triangulation to the IB points, the Neumann condition at each IB point becomes $\nabla \psi(\bm x_{\text{IBP},i}) \cdot \bm n_{\text{IB},i} = (\bm g_{N,\text{IB}})_i$.

The WLS minimization problem slightly changes to impose a gradient value at the IB point. The optimal coefficients $\bm\beta^{i,\star}$ are then used to obtain an approximation of the gradient at IB external faces $f \in \Gamma_\text{IB,ext}$, which is where the Neumann condition is actually imposed at the algebraic level. For each face $f = F_{ij} \in \Gamma_\text{IB,ext}$, with barycenter $\bm f_{ij}$ and face normal $\bm n_{ij}$, linear combination coefficients $\bm s^{ij}$ can be computed, such that
\begin{equation}
    \label{eq:ch3-NeumannIBInterpDatum}
    g_{N,F_{ij}} = S^{ij}_\text{IB}(\bm g_{N,\text{IB}}, \bm \Psi) = s^{ij}_\text{IB} \ (g_{N,\text{IB}})_i + \sum_{k = 1}^n s_k^{ij} \psi_k. 
\end{equation}

\begin{remark}
    We employ this approach \cite{negrini2025IBM} to impose Dirichlet immersed boundary conditions for the velocity field. At rotating solid walls (screws), which are usually handled with the IB method, a known rigid motion is prescribed at each IB point located at $\bm x_{\text{IBP},i}$.
    On the contrary, homogeneous Neumann conditions are typically imposed for volume fraction at solid walls. This approach is not used to impose the immersed boundary Neumann conditions for pressure, to avoid stability issues arising with anisotropic grids. 
    Instead, the IB constraint is imposed on pressure within the PIMPLE solution algorithm presented in the next section, following the approach proposed in \cite{negrini2023phdthesis,negrini2025IBM,ikeno2007}.
\end{remark}

\subsection{The VOF-IB Numerical Solver}
\label{sec:VOFIB}

We start by deriving the finite volume discretization of Equations \eqref{eq:VOFbiphaseNSequations}, obtained by integrating on control volumes, applying the divergence theorem and approximating volume and surface integrals.

Concerning the time discretization, let $\{ t^n \}_{n=1,\dots,N_T}$ be a sequence of time steps in the time-interval $[0,T]$, with $t^{n+1} - t^n = \Delta t$ for $n=0,\dots,N_T - 1$ and let $\{t^k\}_{k=0,\dots,N_\text{sc}-1}$ be the discretization of the sub-interval $[t^n,t^{n+1}]$, where $t^{k+1} - t^k = \Delta t_{\text{sc}}=\frac{\Delta t}{N_\text{sc}}$, needed for the sub-cycling loop performed for the solution of the advection equation \eqref{eq:VOFtransport}, required to satisfy stability constraints on the local interface Courant number. 

The Courant number, also known as the Courant-Friedrichs-Lewy (CFL) number \cite{courant1928partiellen} is defined as $\text{Co} = \frac{U \Delta t}{h}$, where $U$ is the characteristic speed of the flow, that can be taken as $U = \norm{\bm u}_{L^\infty(\Omega)}$ \cite{galusinski2008stability}, and $h$ is the characteristic mesh size. The above definition arises from stability analysis of numerical methods for hyperbolic partial differential equations, particularly concerning explicit finite difference schemes, where typically the requirement is to have $\text{Co} \leq 1$.

At the discrete level, the local Courant number is computed in the FV context as
\begin{equation}
    \label{eq:courantNumberOF}
    \text{Co}_h = \max_{K_i \in \mathcal{T}_h} \frac{\Delta t}{\abs{K_i}} \sum_{f \in \mathcal{F}_i} \abs{\phi_f} 
\end{equation}
with $\phi_f = \bm u_f \cdot \bm S_f$ denoting the flux across faces.

Similarly, a local interface Courant number is defined, restricted to the interface region:
\begin{equation}
    \label{eq:ch3-alphaCourantNumberOF}
    \text{Co}_{\alpha,h} = \max_{K_i \in \mathcal{T}_h} \xi_i \frac{\Delta t}{\abs{K_i}} \sum_{f \in \mathcal{F}_i} \abs{\phi_f} 
\end{equation}
where for each $K_i \in \mathcal{T}_h$:
\begin{equation*}
    \xi_i = 
    \begin{cases}
        1 & \text{if} \quad 0.01 < \alpha_i < 0.99, \\
        0 & \text{if} \quad \alpha_i < 0.01 \vee \alpha_i > 0.99.
    \end{cases}
\end{equation*}
The interface Courant number is employed in OpenFOAM to control the stability of the interface transport equation. It is a heuristic CFL-like constraint, computed as the standard Courant number restricted by the mask $\xi_i$ to the interface region, where the nonlinear compression term is active.

\subsubsection{Non-conforming VOF advection equation}
\label{sec:VOFIB-alphaEq}

The numerical techniques employed for the solution of the advection equation \eqref{eq:VOFtransport} in OpenFOAM \cite{rusche2002phdthesis,Weller1998OpenFOAM,Deshpande2012} rely on a combination of high-resolution interpolation schemes, like van Leer \cite{vanLeer1974towards}, and a specialized flux-limiting algorithm known as MULES (Multidimensional Universal Limiter with Explicit Solution) \cite{Deshpande2012}, that allow to maintain interface sharpness and physical boundedness of $\alpha$ between 0 and 1.

Equation \eqref{eq:VOFtransport} is semi-discretized in time using the sub-time-step size $\Delta t_\text{sc}$, by means of a first order scheme and it is integrated on $K_i \in \mathcal{T}_h$. Therefore, at each $k-$th iteration of the sub-cycling procedure, the following holds: 
\begin{equation*}
    \int_{K_i} \frac{\alpha^{k+1} - \alpha^k}{\Delta t_\text{sc}} + \int_{\partial K_i} \alpha^\star \bm u^n \cdot \bm n \ dS + \int_{\partial K_i} \alpha^k (1 - \alpha^k) \bm u^n_c \cdot \bm n \ dS = 0.
\end{equation*}
The velocity field is taken from the previous global time step $t^n$, so $\bm u^k = \bm u^n$ for all $k = 0,\dots,N_\text{sc} - 1$.

Approximation of the volume integral using the mid-point quadrature formula and of the surface integrals by summing face fluxes across all faces of $K_i$ gives:
\begin{equation}
    \label{eq:ch3-FVVOFEquation-general}
    \alpha^{k+1}_{K_i} + \frac{\Delta t_\text{sc}}{\abs{K_i}} \sum_{f \in \mathcal{F}_i} \alpha_f^\star\phi_f^n
    = \alpha_{K_i}^k  
    - \frac{\Delta t_\text{sc}}{\abs{K_i}} \sum_{f \in \mathcal{F}_i} \alpha^k_f (1 - \alpha^k_f) \phi^n_{cf},
\end{equation}
where $\phi_f = \bm u_f \cdot \bm S_f$ is the flux of velocity through face $f$, while $\phi_{cf}$ denotes the interpolated \textit{compression flux} at face $f$, computed as 
\begin{equation}
    \label{eq:ch3-compression-flux}
    \phi_{cf} = \bm u_{c,f} \cdot \bm S_f = \min \left\{ c_\alpha \dfrac{|\phi_f|}{\| \bm S_f \|}, \max\left(\dfrac{|\phi_f|}{\| \bm S_f \|}\right) \right\} (\bm n_{\Sigma f} \cdot \bm S_f), 
\end{equation}
with $\bm u_{c,f}$ face-interpolated compression velocity \eqref{eq:compressionvelocityVOF} and $\bm n_{\Sigma f}$ interface unit normal vector interpolated at face $f$, which is given by
\begin{equation}
    \label{eq:ch3-discreteInterfaceNormal}
    \bm n_{\Sigma f} = \dfrac{(\nabla \alpha)_f}{\|{(\nabla \alpha)_f}\|}
\end{equation}
and $(\nabla\alpha)_f$ is determined using a central differencing (CD) scheme.

The approximation of face fluxes pertaining to the advection terms is carried out by means of the MULES approach \cite{Deshpande2012}, which is used to reduce the numerical diffusion induced by the upwind (UD) scheme at the interface by introducing higher order schemes for face-interpolation, localized at the interfacial region to keep a low computational cost. Specifically, the advection fluxes $\Phi_{f,\text{up}}$ and $\Phi_{f,\text{c}}$ are defined as:
\begin{subequations}
    \begin{align}
    \Phi^\star_{f,\text{up}} &= (\alpha^\star_f)_\text{up} \phi^n_f, \label{eq:ch3-upwindFlux} \\
    \Phi^\star_{f,\text{c}} &= \alpha^\star_f \phi^n_f - (\alpha^\star_f)_\text{up} \phi^n_f + \alpha^\star_f (1 - \alpha^\star_f) \phi^n_{cf}, \label{eq:ch3-corrFlux}
    \end{align}
\end{subequations}
where we denote by $(\alpha^\star_f)_\text{up}$ the face-interpolated value of $\alpha$ obtained by means of an upwind scheme and by $\alpha^\star_f$ the face-interpolated volume fraction computed as a blending of UD and CD schemes \cite{Deshpande2012,Berberovic2010} to achieve a compromise between boundedness and accuracy, as follows:
\begin{equation}
    \label{eq:ch3-alphaF}
    \alpha^\star_f = (1 - \lambda_\alpha) (\alpha^\star_f)_{\text{up}} + \lambda_\alpha (\alpha^\star_f)_{\text{cd}}.
\end{equation}
Various high resolution schemes are implemented in OpenFOAM for the computation of the blending factor $\lambda_\alpha \in  [0,1]$, the most commonly used being the van Leer scheme \cite{vanLeer1974towards}.

Face fluxes $\phi_f$ and $\phi_{cf}$ are computed at first global time iteration using central differencing for the face interpolation of the initial velocity condition, while at subsequent time instants the value of velocity fluxes is directly inherited from the solution of the pressure-velocity coupled problem (see Section \ref{sec:VOFIB-PIMPLE}).   

An explicit Forward Euler (FE) scheme is used for the solution of Equation \eqref{eq:ch3-FVVOFEquation-general}.
\begin{equation}
    \label{eq:ch3-FVVOFEquationExplicit}
    \begin{aligned}
        \alpha^{k+1}_{K_i} = \alpha_{K_i}^k  
        - \frac{\Delta t_\text{sc}}{\abs{K_i}} \sum_{f \in \mathcal{F}_i} \left[ \Phi^k_{f,\text{up}} + \lambda_M \Phi^k_{f,\text{c}} \right],
    \end{aligned}    
\end{equation}
The limiter parameter $\lambda_M$ is defined to be equal to 1 in the vicinity of the interface, while it is 0 elsewhere.

At the end of each sub-cycle, the mixture properties are updated. Then, the mass ﬂux through cell-faces is computed to obtain the total mass ﬂux corresponding to the global time step as the time average of the sub-cycle mass fluxes, that will be subsequently needed for the solution of the momentum equation:
\begin{equation}
    \label{eq:ch3-massFluxVOF}
    \begin{aligned}
    (\rho_f\phi_f)^{n+1} &
    = \sum_{k=0}^{N_\text{sc} - 1} \frac{\Delta t_\text{sc}}{\Delta t} \left[ (\alpha_f\phi_f)^{k+1} (\rho_1 - \rho_2) + \phi_f^n \rho_2 \right] \\
    &= \frac{1}{N_\text{sc}} \sum_{k=0}^{N_\text{sc} - 1} (\Phi^k_{f,\text{up}}  + \lambda_M \Phi_{f,\text{c}}^k) (\rho_1 - \rho_2) + \phi_f^n \rho_2.
\end{aligned}
\end{equation}

\begin{remark}
    An inner loop, denoted as $\alpha-$correction loop, is performed within each temporal sub-cycle to handle the nonlinearity introduced by the compression term, whose explicit treatment might not guarantee mass conservation if the equation was solved only once within each sub-time step. 
\end{remark}

At the discrete level, the volume fraction equation needs to be equipped with boundary conditions, not only on the inflow boundary $\Gamma^\text{in}$, but also on all other boundaries in $\partial\Omega \smallsetminus \Gamma^\text{in}$.  A common choice is prescribing homogeneous Neumann boundary condition for the volume fraction at solid boundaries. This choice is equivalent to the common and convenient neutral-wetting approximation at the wall, typically used in VOF computational solvers, where a static contact angle $\theta = 90^\circ$ is imposed at the contact line. 
This is appropriate for the present polymer-processing focus since global filling and transport are of primary interest, while a detailed characterization of polymer wettability is beyond the scope of this study. On no-slip walls, moreover, a moving contact line is singular \cite{huh1971}: it advances only through the numerical slip introduced by the finite-volume discretization near the wall, so that its
motion depends on the mesh resolution and on the representation of the wall.  A Navier-slip condition regularizes this singularity.

Since the IB surface typically identifies physical walls, homogeneous Neumann IB conditions are also prescribed for $\alpha$ on $\Sigma_\text{IB}$, namely
\begin{equation*}
     \nabla \alpha \cdot \bm n |_{\Sigma_\text{IB}} = 0.
\end{equation*}

The FE scheme consists of a unique explicit time advancement, thus, no linear system is solved, since the new solution is just a correction of the volume fraction coming from the previous iteration. 

The update of $\alpha$ is performed using \eqref{eq:ch3-FVVOFEquationExplicit} without any change with respect to the conforming case. Then, the resulting field $\bm \alpha$ is forced to satisfy the Neumann IB condition by modifying its value at solid-cell DOFs.
Specifically, since the IB Neumann condition is imposed at IB external faces, given a solid cell $K_j \in \mathcal{T}_S$ and an IB cell $K_i \in \mathcal{T}_\text{IB}$, the following relation holds for face $F_{ij} \in \Gamma_{\text{IB,ext}}$:
\begin{equation}
    \label{eq:ch3-FValphaCorr-IB}
    \alpha^{k+1}_j = \alpha^\star_i \pm g_{N,F_{ij}} \| \bm d_{ij}\|,
\end{equation}
with $g_{N,F_{ij}}$ computed using \eqref{eq:ch3-NeumannIBInterpDatum}.

\subsubsection{Non-conforming momentum equation}
\label{sec:VOFIB-momEq}

The semi-discretization in time of the momentum equation \eqref{eq:VOFmomentumbalance} is obtained applying a first-order backward Euler scheme, linearizing the convective term by lagging the advection velocity. The transposed velocity gradient in the viscous term is typically treated explicitly as well, allowing to decouple the equations for the three velocity components.
The coupling with the advection equation for $\alpha$ is handled in a segregated manner, using the updated field $\alpha^{n+1}$ into the discretized momentum equation. 

Integration of the semi-discretized equation over control volume $K_i \in \mathcal{T}_h$ and application of the divergence theorem yields:
\begin{equation}
    \label{eq:ch3-CVIntMomentumEquation}
    \begin{aligned}
        \int_{K_i} \rho^{n+1} \dfrac{\bm u^{n+1} - \bm u^n}{\Delta t} \ dV
        + \int_{\partial K_i} \rho^{n+1} & (\bm u^{n+1} \otimes \bm u^n) \bm n \ dS 
        - \int_{\partial K_i} \mu^{n+1} (\nabla \bm u^{n+1} + \nabla^{\intercal} \bm u^n) \bm n \ dS \\
        &+ \int_{\partial K_i} p^{n+1}_{rgh} \bm n \ dS = 
        \int_{K_i} - \nabla \rho^{n+1} (\bm g \cdot \bm x)  + \sigma\kappa^{n+1}\nabla\alpha^{n+1} \ dV,
    \end{aligned}
\end{equation}
where the superscript $n+1$ is explicitly written in $\rho^{n+1}$, $\mu^{n+1}$ and $\kappa^{n+1}$ to highlight their dependence on $\alpha^{n+1}$.

Approximating the volume integrals using the mid-point quadrature formula and the surface integrals by summing face fluxes across all faces of $K_i$, we obtain:
\begin{equation}
    \label{eq:ch3-FViMomentumEquation}
    \begin{aligned}
        \dfrac{\rho^{n+1}_i \abs{K_i}}{\Delta t} &\bm u^{n+1}_{i} 
        + \sum_{f \in \mathcal{F}_i} (\rho_f \phi_f)^{n+1} \bm u^{n+1}_f 
        - \sum_{f \in \mathcal{F}_i} \mu^{n+1}_f (\widetilde\nabla^\perp_f \bm u^{n+1}) \cdot \bm S_f
        + \sum_{f \in \mathcal{F}_i} p^{n+1}_{rgh,f} \bm S_f \\
        &= \abs{K_i} \mathcal{R}\left( \left\{(\bm g \cdot \bm x_f) \widetilde\nabla_f^\perp \rho^{n+1} \right\}_{f \in \mathcal{F}_i} \right) 
        + \abs{K_i} \mathcal{R}\left( \left\{ \sigma \kappa^{n+1}_f \widetilde\nabla_f^\perp \alpha^{n+1}\right\}_{f \in \mathcal{F}_i}\right)\\
        &+ \dfrac{\rho^{n+1}_i \abs{K_i}}{\Delta t} \bm u_{i}^n 
        + \sum_{f \in \mathcal{F}_i} \mu^{n+1}_f (\nabla^{\intercal}_h \bm u^n)_f \cdot \bm S_f.
    \end{aligned}
\end{equation}
The mass flux $(\rho_f\phi_f)^{n+1}$ in the convective term is computed after solving the advection equation by means of \eqref{eq:ch3-massFluxVOF}. 

The face-interpolated normal gradient of velocity is approximated as \cite{jasak1996phdthesis}:
\begin{equation}
    \label{eq:ch3-nonOrthNormalGrad}
    \widetilde{\nabla}_f^{\perp} \bm u = \frac{\bm u_{j} - \bm u_{i}}{\norm{\bm d_f \cdot \bm n_f}} + (\nabla_h \bm \psi^n)_f \cdot (\bm n_f - \bm n_\perp), 
\end{equation}
where the second term accounts for mesh non-orthogonality, with $\bm n_\perp$ component of the face unit normal vector $\bm n_f$ parallel to $\bm d_{ij}$ and $(\nabla_h \bm \psi)_f$ explicitly computed as
\begin{equation}
    \label{eq:ch3-faceLinInterpGrad}
    (\nabla_h \bm \psi^n)_f = w_{ij} (\nabla_{h,i} \psi^n) + (1 - w_{ij}) (\nabla_{h,j} \psi^n),
\end{equation}
with $w_{ij}$ interpolation weights \cite{jasak1996phdthesis}. 
Definition \eqref{eq:ch3-nonOrthNormalGrad} is used also to compute the gradient of density and volume fraction fields.
In addition, the transposed gradient of velocity is explicitly computed using \eqref{eq:ch3-faceLinInterpGrad}.

The source terms related to surface tension and hydrostatic pressure are originally computed as numerical fluxes stored at face centers. However, cell-centered values are required in the momentum equations \cite{tolle2020saample}. Therefore, in OpenFOAM VOF solvers a reconstruction operator $\mathcal{R}$ is employed to obtain cell-centered values of the source term \cite{weller2014curl}, namely 
\begin{equation}
    \label{eq:ch3-reconstructOp}
    \bm\psi_i \approx \mathcal{R}\left(\{ \bm\psi_f \}_{f \in \mathcal{F}_i}\right) = \left( \sum_{f \in \mathcal{F}_i} \bm n_f \otimes \bm S_f \right)^{-1} \left( \sum_{f \in \mathcal{F}_i} \bm n_f \otimes \bm\psi_f\right).
\end{equation}

Equation \eqref{eq:ch3-FViMomentumEquation} is valid for each $K_i \ \in \mathcal{T}_h$ and it can be compactly written in matrix form as
\begin{equation}
    \label{eq:ch3-FVMomentumEquation}
    A \bm U^{n+1} + B^{\intercal} \bm p^{n+1} = \bm F + \bm F_{\text{VOF}},
\end{equation}
with $\bm U^{n+1} \in \mathbb{R}^{3N}$ and $\bm p^{n+1} \in \mathbb{R}^N$ denoting the cell-centered FV variables associated with $\bm u^{n+1}$ and $p^{n+1}_{rgh}$ respectively; $A \in \mathbb{R}^{3N\times 3N}$ is the matrix containing coefficients coming from discretization of temporal, convective and viscous diffusion terms and $B \in \mathbb{R}^{N\times 3N}$ stores the coefficients related to the pressure gradient discretization.
The source terms $\bm F \in \mathbb{R}^{3N}$ and $\bm F_{\text{VOF}} \in \mathbb{R}^{3N}$ are given by:
\begin{align*}
    F_i &= \dfrac{\rho^{n+1}_i \abs{K_i}}{\Delta t} \bm u_{i}^n 
        + \sum_{f \in \mathcal{F}_i} \mu^{n+1}_f (\nabla_h^{\intercal} \bm u^n)_f \cdot \bm S_f; \\
    F_{\text{VOF},i} &= \abs{K_i} \mathcal{R}\left( \left\{(\bm g \cdot \bm x_f) \widetilde\nabla_f^\perp \rho^{n+1} \right\}_{f \in \mathcal{F}_i} \right) 
        + \abs{K_i} \mathcal{R}\left( \left\{ \sigma \kappa_f \widetilde\nabla_f^\perp \alpha^{n+1}\right\}_{f \in \mathcal{F}_i}\right).
\end{align*}

\begin{remark}
Within the source term $\bm F$ we include also all correction terms for skewness and non-orthogonality, which are usually treated explicitly. 
\end{remark}

Now, consider an IB surface $\Sigma_\text{IB}$ representing a physical wall such that $\bm u|_{\Sigma_\text{IB}} = \bm g_D$.
The IB interpolation operator is computed such that \eqref{eq:ch3-IBCorrVariablei} holds for the IB constrained velocity $\widetilde{\bm U}$. If $K_i \in \mathcal{T}_F$, then $\widetilde{\bm U}_i = \bm U_i$, therefore all the rows of the system associated with fluid DOFs remain unchanged.
If $K_i \in \mathcal{T}_S \cup \mathcal{T}_\text{IB}$ (solid or IB cell) the matrix coefficients are changed to satisfy $\widetilde{\bm U}_i = S^i_\text{IB}(\bm g, \bm U)$.

We introduce the IB constrained block-diagonal (see Section \ref{sec:VOFIB-highViscIssues}) matrix $\widetilde A \in \mathbb{R}^{3N \times 3N}$ and source term $\widetilde{\bm F} \in \mathbb{R}^{3N}$, such that the IB constrained momentum equation would read
\begin{equation}
    \label{eq:ch3-momentumVOFIBconstrained}
    \widetilde A \bm U + \chi B^\intercal \bm p = \widetilde{\bm F}+ \chi \bm F_\text{VOF},
\end{equation}
denoting by $\chi$ the IB mask such that
\begin{equation}
    \label{eq:ch3-IBMask}
    \chi_i = 
    \begin{cases}
    1 & \text{if } K_i \in \mathcal{T}_F \\
    0 & \text{if } K_i \in \mathcal{T}_\text{IB} \cup \mathcal{T}_S
    \end{cases}.
\end{equation}
Each block $\widetilde A^k, \ k = x,y,z$ of matrix $\widetilde A$ is constructed by imposing
\begin{equation}
    \label{eq:ch3-IBConstrainedA}
    \widetilde A^k_{ij} =
    \begin{cases}
        A^k_{ii} & \text{if } i = j, \\
        0 & \text{if } i \neq j \text{ and } \left(K_i \in \mathcal{T}_\text{IB} \cup \mathcal{T}_S \ \text{or} \ K_j \in \mathcal{T}_\text{IB} \cup \mathcal{T}_S\right), \\
        A^k_{ij} & \text{otherwise},
    \end{cases}
    \quad k = x,y,z \text{ and } i,j = 1,\dots,N
\end{equation}
and the corresponding source term is given by
\begin{equation}
    \label{eq:ch3-IBConstrainedF}
    {\widetilde{F}}^k_i =
    \begin{cases}
        A^k_{ii} S^i_\text{IB}(\bm g_k, \bm U_k) & \text{if } K_i \in \mathcal{T}_\text{IB} \cup \mathcal{T}_S, \\
        {F}^k_i - \displaystyle\sum_{K_j \in \mathcal{K}_i \cap (\mathcal{T}_\text{IB} \cup \mathcal{T}_S)}A^k_{ij}S^j_\text{IB}(\bm g_k, \bm U_k)  & \text{otherwise},
    \end{cases}
    \quad k = x,y,z \text{ and } i = 1,\dots,N.
\end{equation}
Each constrained row, i.e.\ each row with $K_i \in \mathcal{T}_\text{IB} \cup \mathcal{T}_S$, reduces to $A^k_{ii} (\widetilde U_k)_i = A_{ii} S^i_\text{IB}(\bm g_k, \bm U_k)$, namely $(\widetilde U_k)_i = S^i_\text{IB}(\bm g, \bm U)$. In the rows of fluid cells $K_i \in \mathcal{T}_F$, the contributions of the constrained neighbors $K_j \in \mathcal{K}_i \cap (\mathcal{T}_\text{IB} \cup \mathcal{T}_S)$ are moved to the right-hand side. The rows of fluid cells with no constrained neighbor, i.e.\ with $\mathcal{K}_i \cap (\mathcal{T}_\text{IB} \cup \mathcal{T}_S) = \varnothing$, are unchanged: $\widetilde A^k_{ij} = A^k_{ij}$ and $\widetilde F^k_i = F^k_i$.

\subsubsection{IB constrained pressure-velocity problem}
\label{sec:VOFIB-pUCoupling}

The continuity equation \eqref{eq:VOFmassbalance} is also integrated over $K_i$, the divergence theorem is applied and the resulting surface integral is approximated, leading to:
\begin{equation*}
    \int_{K_i} \nabla\cdot\bm u \ dV = \int_{\partial K_i} \bm u \cdot \bm n \ dS \approx \sum_{f \in \mathcal{F}_i} \bm u_f \cdot \bm S_f = 0,
\end{equation*}
which is compactly written in matrix form as:
\begin{equation}
    \label{eq:ch3-FVContinuityEquation}
    B \bm U = \bm 0.
\end{equation}

Putting \eqref{eq:ch3-FVMomentumEquation} and \eqref{eq:ch3-FVContinuityEquation} together we obtain the conforming pressure-velocity coupled algebraic system:
\begin{equation}
    \label{eq:ch3-FVnavierstokes}
    \begin{bmatrix}
        A & B^{\intercal} \\
        B & 0 
    \end{bmatrix}
    \begin{bmatrix}
        \bm U \\
        \bm p
    \end{bmatrix}
    =
    \begin{bmatrix}
        \bm F + \bm F_{\text{VOF}}\\
        \bm 0
    \end{bmatrix}.
\end{equation}

\begin{remark}
    The present work is based on the collocated finite volume method implemented in OpenFOAM, meaning that all variables are stored at cell centers, unlike staggered grids. However, the collocated approach is known to suffer from spurious pressure modes, often denoted as \textit{checkerboard} oscillations, where unphysical pressure fluctuations occur due to a weak numerical coupling between velocity and pressure fields \cite{FerzigerPeric2020}. The typical technique employed to stabilize the problem is the \textit{Rhie--Chow} (RC) interpolation \cite{moukalled2016FV,zhang2014generalized}. For the Stokes problem discretized with the Box method on Voronoi dual meshes, the RC stabilization has been shown to yield a discrete inf-sup condition, under assumptions whose validity was assessed numerically \cite{negrini2024convergence}. The same stabilization has proved robust when solving the IB-constrained Navier--Stokes equations for single-phase
    non-Newtonian flows in industrial mixing devices \cite{negrini2025IBM}. A theoretical result for the two-phase, variable-coefficient, IB-constrained system considered here is not available, and its stability is assessed numerically in Section \ref{sec:results}.

    At the algebraic level, the stabilized system becomes:
    \begin{equation}
    \label{eq:ch3-FVnavierstokes-RC}
    \begin{bmatrix}
        A & B^{\intercal} \\
        B & -C 
    \end{bmatrix}
    \begin{bmatrix}
        \bm U \\
        \bm p
    \end{bmatrix}
    =
    \begin{bmatrix}
        \bm F + \bm F_{\text{VOF}}\\
        \bm 0
    \end{bmatrix}.
    \end{equation}
with the Rhie-Chow stabilization matrix $C$ given by
\begin{equation}
    \label{eq:ch3-RhieChowMatrix}
    C = R(D^{-1}) - BD^{-1}B^\intercal
\end{equation}
where $D$ denotes the diagonal part of $A$ and $R(D^{-1})$ is a short hand to indicate a FV discretized Laplacian operator with diffusivity coefficient $D^{-1}$, following the notation of \cite{negrini2024convergence}, where further details on the assembly of matrix $C$ are reported.

From now on, we will refer to \eqref{eq:ch3-FVnavierstokes-RC} as the FV algebraic formulation associated with the momentum and continuity Equations \eqref{eq:VOFmomentumbalance} and \eqref{eq:VOFmassbalance}.

\end{remark}

The imposition of the IB constraint for velocity on the momentum equation, leads to the non-conforming pressure-velocity problem, that reads:
    \begin{equation}
    \label{eq:ch3-FVnavierstokes-IB}
    \begin{bmatrix}
        \widetilde A & \chi B^{\intercal} \\
        B & -C 
    \end{bmatrix}
    \begin{bmatrix}
        \bm U \\
        \bm p
    \end{bmatrix}
    =
    \begin{bmatrix}
        \widetilde{\bm F} + \chi \bm F_{\text{VOF}}\\
        \bm 0
    \end{bmatrix}.
\end{equation}

\subsubsection{The VOF-IB PIMPLE Algorithm}
\label{sec:VOFIB-PIMPLE}

The VOF-IB PIMPLE solution Algorithm \ref{alg:ch3-VOFIBPIMPLE} is formulated to obtain the non-conforming VOF-IB \texttt{interIbFoam} solver, starting from the reference conforming VOF \texttt{interFoam} solver \cite{Deshpande2012} of OpenFOAM-10 and employing a similar procedure to the one proposed in \cite{negrini2023phdthesis} for  single-phase non-conforming solvers.

The pressure-velocity coupling is handled by the majority of the CFD community using projection schemes, that consist of segregated solution algorithms of \textit{predictor-corrector} kind that allow to solve velocity and pressure fields sequentially, thereby avoiding the solution of the expensive monolithic system.
The basic idea is to determine a pressure field that ensures mass conservation, by re-writing the system of equations in terms of a momentum equation for velocity and a Poisson equation for pressure and solving them sequentially \cite{moukalled2016FV,FerzigerPeric2020}. 

OpenFOAM solvers are based on SIMPLE (Semi Implicit Method for Pressure Linked Equations) algorithm \cite{patankar1972,patankar1980numerical}, formulated for the steady-state Navier--Stokes equations, and its transient version, denoted as PISO (Pressure Implicit with Splitting of Operators) \cite{issa1986solution}.
PIMPLE combines the outer (SIMPLE-type) iterations with inner PISO corrections, which allows larger time steps than PISO alone.

First, the matrix $\widetilde A$ appearing in Equation \eqref{eq:ch3-momentumVOFIBconstrained} is split into its diagonal $\widetilde D$ and off-diagonal $\widetilde H$ components such that
\begin{equation*}
    [\widetilde H (\bm U)]_P = \widetilde{\bm F}_P - \sum_{N \in \mathcal{K}_P} \widetilde A_N \bm U_N,
\end{equation*}
so that the momentum equation can be written as
\begin{equation}
    \label{eq:ch3-splittingTwoPhaseMomEq}
    \widetilde D \bm U = \widetilde H(\bm U) + \chi \left( \bm F_{\text{VOF}} - B^{\intercal} \bm p\right)
\end{equation}
using the continuity equation $B\bm U^{m+1} - C\bm p^{m+1} = 0$, together with \eqref{eq:ch3-splittingTwoPhaseMomEq}, we obtain:
\begin{equation}
    \label{eq:ch3-SIMPLEpressurePoissonIB-incomplete}
    R(\widetilde D^{-1}) \bm p = B (\widetilde D^{-1} \widetilde H(\bm U) + \widetilde D^{-1} \bm F_\text{VOF}).
\end{equation}
Then, we would have a corrected velocity 
\begin{equation}
    \label{eq:correctedUconf} 
    \bm U = \widetilde D^{-1} \widetilde H(\bm U) + \widetilde D^{-1} \bm F_{\text{VOF}} - \widetilde D^{-1} B^{\intercal} \bm p,
\end{equation}
satisfying the incompressibility constraint. However, the velocity field has to be consistent with the IB condition:
\begin{equation}
    \label{eq:IBConstrainedU}
    \bm{U} = S_{\text{IB}}(\bm g, \bm{U}) 
           = S_g \bm g + S \bm{U},  
\end{equation}
thus, we substitute expression \eqref{eq:correctedUconf} for the unconstrained velocity into the right-hand side of \eqref{eq:IBConstrainedU}:
\begin{align}
    \label{eq:correctedIBConstrainedU}
    \bm{U} = S_g \bm g 
                + S \widetilde D^{-1}\widetilde H(\bm U)
                + S \widetilde D^{-1} \bm F_{\text{VOF}} 
                - S \widetilde{D}^{-1} B^{\intercal} \bm{p}.
\end{align}
Following \cite{negrini2025IBM}, the intermediate velocity $\widetilde D^{-1}\widetilde H(\bm U)$ is itself required to
satisfy the IB condition, i.e.\ the IB interpolation is applied to it before the pressure equation is assembled:
\[
    S_{\text{IB}}(\bm g, \widetilde D^{-1}\widetilde H(\bm{U})) = S_g \bm g + S \widetilde D^{-1}\widetilde H(\bm U) =  \widetilde D^{-1}\widetilde H(\bm U).
\]
Substituting this relation into the first two terms on the right-hand side of \eqref{eq:correctedIBConstrainedU} leads to
\begin{equation}
    \label{eq:ch3-PIMPLEVOFIB-IBconstrVel}
    \bm{U} = \widetilde D^{-1}\widetilde H(\bm U)
                + S \widetilde D^{-1} \bm F_{\text{VOF}} 
                - S \widetilde{D}^{-1} B^{\intercal} \bm{p},
\end{equation}
which, by linearity of $S_\text{IB}$, satisfies the IB constraint exactly at each corrector step.
Finally, we impose again the validity of the incompressibility constraint:
\begin{align*}
    0 = B\bm{U} - C\bm p 
    &= B \widetilde D^{-1}\widetilde H(\bm U)
    + B S \widetilde D^{-1} \bm F_{\text{VOF}} 
    - B S \widetilde{D}^{-1} B^{\intercal} \bm{p} - C\bm p\\
    &= B \widetilde D^{-1}\widetilde H(\bm U)
    + B S \widetilde D^{-1} \bm F_{\text{VOF}}  - R(S\widetilde D^{-1}) \bm p,
\end{align*}
obtaining a modified pressure equation that is consistent with the IB constraint:
\begin{equation}
    \label{eq:ch3-PIMPLEVOFIB-pressure}
    R(S\widetilde D^{-1}) \bm p = B (\widetilde D^{-1}\widetilde H(\bm U)
    + S \widetilde D^{-1} \bm F_{\text{VOF}}).
\end{equation}

Within the PIMPLE projection scheme that handles the pressure-velocity coupling, Equation \eqref{eq:ch3-PIMPLEVOFIB-pressure} provides a pressure field that ensures a corrected velocity field, computed using \eqref{eq:ch3-PIMPLEVOFIB-IBconstrVel} that satisfies both incompressibility and IB constraint. The resulting VOF-IB PIMPLE solution procedure is reported in Algorithm \ref{alg:ch3-VOFIBPIMPLE}.

\begin{algorithm}[h!]
\caption{VOF-IB-PIMPLE}
\label{alg:ch3-VOFIBPIMPLE}
\DontPrintSemicolon
\small
Initialize with $\mathbf{U}^0, \mathbf{p}^0_{rgh}, \bm\alpha^0$ \;
\CenterComment{TIME Loop}
\For{$n = 0,\dots,N_T - 1$ }{
    \CenterComment{PIMPLE Loop}
    \For{$l = 0, \dots,N_{\text{outer}} - 1$}{
        \tcp{\textcolor{keywordblue}{1. Volume Fraction Equation}}
        \CenterComment{SUB-CYCLE Loop}
        \For{$k = 0, \dots,N_{\text{sc}} - 1$, \text{with} $\bm\alpha^0 = \bm\alpha^{l}$}{
            \CenterComment{$\alpha-$Correction Loop}
            \For{$q = 0, \dots, N_{\alpha, \text{corr}} - 1$, \text{with} $\alpha^0 = \alpha^{k}$}{
                Compute \quad $\alpha_i^{q+1} = \alpha_i^q  
                    - \frac{\Delta t_\text{sc}}{\abs{K_i}} \sum_{f \in \mathcal{F}_i} \left[ \Phi^q_{f,\text{up}} + \lambda_M \Phi^q_{f,\text{c}} \right]
                    $\;
                Impose IB constraint on $K_j \in \mathcal{T}_S \cap \mathcal{K}_i$, $K_i \in \mathcal{T}_\text{IB}$: $\alpha_j^{q+1} = \alpha_i^{q+1} \pm g_{N,F_{ij}}\|\bm d_{ij}\|$;
            }
              $\bm\alpha^{k+1} \gets \bm\alpha^{N_{\alpha,\text{corr}}}$\;
              Update \quad $\bm\rho^{k+1} = \bm\alpha^{k+1} \rho_1 + (1 - \bm\alpha^{k+1})\rho_2$\;
              Update \quad $\bm\mu^{k+1}=\bm\alpha^{k+1} \mu_1 + (1 - \bm\alpha^{k+1})\mu_2$\;
              Store mass flux \quad $(\rho_f\phi_f)^{k+1} = (\alpha_f\phi_f)^{k+1} (\rho_1 - \rho_2) + \phi_f^l\rho_2$\;
        }
          $\bm\alpha^{l+1} \gets \bm\alpha^{N_{\text{sc}}}$\;
          Compute global mass flux \quad $(\rho_f\phi_f)^{l+1} = \frac{1}{N_\text{sc}} \sum_k (\rho_f\phi_f)^{k+1}$ \;
        \tcp{\textcolor{keywordblue}{2. Momentum Predictor (\textit{optional})}}
        Impose IB constraint on \eqref{eq:ch3-FVMomentumEquation}: $A \to \widetilde A$ and $\bm F \to \widetilde{\bm F}$ \;
        Solve \quad $\widetilde A \bm{U}^{l+\frac{1}{2}} = \widetilde{\bm F} +\chi \left(\bm F_{\text{VOF}} - B^{\intercal} \bm p^l\right)$\;
        \CenterComment{PISO Loop}
        \For{$m = 0,\dots,N_{\text{inner}} - 1$, \text{with} $\bm U^0 = \bm U^l$ or $\bm U^0 = \bm U^{l+\frac{1}{2}}$}{
        \tcp{\textcolor{keywordblue}{3. Pressure equation}}
              Solve \quad $R(S \widetilde D^{-1}) \bm p^{m+1} = B (\widetilde D^{-1} \widetilde H(\bm U^{m}) + S \widetilde D^{-1} \bm F_\text{VOF})$ \;
        \tcp{\textcolor{keywordblue}{4. Pressure relaxation}}
              Compute \quad $\bm{p}^{m+1} \gets (1 - \beta_p)\bm{p}^m + \beta_p \bm{p}^{m+1}$ \;
        \tcp{\textcolor{keywordblue}{5. Velocity correction}}
              Compute \quad $\bm{U}^{m+1} = \widetilde D^{-1}\widetilde H(\bm U^{m}) - S \widetilde D^{-1}B^\top \bm{p}^{m+1} + S \widetilde D^{-1}\bm{F}_{\text{VOF}}$ \;
        }
          $\bm{U}^{l+1} \gets \bm{U}^{N_\text{inner}}$ \;
          $\bm{p}^{l+1} \gets \bm{p}^{N_\text{inner}}$ \;
    }
    
      $\bm{U}^{n+1} \gets \bm{U}^{N_\text{outer}}$, $\mathbf{p}^{n+1} \gets \bm p^{N_\text{outer}}$, $\bm\alpha^{n+1} \gets \bm\alpha^{N_\text{outer}}$\;
}
\end{algorithm}

\subsubsection{Issues with high-viscosity contrasts}
\label{sec:VOFIB-highViscIssues}

The standard VOF solver of OpenFOAM implements a segregated approach not only for the pressure-velocity coupling, handled by means of the PIMPLE projection scheme, but also for the momentum equation itself that couples the three velocity components. The viscous diffusion term is split into an implicit component pertaining to the gradient of velocity and an explicit one concerning the transposed gradient. This semi-implicit treatment allows each velocity component equation to be solved sequentially and independently of the others.
The structure of matrix $A \in \mathbb{R}^{3N \times 3N}$ appearing in the pressure-velocity problem \eqref{eq:ch3-FVnavierstokes-RC} is block-diagonal, such that:
\begin{equation}
    \label{eq:ch4-FVnavierstokesSegregated}     
    \begin{bmatrix}
        A_x & 0   & 0   & B_x \\
        0   & A_y & 0   & B_y \\
        0   & 0   & A_z & B_z \\
        B_x & B_y & B_z & -C
    \end{bmatrix}
    \begin{bmatrix}
        \bm U_x \\
        \bm U_y \\
        \bm U_z \\
        \bm p
    \end{bmatrix}
    =
    \begin{bmatrix}
        (\bm F + \bm F_{VOF})_x \\
        (\bm F + \bm F_{VOF})_y \\
        (\bm F + \bm F_{VOF})_z \\
        \bm 0
    \end{bmatrix}.
\end{equation}
Therefore, the momentum predictor step consists of the solution of three smaller systems of dimension $N \times N$ sequentially. The advantage of this approach is its easy implementation as well as the lower computational cost for the solution of the resulting finite-volume algebraic system. 

Treating part of the viscous term explicitly, however, restricts the time-step size for which a stable and accurate solution is obtained. When viscous diffusion is integrated explicitly, the time step on a uniform three-dimensional mesh of size $h$ is limited by \cite{dodd2014fast}
\begin{equation}
    \label{eq:ch4-deltaTNu}
    \Delta t \leq \Delta t_\nu = \text{Re}\, \frac{h^2}{6},
\end{equation}
where $\Delta t$ and $h$ are made non-dimensional with $L/U$ and $L$, respectively, $\text{Re} = \rho_1 U L/\mu_1$ is the Reynolds number based on the properties of the polymer (fluid 1), and $U$ and $L$ are a characteristic velocity and length. In dimensional form, \eqref{eq:ch4-deltaTNu} reads $\Delta t \le h^2/(6\nu_1)$, with $\nu_1 = \mu_1/\rho_1$ the kinematic viscosity of the polymer, the largest in the system. The restriction depends quadratically on $h$ and becomes particularly severe at low Reynolds numbers ($\text{Re} \ll 1$), where viscous effects dominate, as in polymer mixing, and on fine meshes.

In the segregated solver, only the transposed-gradient part $\nabla \cdot (\mu \nabla^\intercal \bm u)$ is explicit. For incompressible flows this term vanishes where the viscosity is uniform, since $\nabla \cdot (\mu \nabla^\intercal \bm u) = \mu \nabla (\nabla \cdot \bm u) = 0$; it is therefore active
only where $\nabla \mu \neq \bm 0$, i.e.\ across the two-phase interface and, for shear-thinning materials, in regions of strong shear-rate gradients. The global bound \eqref{eq:ch4-deltaTNu} is thus not attained, but a restriction of the same nature arises where the viscosity varies sharply. Multiphase flows with high viscosity contrasts are indeed known to be prone to instabilities \cite{ouafa2021monolithic,rajendran2022new}, owing to the strong coupling between velocity components introduced by the transposed-gradient term near the interface \cite{galusinski2008stability,rajendran2022new}. Dodd and Ferrante \cite{dodd2014fast} observed that the restriction caused by large density and viscosity gradients at the interface can be more severe than $\Delta t_\nu$, independently of the pressure-correction method. This is expected to be the case in polymer mixing, where kinematic viscosity ratios can reach $10^8$. 
Moreover, the large jumps in material properties across the interface make the matrix $A$, arising from the discretization of the unsteady, convective and viscous terms, severely ill-conditioned.

The severity of this restriction can be illustrated by considering a region where the viscous phase moves almost rigidly, as occurs for the polymer adjacent to a rotating screw, so that $\bm u \approx  \omega r \bm e_\theta$, with $\omega$ the angular velocity of the screw.. The velocity gradient of a rigid rotation is skew-symmetric, so that the rate of strain and the viscous stress vanish, $\nabla \bm u + \nabla^\intercal \bm u \approx \bm 0$; the two contributions to the viscous term, however, do not vanish separately where the viscosity varies. If $\mu$ varies in the radial direction, e.g.\ across an interface of thickness $\delta$, then, since $\Delta \bm u = \bm 0$ for a rigid rotation,
\begin{equation*}
    \nabla \cdot (\mu \nabla \bm u)
    = (\nabla \mu \cdot \nabla)\, \bm u + \mu \Delta \bm u
    = \mu'(r)\, \frac{\partial \bm u}{\partial r}
    = \mu'\, \omega\, \bm e_\theta
    = -\nabla \cdot (\mu \nabla^\intercal \bm u),
    \qquad |\mu'| \sim \frac{|\mu_1 - \mu_2|}{\delta}.
\end{equation*}
When the viscosity ratio is large, the two terms are therefore very large, opposite, and must cancel. In the segregated solver, $\nabla\cdot(\mu \nabla\bm u)$ is evaluated implicitly and $\nabla\cdot(\mu \nabla^\intercal\bm u)$ explicitly, with different discrete operators, so that the cancellation is only approximate: it leaves a residual that scales with $|\mu'|$ and arises both from the change of velocity over a time step and from the mismatch between the two discrete operators, producing spurious velocities or instabilities. In the block-coupled approach introduced in Section \ref{sec:methods-BC-VOF-IB}, both contributions are treated implicitly within the same system, and the cancellation is enforced at each iteration.


\section{A Block-Coupled VOF-IB Solver}
\label{sec:methods-BC-VOF-IB}

Several remedies can be considered to mitigate the instabilities appearing when simulating multiphase flows with high viscosity contrasts. Reducing the time step until the explicit term is stable is the simplest option, but the resulting cost is prohibitive for the transient, three-dimensional simulations of interest here (Section \ref{sec:results}). At the opposite end, a fully coupled (monolithic) solution of velocity and pressure can be considered. El Ouafa et al. \cite{ouafa2021monolithic} showed that monolithic solvers are considerably more robust than projection methods at density and viscosity
ratios up to $10^5$ and $10^{10}$, respectively, allowing larger time steps at a lower overall cost. These solvers, however, require dedicated block preconditioners for the saddle-point system and were developed on staggered Cartesian grids. Their extension to collocated unstructured meshes \cite{darwish2009coupled} would require pressure stabilization within the saddle-point system, and, since in the present method the IB constraint is embedded in the segregated pressure equation~\eqref{eq:ch3-PIMPLEVOFIB-pressure}, the IB constraint would also have to be reformulated for the coupled velocity-pressure system.

Here we adopt an intermediate strategy, denoted as \textit{block-coupled} approach: the momentum equation is solved implicitly for the three velocity components simultaneously, including the full viscous stress $\nabla \cdot \left[\mu (\nabla \bm u + \nabla^{\intercal} \bm u)\right]$, while the pressure-velocity coupling is kept segregated within the PIMPLE algorithm. To this end, we adapt the block-coupled finite volume discretization proposed by Cardiff et al. \cite{cardiff2016block} for solid mechanics, where an analogous coupling between displacement components arises from the transposed gradient of the displacement within the Cauchy stress tensor and limits the performance of segregated solvers. Treating the transposed-gradient term implicitly removes the lagged explicit dependence on the local viscosity contrast, so that large time steps can be used without degrading the solution. 
Nonetheless, the resulting system remains ill-conditioned because of the property jumps and  it is solved with a preconditioned Krylov method.

\subsection{Derivation of the conforming BC-VOF method}
\label{sec:methods-BC-derivation}

We write the semi-discretized momentum equation, integrated over control volume $K_i \in \mathcal{T}_h$:
\begin{equation}
    \label{eq:ch4-CVIntMomentumEquation-compi}
    \begin{aligned}
        \int_{K_i} \rho^{n+1} \dfrac{u_i^{n+1} - u_i^n}{\Delta t} \ dV
        + \int_{\partial K_i}  \sum_{j=1}^3 \rho^{n+1} & u_i^{n+1} u_j^n n_j \ dS 
        - \int_{K_i} \sum_{j=1}^3 \dfrac{\partial}{\partial x_j}\left( \mu^{n+1} \frac{\partial u_i^{n+1}}{\partial x_j}  + \mu^{n+1} \frac{\partial u_j^{n+1}}{\partial x_i}  \right) \ dV \\
        &+ \int_{\partial K_i} p^{n+1}_{rgh} \bm n_i \ dS = 
        \int_{K_i} - (\bm g \cdot \bm x) \frac{\partial \rho}{\partial x_i}  + \sigma\kappa^{n+1}\frac{\partial \alpha^{n+1}}{\partial x_i} \ dV,
    \end{aligned}
\end{equation}
associated with the $i-$th velocity component, so that the inter-component couplings introduced by the implicit treatment of the transposed gradient term are clearly visible: the $i-$th equation valid on CV $K_i$ \eqref{eq:ch4-CVIntMomentumEquation-compi} cannot be decoupled from the other two components, since there is an explicit dependence on the unknowns $u_j^{n+1}, j \neq i$ at time $t^{n+1}$.

Focusing only on the viscous diffusion term, in \cite{cardiff2016block} the following relation is proved to hold:
\begin{equation}
\label{eq:ch4-viscousStressDecomposition}
    \int_{K_i} \nabla \cdot \left( \mu \nabla \bm u + \mu \nabla ^\intercal \bm u \right) \ dV = \int_{\partial K_i} \underbrace{2 \mu (\nabla \bm u_n) \bm n }_{\bm{T}_n} + \underbrace{\mu (\nabla \bm u_t)\bm n + \mu \nabla_t u_n}_{\bm{T}_t} \ dS,
\end{equation}
where $\bm u_n = (\bm n \otimes \bm n) \bm u$ and $\bm u_t = (\bm I - \bm n \otimes \bm n) \bm u$ are the normal and tangential components to $\partial K_i$, respectively, and $\nabla_t = (\bm I - \bm n \otimes \bm n) \nabla$ is the tangential gradient operator.

The last term requires particular care:
\begin{equation}
    \label{eq:ch4-tangentialTraction2}
    \int_{\partial K_i} \mu (\nabla_t u_n) \ dS \approx
     \sum_{f \in \mathcal{F}_i} \mu_f (\nabla_t u_n)_f \cdot \bm S_f 
     \approx \sum_{f \in \mathcal{F}_i} \mu_f \left(\dfrac{1}{\abs{\bm S_f}} \sum_{e \in \mathcal{E}_f} \bm m_e (u_n)_e L_e \right) \cdot \bm S_f
\end{equation}
where the face-interpolated tangential gradient is computed by means of a Finite Area (FA) method \cite{cardiff2016block}.
We denote by $\mathcal{E}_f$ the set of edges of face $f$, by $L_e$ the length of edge $e$ and by $\hat{\bm e}$ the unit vector directed along edge $e$, so that the unit bi-normal vector can be computed as $\bm m_e = \hat{\bm e} \times \bm n_f$ (see Figure \ref{fig:ch4-edge-and-vertices-cellKi}).

\begin{figure}[ht!]
    \centering

\begin{tikzpicture}[
    scale=2.5, 
    line join=round, 
    line cap=round, 
    >=stealth,
    dot/.style={circle, fill=black, inner sep=1.2pt},
    pdot/.style={circle, fill=newgreen, inner sep=1.5pt},
    edot/.style={circle, fill=newblue, inner sep=1.2pt}
]

    \coordinate (A) at (0,0,0);
    \coordinate (B) at (1,0,0);
    \coordinate (C) at (1,0,1);
    \coordinate (D) at (0,0,1);

    \coordinate (At) at (0,1,0);
    \coordinate (Bt) at (1,1,0);
    \coordinate (Ct) at (1,1,1);
    \coordinate (Dt) at (0,1,1);

    \coordinate (Ki) at (0.5, 0.5, 0.5);
    \coordinate (Fcenter) at (1, 0.5, 0.5);
    \coordinate (E) at (1, 0.5, 0); 
    
    \fill[newred!20] (B) -- (Bt) -- (Ct) -- (C) -- cycle;
    \node at (0.85,0.5,0.2) {\color{newred} $f$};

    \draw[dashed, thick] (A) -- (B);
    \draw[dashed, thick] (A) -- (D);
    \draw[dashed, thick] (A) -- (At);

    \draw[thick] (B) -- (C) -- (D) -- (Dt) -- (At) -- (Bt) -- cycle;
    \draw[thick] (Bt) -- (Ct) -- (Dt);
    \draw[thick] (C) -- (Ct);
    \draw[ultra thick, newblue] (B) -- (Bt); 
    
    \node[dot, label=left:{$K_i$}] at (Ki) {};

    \draw[->, ultra thick] (Fcenter) -- ++(0.4,0,0) node[right,black] {\small $\bm{n}_f$};
    \node[dot,newred] at (Fcenter) {};
    
    \node[pdot, label=below right:{\small \color{newgreen}$p_1$}] at (B) {};
    \node[pdot, label=above right:{\small \color{newgreen}$p_2$}] at (Bt) {};

    \draw[->, ultra thick] (E) -- ++(0,0.25,0) node[above right] {\small $\bm{\hat{e}}$};
    \draw[->, ultra thick] (E) -- ++(0.2,0.15,-0.2) node[right] {\small $\bm{m}_e$};
    \node[edot] at (E) {};
    \node[right,yshift=-5pt] at (E) {\color{newblue}$e$};

\end{tikzpicture}

\caption{Geometric representation of a polyhedral cell $K_i \in \mathcal{T}_h$, showing face $f \in \mathcal{F}_i$ with its unit normal vector $\bm n_f$, and an edge $e \in \mathcal{E}_f$ having vertices $p_1$ and $p_2$, tangent vector $\hat{\bm e}$ and bi-normal vector $\bm m_e$.}
\label{fig:ch4-edge-and-vertices-cellKi}
\end{figure}

The evaluation of $u_n$ at the edge middle point, indicated as $(u_n)_e$, is obtained as a linear combination of the values in the edge nodes:
\begin{equation}
    \label{eq:ch4-linEdgeInterp}
    \bm u_e = \dfrac{\bm u_{p_1} + \bm u_{p_2}}{2}
\end{equation}
where the subscripts $p_1$ and $p_2$ identify the extrema (nodes) of edge $e$.

In turn, $\bm u_{p_i}$ needs to be expressed in terms of cell-centered values, which are the actual DOFs of collocated finite volume method. For this purpose, Cardiff et al. \cite{cardiff2016block} utilize a WLS method to approximate the point field values with a linear fit. The WLS interpolation stencil is defined for each vertex $p_i$ of the computational grid, consisting of all point-neighbors, i. e. all cells containing vertex $p_i$, whose set we denote by $\mathcal{P}_{p_i}$. Additionally, we define the set of point-cell neighbors of the two vertices of edge $e$ as $\mathcal{P}_e = \mathcal{P}_{p_1} \cup \mathcal{P}_{p_2}$, as reported in Figure \ref{fig:ch4-point-cell-neighbors}. If a node belongs to the boundary of the domain, boundary face-centers are included in the stencil instead of the neighbor cell centers.

\begin{figure}[ht]
\centering

\begin{tikzpicture}[
    >=Stealth,
    scale=1.2,             
    transform shape,       
    cell/.style={rectangle, draw=black, minimum size=1cm, inner sep=0pt},
    stencilp1/.style={fill=newblue!40!white},
    stencilp2/.style={fill=newred!40!white},
    stencile/.style={fill=newred!40!white!50!newblue!40!white},
    solid/.style={fill=wallgrey!20!white}
]

    \def\shift{0.5}

    \foreach \x in {0,1,2} {
        \foreach \y in {0,1,2} {
            \pgfmathsetmacro{\cellstyle}{
                (\y < 1) ? "solid" : (
                (\x < 1 ) ? "stencilp1" :  (
                (\x > 1) ? "stencilp2" : "stencile"  ))
            }
            
            \node[\cellstyle, cell] (c-\x-\y) at (\x, \y) {};
            \fill[black] (\x, \y) circle (1.5pt); 
        }
    }

    \fill[pattern=north west lines, pattern color=black!45] 
      (0-\shift,1-\shift) rectangle (2-\shift,3-\shift);
    \draw[newblue,ultra thick,dashed] (0-\shift,1-\shift) rectangle (2-\shift,3-\shift);

    \fill[pattern=north east lines, pattern color=black!45] 
      (1-\shift,1-\shift) rectangle (3-\shift,3-\shift);
    \draw[newred, ultra thick,dashed] (1-\shift,1-\shift) rectangle (3-\shift,3-\shift);

    \draw[black, ultra thick] (1-\shift, 2-\shift) -- (2-\shift,2-\shift);
    \fill[newblue] (1-\shift, 2-\shift) circle (2.5pt);
    \fill[newred] (2-\shift, 2-\shift) circle (2.5pt);

    \foreach \x in {0,1,2} {
        \foreach \y in {0,1,2} {
            \fill[black] (\x, \y) circle (2pt); 
        }
    }

    \matrix[draw=none, anchor=north west] at (3, 2.6) {
        \draw[newblue, thick,postaction={pattern=north west lines, pattern color=black!60},fill=newblue!40] (0,0) rectangle (0.4,0.4); 
        \node[right] at (0.4,0.1) {\small $\mathcal{P}_{p_1}$}; \\ 
        \draw[newred, thick,postaction={pattern=north east lines, pattern color=black!60},fill=newred!40] (0,0) rectangle (0.4,0.4); 
        \node[right] at (0.4,0.1) {\small $\mathcal{P}_{p_2}$};  \\
        \draw[fill=newblue, draw=newblue] (0.2, 0.1) circle (2.5pt);
        \node[right] at (0.4,0.1) {\small $p_1$}; \\
        \draw[fill=newred, draw=newred] (0.2, 0.1) circle (2.5pt);
        \node[right] at (0.4,0.1) {\small $p_2$}; \\
        \draw[black, ultra thick] (0,0) -- (0.4,0);
        \node[right] at (0.4,0) {\small $e$};\\
    };

\end{tikzpicture}

\caption{Sketch of point-cell neighbors sets associated to vertices $p_1$ and $p_2$ of edge $e$.}
\label{fig:ch4-point-cell-neighbors}

\end{figure}

The result is that, for a given control volume $K_i \in \mathcal{T}_h$, in order to approximate the tangential gradient at each face $f \in \mathcal{F}_i$, the \textit{extended stencil} of all point-neighbors of cell $K_i$ is considered, consisting of all cells $K_j$ (or boundary faces if $K_i$ is a boundary cell) sharing a vertex with $K_i$, that we denote as $\mathcal{K}_i^p$ (see Figure \ref{fig:ch4-extendedStencil}). We call $\mathcal{K}_i^p$ the \textit{extended stencil} associated with cell $K_i$ as opposed to the \textit{compact stencil} $\mathcal{K}_i$ (Figure \ref{fig:ch4-compactStencil}) collecting face-neighbors of cell $K_i$, which is usually employed in OpenFOAM finite volume schemes.

\tikzset{
    cell/.style={draw=newblue, minimum size=1cm, fill=newblue!20, font=\Large\itshape},
    cellP/.style={draw=newred, minimum size=1cm, fill=newred!20, font=\Large\itshape},
    empty/.style={draw=black,minimum size=1cm,fill=wallgrey!20}
}

\begin{figure}[h!]
\centering
\begin{subfigure}[t]{0.45\textwidth}
\centering

\begin{tikzpicture}[
    scale=1.5,
    >=Stealth,
    cell/.style={draw=black, minimum size=1.5cm, fill=newblue!20, font=\Large\itshape},
    cellP/.style={draw=black, minimum size=1.5cm, fill=newred!20, font=\Large\itshape},
    empty/.style={draw=black,minimum size=1.5cm,fill=wallgrey!20}
]
            \path[use as bounding box] (-0.7,-2.7) rectangle (2.7,0.7);
            
            \def\shift{0.5}
            
            \foreach \x in {0,1,2}{
                \foreach \y in {0,1,2}{
                    \pgfmathtruncatemacro{\indexi}{\x + 3*\y + 1}
                    \pgfmathsetmacro{\indexi}{
                    (\indexi == 4) ? 6 : (
                    (\indexi == 6) ? 4 : \indexi
                    )}

                    \pgfmathtruncatemacro{\isCenter}{\indexi==5}
                    \pgfmathtruncatemacro{\isCross}{(\indexi==2) || (\indexi==4) || (\indexi==6) || (\indexi==8)}
                    
                    \ifnum\isCenter=1
                        \node[cellP] at (\x,-\y) {};
                        \fill[newred] (\x, -\y) circle (2pt) node[black,xshift=7pt,yshift=7pt] {\footnotesize\indexi};
                    \else
                        \ifnum\isCross=1
                            \node[cell] at (\x,-\y) {};
                            \fill[newblue] (\x, -\y) circle (2pt) node[black,xshift=7pt,yshift=7pt] {\footnotesize\indexi};
                        \else
                            \node[empty] at (\x,-\y) {};
                            \fill[black] (\x, -\y) circle (2pt) node[black,xshift=7pt,yshift=7pt] {\footnotesize\indexi};
                        \fi
                    \fi

                }
            }

            \foreach \x in {0,1,2}{
                \foreach \y in {0,1,2}{
                    \ifnum\x=2

                    \else
                        \fill[black] (\x+\shift, -\y) circle (1.5pt);
                        \ifnum \y=1
                            \draw[->,thick,black] (\x+\shift, -\y) -- (\x+0.2, -\y);
                        \else
                            \draw[->,thick,black] (\x+\shift, -\y) -- (\x+0.8, -\y); 
                        \fi
                    \fi  
                                        
                    \ifnum\y=2

                    \else
                        \fill[black] (\x, -\y-\shift) circle (1.5pt);
                        \draw[->,thick,black] (\x, -\y-\shift) -- (\x, -\y-0.8); 
                    \fi   
                }
            }
        
            \draw[black,thick] (0-\shift,-2-\shift) -- (2+\shift,-2-\shift) -- (2+\shift,0+\shift) -- (0-\shift,0+\shift) -- (0-\shift,-2-\shift);
\end{tikzpicture}

\caption{Compact stencil $\mathcal{K}_5$.}
\label{fig:ch4-compactStencil}
\end{subfigure}
\begin{subfigure}[t]{0.45\textwidth}

\centering

\begin{tikzpicture}[
    scale=1.5,
    >=Stealth,
    cell/.style={draw=black, minimum size=1.5cm, fill=newblue!20, font=\Large\itshape},
    cellP/.style={draw=black, minimum size=1.5cm, fill=newred!20, font=\Large\itshape},
    empty/.style={draw=black,minimum size=1.5cm,fill=wallgrey!20}
]
            \path[use as bounding box] (-0.7,-2.7) rectangle (2.7,0.7);
            \def\shift{0.5}
            
            \foreach \x in {0,1,2}{
                \foreach \y in {0,1,2}{
                    \pgfmathtruncatemacro{\indexi}{\x + 3*\y + 1}
                    \pgfmathsetmacro{\indexi}{
                    (\indexi == 4) ? 6 : (
                    (\indexi == 6) ? 4 : \indexi
                    )}

                    \pgfmathtruncatemacro{\isCenter}{\indexi==5}
                    
                    \ifnum\isCenter=1
                        \node[cellP] at (\x,-\y) {};
                        \fill[newred] (\x, -\y) circle (2pt) node[black,xshift=7pt,yshift=7pt] {\footnotesize\indexi};
                    \else
                        \node[cell] at (\x,-\y) {};
                        \fill[newblue] (\x, -\y) circle (2pt) node[black,xshift=7pt,yshift=7pt] {\footnotesize\indexi};
                    \fi
                }
            }

            \foreach \x in {0,1,2}{
                \foreach \y in {0,1,2}{
                    \ifnum\x=2
                    \else
                        \fill[black] (\x+\shift, -\y) circle (1.5pt);
                        \ifnum \y=1
                            \draw[->,thick,black] (\x+\shift, -\y) -- (\x+0.2, -\y);
                        \else
                            \draw[->,thick,black] (\x+\shift, -\y) -- (\x+0.8, -\y); 
                        \fi
                    \fi  

                    \ifnum\y=2
                    \else
                        \fill[black] (\x, -\y-\shift) circle (1.5pt);
                        \draw[->,thick,black] (\x, -\y-\shift) -- (\x, -\y-0.8); 
                    \fi   
                }
            }

            \draw[black,thick] (0-\shift,-2-\shift) -- (2+\shift,-2-\shift) -- (2+\shift,0+\shift) -- (0-\shift,0+\shift) -- (0-\shift,-2-\shift);

            \foreach \x [count=\i from 10] in {0,1,2}
                \fill[black] (\x, \shift) circle (1.5pt) node[yshift=8pt,black] {\footnotesize\i};

            \foreach \y [count=\i from 13] in {0,1,2}
                \fill[black] (2+\shift, -\y) circle (1.5pt) node[xshift=8pt,black] {\footnotesize\i};

            \foreach \x [count=\i from 16] in {2,1,0}
                \fill[black] (\x, -2-\shift) circle (1.5pt) node[yshift=-8pt,black] {\footnotesize\i};

            \foreach \y [count=\i from 19] in {2,1,0}
                \fill[black] (-\shift, -\y) circle (1.5pt) node[xshift=-8pt,black] {\footnotesize\i};

\end{tikzpicture}

\caption{Extended stencil $\mathcal{K}_5^p$.}
\label{fig:ch4-extendedStencil}
\end{subfigure}

\caption{Comparison between compact and extended stencils for a simple $3 \times 3$ two-dimensional grid (adapted from \cite{cardiff2016block}).}
\label{fig:ch4-compactExtendedStencil}

\end{figure}

Once the WLS approximator is obtained, using field values $u_{K_i}$ for $K_i \in \mathcal{P}_{p_i}$ as observation data, each point field value $\bm u_{p_i}$ can be expressed as
\begin{equation}
    \label{eq:ch4-pointFieldValueApprox}
    \bm u_{p_i} = \sum_{j = 1}^{N_i} \gamma_{j} \bm u_j,
\end{equation}
where $N_i = \text{card}(\mathcal{P}_{p_i})$ and $\gamma_j$ are interpolation weights, whose values can be obtained after solving the WLS minimization problem.

Therefore, the discretization of the tangential gradient term, starting from \eqref{eq:ch4-tangentialTraction2}, becomes:
\begin{equation}
    \label{eq:ch4-tangentialTraction2-FAWLS}
    \int_{\partial K_i} \mu (\nabla_t u_n) \ dS \approx
     \sum_{f \in \mathcal{F}_i} \mu_f (\nabla_t u_n)_f \cdot \bm S_f = \sum_{f \in \mathcal{F}_i} \sum_{e \in \mathcal{E}_f} \sum_{P \in \mathcal{P}_e} \dfrac{\mu _f}{2} L_e \gamma_{P} (\bm m_e \otimes \bm n_f) \bm u_P.
\end{equation}
Consequently, the FV approximation of the implicit viscous term \eqref{eq:ch4-viscousStressDecomposition}, under the simplifying assumption of orthogonal mesh, reads:
\begin{equation}
\label{eq:ch4-blockCoupledViscousTerm}
\begin{aligned}
    \int_{K_i} &\nabla \cdot \left( \mu \nabla \bm u + \mu \nabla ^\intercal \bm u \right) \ dV = \int_{\partial K_i} \underbrace{2 \mu (\nabla \bm u_n) \bm n }_{\bm{T}_n} + \underbrace{\mu (\nabla \bm u_t)\bm n + \mu \nabla_t u_n}_{\bm{T}_t} \ dS \\
    \approx& \sum_{f \in \mathcal{F}_i} 2 \mu_f \abs{\bm S_f} (\bm n_f \otimes \bm n_f) \frac{\bm u_N - \bm u_{P}}{\|\bm d_f \|} 
    + \sum_{f \in \mathcal{F}_i} \mu_f \abs{\bm S_f} (\bm I - \bm n_f \otimes \bm n_f) \frac{\bm u_{N} - \bm u_{P}}{\|\bm d_f\|} 
    + \sum_{f \in \mathcal{F}_i} \sum_{e \in \mathcal{E}_f} \sum_{P \in \mathcal{P}_e} \dfrac{\mu _f}{2} L_e \gamma_{P} (\bm m_e \otimes \bm n_f) \bm u_P. 
\end{aligned}   
\end{equation}
The first two terms in \eqref{eq:ch4-blockCoupledViscousTerm} correspond to the normal gradient discretization, using standard FV schemes that only include the DOFs belonging to the compact stencil $\mathcal{K}_i$ (Figure \ref{fig:ch4-compactStencil}).
The last row of \eqref{eq:ch4-blockCoupledViscousTerm} is the tangential gradient discretized by means of the FA-WLS approach \cite{cardiff2016block}, where the three encapsulated sums involve the DOFs belonging to every point-neighbor $K_j$ of cell $K_i$ belonging to the extended stencil $\mathcal{K}_i^p$ (Figure \ref{fig:ch4-extendedStencil}).
    
Boundary-face centers are point neighbors of boundary cells and therefore become additional unknowns; one equation per boundary face is appended to the system (see \cite{cardiff2016block}), as reported in Equation \eqref{eq:ch4-FVnavierstokesCoupled}.

\begin{remark}
    We have assumed an orthogonal mesh, but when working with real industrial applications, the orthogonality of the computational grid is almost never guaranteed. Non-orthogonality corrections are actually accounted for in the discretization of normal gradients, leading to additional tangential gradient contributions that are handled with the same FA-WLS method employed to obtain \eqref{eq:ch4-tangentialTraction2-FAWLS}. Further details can be found in \cite{cardiff2016block}.
\end{remark}

The unsteady and convective contributions to the discrete momentum balance on each control volume $K_i$ are the same as in \eqref{eq:ch3-FViMomentumEquation}. The fully implicit treatment of viscous diffusion \eqref{eq:ch4-blockCoupledViscousTerm}, however, couples the three velocity components. The momentum matrix is therefore no longer block-diagonal, with one identical scalar block per
velocity component as in \eqref{eq:ch4-FVnavierstokesSegregated}.

In the present implementation, the block-coupled discretization is used only in the momentum predictor of the PIMPLE algorithm, while the pressure equation is unchanged. Moreover, unlike in Algorithm \ref{alg:ch3-VOFIBPIMPLE}, the volume-fraction sub-cycles and the momentum predictor are performed once per time step, at the first outer iteration ($l = 0$), so that the subsequent outer iterations only contain the pressure correctors. Using the notation of Algorithm \ref{alg:ch3-VOFIBPIMPLE}, the predicted velocity is obtained by solving
\begin{equation}
    \label{eq:ch4-FVnavierstokesCoupled}
    \begin{bmatrix}
		A^c_I & A^c_{I,\text{BF}} \\
		A^c_{\text{BF},I}   & A^c_\text{BF}
	\end{bmatrix}
    \begin{bmatrix}
    	\bm U_I^{l+\frac{1}{2}}  \\
    	\bm U_\text{BF}^{l+\frac{1}{2}}
    \end{bmatrix}
    = 
    \begin{bmatrix}
    	\bm F + \bm F_\text{VOF} - B^\intercal \bm p^l  \\
    	\bm F_\text{BF}
    \end{bmatrix} \quad \text{for } l = 0 \ (\text{only at }1^\text{st} \text{ outer iteration}).
\end{equation}
Here $A^c \in \mathbb{R}^{3N_E \times 3N_E}$ denotes the block-coupled momentum matrix. Let $\mathcal{T}^E_h  = \mathcal{T}_h \cup \mathcal{F}_{h,B}$ be the set of the extended-stencil unknowns, collecting cells and boundary faces, and $N_E = \text{card}(\mathcal{T}^E_h) = N + N_\text{BF}$, where $N = \text{card}(\mathcal{T}_h)$ is the number of cells and
$N_\text{BF} = \text{card}(\mathcal{F}_{h,B})$ the number of boundary faces. Accordingly, the velocity
unknowns are split into cell-centered values $\bm U_I \in \mathbb{R}^{3N}$ and boundary-face values
$\bm U_\text{BF} \in \mathbb{R}^{3N_\text{BF}}$, and $A^c$ is partitioned into
$A^c_I \in \mathbb{R}^{3N \times 3N}$,
$A^c_{I,\text{BF}} \in \mathbb{R}^{3N \times 3N_\text{BF}}$,
$A^c_{\text{BF},I} \in \mathbb{R}^{3N_\text{BF} \times 3N}$ and
$A^c_\text{BF} \in \mathbb{R}^{3N_\text{BF} \times 3N_\text{BF}}$. Each sub-matrix consists of $3 \times 3$ blocks, which are non-zero only if $K_j$ belongs to the extended stencil $\mathcal{K}_i^p$; the off-diagonal entries of each block couple different velocity components through the transposed-gradient term.
The second block row of Equation \eqref{eq:ch4-FVnavierstokesCoupled} contains one vector equation per boundary face, which imposes the velocity
boundary condition, whose corresponding boundary data are collected in $\bm F_\text{BF} \in \mathbb{R}^{3N_\text{BF}}$.
As in the segregated case, $B^\intercal \in \mathbb{R}^{3N \times N}$ is the discrete pressure gradient.

The pressure equation is then assembled as in the original VOF solver, using the diagonal $D$ and the operator $H$ of the segregated matrix $A$ in \eqref{eq:ch4-FVnavierstokesSegregated} evaluated at $\bm U_I^{l+\frac{1}{2}}$:
\begin{equation}
\label{eq:ch3-PIMPLEBCVOF-pressure}
    R(D^{-1}) \bm p^{m+1} = B (D^{-1} H(\bm U^m) + D^{-1} \bm F_\text{VOF}), \qquad \bm U^0 = \bm U_I^{l+\frac{1}{2}} \text{ for } l = 0 \text{ and } \bm U^0 = \bm U_I^l \text{ for } l > 0
\end{equation}
where $m$ is the inner-corrector index.
The transposed-gradient term still enters $H$ explicitly, but it is evaluated at the implicitly predicted velocity. Note that the predictor and the pressure correction therefore use different discretizations of the transposed-gradient term, implicit FA-WLS in \eqref{eq:ch4-FVnavierstokesCoupled} and compact explicit within $H$ in \eqref{eq:ch3-PIMPLEBCVOF-pressure}; a consistent formulation, with $D$ and $H$ derived from the block-coupled matrix, is left for future work. The boundary-face unknowns $\bm U_\text{BF}$ appear only in the momentum predictor, while the boundary fluxes in the pressure equation are computed from the velocity boundary conditions, as in the segregated solver.

\subsection{Integration with the IB method}
\label{sec:methods-BC-IB-integration}

Because $D$ and $H$ are taken from the segregated system, the IB-constrained pressure equation \eqref{eq:ch3-PIMPLEVOFIB-pressure} of Section \ref{sec:VOFIB-pUCoupling} carries over to the BC-VOF-IB solver without modification, with $\widetilde D$ and $\widetilde H$ obtained from the IB-modified segregated matrix $\widetilde A$ \eqref{eq:ch3-IBConstrainedA}.
The IB constraint must therefore be imposed only on the block-coupled momentum predictor \eqref{eq:ch4-FVnavierstokesCoupled}, which is replaced by
\begin{equation}
    \label{eq:ch4-blockPredictorIB}
    \widetilde A^c \bm U_C^{l+\frac{1}{2}} = \widetilde{\bm F}^c + 
    \begin{bmatrix}
        \chi\left(\bm F_\text{VOF} - B^\intercal \bm p^l\right)\\
        \bm 0
    \end{bmatrix} \quad \text{for } l = 0,
\end{equation}
where $\bm U_C^{l+\frac{1}{2}} = \big(\bm U_I^{l+\frac{1}{2}},\ \bm U_\text{BF}^{l+\frac{1}{2}}\big)^\intercal$ and
$\bm F^c = \big(\bm F,\ \bm F_\text{BF}\big)^\intercal$ denotes the right-hand side of \eqref{eq:ch4-FVnavierstokesCoupled} without pressure-gradient, gravity and surface tension contributions. 

The unknowns are numbered so that $i = 1, \dots, N$ correspond to cells and $i = N+1, \dots, N_E$ to
boundary faces. In analogy with \eqref{eq:ch3-IBConstrainedA}, the blocks $\widetilde A^c_{ij} \in \mathbb{R}^{3 \times 3}, \ i,j = 1,\dots, N_E$ of the constrained matrix $\widetilde A^c$ are obtained by modifying $A^c$ at the IB and solid unknowns:
\begin{equation}
    \label{eq:ch4-IBConstrainedAc}
    \widetilde{A}^c_{ij} =
    \begin{cases}
        A^c_{ii} & \text{if } i = j \\
        0        & \text{if } i \neq j \text{ and } \quad
                \begin{aligned}[t]
                    & K_i \in \mathcal{T}_\text{IB} \cup \mathcal{T}_S \ (i \leq N) \text{ or }\\
                    & K_j \in \mathcal{T}_\text{IB} \cup \mathcal{T}_S \ (j \leq N) \text{ or } \\
                    & F_i \in \Gamma_\text{IB,ext} \cup \Gamma_S \ (i > N) \text{ or } \\
                    & F_j \in \Gamma_\text{IB,ext} \cup \Gamma_S \ (j > N)
                \end{aligned}      \\
        A^c_{ij} & \text{otherwise}
    \end{cases}
    \qquad i,j = 1,\dots,N_E.
\end{equation}
To shorten the notation, we denote by $\mathcal{K}_{i,\text{IB}}^p =  \mathcal{K}_i^p \cap (\mathcal{T}_\text{IB} \cup \mathcal{T}_S)$ and $\mathcal{K}_{i,\text{IB},\Gamma}^p =  \mathcal{K}_i^p \cap (\Gamma_\text{IB,ext} \cup \Gamma_S)$ the sets of IB-constrained cells and boundary faces, respectively. In analogy with \eqref{eq:ch3-IBConstrainedF}, the source term is given by
\begin{equation}
\label{eq:ch4-IBConstrainedFc}
    \widetilde{\bm F}^c_i =
    \begin{cases}
        A^c_{ii}\, S^i_\text{IB}(\bm g, \bm U) &\text{if }
        \begin{aligned}[t]
            &K_i \in \mathcal{T}_\text{IB} \cup \mathcal{T}_S \ (i \leq N)\\
            &F_i \in \Gamma_\text{IB,ext} \cup \Gamma_S \ (i > N)
        \end{aligned}\\[3pt]
        \bm F_i - \displaystyle\sum_{\substack{K_j \in \mathcal{K}_{i,\text{IB}}^p \\ F_j \in \mathcal{K}_{i,\text{IB},\Gamma}^p}} A^c_{ij} S^j_\text{IB}(\bm g, \bm U) & \text{if } K_i \in \mathcal{T}_F \ (i \leq N), \\[3pt]
        \bm F_{\text{BF},i} & \text{if } F_i \in \Gamma_F \ (i > N),
    \end{cases}
    \qquad i = 1,\dots,N_E.
\end{equation}
As in the segregated case, each constrained row reduces to
$A^c_{ii} \bm U_{C,i} = A^c_{ii} S^i_\text{IB}(\bm g, \bm U)$, namely $\bm U_{C,i} = S^i_\text{IB}(\bm g, \bm U)$, for both cell and boundary-face unknowns.
A boundary face can be constrained only if its owner cell is an IB or a solid cell. Since boundary faces
never belong to $\Gamma_\text{IB,int}$, the boundary faces of IB cells lie in $\Gamma_\text{IB,ext}$,
i.e.\ on the solid side of the immersed boundary; such faces exist where the background grid contains no solid cell beyond the IB cell. On all constrained boundary faces, $F_i \in \Gamma_\text{IB,ext} \cup \Gamma_S$, the solid datum is therefore imposed, $S^i_\text{IB}(\bm g, \bm U) = \bm g_{S,i}$, and no WLS reconstruction is required.

In the rows of fluid cells $K_i \in \mathcal{T}_F$, the contributions of the constrained cells and boundary faces in the stencil, if any, are moved to the right-hand side. The rows of unconstrained boundary faces $F_i \in \Gamma_F$ (because $\Gamma_\text{IB,int} \cap \mathcal{F}_{h,B} = \varnothing$) are unchanged. This holds because the boundary conditions imposed at conforming boundaries involve only the face-normal gradient, without tangential-gradient or non-orthogonal correction terms; each unconstrained boundary-face row therefore couples the face only with its fluid owner cell, and never with point neighbors that might be constrained.

Two differences arise with respect to~\eqref{eq:ch3-IBConstrainedA}--\eqref{eq:ch3-IBConstrainedF}.
First, the sum in~\eqref{eq:ch4-IBConstrainedFc} runs over the extended stencil of point neighbors $\mathcal{K}_i^p$ instead of the face neighbors $\mathcal{K}_i$, so that fluid cells sharing only a vertex with a constrained cell or boundary face also receive IB contributions. Second, the additional rows of boundary-face unknowns are modified in the same way as the rows of cells unknowns.

\begin{remark}
As anticipated at the beginning of this section, both $\widetilde A$, which provides $\widetilde D$ and $\widetilde H$ for the pressure equation, and $\widetilde A^c$, which defines the predictor, are assembled at each outer iteration.
\end{remark}


\section{Numerical Results}
\label{sec:results}

The present section reports the numerical results obtained to assess the performance and reliability of the proposed modeling approach. As a preliminary step, the methodology is first assessed on the dog-bone injection-molding test case \cite{de2023analysis}, conducting a solver comparison between conforming and IB approaches, as well as segregated and BC schemes (Section \ref{sec:dogbone}). A physical demonstration of the methods capabilities is then carried out, extending the analysis to more complex and realistic configurations, focusing on continuous mixing devices of industrial interest (Section \ref{sec:industrial-app}).
Surface tension is neglected in the present applications by setting $\sigma = 0$, since in polymer processing viscous forces dominate over surface tension, which corresponds to large capillary numbers.

In all simulations, time derivatives are discretized with the first-order implicit Euler scheme. Gradients are computed
with the Gauss scheme and linear interpolation, the convective term of the momentum equation is discretized with a limited linear TVD scheme, the volume-fraction flux with the van Leer limiter \cite{vanLeer1974towards}, and the compressive flux with the interface-compression scheme of OpenFOAM.
In the segregated solvers, the Laplacian and explicit transposed-gradient terms are discretized with the Gauss scheme with linear interpolation and explicit non-orthogonal correction, whereas in the block-coupled solvers the full viscous term is discretized with the point-based least-squares operators \cite{cardiff2016block} of Section \ref{sec:methods-BC-VOF-IB}. The pressure equation is solved with the conjugate gradient method preconditioned by algebraic multigrid (GAMG, with DIC Gauss--Seidel smoothing), and the coupled momentum system with GMRES preconditioned by LU-SGS (Lower Upper Symmetric Gauss--Seidel), with 8 Krylov directions and at most 20 iterations; tolerances are reported in Tables~\ref{tab:ch4-dogbone-setup}, \ref{tab:ch5-SSE-setup} and~\ref{tab:ch5-TSE-setup}.

\subsection{Dog-bone injection molding test case}
\label{sec:dogbone}
The dog-bone injection molding test case \cite{de2023analysis} is considered to assess the performance of segregated and block-coupled solvers, as well as to compare the solution obtained with either conforming or IB approaches.

The test case consists of a three-dimensional dog-bone shaped channel (Figure \ref{fig:ch3-dogbone-domain}), initially filled with air, into which a generalized Newtonian material is injected, so that a two-phase front advances through the domain.

\input{img/tikz/ch3-dogbone-domain}

Both conforming and non-conforming grids are generated, which are reported in Figure \ref{fig:ch3-dogbone-conf-IB-grids-60K} together with mesh indicators in Table \ref{tab:ch4-dogbone-grids}, highlighting the better mesh quality allowed in the non-conforming case.
The background mesh for the IB case is a rectangular prism and two triangulated surfaces are constructed representing the two curved walls as immersed solid bodies. The IB grid is refined so that the two grids have the same mesh size in the restriction of the channel. Two refinement levels,
$h_1 \sim 0.5 \ \unit{mm}$ and $h_2 \sim 0.25 \ \unit{mm}$, are used to assess the reduction of the discrepancy between conforming and IB results.

\input{img/tikz/ch3-dogbone-conf-IB-grids-60K}

\begin{table}[h!]
    \renewcommand{\arraystretch}{1.3}
    \setlength{\tabcolsep}{7.5pt}
    \small
    \centering
    \begin{NiceTabular}{|c|c|c|c|c|c|c|c|} 
        \noalign{\hrule height 1.2pt}
        \Block{2-1}{Grid} &  \Block{2-1}{$h \ [\unit{mm}]$} & \Block{2-1}{$\#$DOFs} & \multicolumn{2}{c}{Non-orthogonality} & Skewness & \multicolumn{2}{c}{Cell volume $[\unit{m}^3]$} \\
        \cline{4-8}  
        & & & $\max(\tau_{ij})$ & $\text{avg}(\tau_{ij})$ & $\max(\psi_{ij})$ & $\displaystyle \min
        |K_i|$ & $\displaystyle \max|K_i|$\\
        \noalign{\hrule height 0.8pt}
        \Block{2-1}{Conforming}  & $\sim 0.50$  & \num{64704} & 35.21 & 4.35 & $9.01 \cdot 10^{-1}$ & $1.05\cdot 10^{-10}$ & $2.11 \cdot 10^{-10}$ \\
        \cline{2-8}
        & $\sim 0.25$ & \num{530768} & 36.46 & 4.35 & $9.01 \cdot 10^{-1}$ & $1.27\cdot 10^{-11}$ & $2.56 \cdot 10^{-11}$ \\
        \noalign{\hrule height 0.6pt}
        \Block{2-1}{IB}  & $\sim 0.50$ &  \num{126720} & 0 & 0 & $3.23\cdot 10^{-13}$ & $1.07\cdot 10^{-10}$ & $1.07\cdot 10^{-10}$ \\
        \cline{2-8}
        & $\sim 0.25$ &  \num{1306144} & 0 & 0 & $1.09\cdot 10^{-12}$ & $1.04\cdot 10^{-11}$ & $1.04\cdot 10^{-11}$ \\
        \noalign{\hrule height 1.2pt}
    \end{NiceTabular}    
    \caption{Mesh size and key quality indicators for the dog-bone benchmark case.}
    \label{tab:ch4-dogbone-grids}
\end{table}

The boundary conditions imposed to the system of two-phase Navier--Stokes equations \eqref{eq:VOFbiphaseNSequations} take the following form:
\begin{equation*}
    \begin{cases}
        \bm u = \bm u_{\text{in}}, \ \nabla p_{rgh} \cdot \bm n = 0, \ \alpha = 1 & \quad  \text{at } x = 0, \ y \in [-0.01,0.01], \ z \in [-0.004,0], \\
        (\nabla \bm u) \bm n = \bm 0, \ p_{rgh} = 0, \ \nabla \alpha \cdot \bm n = 0 & \quad \text{at } x = 0.17, \ y \in [-0.01,0.01], \ z \in [-0.004,0], \\
        \bm u = \bm 0, \ \nabla p_{rgh} \cdot \bm n = 0, \ \nabla \alpha \cdot \bm n = 0 & \quad \text{elsewhere}.
    \end{cases}
\end{equation*}
where a uniform inlet velocity profile is prescribed, such that
$\bm u_\text{in} = (u_{\text{in},x}, 0, 0)$ with $u_{\text{in},x} = 0.12 \ \unit{m/s}$. 

A highly viscous non-Newtonian material is considered, following a power-law model, whose parameters are calibrated to describe the rheology of a low density polyethylene \cite{kim2006numerical}:
\begin{itemize}
    \item \textit{Fluid 1}: $\mu_1 = K \dot\gamma^{n-1}$, with $K = \num{16630} \ \unit{Pa.s}^n$, $n = 0.334$ and limiting kinematic viscosities $\nu_{\max} = 5 \cdot 10^3 \ \unit{m}^2/\unit{s}$ and $\nu_{\min} = 10^{-2} \ \unit{m}^2/\unit{s}$, while density is set to $\rho_1 = 785 \ \unit{kg/m}^3$.
    \item \textit{Fluid 2 (air):} $\mu_2 = 2.6 \cdot 10^{-5} \ \unit{Pa.s}$, $\rho_2 = 0.73 \ \unit{kg/m}^3$. These values correspond to air at about $210 \ ^\circ\mathrm{C}$, a typical processing temperature at which the isothermal simulations are performed.
\end{itemize}
The kinematic viscosity ratio between the two fluids reaches the order of $10^8$ when $\nu_1 = \nu_{\max}$, corresponding to the limiting value for $\dot\gamma \to 0$. 

Three conforming solvers are compared on the coarse grid (Figure \ref{fig:ch3-dogbone-conf-grid-60K}): the standard segregated solver of OpenFOAM (\texttt{interFoam}), denoted by VOF; a segregated variant with the same algorithmic structure as the block-coupled solver, denoted by VOF$^\ast$, in which the volume-fraction sub-cycles and the momentum predictor are performed only at the first outer iteration (see Section \ref{sec:methods-BC-derivation}); and the block-coupled solver BC-VOF (\texttt{UCInterFoam}). VOF$^\ast$ and BC-VOF therefore differ only in the discretization of the viscous term. Afterwards, the non-conforming BC-VOF-IB solver (\texttt{UCInterIbFoam}), formulated in Section \ref{sec:methods-BC-IB-integration}, is compared with its conforming counterpart BC-VOF. 

Adaptive time stepping is enabled, setting $\Delta t_{\max} = 0.01 \ \unit{s}$ and initial time-step size $\Delta t^0 = 0.001 \ \unit{s}$, so that $\Delta t$ varies during the simulation while the Courant numbers $\text{Co}_h$ and $\text{Co}_{\alpha,h}$ remain below user-defined bounds. 
Two pairs of bounds are considered, $(\max(\text{Co}_h), \max (\text{Co}_{\alpha,h})) = (1, 0.4)$ and $(50, 2)$. Table \ref{tab:ch4-dogbone-setup} lists all simulation settings of the segregated and block-coupled solvers when using the coarse grids.
A semi-implicit treatment of the volume-fraction transport (\texttt{MULESCorr}), available in OpenFOAM, was also tested with the segregated solver, but it did not improve stability or allow larger time steps; it is therefore not used in any of the solvers, which also isolates the effect of the momentum discretization.
All runs are performed on computing nodes of the same type with 4 MPI processes, and the reported computing times are wall-clock times for $t \in (0, 1.1) \ \unit{s}$.\footnote{HPC (High Performance Computing) cluster of the Department of Mathematics (DMAT) of Politecnico di Milano, composed of 40 computing nodes having 96 CPUs (AMD EPYC 7413 24-Core Processor) each and 512 GB RAM per node.}

\begin{table}[h!]
    \renewcommand{\arraystretch}{1.3}
    \setlength{\tabcolsep}{5pt}  
    \small
    \centering
    \begin{NiceTabular}{|c|c|c|c|c|}
        \noalign{\hrule height 1.2pt}
        Parameter & VOF & VOF$^\ast$ & BC-VOF & BC-VOF-IB \\
        \noalign{\hrule height 0.8pt}
        $\alpha$ sub-cycles and momentum predictor & every outer it. & $1^\text{st}$ outer it. & $1^\text{st}$ outer it. & $1^\text{st}$ outer it. \\
        \hline
        $c_\alpha$ & 1 & 1 & 1 & 1  \\
        \hline
        \texttt{MULESCorr} & no & no & no & no \\
        \hline
        $N_\text{sc}$ & 40 & 40 & 40 & 40 \\        
        \hline
        $N_{\alpha,\text{corr}}$ & 1 & 1 & 1 & 1 \\
        \hline
        $N_\text{inner}$ & 10 & 10 & 10 & 20 \\
        \hline
        $N_\text{outer, max}$ & 20 & 20 & 20 & 15 \\
        \cline{2-5}
        \Block{1-1}{Outer tolerances} & \Block{1-4}{$\bm u$: $5 \cdot 10^{-4}$, \quad $p_{rgh}$: $10^{-5}$ \quad (absolute, initial residuals)} \\
        \hline
        $N_\text{non-orth}$ & 2 & 2 & 2 & 0 \\
        \noalign{\hrule height 0.8pt}
        \Block{2-1}{$p_{rgh}$ solver} & PCG (GAMG) & PCG (GAMG) & PCG (GAMG) & PCG (GAMG) \\
        & $tol= 10^{-7}$ & $tol= 10^{-7}$ & $tol= 10^{-7}$ & $tol= 10^{-7}$ \\
        \hline
        \Block{2-1}{$\bm u$ solver} & \Block{2-1}{no} & \Block{2-1}{no} & GMRES (LUSGS) & GMRES (LUSGS) \\
        & & & $tol = 10^{-4}$ & $tol = 10^{-4}$ \\
        \noalign{\hrule height 0.8pt} 
        $\text{Co}_{h,\max}$ & 1 \,/\, 50 & 1 \,/\, 50 & 50 & 50 \\
        \hline
        $\text{Co}_{\alpha,h,\max}$ & 0.4 \,/\, 2 & 0.4 \,/\, 2 & 2 & 2 \\
        \hline
        $\Delta t^0 \ [\unit{s}]$ & 0.001 & 0.001 & 0.001 & 0.001 \\
        \hline
        $\Delta t_{\max} \ [\unit{s}]$ & 0.01 & 0.01 & 0.01 & 0.01 \\
        \hline
        $T_f \ [\unit{s}]$ & 1.1 & 1.1 & 1.1 & 1.1 \\
        \noalign{\hrule height 1.2pt}
    \end{NiceTabular}
    \caption{Numerical settings for the dog-bone case, using the coarsest grid refinement $h_1 \sim 0.5 \ \unit{mm}$.  For the segregated solvers, the two values of the Courant-number bounds refer to the runs with bounds
    $(1, 0.4)$ and $(50, 2)$, respectively; the VOF run with bounds $(50, 2)$ diverges at $t \approx 0.13~\unit{s}$.
    The momentum predictor of the segregated solvers is disabled.}
    \label{tab:ch4-dogbone-setup}
\end{table}

\subsubsection{Segregated and block-coupled solvers}
\label{sec:dogbone-seg-bc}

We first consider the momentum predictor of the segregated solvers. When it is enabled, VOF$^\ast$ diverges at $t \approx 0.35 \ \unit{s}$ with the bounds $(1, 0.4)$, after the maximum velocity has grown to about $3.3 \ \unit{m/s}$, and produces velocities of about $200 \ \unit{m/s}$ near the second restriction with the bounds $(50, 2)$. The momentum predictor is therefore disabled in all segregated runs reported in the following.

\begin{figure}[h!]
\centering

\begin{subfigure}{\textwidth}
\centering
\includegraphics[width=0.7\textwidth]{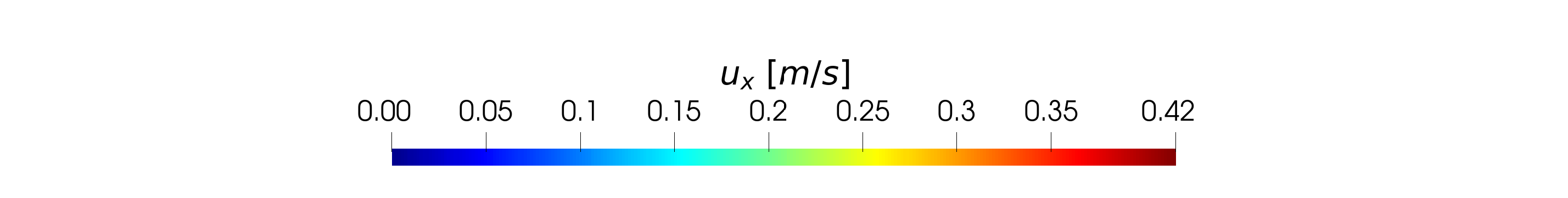} \\
\includegraphics[width=0.32\textwidth]{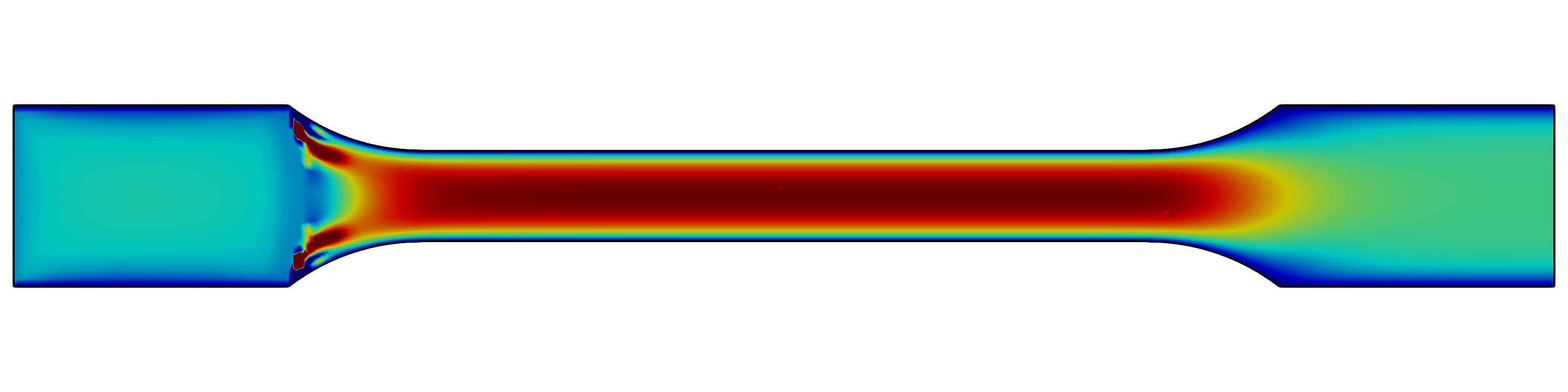}
\includegraphics[width=0.32\textwidth]{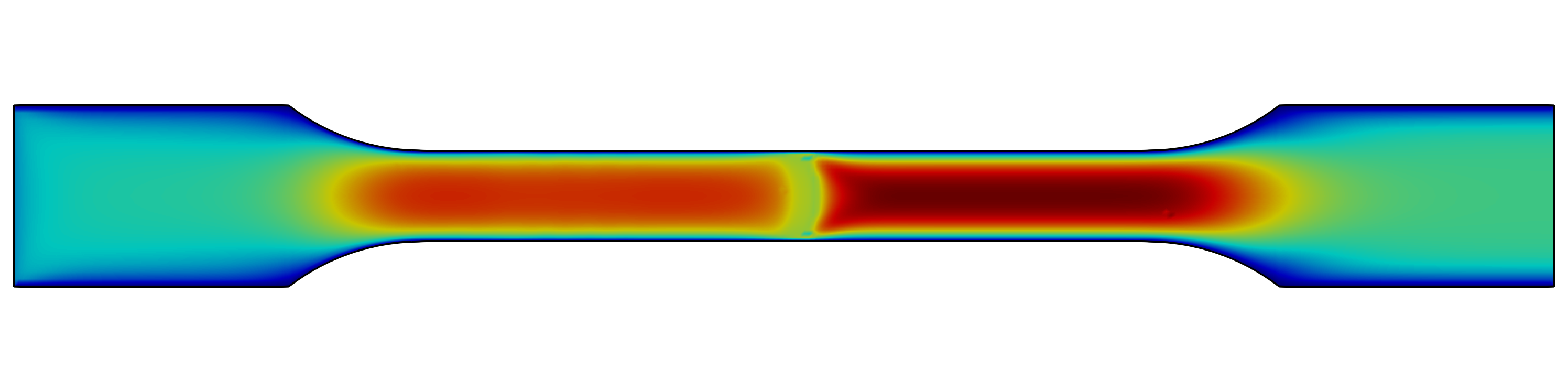}
\includegraphics[width=0.32\textwidth]{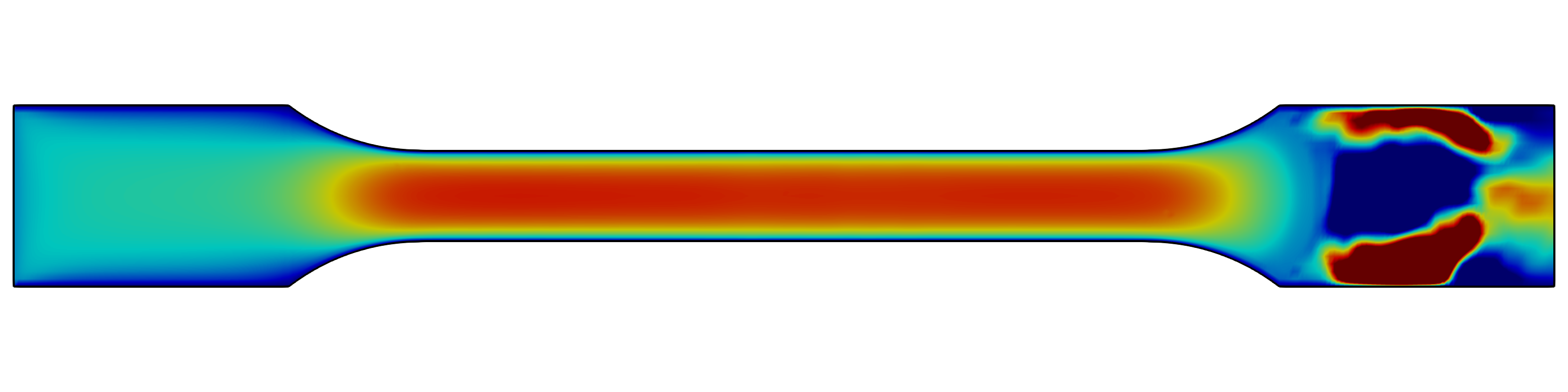}
\caption{Standard segregated VOF solution with (1, 0.4) Courant number bounds.}
\label{fig:ch4-dogbone-non-newtonian-high-interFoam}
\end{subfigure}

\begin{subfigure}{\textwidth}
\centering
\includegraphics[width=0.32\textwidth]{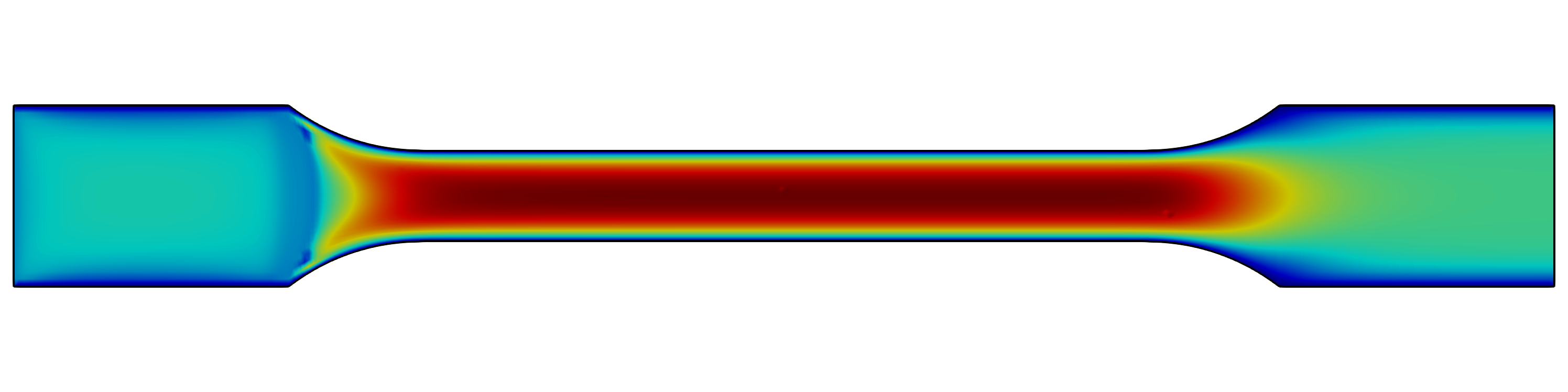}
\includegraphics[width=0.32\textwidth]{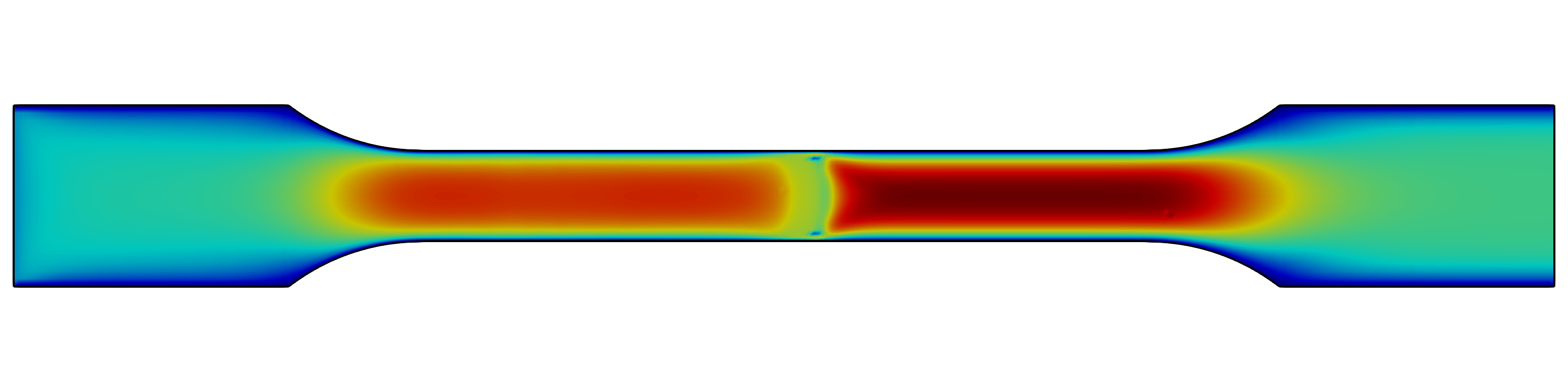}
\includegraphics[width=0.32\textwidth]{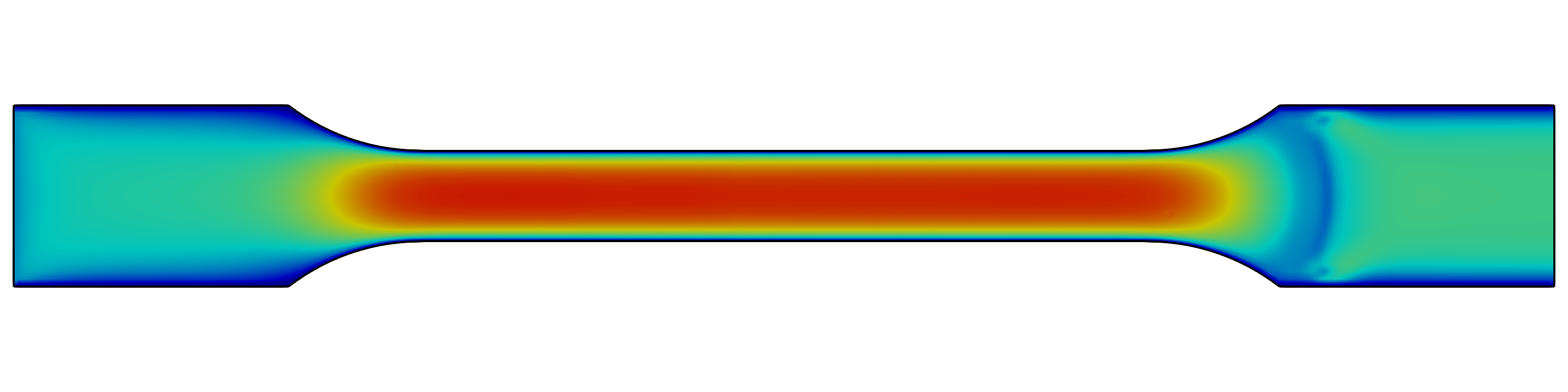}
\caption{Segregated VOF$^\ast$ solution with (1, 0.4) Courant number bounds.}
\label{fig:ch4-dogbone-non-newtonian-high-interFoam-star}
\end{subfigure}

\begin{subfigure}{\textwidth}
\centering
\includegraphics[width=0.32\textwidth]{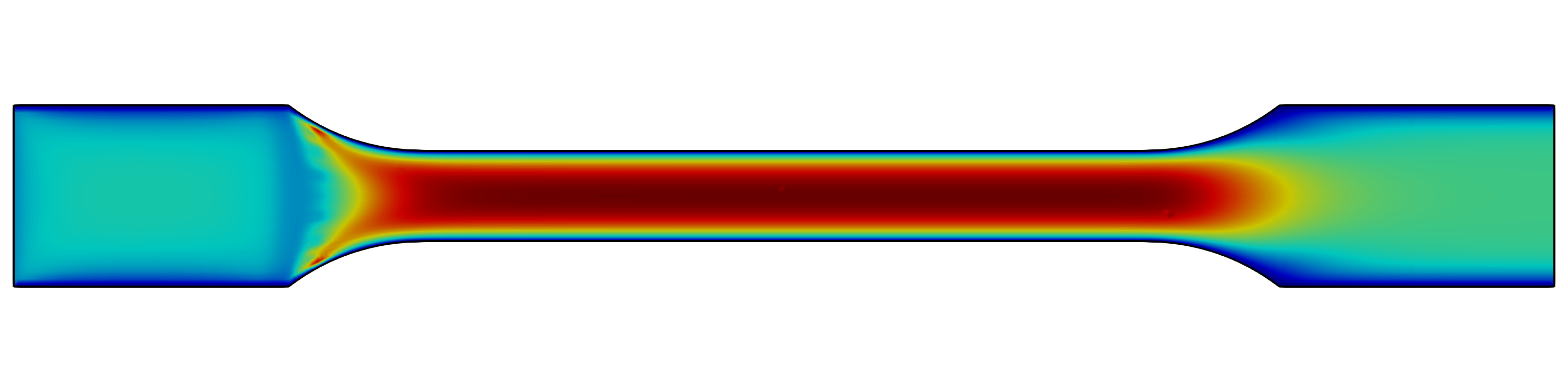}
\includegraphics[width=0.32\textwidth]{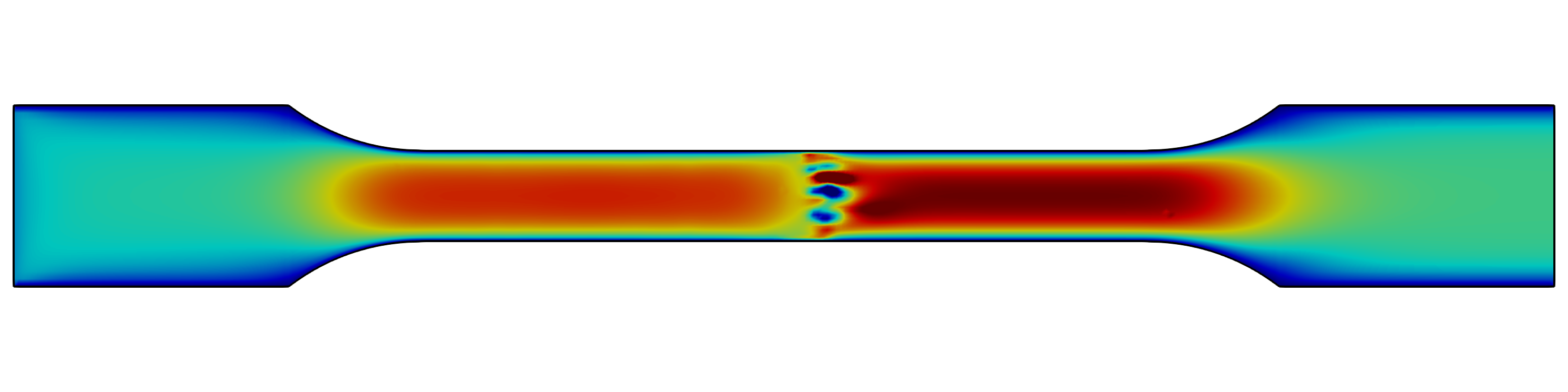}
\includegraphics[width=0.32\textwidth]{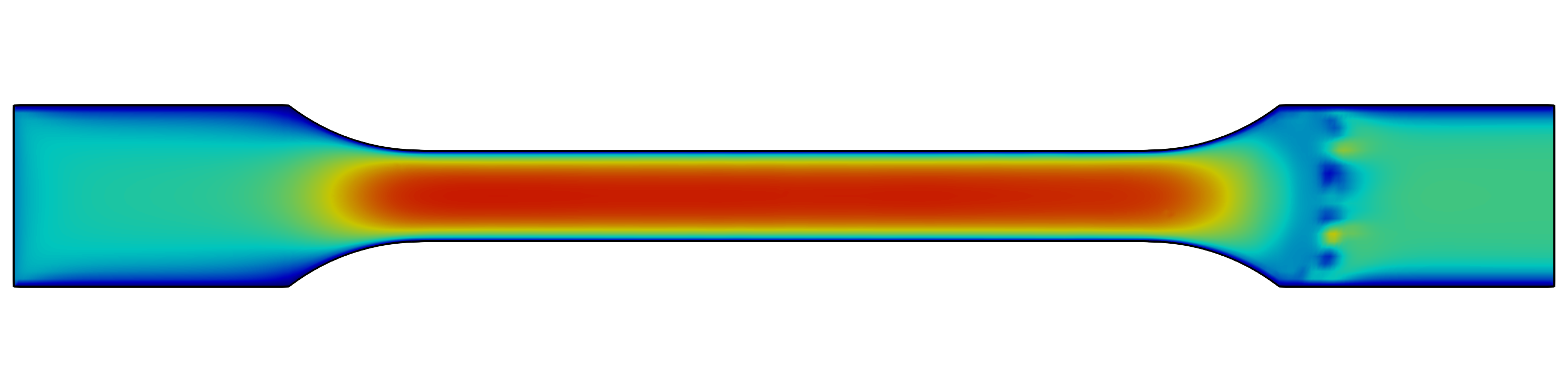}
\caption{Segregated VOF$^\ast$ solution with (50, 2) Courant number bounds.}
\label{fig:ch4-dogbone-non-newtonian-high-interFoam-star-highCo}
\end{subfigure}

\begin{subfigure}{\textwidth}
\centering
\includegraphics[width=0.32\textwidth]{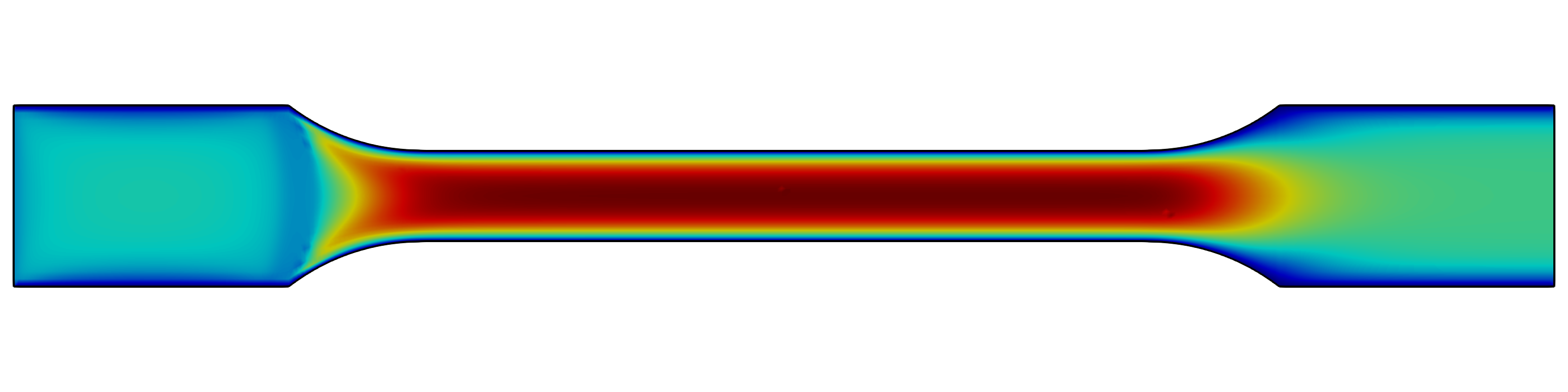}
\includegraphics[width=0.32\textwidth]{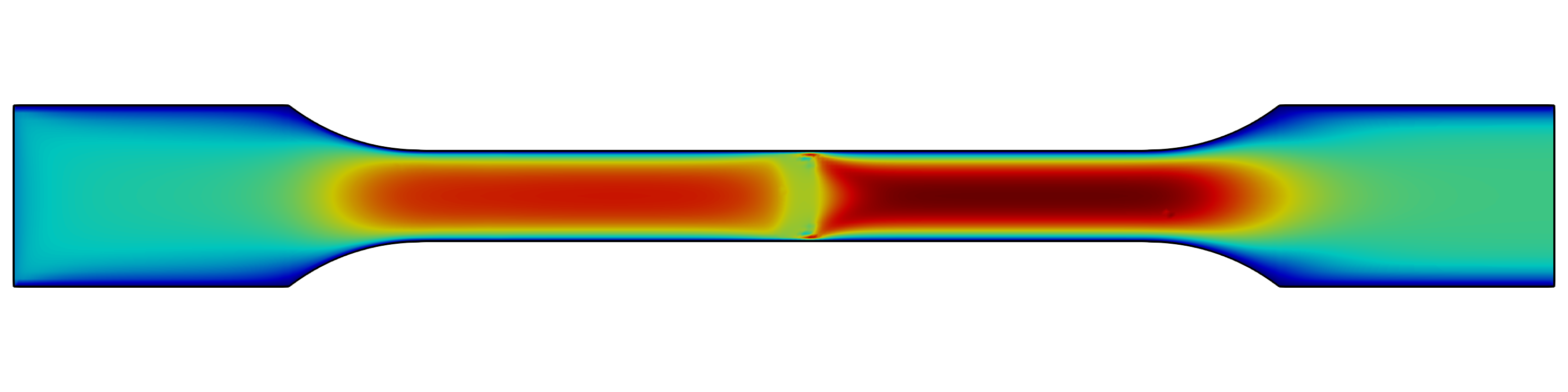}
\includegraphics[width=0.32\textwidth]{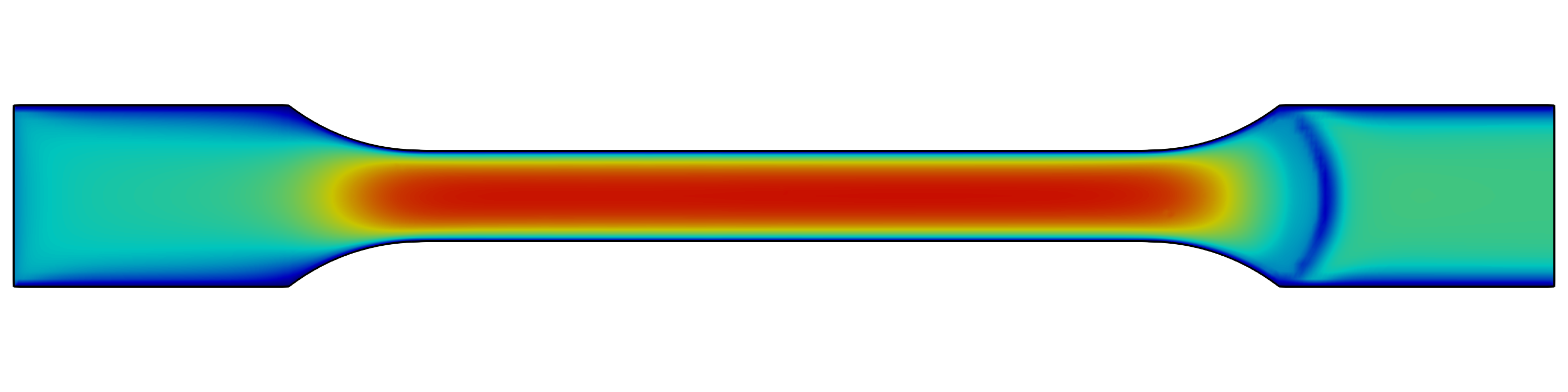}
\caption{BC-VOF solution with (50, 2) Courant number bounds.}
\label{fig:ch4-dogbone-non-newtonian-high-UCInterFoam}
\end{subfigure}

\caption{Axial velocity for the dog-bone case obtained with VOF, VOF$^\ast$ and BC-VOF solvers at different Courant number bounds. The velocity field is plotted on a slice in the $x-y$ plane for both fluid phases at three successive time-instants, $t = 0.25 \ \unit{s}$, $t = 0.5 \ \unit{s}$ and $t = 0.75 \ \unit{s}$.}
\label{fig:ch4-dogbone-non-newtonian-high}

\end{figure}

A qualitative comparison of the velocity fields obtained with VOF, VOF$^\ast$ and BC-VOF solvers is illustrated in Figure \ref{fig:ch4-dogbone-non-newtonian-high}.
The solution of the standard VOF, with the bounds $(50, 2)$, is not reported in the figure since the solver diverges after 42 time steps ($t \approx 0.13 \ \unit{s}$), with Courant numbers that grow rapidly before the simulation crashes. VOF$^\ast$ completes the simulation with the same bounds, but it produces spurious velocities in the air phase (Figure \ref{fig:ch4-dogbone-non-newtonian-high-interFoam-star-highCo}): the maximum velocity magnitude is $2.31 \ \unit{m/s}$, reached at time $t = 0.7$ s, against $1.20 \ \unit{m/s}$ for BC-VOF at $ t = 0.3$ s (Figure \ref{fig:ch4-dogbone-non-newtonian-high-UCInterFoam}). This behavior reflects a more general difference between the two approaches. In the segregated solutions, spurious velocities develop along the entire interface, often spreading into the air phase, and persist even with a Navier-slip condition \eqref{eq:navierSlipBC}. On the contrary, in the block-coupled simulations, both conforming and IB, they are confined to the channel edges, where they originate from the moving contact line singularity, and can be easily suppressed by introducing partial slip.

The bounds $(1, 0.4)$ are necessary to obtain physically meaningful results with the segregated approach. With these bounds, VOF$^\ast$ yields a physical solution (Figure \ref{fig:ch4-dogbone-non-newtonian-high-interFoam-star}), apart from a localized velocity peak at the edges of the interface near the restrictions, similar to the one observed with BC-VOF over the same time interval.
The standard VOF solver, instead, still produces a degenerate velocity field in the air phase (Figure \ref{fig:ch4-dogbone-non-newtonian-high-interFoam}): the maximum velocity magnitude, $16.68 \ \unit{m/s}$ at $t = 0.25$ s, is reached along the interface edges at the first restriction, while at the second restriction the peak is milder but the velocity field becomes completely unphysical.

Since VOF$^\ast$ and BC-VOF share the same algorithmic structure, mesh, viscosity averaging, volume-fraction transport and pressure equation, the different behavior observed with the bounds $(50, 2)$ can be attributed to the treatment of the transposed-gradient term. This is consistent with the analysis of Section \ref{sec:VOFIB-highViscIssues}: with the explicit treatment, the cancellation between the two parts of the viscous term near the interface is only approximate, at large time steps the residual produces spurious velocities when the momentum predictor is disabled and instabilities when it is
enabled.

The performance of segregated and block-coupled solvers is presented in Table \ref{tab:ch4-dogbone-non-newtonian-high-deltaT-Co} concerning the simulations on the coarse grid. The finer grids are only employed in Section \ref{sec:dogbone-conf-ib} where BC-VOF and BC-VOF-IB results are presented, because running the segregated solvers to completion on the finer mesh was computationally infeasible.

The fair comparison is between VOF$^\ast$ with bounds $(1, 0.4)$ and BC-VOF, i.e.\ between the fastest segregated run that remains physical and the block-coupled run. BC-VOF allows average time steps about 4.8 times larger, but each time step is about 1.8 times more expensive (5.05 s against 2.81 s), owing to the solution of the coupled momentum system and to the larger
number of outer correctors. The net reduction of the computing time is about a factor of 2.7, from 1 h 4 min to 24 min. The standard VOF solver (\texttt{interFoam}), instead, requires 3 h with bounds $(1, 0.4)$, without yielding a physical solution.

\begin{center}
\begin{table}[h!]
    \renewcommand{\arraystretch}{1.3}
    \setlength{\tabcolsep}{4pt}  
    \small
    \centering
    \begin{tabular}{|c|c|c|c|c|c|c|c|}
        \noalign{\hrule height 1pt}
        Solver & Bounds & $\min /\ \text{avg} \ \Delta t \ [\unit{s}]$ & $\max /\ \text{avg} \ \text{Co}_h$ & $\max /\ \text{avg} \ \text{Co}_{\alpha,h}$ & $\max\|\bm u\|$ [m/s] & WCT$_\text{tot}$ & $N_T$ \\
        \noalign{\hrule height 0.7pt}
        VOF & (1, 0.4) & $9.23 \cdot 10^{-6} $   /  $2.26 \cdot 10^{-4} $  & 8.09 / 0.96 & 0.49 / 0.12 & 16.68 ($t = 0.25$ s)&  3 h & 4872 \\
        \cline{3-8}
        VOF & (50, 2)& \multicolumn{6}{c|}{diverges at $t \approx 0.13~\unit{s}$, when $\max(\text{Co}) \sim 10^{11}$, $\max(\text{Co}_{\alpha,h}) \sim 10^8$ and $\Delta t \sim 10^{-13}$ s}  \\
        \hline
        VOF$^\ast$ & (1, 0.4) & $3.27\cdot10^{-4}$ / $8.01\cdot10^{-4}$ & 2.13 / 0.85 & 0.56 / 0.37 & 1.04 ($t = 0.3$ s) & 1 h 4 min & 1373 \\
        VOF$^\ast$ & (50, 2)  & $1.01\cdot10^{-3}$ / $2.56\cdot10^{-3}$ & 10.5 / 3.97 & 3.97 / 1.71 & 2.31 ($t = 0.7$ s) & 20 min & 430 \\
        \hline
        BC-VOF & (50, 2) & $1.19\cdot10^{-3}$ / $3.83\cdot10^{-3}$     & 9.82 / 4.17 & 3.08 / 1.73 & 1.20 ($t = 0.3$ s) & 24 min & 287 \\
        \hline
        BC-VOF-IB & (50, 2) & $9.73 \cdot 10^{-4} $ / $2.48 \cdot 10^{-3}$  & 11.18 / 2.45 & 6.12 / 1.78 & 1.23 ($t = 0.9$ s) & 1 h & 443 \\
        \noalign{\hrule height 1pt}
    \end{tabular}
    \caption{Performance of the segregated and block-coupled solvers for the dog-bone case on the coarse grid, for $t \in (0, 1.1) \ \unit{s}$: Courant-number bounds, time-step sizes, maximum and average Courant numbers, maximum velocity, wall-clock time on 4 MPI processes and total number of time iterations $N_T$.}
\label{tab:ch4-dogbone-non-newtonian-high-deltaT-Co}
\end{table}
\end{center}

\begin{figure}[h!]
\centering
\begin{subfigure}{\textwidth}
\centering
\includegraphics[width=\textwidth]{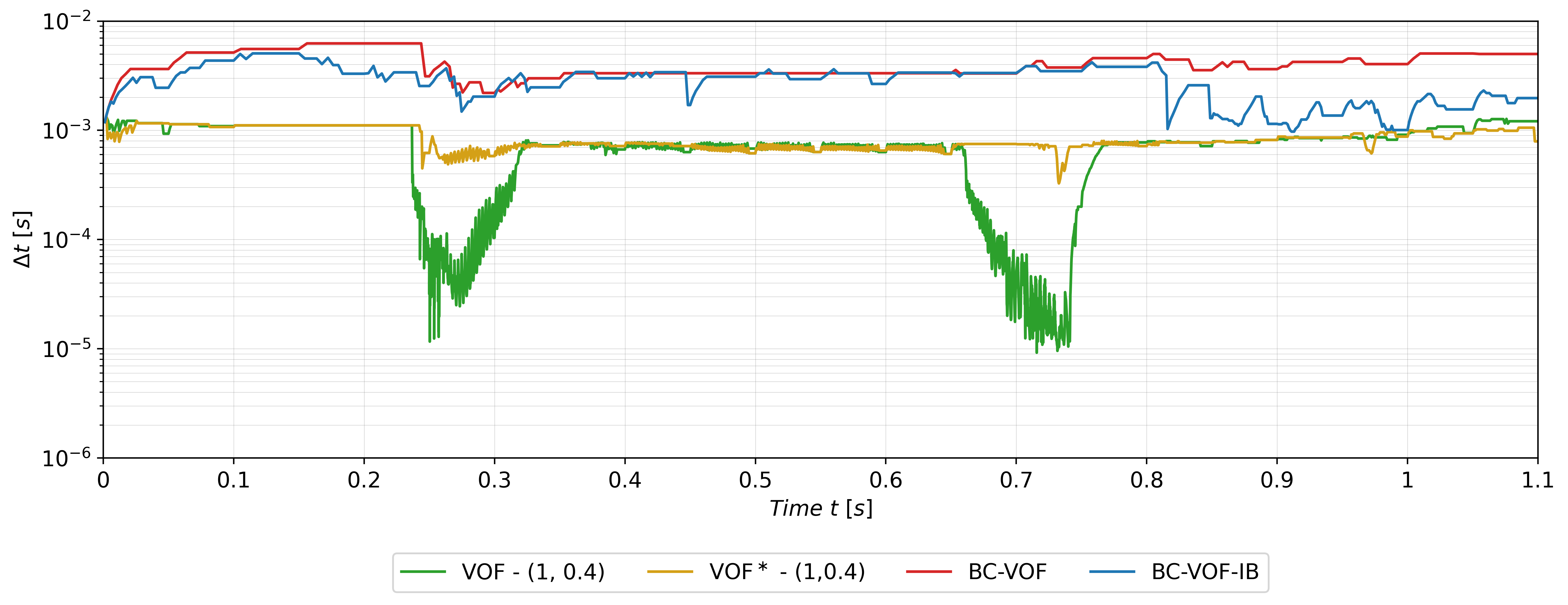}
\caption{Trend of $\Delta t$ values throughout the simulation. The $y-$axis is represented in logarithmic scale.}
\label{fig:ch4-dogbone-non-newtonian-high-deltat}
\end{subfigure}

\begin{subfigure}{\textwidth}
\centering
\bigskip
\includegraphics[width=\textwidth]{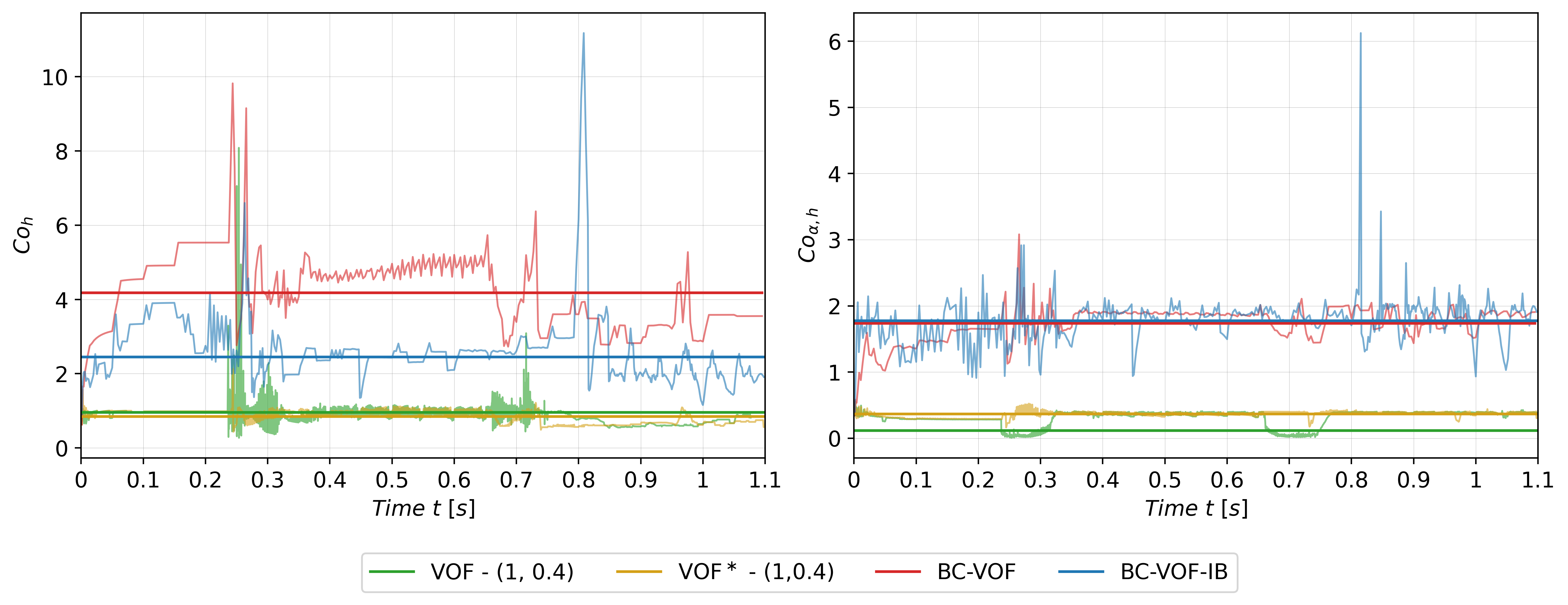}
\caption{Trends of $\max\text{Co}_h$ (left) and $\max\text{Co}_{\alpha,h}$ (right) throughout the simulation.}
\label{fig:ch4-dogbone-non-newtonian-high-Co}
\end{subfigure}

\caption{Time histories of (a) the time-step size and (b) the maximum Courant numbers for the dog-bone case on the coarse grid, obtained with VOF and VOF$^\ast$ with bounds $(1, 0.4)$ and with BC-VOF and BC-VOF-IB with bounds $(50, 2)$.}
\label{fig:ch4-dogbone-non-newtonian-high-deltat-Co}
\end{figure}

The time histories of $\Delta t$ and of the Courant numbers over the whole simulation are reported in Figures \ref{fig:ch4-dogbone-non-newtonian-high-deltat} and  \ref{fig:ch4-dogbone-non-newtonian-high-Co}, respectively. The smallest time steps, corresponding to the Courant number peaks, are attained when the phase front passes through the restriction of the dog-bone domain, for $t \in (0.2,0.3) \ \unit{s}$ and $t \in (0.7, 0.8) \ \unit{s}$, where the shear rate, and hence the viscosity gradient of the power-law melt, is high. Since the adaptive time-stepping strategy of OpenFOAM computes each time step from the Courant numbers of the previous one and limits its growth, sudden increases of the velocity may locally produce Courant numbers above the bounds; the time-averaged values, however, always satisfy them (Table \ref{tab:ch4-dogbone-non-newtonian-high-deltaT-Co}). With BC-VOF, the time step is limited by the interface Courant number, whose average (1.73) is close to its bound, while the average Courant number (4.17) remains well below its own.

\subsubsection{Conforming and IB block-coupled solvers}
\label{sec:dogbone-conf-ib}

The non-conforming BC-VOF-IB solver (\texttt{UCInterIbFoam}) is compared with its conforming counterpart BC-VOF. 
The setup of Table \ref{tab:ch4-dogbone-setup} remains nearly identical during mesh refinement, with the exception of the Courant number limits, which are reduced to $\text{Co}_{h,\max} = 20$ and  $\text{Co}_{\alpha,h,\max} = 1$, as well as $\Delta t_\text{max} = 10^{-4} \ s$, while the number of sub-cycles is increased to $N_\text{sc} = 50$. This choice is motivated by the stricter stability constraints on $\Delta t$ at smaller mesh sizes $h$.

\begin{figure}[!ht]
\centering

\begin{subfigure}{\textwidth}
\centering
\includegraphics[width=0.7\textwidth]{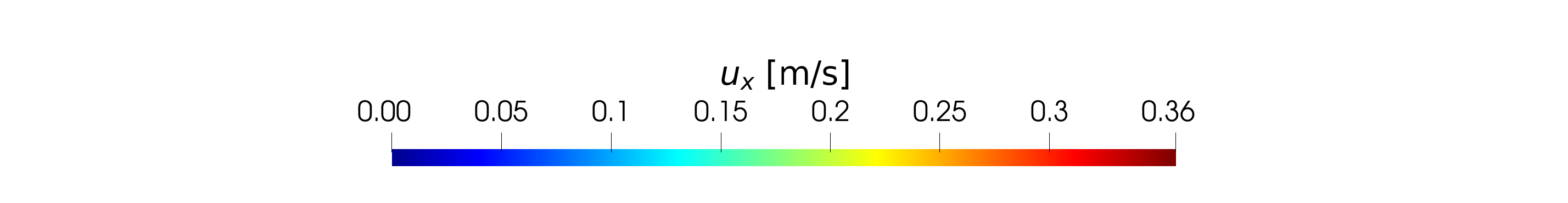} \\
\includegraphics[width=0.32\textwidth]{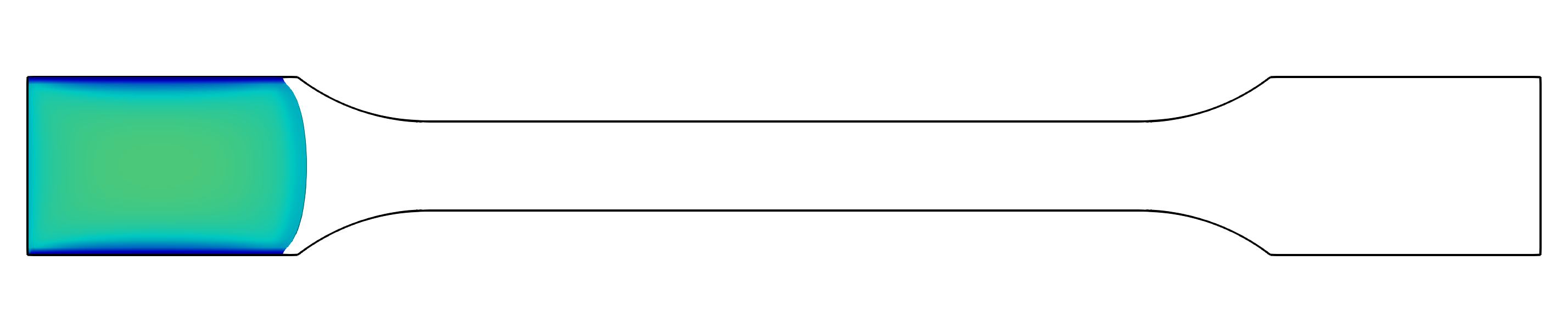}
\includegraphics[width=0.32\textwidth]{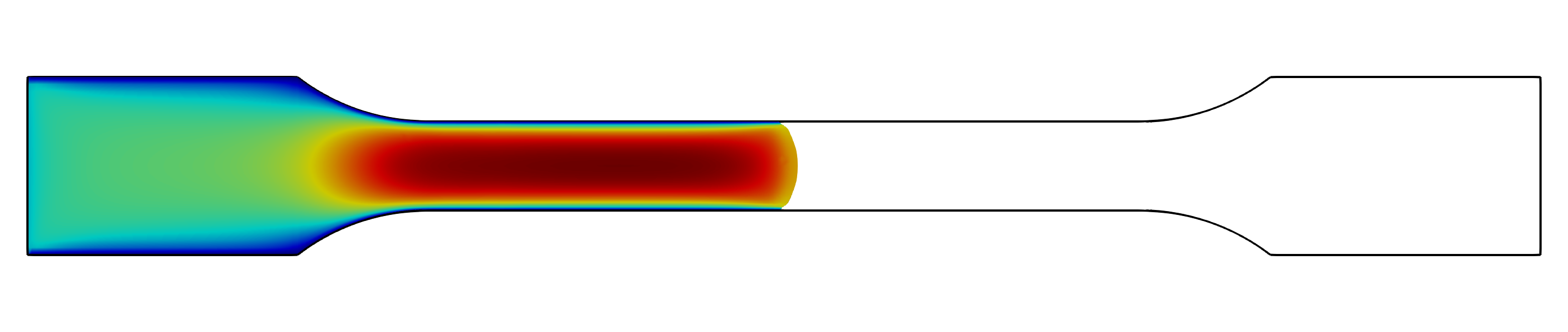}
\includegraphics[width=0.32\textwidth]{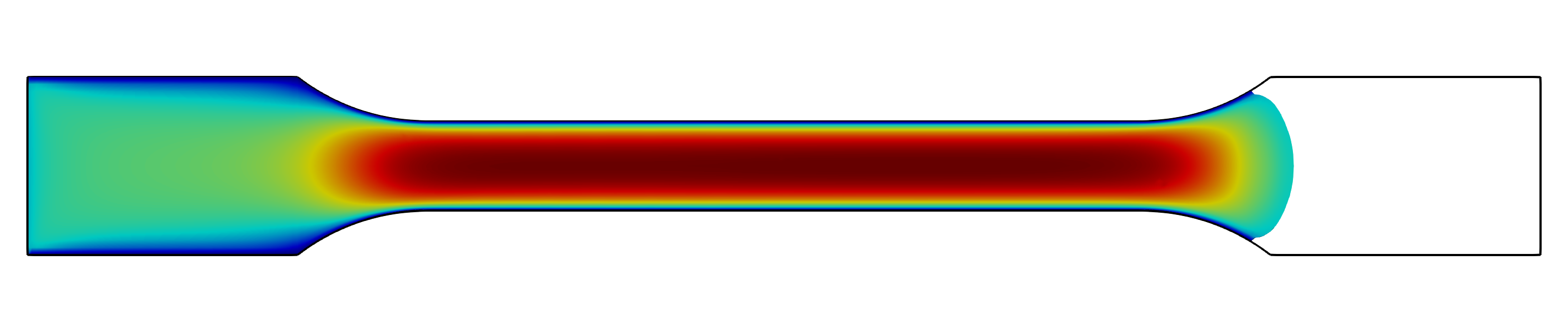}
\caption{BC-VOF solver.}
\label{fig:ch4-dogbone-IB-non-newtonian-high-uclip-UCInterFoam}
\end{subfigure}

\begin{subfigure}{\textwidth}
\centering
\includegraphics[width=0.32\textwidth]{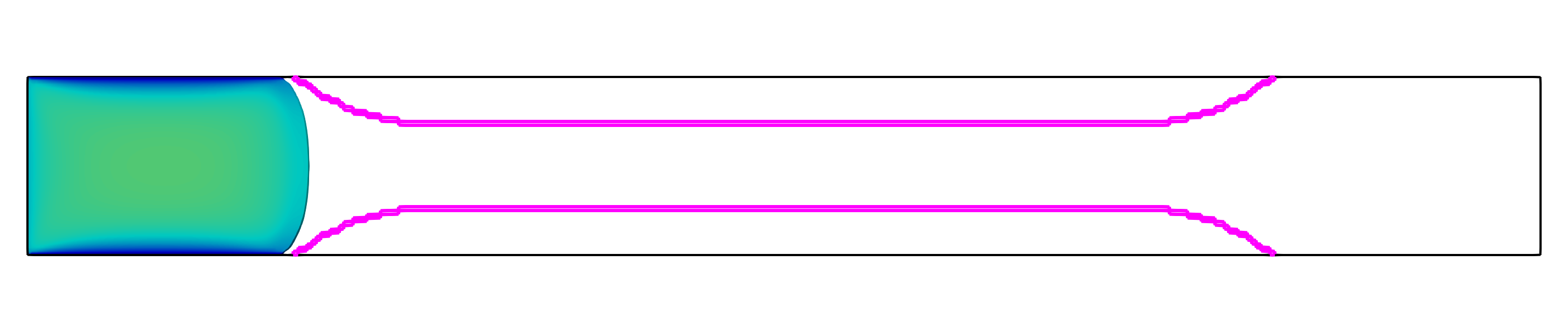}
\includegraphics[width=0.32\textwidth]{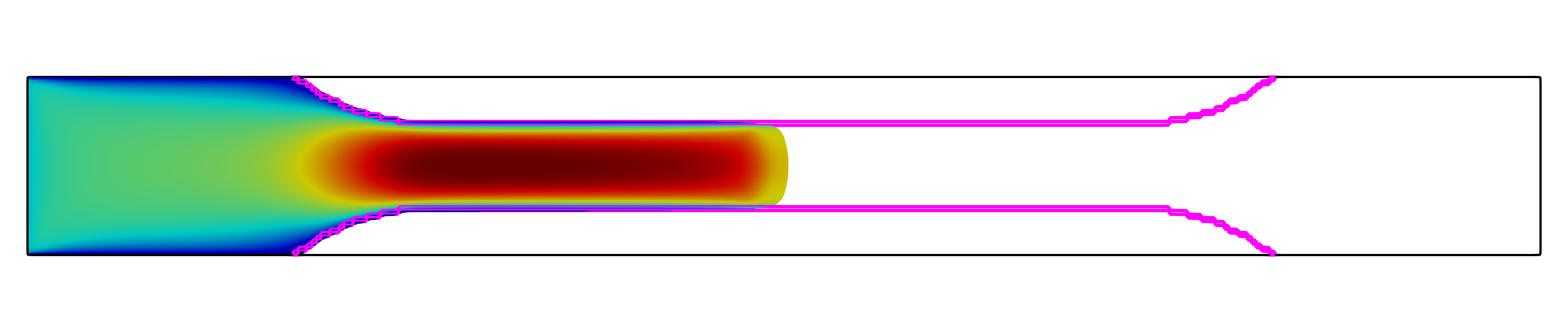}
\includegraphics[width=0.32\textwidth]{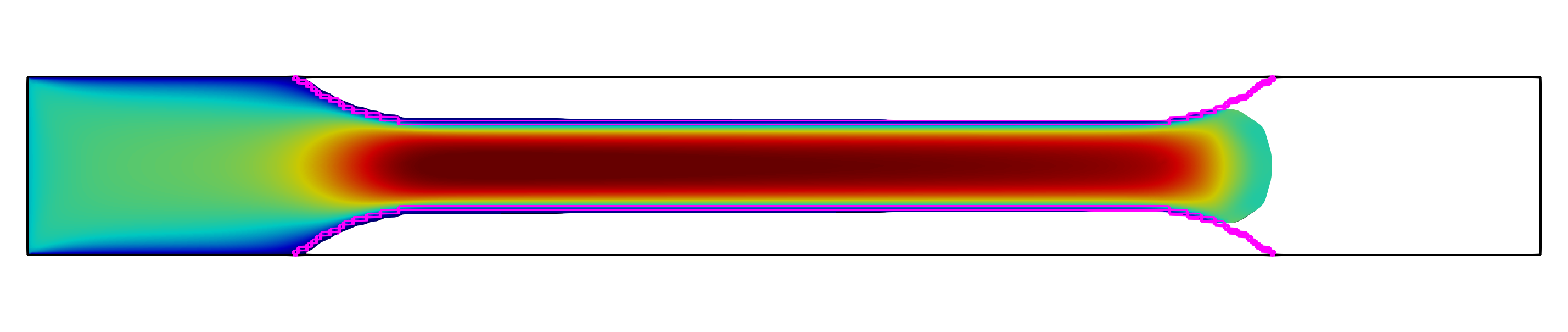}
\caption{BC-VOF-IB solver.}
\label{fig:ch4-dogbone-IB-non-newtonian-high-uclip-UCInterIbFoam}
\end{subfigure}

\caption{Axial velocity $u_x$ for the dog-bone case on the coarse grids ($h_1 \sim 0.5~\unit{mm}$) obtained with (a) BC-VOF and (b) BC-VOF-IB solvers at three successive time-instants, $t = 0.25 \ \unit{s}$, $t = 0.5 \ \unit{s}$ and $t = 0.75 \ \unit{s}$, and plotted on a slice in the $x-y$ plane where $0.5 < \alpha < 1$.}
\label{fig:ch4-dogbone-IB-non-newtonian-high-uclip}

\end{figure}

\begin{figure}[h!]
\centering

\begin{subfigure}{\textwidth}
\centering
\includegraphics[width=0.7\textwidth]{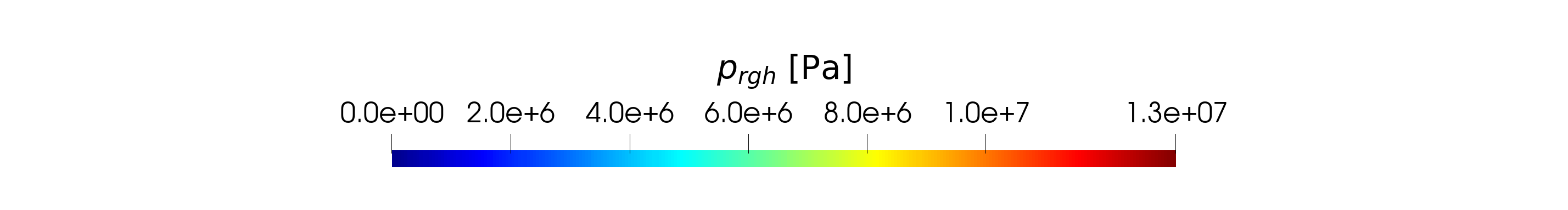} \\
\includegraphics[width=0.32\textwidth]{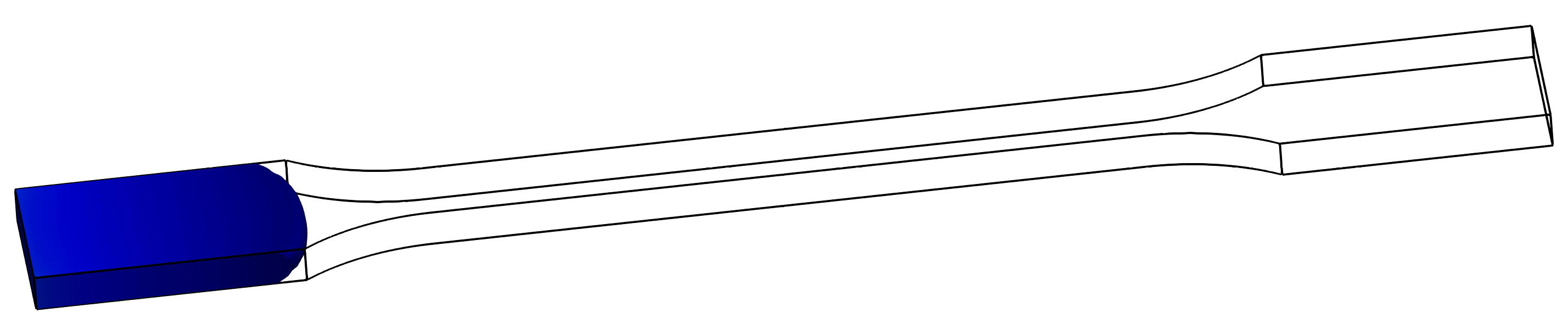}
\includegraphics[width=0.32\textwidth]{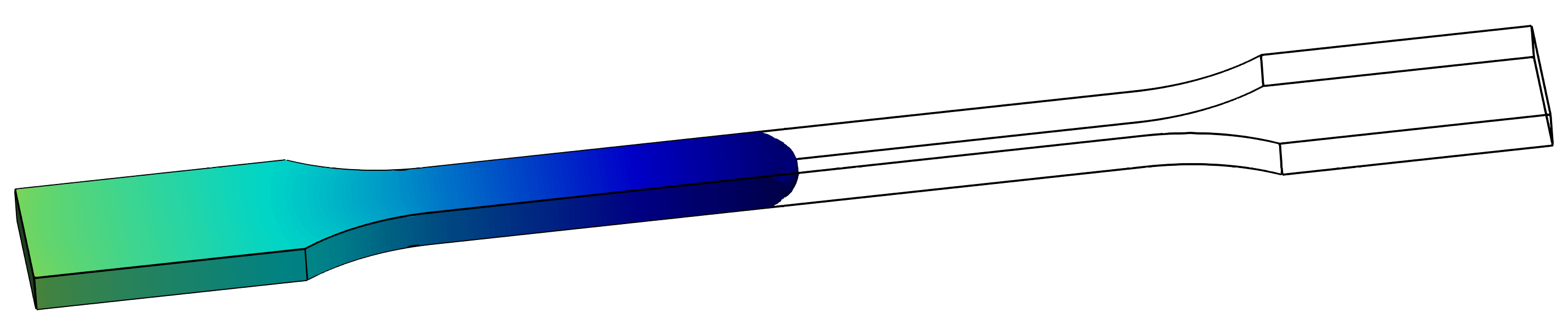}
\includegraphics[width=0.32\textwidth]{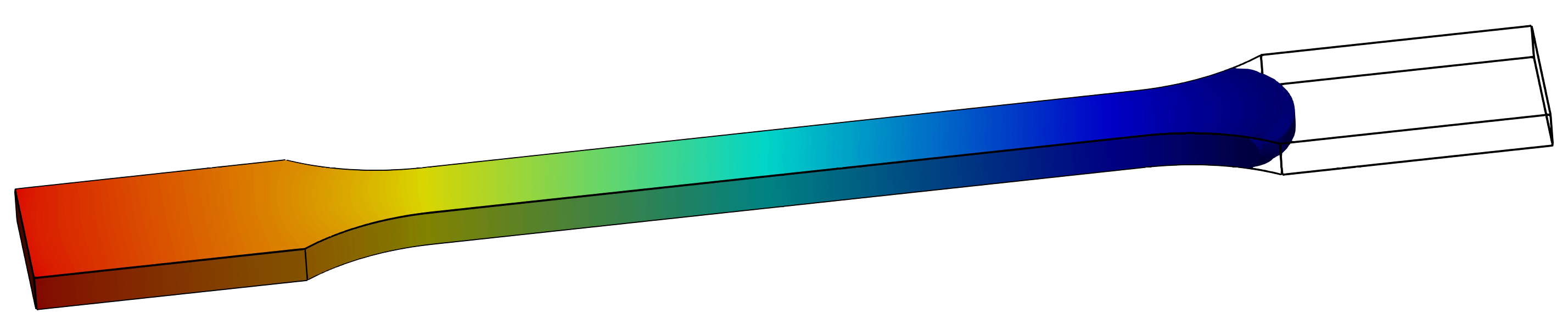}
\caption{BC-VOF solver.}
\label{fig:ch4-dogbone-IB-non-newtonian-high-pclip-UCInterFoam}
\end{subfigure}

\begin{subfigure}{\textwidth}
\centering
\includegraphics[width=0.32\textwidth]{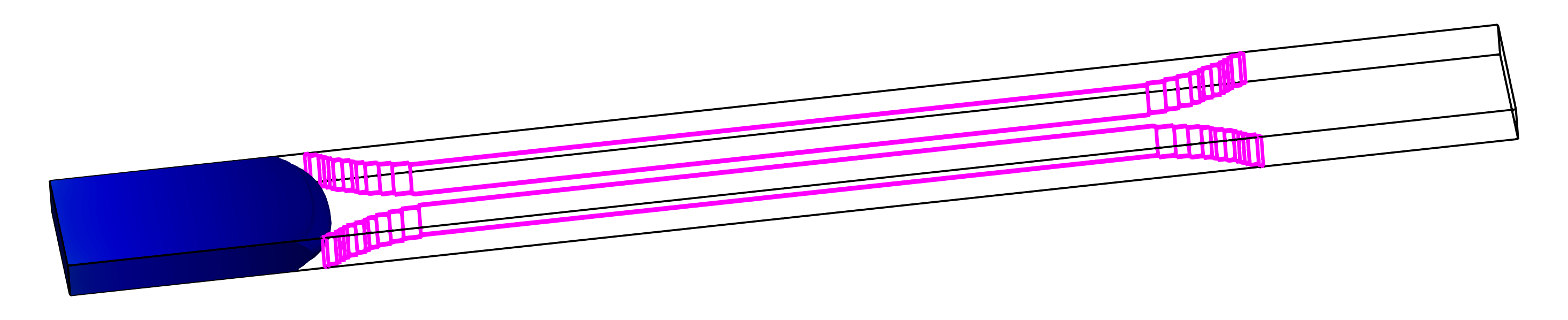}
\includegraphics[width=0.32\textwidth]{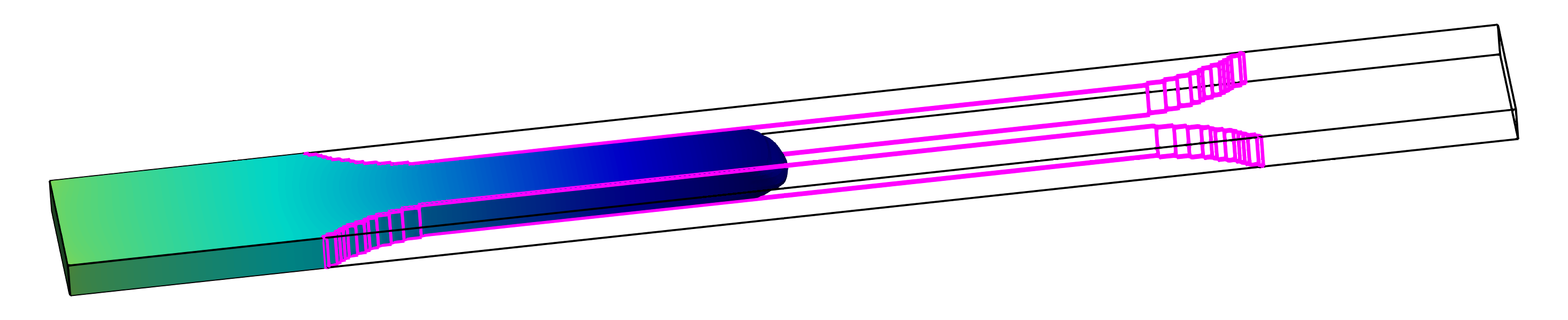}
\includegraphics[width=0.32\textwidth]{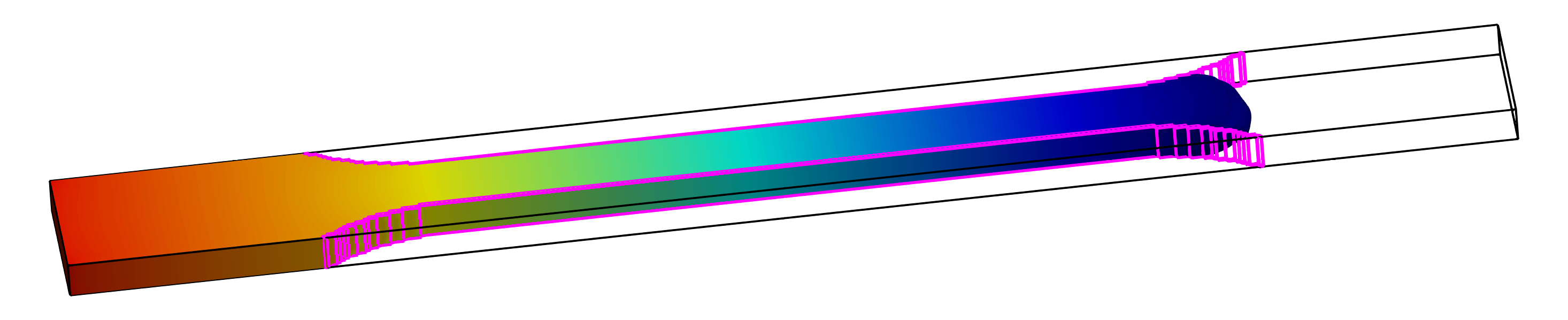}
\caption{BC-VOF-IB solver.}
\label{fig:ch4-dogbone-IB-non-newtonian-high-pclip-UCInterIbFoam}
\end{subfigure}

\caption{Pressure $p_{rgh}$ for the dog-bone case on the coarse grids ($h_1 \sim 0.5~\unit{mm}$) obtained with (a) BC-VOF and (b) BC-VOF-IB solvers at three successive time-instants, $t = 0.25 \ \unit{s}$, $t = 0.5 \ \unit{s}$ and $t = 0.75 \ \unit{s}$, and plotted in the whole domain where $0.5 < \alpha < 1$.}
\label{fig:ch4-dogbone-IB-non-newtonian-high-pclip}

\end{figure}

\begin{figure}[h!]
\centering

\begin{subfigure}{\textwidth}
\centering
\includegraphics[width=0.7\textwidth]{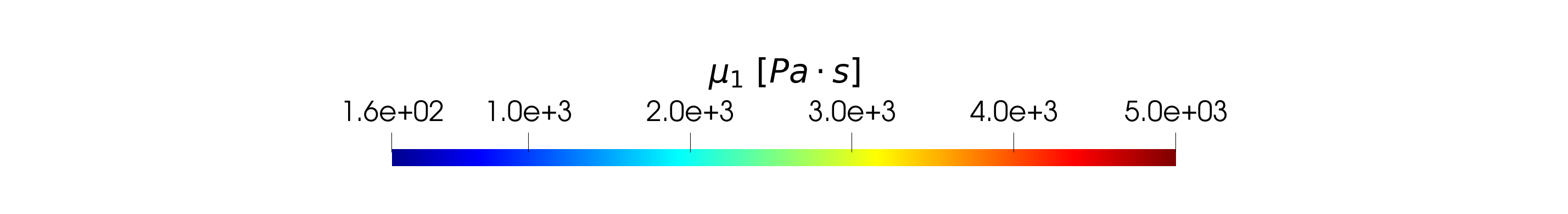} \\
\includegraphics[width=0.32\textwidth]{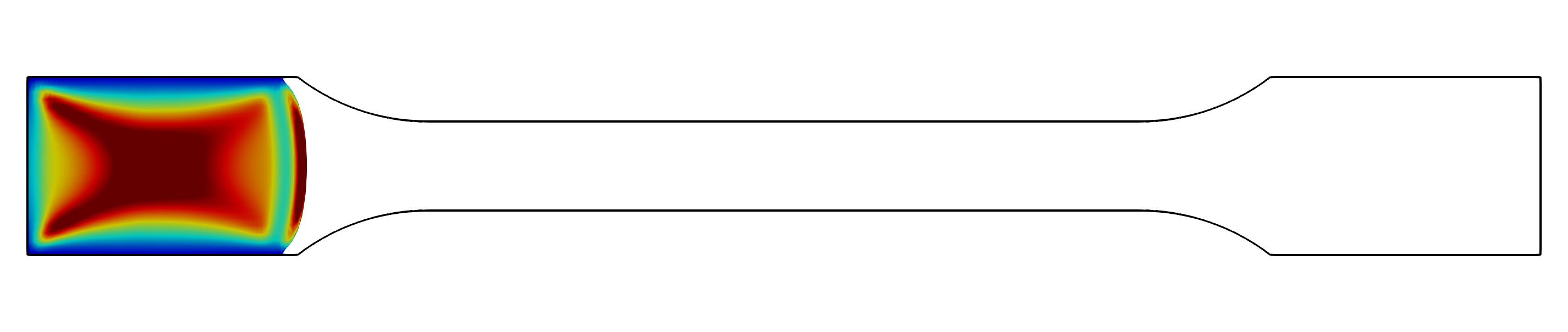}
\includegraphics[width=0.32\textwidth]{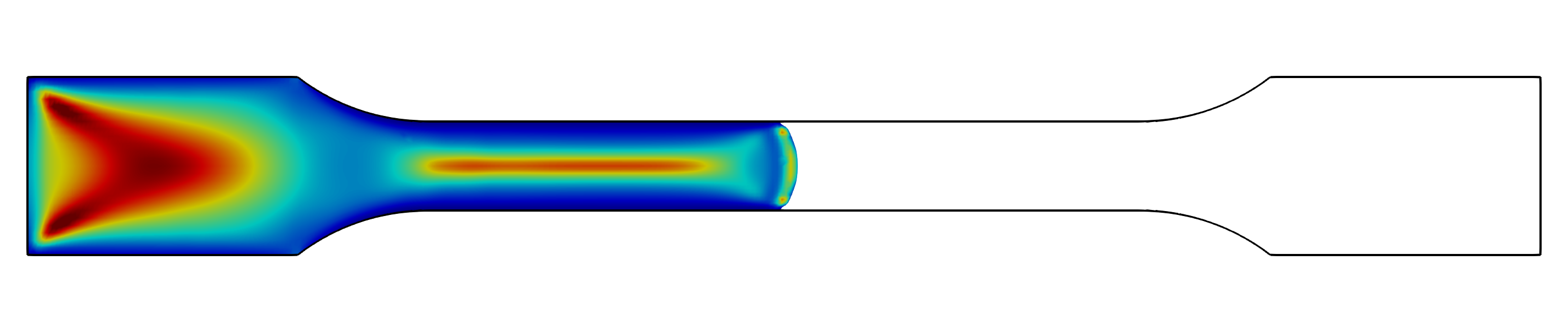}
\includegraphics[width=0.32\textwidth]{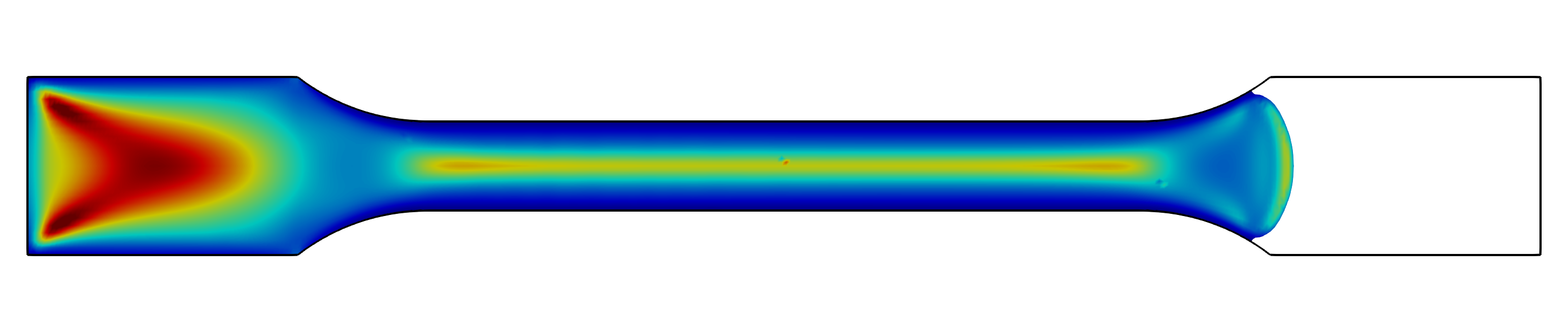}
\caption{BC-VOF solver.}
\label{fig:ch4-dogbone-IB-non-newtonian-high-muclip-UCInterFoam}
\end{subfigure}

\begin{subfigure}{\textwidth}
\centering
\includegraphics[width=0.32\textwidth]{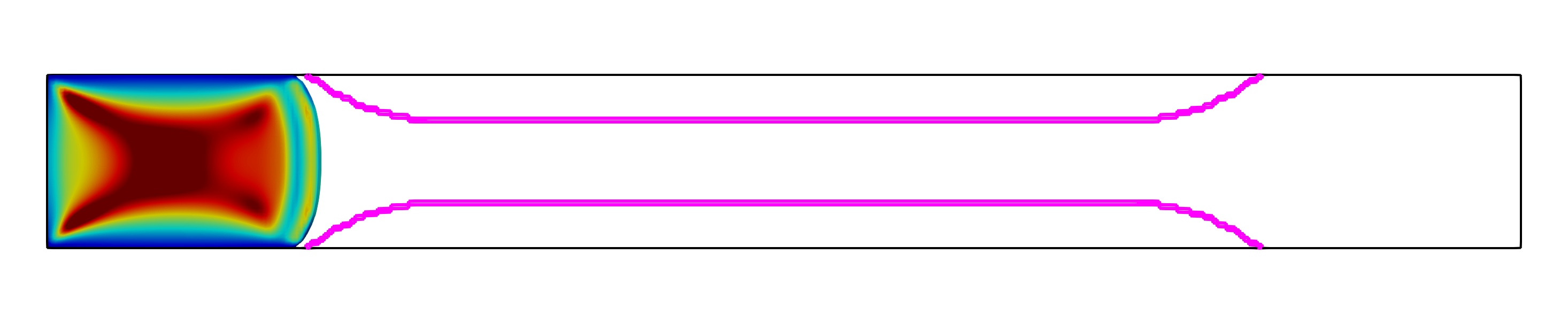}
\includegraphics[width=0.32\textwidth]{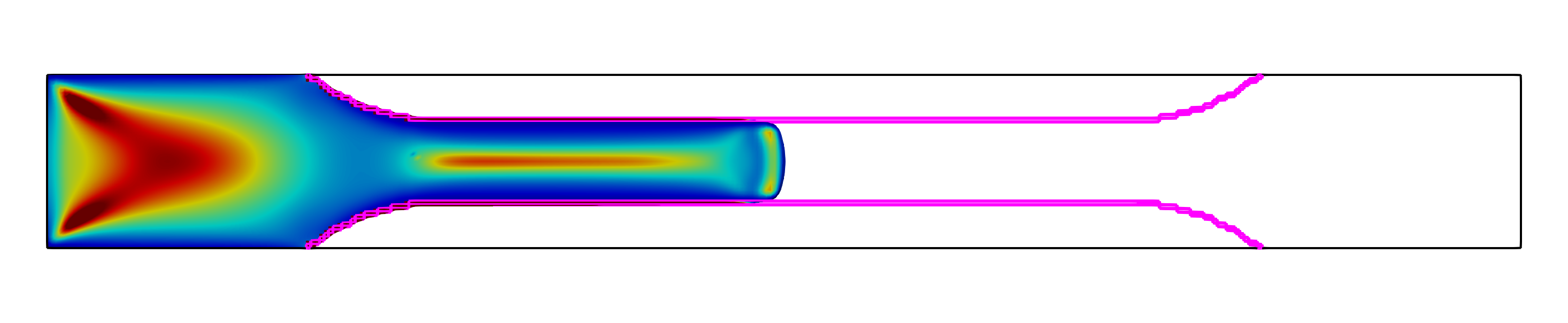}
\includegraphics[width=0.32\textwidth]{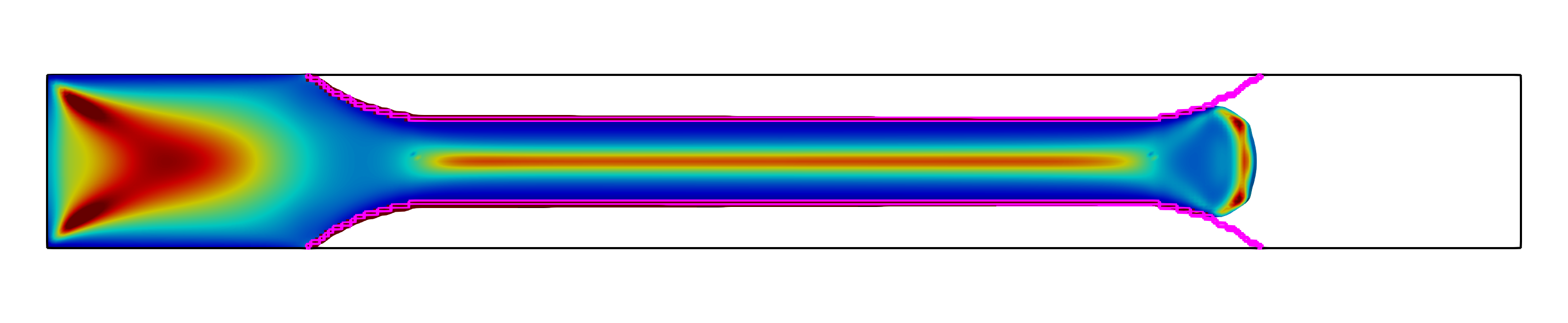}
\caption{BC-VOF-IB solver.}
\label{fig:ch4-dogbone-IB-non-newtonian-high-muclip-UCInterIbFoam}
\end{subfigure}

\caption{Viscosity field for the dog-bone case on the coarse grids ($h_1 \sim 0.5~\unit{mm}$) obtained with (a) BC-VOF and (b) BC-VOF-IB solvers at three successive time-instants, $t = 0.25 \ \unit{s}$, $t = 0.5 \ \unit{s}$ and $t = 0.75 \ \unit{s}$, and plotted on a slice in the $x-y$ plane where $0.5 < \alpha < 1$.}
\label{fig:ch4-dogbone-IB-non-newtonian-high-muclip}

\end{figure}

Figures \ref{fig:ch4-dogbone-IB-non-newtonian-high-uclip}, \ref{fig:ch4-dogbone-IB-non-newtonian-high-pclip} and \ref{fig:ch4-dogbone-IB-non-newtonian-high-muclip} show the numerical solutions of axial velocity, pressure and dynamic viscosity fields, respectively, comparing BC-VOF (\texttt{UCInterFoam}) with BC-VOF-IB (\texttt{UCInterIbFoam}) results on the coarse grids at three time instants. The
interface advancement obtained with the IB solver agrees qualitatively with the conforming one. The largest discrepancies are observed at $t = 0.75 \ \unit{s}$, when the interface passes through the restriction where the lateral walls are curved: there, the interface is slowed down in the IB case, and a viscosity peak appears (Figure~\ref{fig:ch4-dogbone-IB-non-newtonian-high-muclip-UCInterIbFoam}) that is absent in the conforming solution (Figure \ref{fig:ch4-dogbone-IB-non-newtonian-high-muclip-UCInterFoam}).

Table \ref{tab:ch4-dogbone-non-newtonian-high-deltaT-Co} also reports the performance of BC-VOF-IB on the coarse grid. With the same Courant-number bounds as BC-VOF, the IB solver takes 443 time steps instead of 287, with an average time step about
35\% smaller ($2.48 \cdot 10^{-3} \ \unit{s}$ against $3.83 \cdot 10^{-3} \ \unit{s}$). The time histories in Figure \ref{fig:ch4-dogbone-non-newtonian-high-deltat} show that this difference is not uniform in time. Up to $t \approx 0.8 \ \unit{s}$, the
time steps of the two block-coupled solvers are comparable. The difference arises mainly after the phase front has passed the second restriction: for $t \in (0.8, 1.1) \ \unit{s}$, the average time step of BC-VOF-IB is about 2.8 times smaller than that of BC-VOF, and the interface Courant number reaches a value $\text{Co}_{\alpha,h} \approx 6.1$ at $t \approx 0.82 \ \unit{s}$, against a maximum of 3.1 for BC-VOF, which is reached in the first restriction. This behavior is consistent with the slower interface advancement and the viscosity peak observed in the IB solution where the interface crosses the restriction with curved walls Figure \ref{fig:ch4-dogbone-IB-non-newtonian-high-muclip-UCInterIbFoam}), which produce sharper local velocity variations near the interface after the restriction. As for BC-VOF, the time step of BC-VOF-IB is limited by the interface Courant number, whose average (1.78) is close to its bound, while the average Courant number (2.45) remains well below 50.

The cost per time step of BC-VOF-IB is higher than that of BC-VOF for two reasons. First, the IB background grid has
about twice as many cells as the conforming grid; second, the
IB-constrained pressure equation is harder to solve, requiring more iterations of the linear solver on average. Overall, BC-VOF-IB requires about 1 h, i.e.\ about 2.5 times the computing time of BC-VOF.

A quantitative evaluation of the discrepancy between the interface positions obtained in the conforming and IB frameworks is carried out by estimating the distance between the two iso-surfaces localized at $\alpha = 0.5$, obtained at two consecutive mesh refinements. We indicate with $\Sigma_\text{conf}(t)$ and $\Sigma_\text{IB}(t)$ the sets of $N_\text{conf}$ and $N_\text{IB}$ points on the two iso-surfaces. For each point $p_i \in \Sigma_\text{IB}(t)$ the \textit{distance to the nearest point} on $\Sigma_\text{conf}(t)$ is computed as
\begin{equation*}
    d_{\text{conf},i}(t) = \text{dist}(p_i, \Sigma_\text{conf}) = \min_{q \in \Sigma_\text{conf}} \| p_i - q \|_2 \quad \forall i = 1,\dots,N_\text{IB}.
\end{equation*}
Similarly, for each point $p_i \in \Sigma_\text{conf}(t)$ we can define
\begin{equation*}
    d_{\text{IB},i}(t) = \text{dist}(p_i, \Sigma_\text{IB}) = \min_{q \in \Sigma_\text{IB}} \| p_i - q \|_2 \quad \forall i = 1,\dots,N_\text{conf}.
\end{equation*}
Then, we employ the following metrics to evaluate the nearest-neighbor distance between the two iso-surfaces:
\begin{itemize}
    \item \textit{Mean distance} of $\Sigma_\text{IB}$ from $\Sigma_\text{conf}$
    \begin{equation*}
        \bar d(t) = \frac{1}{N_\text{IB}} \sum_{i=1}^{N_\text{IB}} d_{\text{conf},i}(t);
    \end{equation*}
    \item \textit{Root Mean Square (RMS) distance} of $\Sigma_\text{IB}$ from $\Sigma_\text{conf}$
    \begin{equation*}
        d_\text{rms}(t) = \sqrt{\frac{1}{N_\text{IB}} \sum_{i=1}^{N_\text{IB}} d_{\text{conf},i}^2};
    \end{equation*}
    \item \textit{Symmetric Hausdorff distance}
    \begin{equation*}
        d_H(\Sigma_\text{IB},\Sigma_\text{conf}) = \max\left ( \max_i d(p_i,\Sigma_\text{conf}), \max_j d(p_j,\Sigma_\text{IB}) \right),
    \end{equation*}
    which is the worst-case distance.
\end{itemize}
Table \ref{tab:ch4-dogbone-non-newtonian-high-distances} collects the values of the three distances for the two consecutive grid refinements. On the coarse grid, all distances grow with time, as the interface accumulates the effect of the different wall representations along the channel. The mean and RMS distances decrease with mesh refinement at all time instants, except for the mean distance at $t = 0.25 \ \unit{s}$, which is almost unchanged. On the contrary, the symmetric Hausdorff distance does not decrease monotonically with $h$: it increases at $t = 0.25~\unit{s}$ and $t = 0.75~\unit{s}$ and decreases at $t = 0.5~\unit{s}$. The largest distances are located near the walls, where the contact line moves only through numerical slip. Refining the grid makes the moving-contact-line singularity (contact line paradox) more evident, numerically resulting in a more fragmented two-phase front with air bubbles entrapped at the walls.

\begin{table}[h!]
    \renewcommand{\arraystretch}{1.3}
    \setlength{\tabcolsep}{10pt}  
    \small
    \centering
    \begin{NiceTabular}{|c|cc|cc|cc|}
        \noalign{\hrule height 1pt}
        \Block{2-1}{Time \ $[\unit{s}]$} & \multicolumn{2}{c}{$\bar{d}(t) \ [\unit{m}]$} & \multicolumn{2}{c}{$d_\text{rms}(t) \ [\unit{m}]$} & \multicolumn{2}{c}{$d_H(t) \ [\unit{m}]$} \\
        \cline{2-7}
        & $h_1$ & $h_2$ & $h_1$ & $h_2$ & $h_1$ & $h_2$ \\
        \noalign{\hrule height 0.8pt}
        0.25 & $2.65 \cdot 10^{-4}$ & $2.59 \cdot 10^{-4}$ & $2.98 \cdot 10^{-4}$ & $2.84 \cdot 10^{-4}$ & $5.23 \cdot 10^{-4}$ & $1.45 \cdot 10^{-3}$ \\
        \hline
        0.50 & $1.12 \cdot 10^{-3}$ & $2.15 \cdot 10^{-4}$ & $1.18 \cdot 10^{-3}$ & $2.28 \cdot 10^{-4}$ & $2.67 \cdot 10^{-3}$ & $4.85 \cdot 10^{-4}$ \\
        \hline
        0.75 & $2.23 \cdot 10^{-3}$ & $3.38 \cdot 10^{-4}$ & $2.27 \cdot 10^{-3}$ & $4.10 \cdot 10^{-4}$ & $3.39 \cdot 10^{-3}$ & $4.20 \cdot 10^{-3}$\\
        \noalign{\hrule height 1pt}
    \end{NiceTabular}
    \caption{Interface position discrepancy metrics for the dog-bone case between BC-VOF and BC-VOF-IB iso-surfaces at three successive time-instants, $t = 0.25 \ \unit{s}$, $t = 0.5 \ \unit{s}$ and $t = 0.75 \ \unit{s}$, at refinement levels $h_1 \sim 0.5 \ \unit{mm}$ and $h_2 \sim 0.25 \ \unit{mm}$.} 
    \label{tab:ch4-dogbone-non-newtonian-high-distances}
\end{table}

\begin{figure}[h!]
    \centering
    \includegraphics[width=0.8\textwidth]{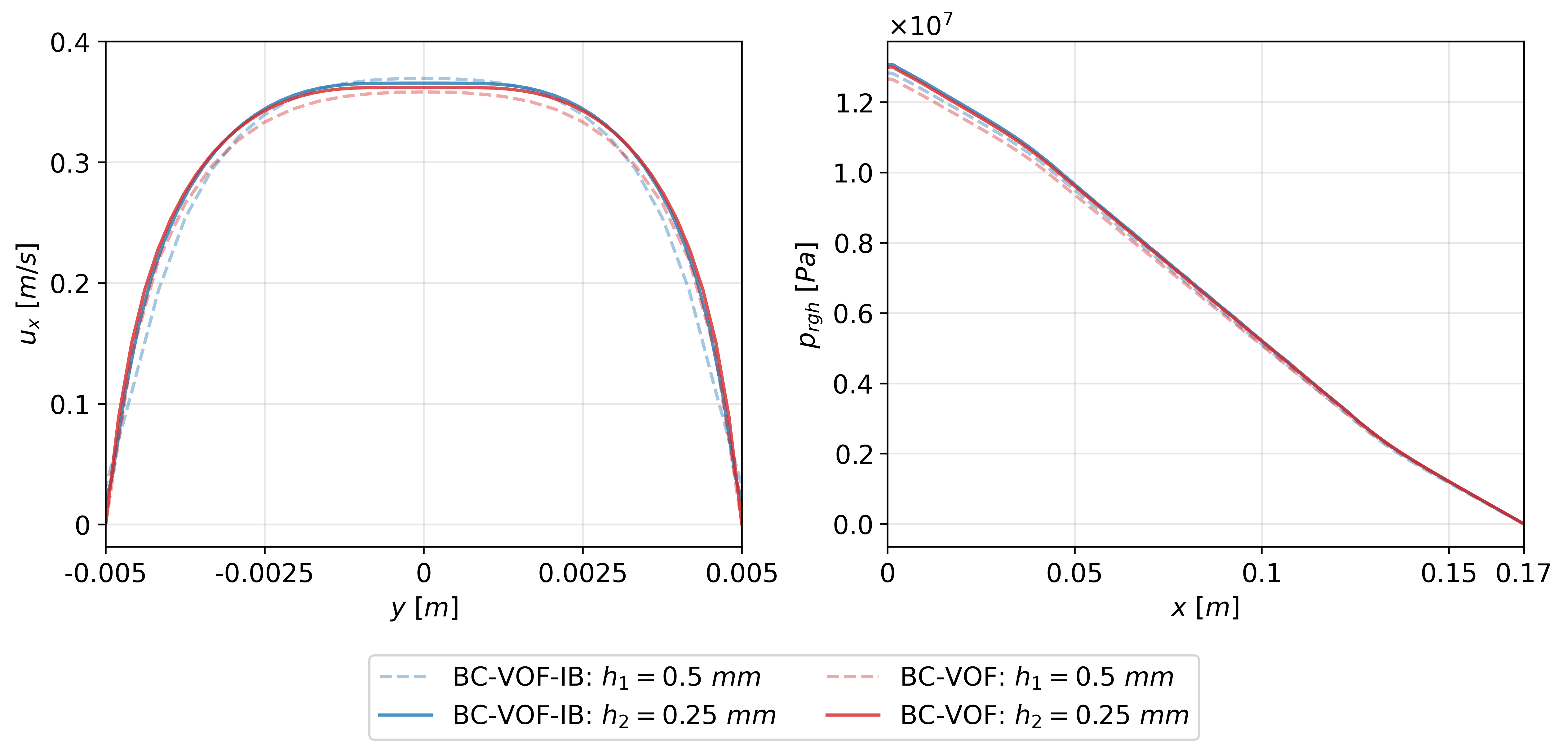}
    \caption{Axial velocity $u_x$ along vertical centerline (left), and pressure $p_{rgh}$ along horizontal centerline (right) for the dog-bone case at final time. The comparison between (a) BC-VOF and (b) BC-VOF-IB is reported, at refinement levels $h_1 \sim 0.5 \ \unit{mm}$ and $h_2 \sim 0.25 \ \unit{mm}$.}
    \label{fig:ch4-dogbone-IB-non-newtonian-high-ux-p-mu}
\end{figure}

We show in Figure \ref{fig:ch4-dogbone-IB-non-newtonian-high-ux-p-mu} the axial velocity profile along the vertical centerline and the pressure field along the horizontal centerline at final time (when the phase front has surpassed the outlet), obtained with the conforming and IB solvers at two consecutive grid refinements. The relative errors are computed for each variable $\psi$ as
\begin{equation*}
    e_r(\psi) = \dfrac{\lVert\psi_\text{IB} - \psi_\text{conf}\rVert}{\lVert\psi_\text{conf}\rVert},
\end{equation*}
with $\psi_\text{IB}$ and $\psi_\text{conf}$ identifying the solutions provided by BC-VOF-IB and BC-VOF, respectively, sampled at the segment center of either the vertical or the horizontal centerline, and the discrete norms $\lVert \psi \rVert_{L^1} = \sum_s \ell_s |\psi_s|$ and $\lVert \psi \rVert_{L^2} = \big(\sum_s \ell_s |\psi_s|^2\big)^{1/2}$, where $\ell_s$ is the length of segment $s$.

Table \ref{tab:ch4-dogbone-non-newtonian-high-errors} reports the relative errors in $L^2$ and $L^1$ discrete norms for the axial velocity and pressure fields along the centerlines, together with the relative error on the final pressure drop $\Delta p_{rgh} = p_{rgh,\text{out}} - p_{rgh,\text{in}}$, computed as $e_r(\Delta p_{rgh}) = |\Delta p_\text{IB} - \Delta p_\text{conf}|/|\Delta p_\text{conf}|$.
When the mesh size is halved, the velocity errors decrease from $4.1 \cdot 10^{-2}$ to $1.4 \cdot 10^{-2}$ in the $L^1$ norm and from $4.8 \cdot 10^{-2}$ to $1.8 \cdot 10^{-2}$ in the $L^2$ norm, i.e.\ by factors of about 2.9 and 2.6. The error on the
pressure drop decreases from 1.4\% to 0.5\%. The pressure error decreases by a factor of about 9 in the $L^1$ norm, but only by a factor of 2.6 in the $L^2$ norm. These reductions correspond to apparent rates between about 1.4 and 1.5 for the velocity and the pressure drop. We remark that these errors measure the  discrepancy between two numerical solutions on grids of the same size, not the error with respect to an exact solution, and that two refinement levels are not sufficient to establish an order of convergence.

\begin{table}[h!]
    \renewcommand{\arraystretch}{1.3}
    \setlength{\tabcolsep}{10pt}  
    \small
    \centering
    \begin{NiceTabular}{|c|cc|cc|c|}
        \noalign{\hrule height 1pt}
        & \multicolumn{2}{c}{Velocity error} & \multicolumn{3}{c}{Pressure error} \\
        \cline{2-6}
        & $e_{L^1,r}(u_x)$ & $e_{L^2,r}(u_x)$ & $e_{L^1,r}(p_{rgh})$ & $e_{L^2,r}(p_{rgh})$ & $e_{r}(\Delta p_{rgh})$  \\
        \noalign{\hrule height 0.8pt}
        $h_1$ & $4.09 \cdot 10^{-2}$ & $4.83 \cdot 10^{-2}$ & $1.30 \cdot 10^{-2}$ & $1.39 \cdot 10^{-2}$ & $1.38 \cdot 10^{-2}$  \\
        \hline
        $h_2$ & $1.41 \cdot 10^{-2}$ & $1.84 \cdot 10^{-2}$ & $1.41 \cdot 10^{-3}$ & $5.34 \cdot 10^{-3}$ & $5.13 \cdot 10^{-3}$ \\
        \noalign{\hrule height 1pt}
    \end{NiceTabular}
    \caption{Relative errors for the dog-bone case at final time, computed for axial velocity and pressure profile along centerlines between IB and conforming solution, at refinement levels $h_1 \sim 0.5 \ \unit{mm}$ and $h_2 \sim 0.25 \ \unit{mm}$.} 
    \label{tab:ch4-dogbone-non-newtonian-high-errors}
\end{table}

Overall, in the polymer phase, the IB solver reproduces the conforming solution with relative discrepancies below 5\% for the velocity and 1.5\% for the pressure on the coarse grid, decreasing to below 2\% and 0.6\%, respectively, on the fine grid. The largest differences concern the interface position near curved walls and in the contact-line region.

An evaluation of the phase-mass conservation properties of both conforming and IB approaches has also been conducted. Under the incompressibility assumption for the two fluid phases, the volume of injected material over time has the analytical expression given by $V_\text{ref}(t) = S_{\text{in}} u_{\text{in},x} t$, with $u_{\text{in},x} = 0.12 \ \unit{m/s}$ axial inlet velocity and $S_\text{in} = 8 \cdot 10^{-5} \ \unit{m}^2$ surface area of the inlet boundary, until the phase front reaches the outlet at $t \approx 0.97~\unit{s}$ 

The injected volume resulting from the numerical simulations is computed in the BC-VOF and BC-VOF-IB frameworks, respectively, as $V_\text{conf}(t) = \int_\Omega \alpha \ dV$ and $V_{\text{IB}}(t) = \int_\Omega \chi^\text{ext} \alpha \ dV$, where $\chi^\text{ext}$ is the external IB mask, that is null on the solid cell set $\mathcal{T}_s$.
Then, the injected volumes are normalized by the total domain volume, $|\Omega|$ and $|\Omega_\text{IB}| = \int_\Omega \chi^\text{ext} \ dV$, obtaining the filling ratios $F_\text{conf} = \frac{V_\text{conf}}{|\Omega|}$ and $F_\text{IB} = \frac{V_\text{IB}}{|\Omega_\text{IB}|}$.

\begin{figure}[h!]
    \centering
    \includegraphics[width=0.8\textwidth]{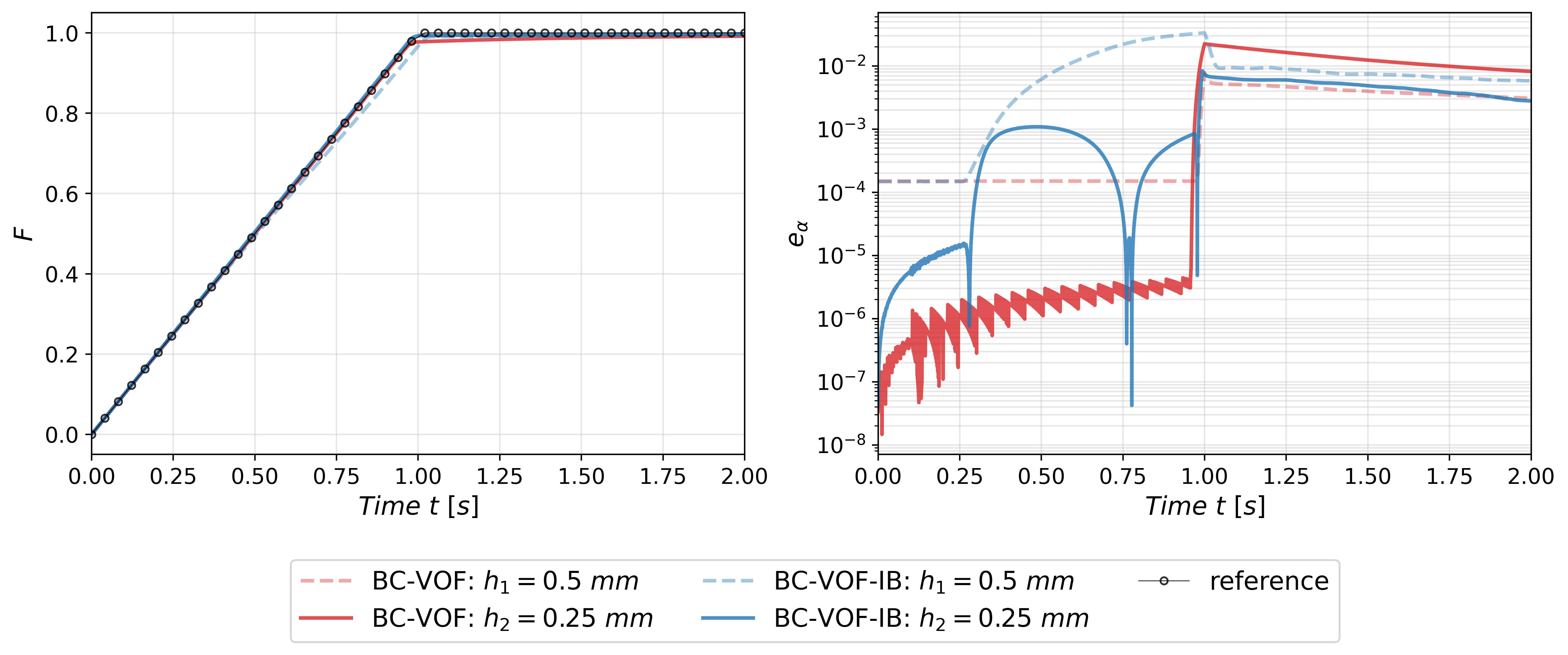}
    \caption{Filling ratio $F$ (left) and magnitude of the phase-mass conservation error $e_\alpha$ (right) of the polymer phase in the dog-bone case, for the BC-VOF and BC-VOF-IB solvers on two grid refinements. The reference filling ratio is $F_\text{ref} = \min(V_\text{ref}/|\Omega|, 1)$. After $t \gtrsim 0.97 \ \unit{s}$ the front has reached the outlet and $1 - F$ measures the residual entrapped air rather than a conservation error.}
    \label{fig:ch4-dogbone-non-newtonian-alpha-errors}
\end{figure}

The phase-mass conservation errors are then computed as
\begin{equation*}
    e_{\alpha,\text{conf}} = \dfrac{|V_\text{conf} - V_\text{ref}|}{|\Omega|} \quad \text{and} \quad e_{\alpha,\text{IB}} = \dfrac{|V_\text{IB} - V_\text{ref}|}{|\Omega_\text{IB}|}. 
\end{equation*}
The comparison of conforming and IB filling ratios with respect to the reference $F_\text{ref} = \frac{V_\text{ref}}{|\Omega|}$, together with phase-mass conservation errors, 
is reported in Figure \ref{fig:ch4-dogbone-non-newtonian-alpha-errors} for two consecutive grid refinements, together with the mass conservation errors.
Before the front reaches the outlet, the conforming solver conserves the injected volume to within $1.5 \cdot 10^{-4}$ on the coarse grid, present from the first time steps, and to within $4 \cdot 10^{-6}$ on the fine grid. The IB error coincides with the conforming one until the front reaches the first restriction, at $t \approx 0.27 \ \unit{s}$; it then grows, up to about $3 \cdot 10^{-2}$ on the coarse grid and $10^{-3}$ on the fine grid, i.e.\ it decreases by a factor of about 30 when the mesh size is halved. This error originates near the immersed walls, where volume fraction can be exchanged with the solid cells, which are excluded by $\chi^\text{ext}$, and where IB cells are partially occupied by the solid even if they are counted as entirely fluid. In all simulations, the volume fraction remains bounded in $[0, 1]$ up to machine precision.

After the front has reached the outlet, material leaves the domain and $V_\text{ref}$ no longer applies: the departure of $F$ from unity then measures the volume of air that remains entrapped in the domain, rather than a conservation error. This residual air fraction is largest for the conforming solver on the fine grid and decreases slowly as bubbles are expelled, consistently with the contact-line behavior, which leads to more air entrapped at the walls on finer grids.

\subsection{Continuous mixing devices}
\label{sec:industrial-app}

The current section is devoted to free-surface numerical simulations obtained with the conforming BC-VOF and the non-conforming BC-VOF-IB solvers, applied to realistic geometries employed in the industrial context of polymer mixing processes.
To the best of our knowledge, the results reported here represent a novel contribution to the literature on polymer mixing simulations, as few works \cite{olofsson2023cfd,Dong2022,Dong2023SPH,Yu2025MPS,Poudyal2019speedratio,Wang2023rubber} currently address such complex configurations under partial filling conditions.

As for the dog-bone case, the simulations are run on the HPC cluster described in Section \ref{sec:dogbone}.

Here, we focus on the family of continuous mixing processes, so called because the material is continuously fed from a feeding hopper with a constant flow rate into the extruder device, consisting of a stationary barrel containing one or more screws that rotate at a predefined angular velocity. The screw rotation allows the material to be mixed and transported towards the outlet. The mechanical forces applied to the polymer by the barrel and the screws cause viscous friction that, together with the heat coming from thermo-regulation of the solid walls, heats and melts the material; the resulting temperature increase lowers its viscosity. The material is continuously discharged at the end of the barrel, and exits the extruder through a die.

We introduce the \textit{drag flow rate}, denoted as $Q_0$, as the flow rate generated solely by the rotation of the screw, that only depends on the screw shape and velocity. It can be computed assuming there is no die at the end of the barrel causing a pressure resistance and in the absence of an imposed inlet flow rate. Different combinations of inlet flow rate $Q_{\text{in}}$, imposed as a parameter of the process, and $Q_0$ lead to either a stationary fully filled or partially filled scenario. In particular, for values of $Q_{\text{in}} > Q_0$ the material is expected to fully fill the extruder domain because the screw cannot move forward enough material as fast as it is fed and a negative pressure gradient is generated from inlet to outlet. On the contrary, when $Q_{\text{in}} < Q_0$ the inlet flow rate is lower than the screw's pumping capacity and the device will be only partially filled.

A strong assumption underlying all simulations that will be presented in the following is the restriction to isothermal processes. This is due to the fact that our two-phase solvers currently do not include the solution of the energy equation. The results obtained on real geometries do not have a quantitative experimental counterpart for comparison, but qualitative evaluation and considerations are carried out based on the outcomes expected experimentally given the working conditions of the process.

\subsubsection{The Single Screw Extruder}
\label{sec:SSE}

A preliminary test of the free-surface flow of a high-viscosity Newtonian material flowing in a simplified SSE geometry is provided.
Since the SSE has only one rotating screw, the process can be simulated with both conforming-mesh and IB approaches.
Specifically, the conforming simulation can be carried out by solving the two-phase Navier--Stokes equations \eqref{eq:VOFbiphaseNSequations} in non-inertial reference frame, solved with respect to the relative velocity of a moving observer that rotates with the screw, i.e.\ using a Single Rotating Frame (SRF) approach.

\begin{figure}[h!]
    \centering
    \includegraphics[width=0.5\textwidth]{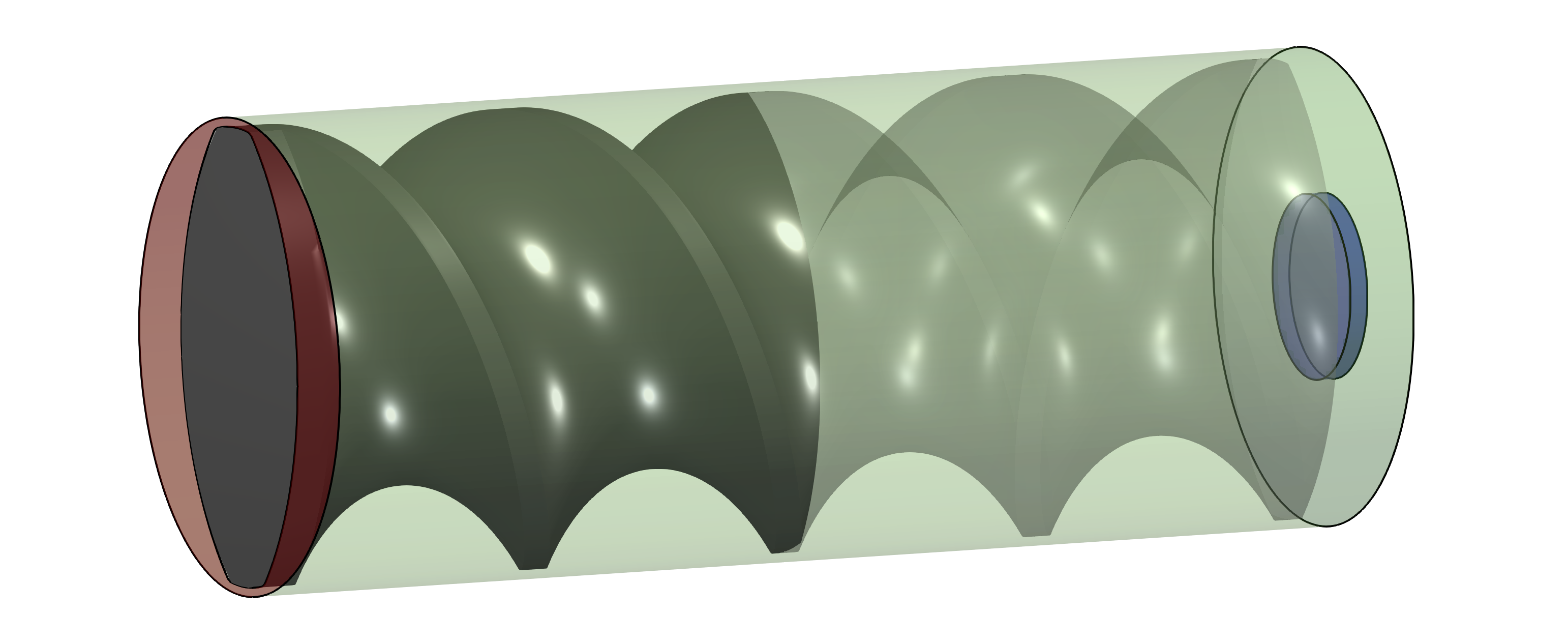}
    \caption{Geometry of the simplified SSE with restricted outlet.}
    \label{fig:ch5-simp-SSE-2L-0.5mmGap-domain}
\end{figure}

The SSE is the simplest continuous mixing device considered in this study, as it consists of a single screw rotating inside a stationary barrel. 
The SSE simplified geometry is displayed in Figure \ref{fig:ch5-simp-SSE-2L-0.5mmGap-domain}. The barrel length and diameter are $L_B = 64 \ \unit{mm}$ and $d_B = 26 \ \unit{mm}$, respectively. The maximum and minimum screw diameters are $d_{S,\max} = 25 \ \unit{mm}$ and $d_{S,\min} = 15 \ \unit{mm}$, resulting in a minimum clearance between screw and barrel equal to $a = \frac{d_B - d_{S,\max}}{2} = 0.5 \ \unit{mm}$. The screw length is $L_S = 62 \ \unit{mm}$ so that a gap of $2 \ \unit{mm}$ is left between the end of the screw and the physical outlet, that does not coincide with the entire barrel section, but it has a reduced diameter $d_\text{out} = 10 \ \unit{mm}$.  

\begin{figure}[h!]
\centering

\begin{subfigure}{\textwidth}
\centering
\includegraphics[width=0.3\textwidth]{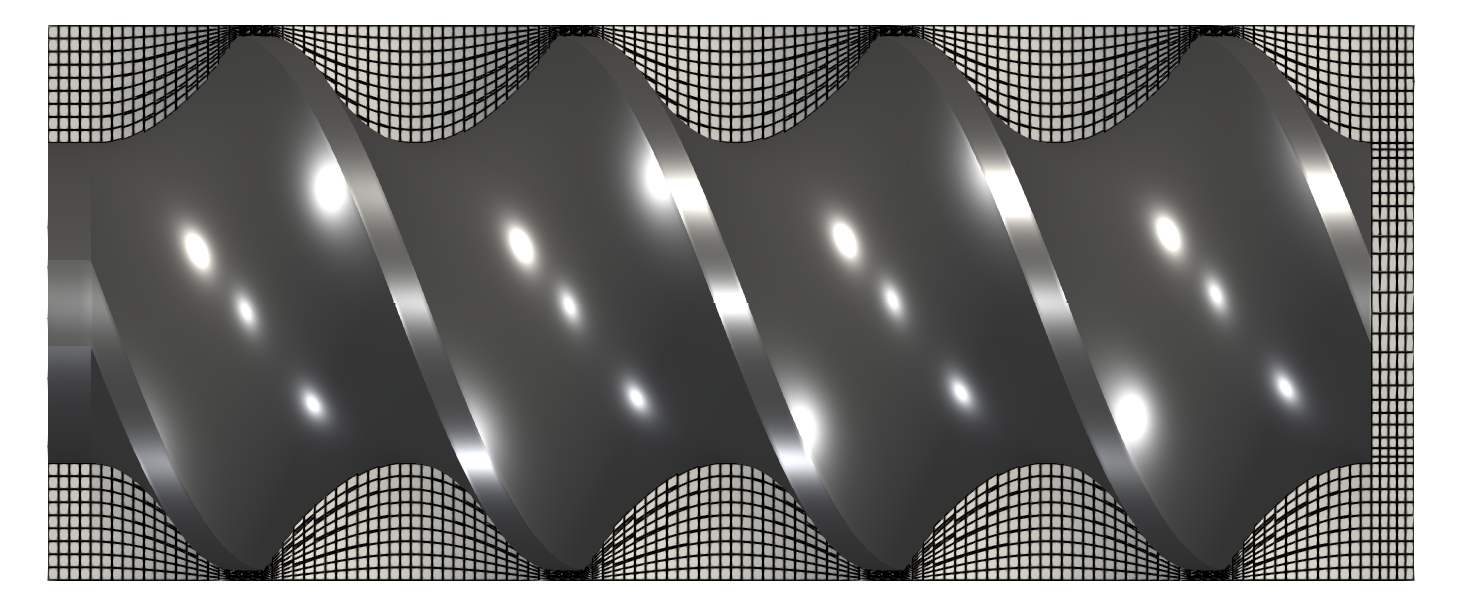}
\includegraphics[width=0.3\textwidth]{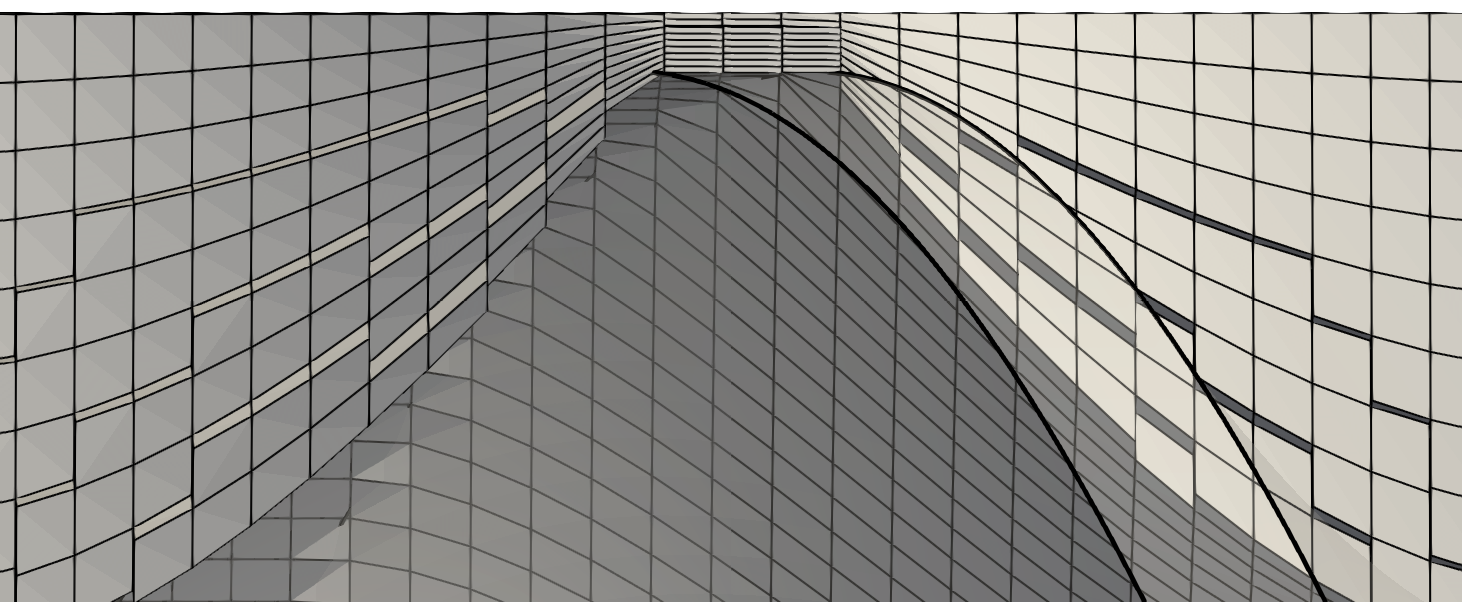}
\caption{Conforming grid.}
\label{fig:ch5-simp-SSE-2L-0.5mmGap-conf-grid}
\end{subfigure}

\begin{subfigure}{\textwidth}
\centering
\includegraphics[width=0.3\textwidth]{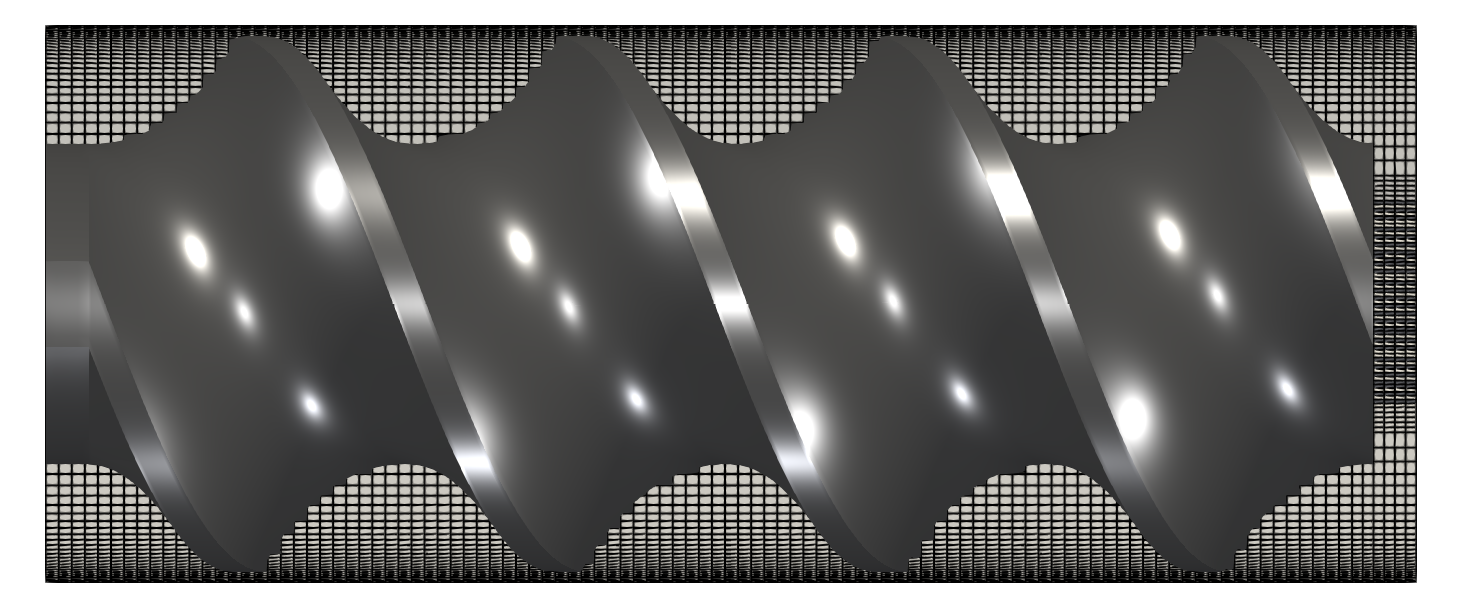}
\includegraphics[width=0.3\textwidth]{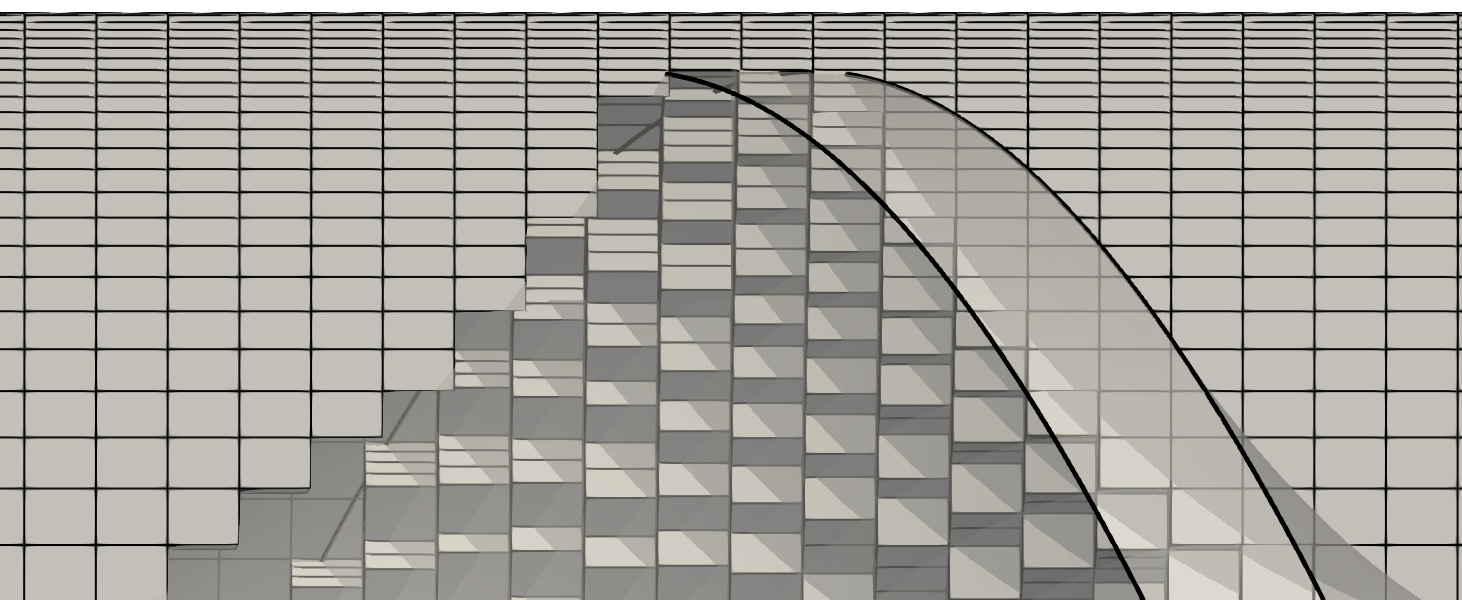}
\caption{IB grid.}
\label{fig:ch5-simp-SSE-2L-0.5mmGap-IB-grid}
\end{subfigure}

\caption{Conforming and non-conforming computational grids for the simplified SSE case.}
\label{fig:ch5-simp-SSE-2L-0.5mmGap-grids}

\end{figure}

\begin{figure}[h!]
\centering
\begin{subfigure}{0.5\textwidth}
\centering
\includegraphics[width=0.8\textwidth]{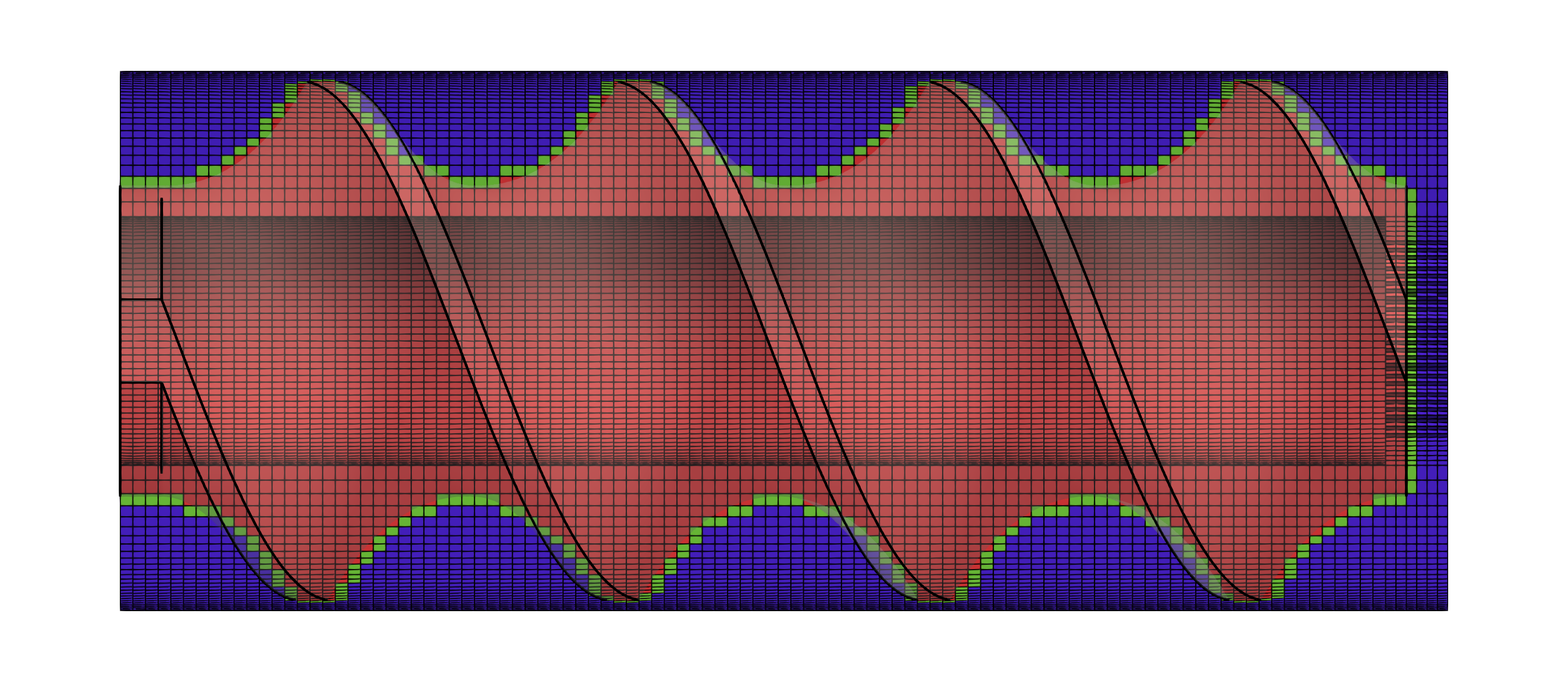} 
\caption{Top view of a slice in the $x-z$ plane.}
\label{fig:ch5-simp-SSE-2L-0.5mmGap-IB-grid-IBmask-xz}
\end{subfigure}
\begin{subfigure}{0.3\textwidth}
\centering
\includegraphics[width=0.7\textwidth]{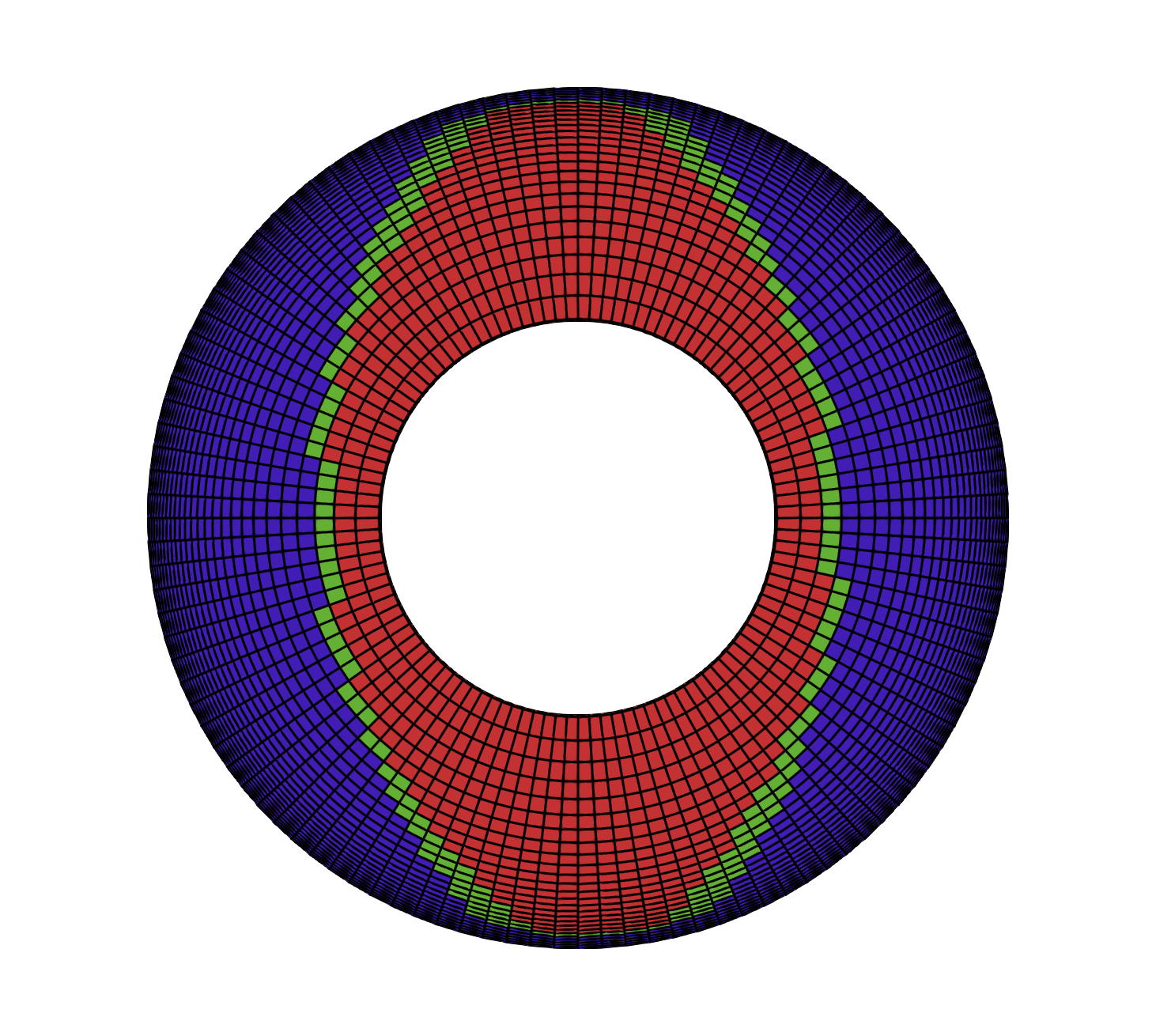}
\caption{Front view of a slice in the $x-y$ plane.}
\label{fig:ch5-simp-SSE-2L-0.5mmGap-IB-grid-IBmask-xy}
\end{subfigure}

\caption{Subdivision of the non-conforming background grid into IB cells (green), fluid cells (blue) and solid cells (red) for the simplified SSE case.}
\label{fig:ch5-simp-SSE-2L-0.5mmGap-IB-grid-IBmask}
\end{figure}

Two computational grids are generated to perform simulations using both conforming and IB approaches. Figure \ref{fig:ch5-simp-SSE-2L-0.5mmGap-grids} reports a qualitative comparison of the two meshes, while mesh quality indicators are collected in Table \ref{tab:ch5-SSE-grids}. 
The conforming grid shown in Figure \ref{fig:ch5-simp-SSE-2L-0.5mmGap-conf-grid} is generated starting from a cylindrical annulus with outer radius $R_{\text{out}} = \frac{d_B}{2}$ and inner radius $R_{\text{in}} = \frac{d_{S,\max}}{2}$. Subsequently, the grid is perturbed to transform the inner circular axial sections into the elliptic screw section.
In contrast, the non-conforming grid is much easier to generate. It is constructed using a cylindrical annulus of outer radius $R_{\text{out}} = \frac{d_B}{2}$ and inner radius $R_{\text{in}} = 7.5 \ \unit{mm} < \frac{d_{S,\min}}{2}$. The triangulated STL surface of the screw is used to generate the IB mask and subdivide the cells into fluid, IB and solid cells (see Figure \ref{fig:ch5-simp-SSE-2L-0.5mmGap-IB-grid-IBmask}). The resulting non-conforming grid of Figure \ref{fig:ch5-simp-SSE-2L-0.5mmGap-IB-grid} is necessarily finer than the body-fitted counterpart. Progressive refinement in the radial direction is adopted to ensure an adequate resolution of the narrow gap between screw and barrel. Because the IB approach does not require the grid to conform to the screw surface, the mesh elements can be more orthogonal and less skewed.

\begin{table}[h!]
    \renewcommand{\arraystretch}{1.3}
    \setlength{\tabcolsep}{7.5pt}
    \small
    \centering
    \begin{NiceTabular}{|c|c|c|c|c|c|c|c|} 
        \noalign{\hrule height 1.2pt}
        \Block{2-1}{Grid} &  \Block{2-1}{$\#$DOFs} & $\#$DOFs across& \multicolumn{2}{c}{Non-orthogonality} & Skewness & \multicolumn{2}{c}{Cell volume $[\unit{m}^3]$} \\
        \cline{4-8}  
        & & screw-barrel gap & $\max(\tau_{ij})$ & $\text{avg}(\tau_{ij})$ & $\max(\psi_{ij})$ & $\displaystyle \min_{K_i \in \mathcal{T}_h}|K_i|$ & $\displaystyle \max_{K_i \in \mathcal{T}_h}|K_i|$\\
        \noalign{\hrule height 0.8pt}
        Conforming               & \num{201146} & 9 & 62.41 & 20.38 & 1.37 & $9.46 \cdot 10^{-12}$ & $1.49 \cdot 10^{-10}$ \\
        \hline
        IB                       &  \num{316736} & 6 & 32.24 & 0.76 & 0.82 & $2.17 \cdot 10^{-11}$ & $1.56 \cdot 10^{-10}$ \\
        \noalign{\hrule height 1.2pt}
    \end{NiceTabular}    
    \caption{Mesh size and key quality indicators for the simplified SSE test case.}
    \label{tab:ch5-SSE-grids}
\end{table}

In the simulations we adopt for fluid 1 the physical properties of a highly viscous Newtonian material, with viscosity $\mu_1 = 1232 \ \unit{Pa.s}$ and density $\rho_1 = 880 \ \unit{kg/m}^3$, while fluid 2 is assigned values of viscosity $\mu_2 = 2.6 \cdot 10^{-4} \ \unit{Pa.s}$ and density $\rho_2 = 0.73 \ \unit{kg/m}^3$.
This choice is not intended to represent a realistic polymeric material. Rather, the objective of the present test is to assess the robustness of the BC-VOF and BC-VOF-IB solvers in handling a two-phase flow with strong dynamic and kinematic viscosity contrasts, $\frac{\mu_1}{\mu_2} \sim 5 \cdot 10^6$ and $\frac{\nu_1}{\nu_2} \sim 4 \cdot 10^3$, respectively, within a domain featuring complex moving boundaries.

A preliminary single-phase simulation is run to compute the \textit{drag flow rate} $Q_0$ associated with a chosen screw velocity of 60 rpm. 
Afterwards, a free-surface simulation is set up reproducing a total filling scenario, by imposing a constant inlet flow rate $Q_\text{in} > Q_0$. The SSE device is initially full of air and the highly viscous material is injected from the inlet patch (the red surface of Figure \ref{fig:ch5-simp-SSE-2L-0.5mmGap-domain}) with a uniform axial velocity $\bar u_{\text{in},z}$ given by $\bar u_{\text{in},z} = \frac{Q_{\text{in}}}{\rho_1 |S_\text{in}|}$, with $|S_\text{in}|$ denoting the surface area of the inflow boundary.
Navier slip conditions are imposed at barrel walls, allowing the viscous material to advance more freely along the solid boundary.

The same simulation settings, concerning volume fraction, pressure and velocity solvers, as well as number of iterations and adaptive time-stepping parameters, are used for both the conforming and IB cases and are reported Table \ref{tab:ch5-SSE-setup}.

\begin{table}[h!]
    \renewcommand{\arraystretch}{1.3}
    \setlength{\tabcolsep}{5pt}  
    \small
    \centering
    \begin{NiceTabular}{|c|cc|}
        \noalign{\hrule height 1.2pt}
        Parameter & BC-VOF & BC-VOF-IB \\
        \noalign{\hrule height 0.8pt}
        $c_\alpha$ & \multicolumn{2}{c|}{1}  \\
        \hline
        \texttt{MULESCorr} & \multicolumn{2}{c|}{no} \\
        \hline
        $N_\text{sc}$ & \multicolumn{2}{c|}{10} \\        
        \hline
        $N_{\alpha,\text{corr}}$ & \multicolumn{2}{c|}{1} \\
        \hline
        $N_\text{inner}$ & \multicolumn{2}{c|}{5} \\
        \hline
        $N_\text{outer}$ & \multicolumn{2}{c|}{5} \\
        \noalign{\hrule height 0.8pt}
        $p_{rgh}$ solver & \multicolumn{2}{c|}{PCG (GAMG) $tol = 10^{-8}$} \\
        \hline
        $\bm u$ solver & \multicolumn{2}{c|}{GMRES (LUSGS) $tol = 10^{-5}$} \\
        \noalign{\hrule height 0.8pt} 
        $N_T$ & \num{4996} & \num{5034} \\
        \hline
        $\text{Co}_{h,\max}$ & \multicolumn{2}{c|}{50} \\
        \hline
        $\text{Co}_{\alpha,h,\max}$ & \multicolumn{2}{c|}{2} \\
        \hline
        $\Delta t^0 \ [\unit{s}]$ & \multicolumn{2}{c|}{$10^{-5}$} \\
        \hline
        $\Delta t_{\max} \ [\unit{s}]$ & \multicolumn{2}{c|}{0.002} \\
        \hline
        $T_f \ [\unit{s}]$ & \multicolumn{2}{c|}{10} \\
        \noalign{\hrule height 1.2pt}
    \end{NiceTabular}
    \caption{Numerical settings for the simplified SSE test case.}
    \label{tab:ch5-SSE-setup}
\end{table}

\begin{figure}[h!]
\centering

\begin{subfigure}{\textwidth}
\centering
\includegraphics[width=0.4\textwidth]{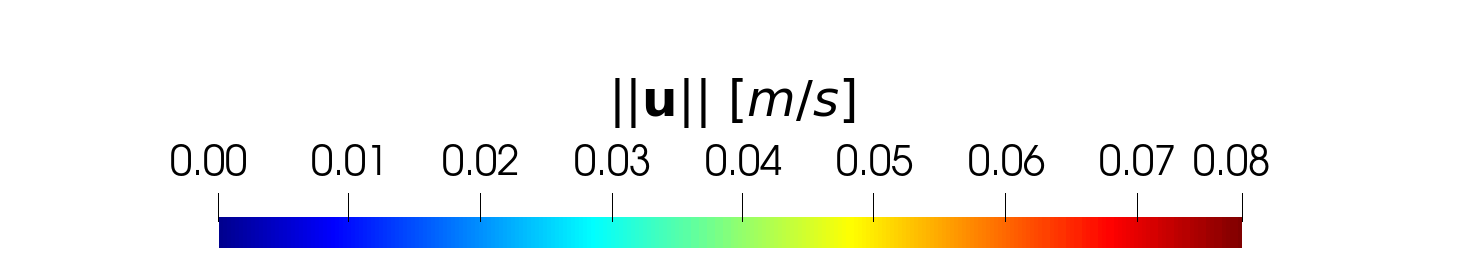} \\
\vspace{0.2cm}
\includegraphics[width=0.24\textwidth]{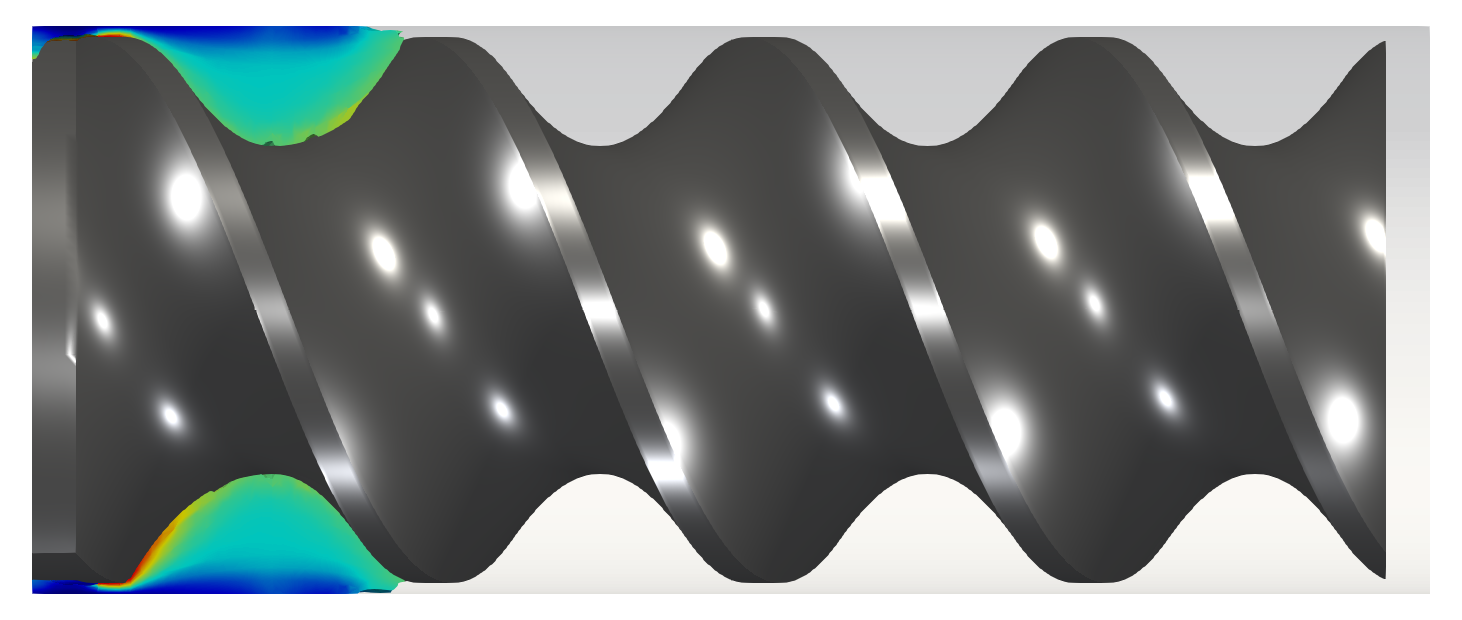}
\includegraphics[width=0.24\textwidth]{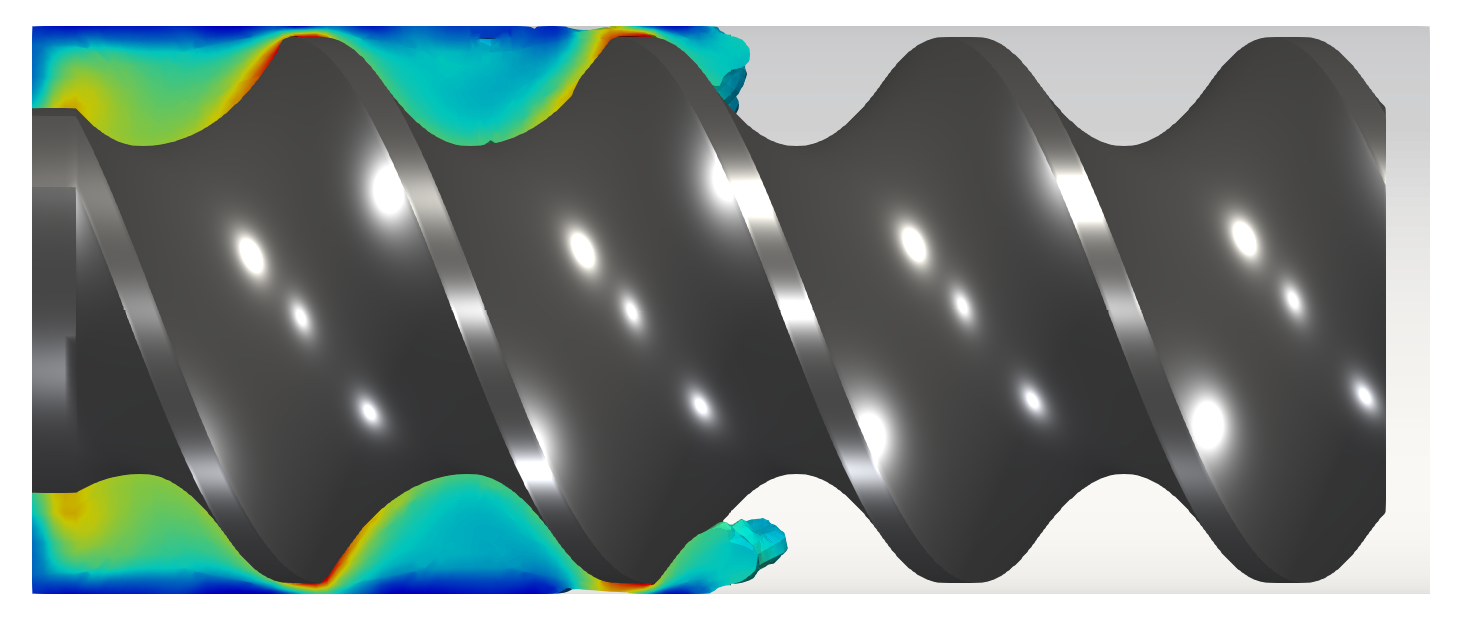}
\includegraphics[width=0.24\textwidth]{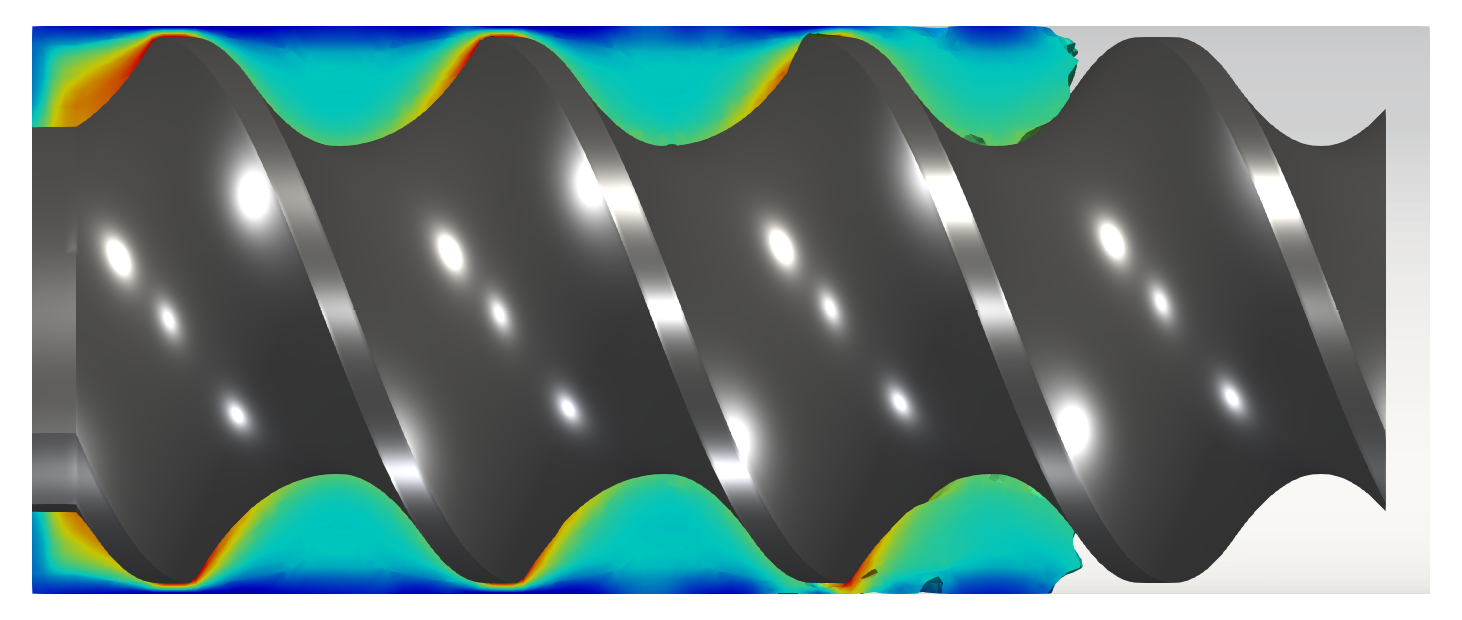}
\includegraphics[width=0.24\textwidth]{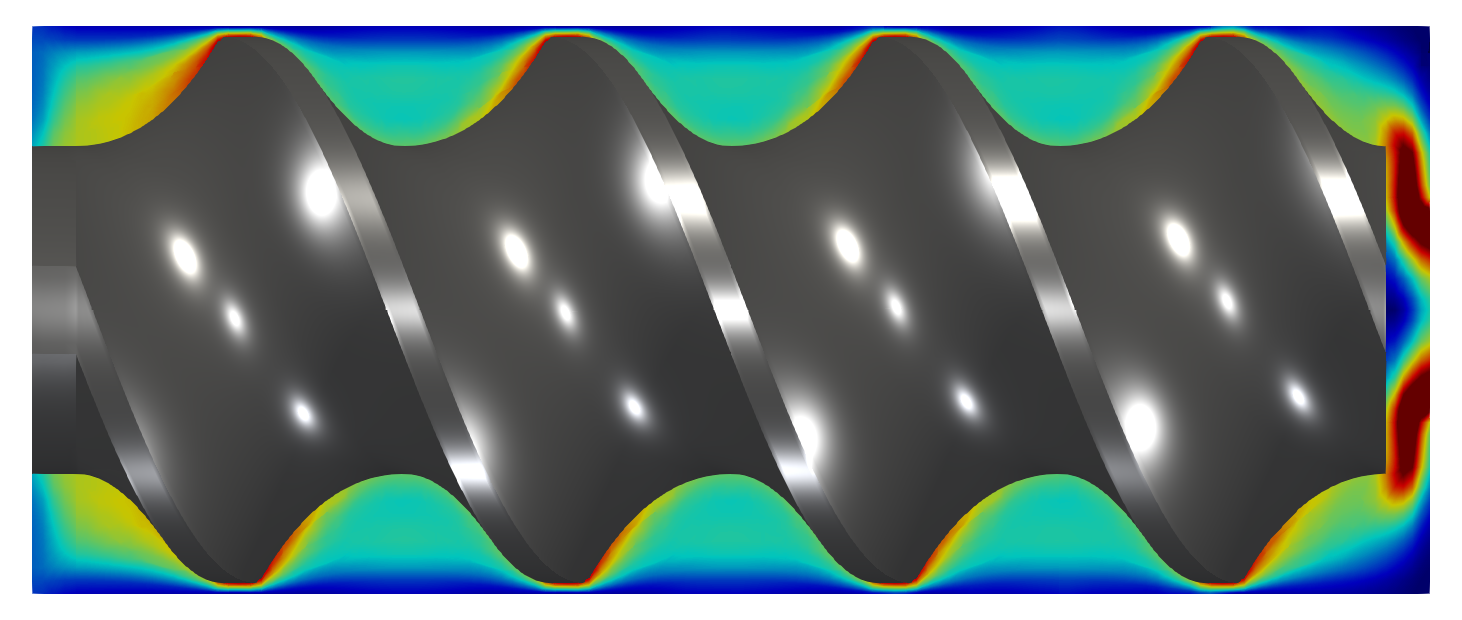}
\caption{BC-VOF solver using SRF tools.}
\label{fig:ch-5simp-SSE-2L-0.5mmGap-60RPM-2Q0-conf-high-newtonian-u}
\end{subfigure}
\hfill
\begin{subfigure}{\textwidth}
\centering
\includegraphics[width=0.24\textwidth]{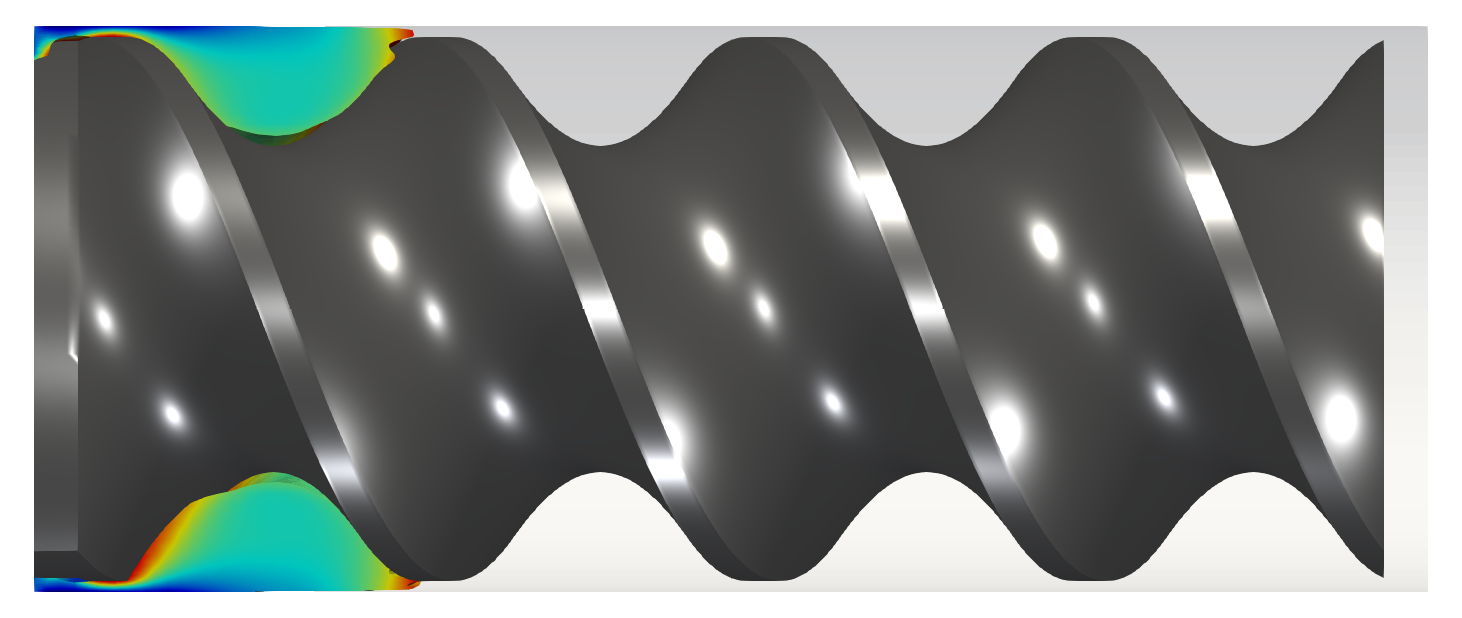}
\includegraphics[width=0.24\textwidth]{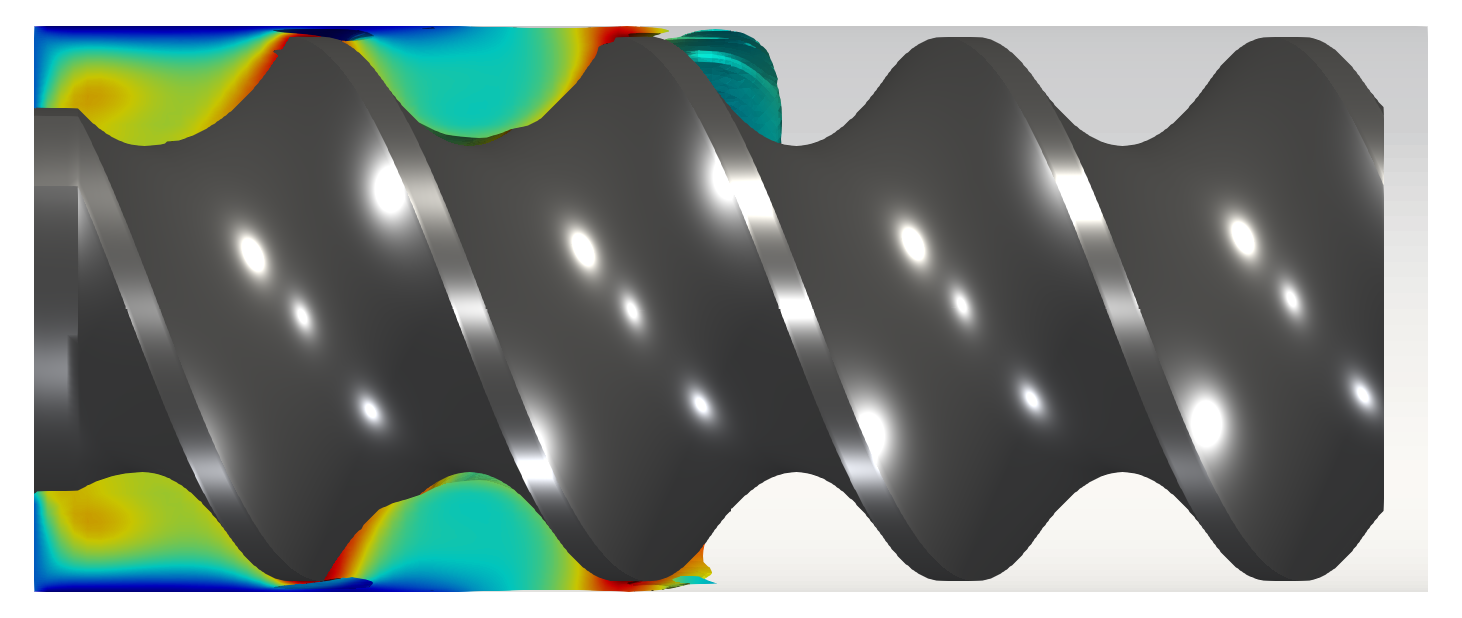}
\includegraphics[width=0.24\textwidth]{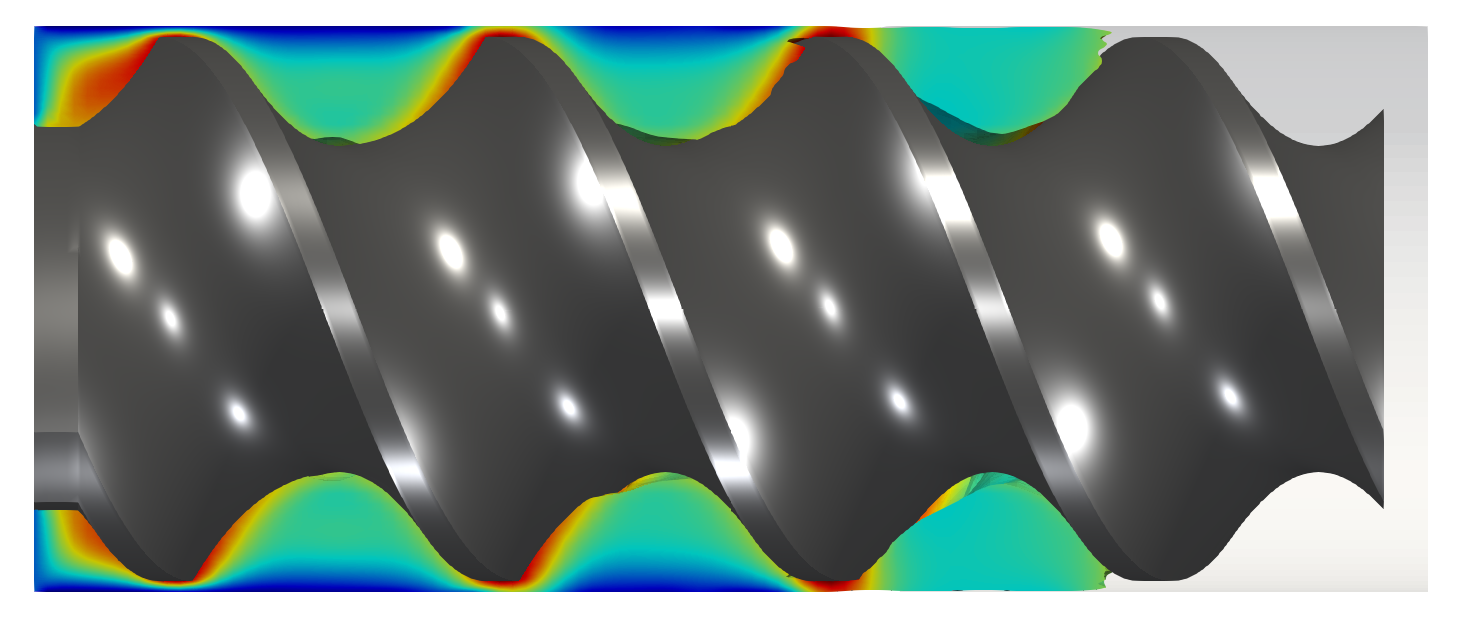}
\includegraphics[width=0.24\textwidth]{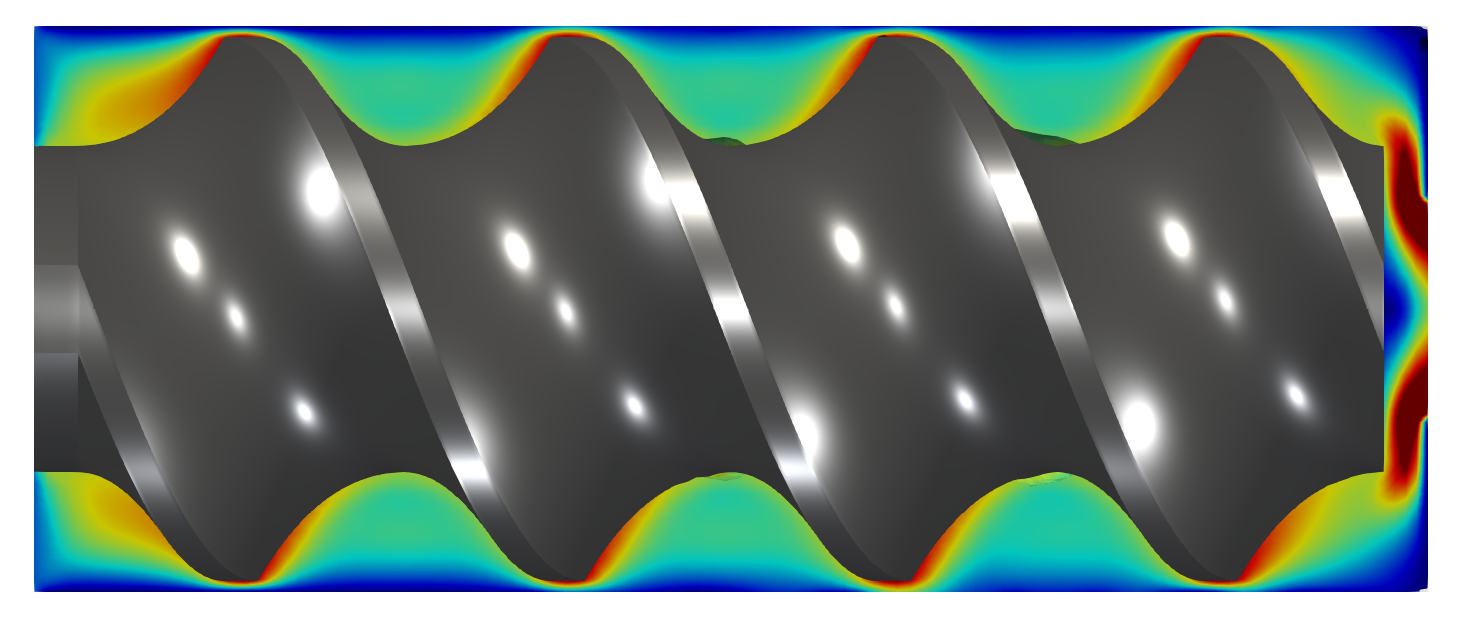}
\caption{BC-VOF-IB solver using SRF tools.}
\label{fig:ch5-simp-SSE-2L-0.5mmGap-60RPM-2Q0-IB-high-newtonian-u}
\end{subfigure}

\caption{Velocity magnitude of fluid 1 (highly viscous Newtonian material) obtained with BC-VOF and BC-VOF-IB solvers reported on a slice along the $x-z$ plane (view from above) at time instants $t = 0.8 \ \unit{s}, \ 1.6 \ \unit{s}, \ 2.4 \ \unit{s}, \ 5 \ \unit{s}$.}
\label{fig:ch5-simp-SSE-2L-0.5mmGap-60RPM-2Q0-high-newtonian-u}

\end{figure}

\begin{figure}[h!]
\centering

\begin{subfigure}{\textwidth}
\centering
\includegraphics[width=0.4\textwidth]{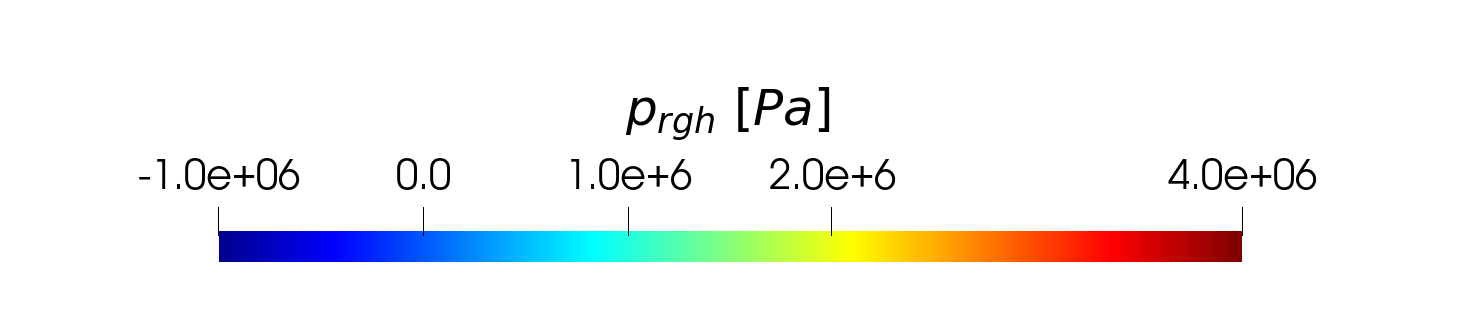} \\
\vspace{0.2cm}
\includegraphics[width=0.24\textwidth]{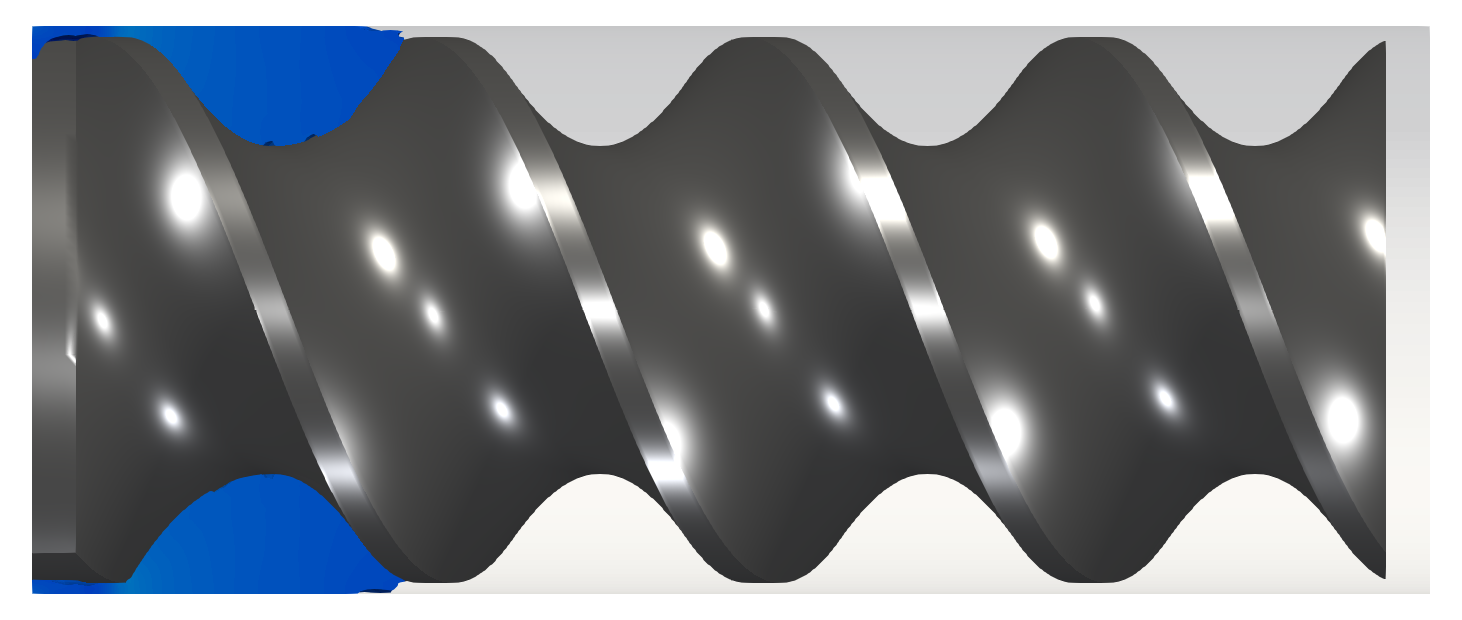}
\includegraphics[width=0.24\textwidth]{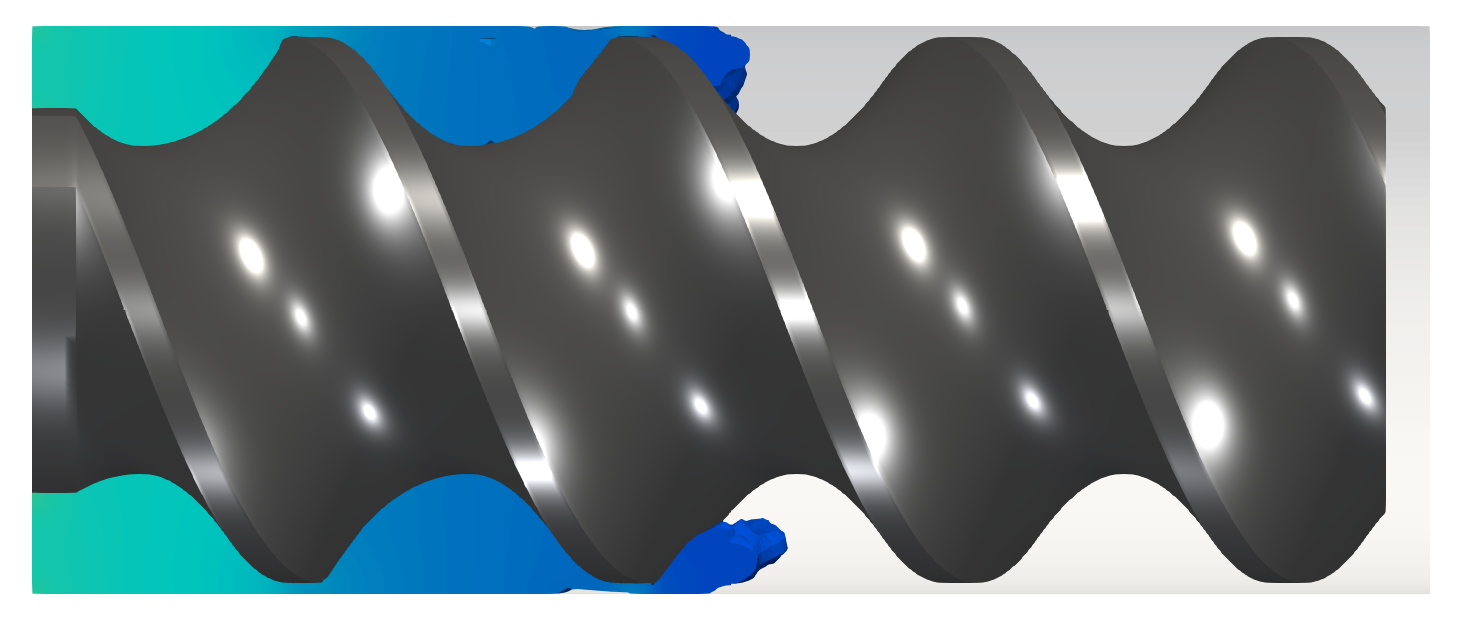}
\includegraphics[width=0.24\textwidth]{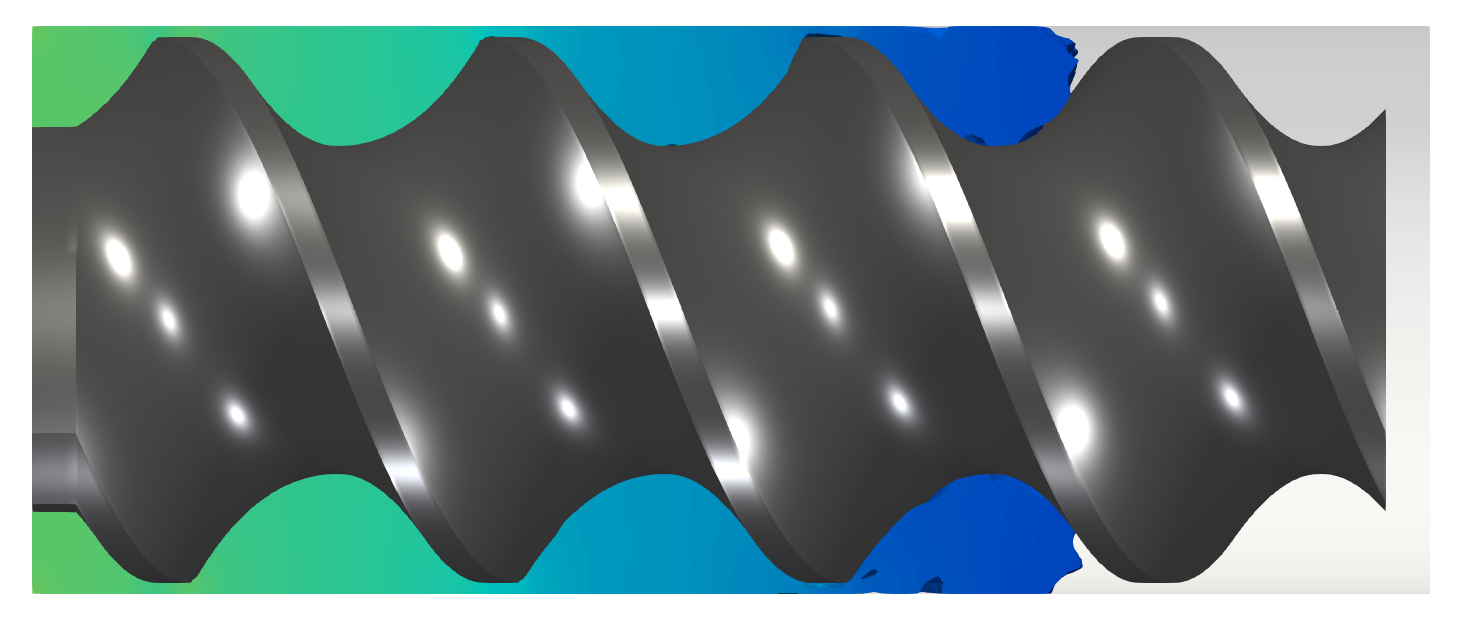}
\includegraphics[width=0.24\textwidth]{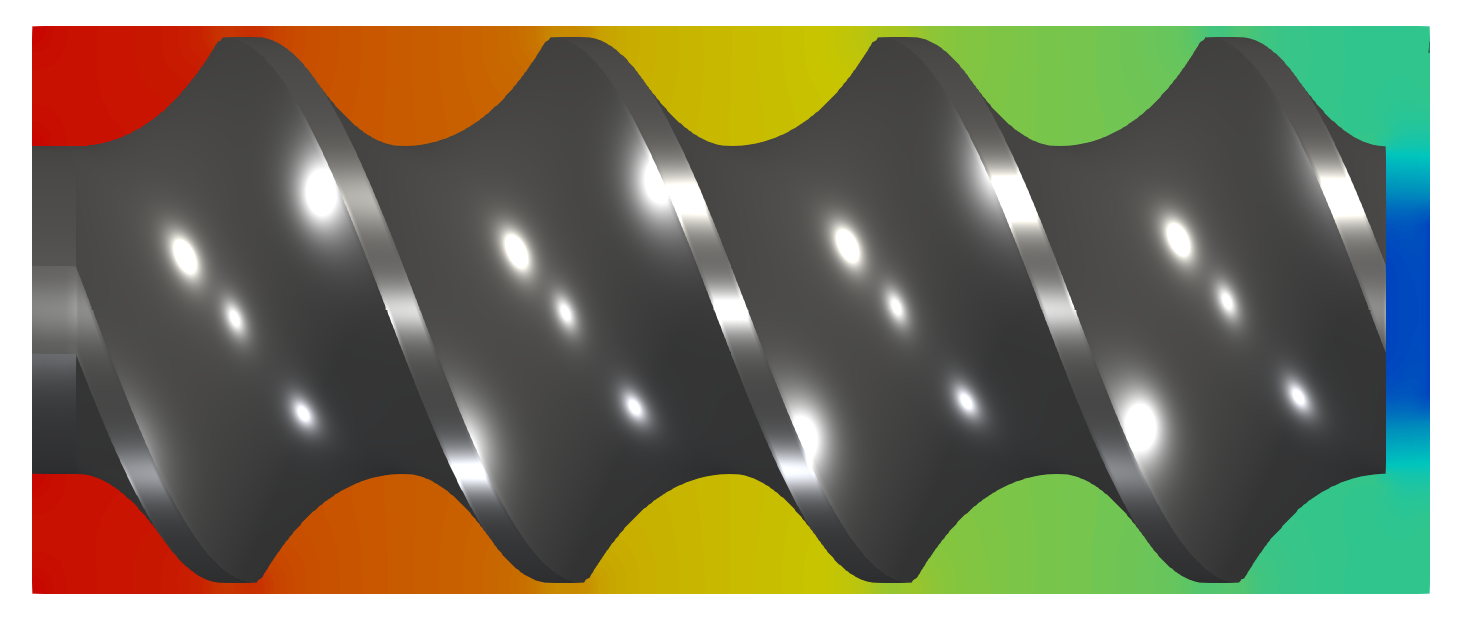}
\caption{BC-VOF solver using SRF tools.}
\label{fig:ch5-simp-SSE-2L-0.5mmGap-60RPM-2Q0-conf-high-newtonian-p}
\end{subfigure}
\hfill
\begin{subfigure}{\textwidth}
\centering
\includegraphics[width=0.24\textwidth]{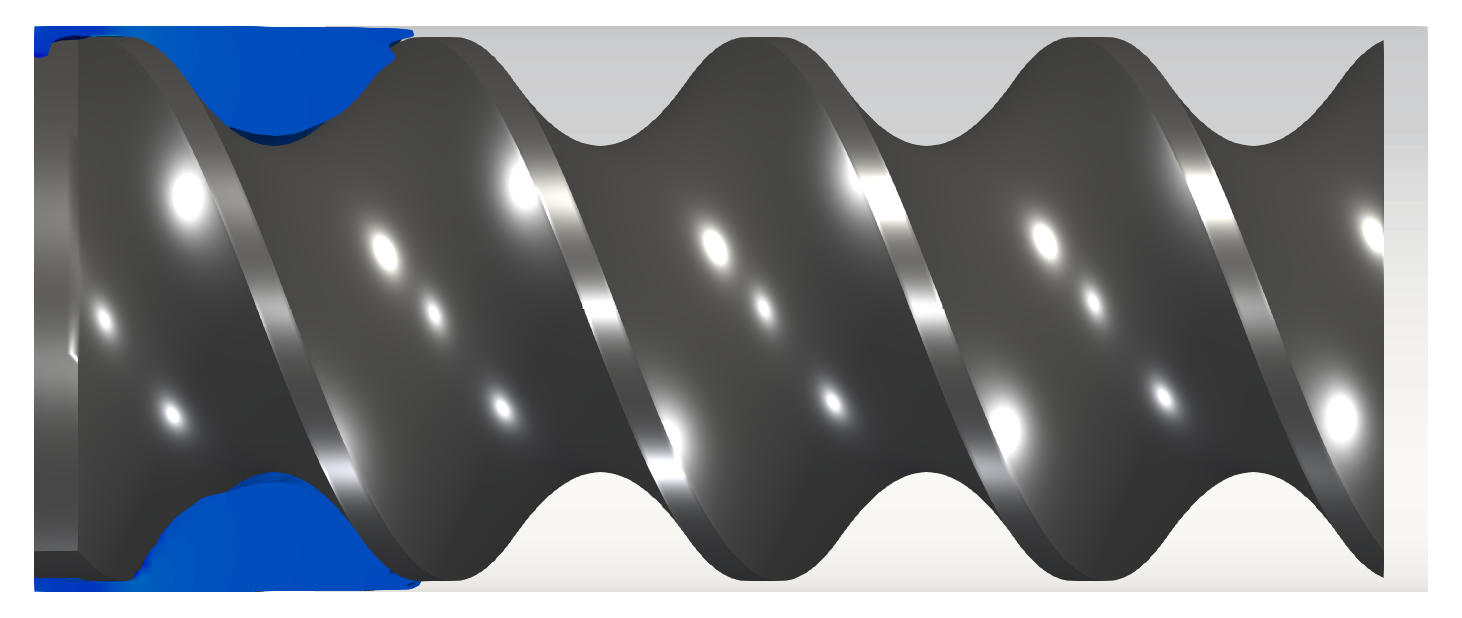}
\includegraphics[width=0.24\textwidth]{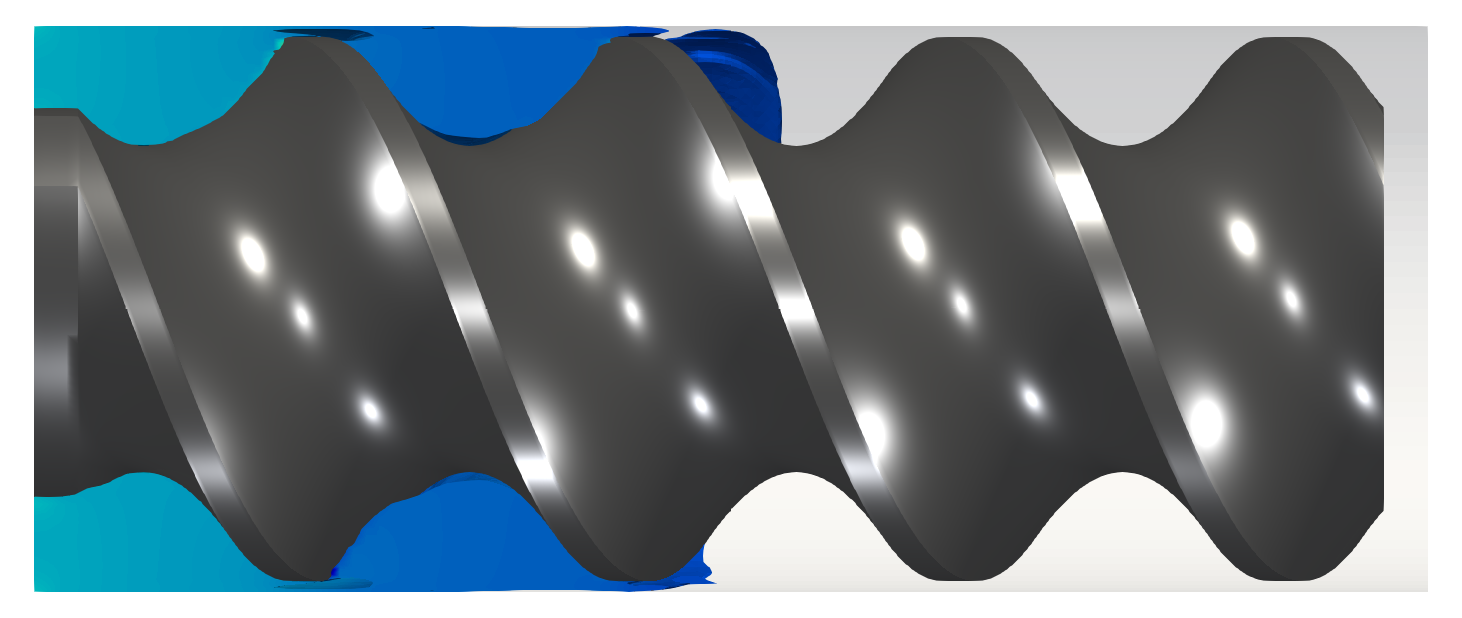}
\includegraphics[width=0.24\textwidth]{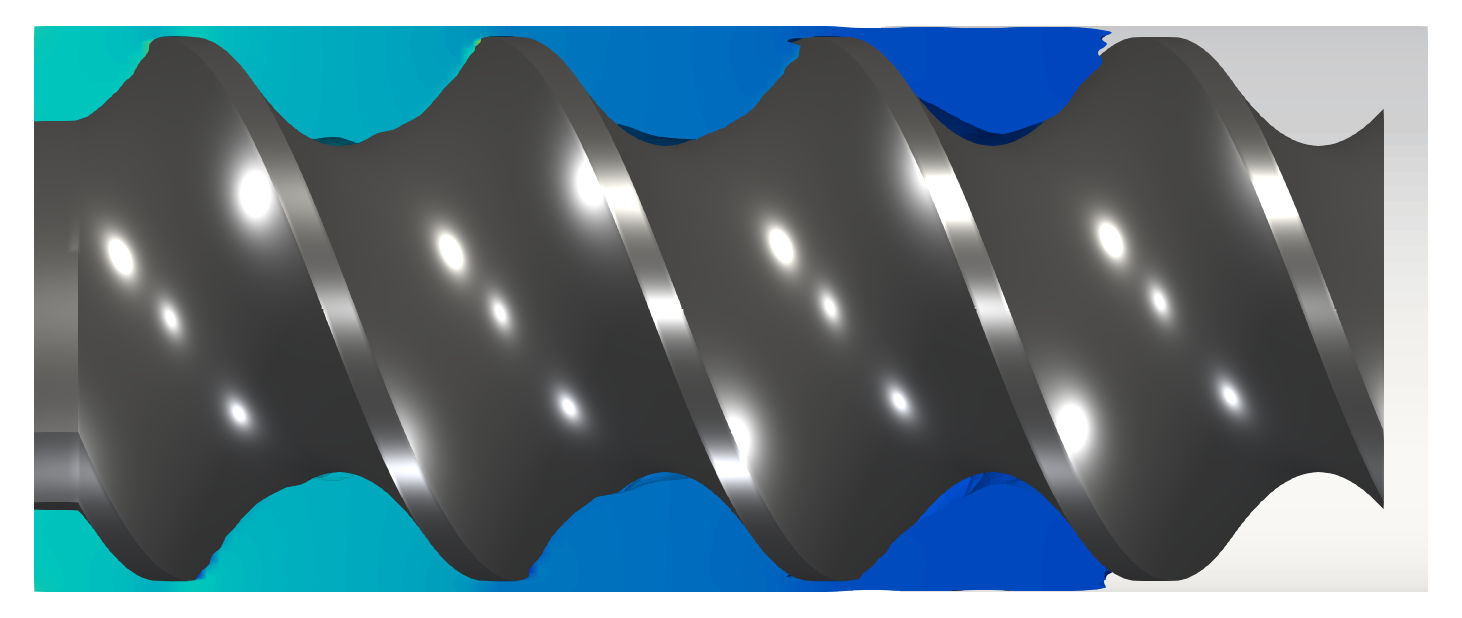}
\includegraphics[width=0.24\textwidth]{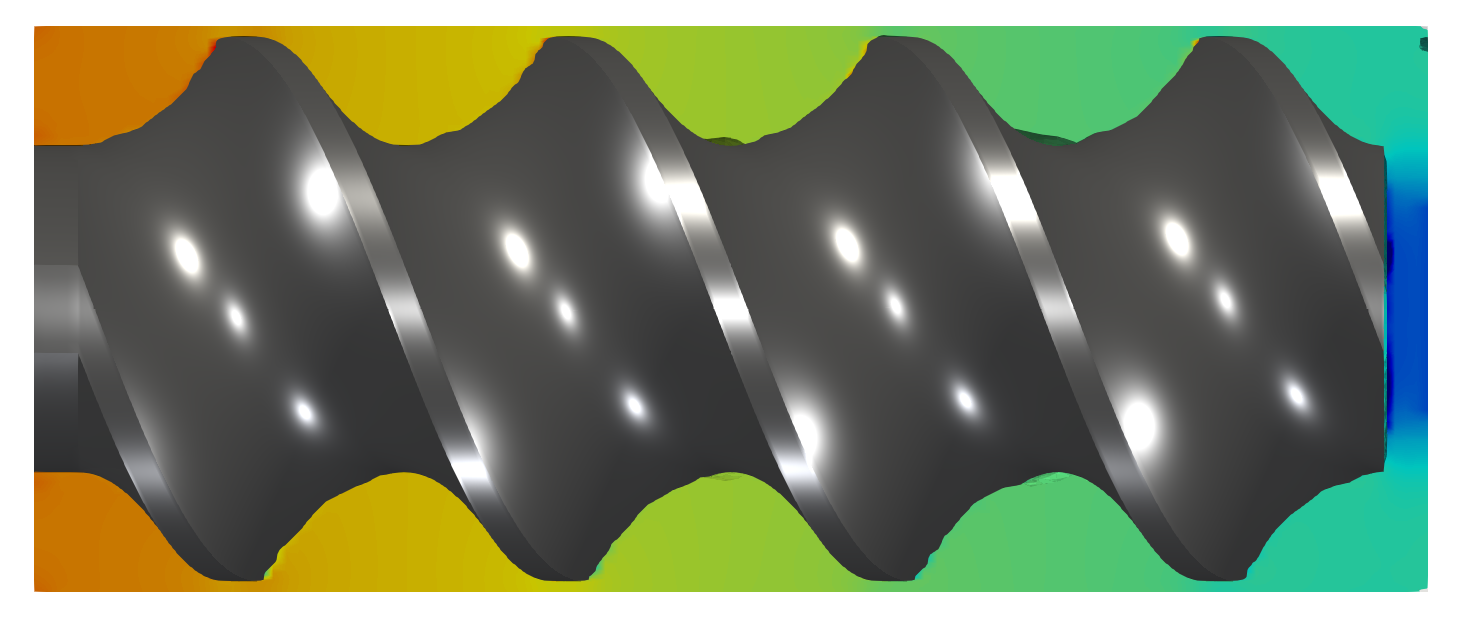}
\caption{BC-VOF-IB solver using SRF tools.}
\label{fig:ch5-simp-SSE-2L-0.5mmGap-60RPM-2Q0-IB-high-newtonian-p}
\end{subfigure}

\caption{Pressure $p_{rgh}$ of fluid 1 (highly viscous Newtonian material) obtained with BC-VOF and BC-VOF-IB solvers reported on a slice along the $x-z$ plane (view from above) at time instants $t = 0.8 \ \unit{s}, \ 1.6 \ \unit{s}, \ 2.4 \ \unit{s}, \ 5 \ \unit{s}$.}
\label{fig:ch5-simp-SSE-2L-0.5mmGap-60RPM-2Q0-high-newtonian-p}

\end{figure}

\begin{figure}[h!]
\centering

\begin{subfigure}{\textwidth}
\centering
\includegraphics[width=0.24\textwidth]{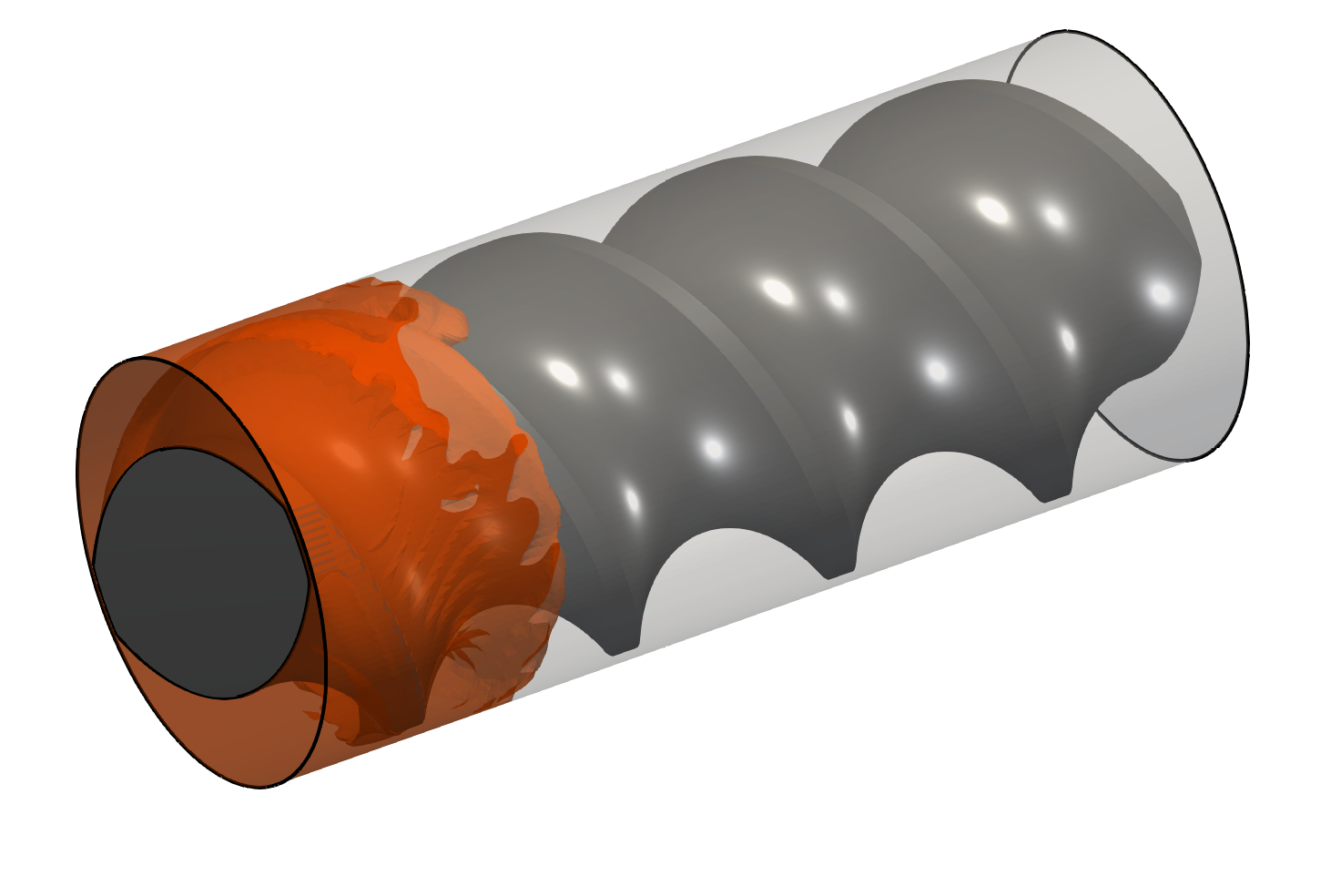}
\includegraphics[width=0.24\textwidth]{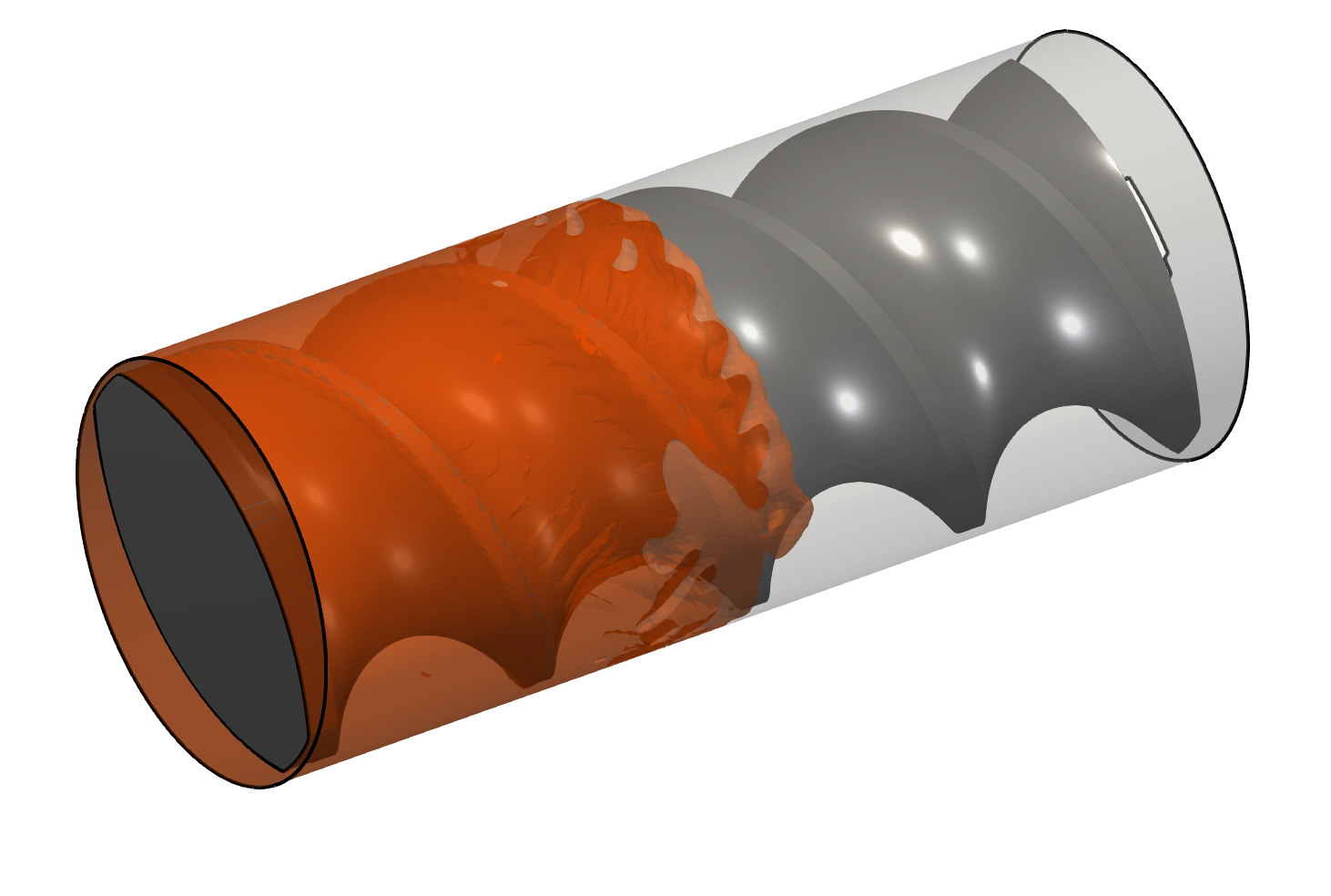}
\includegraphics[width=0.24\textwidth]{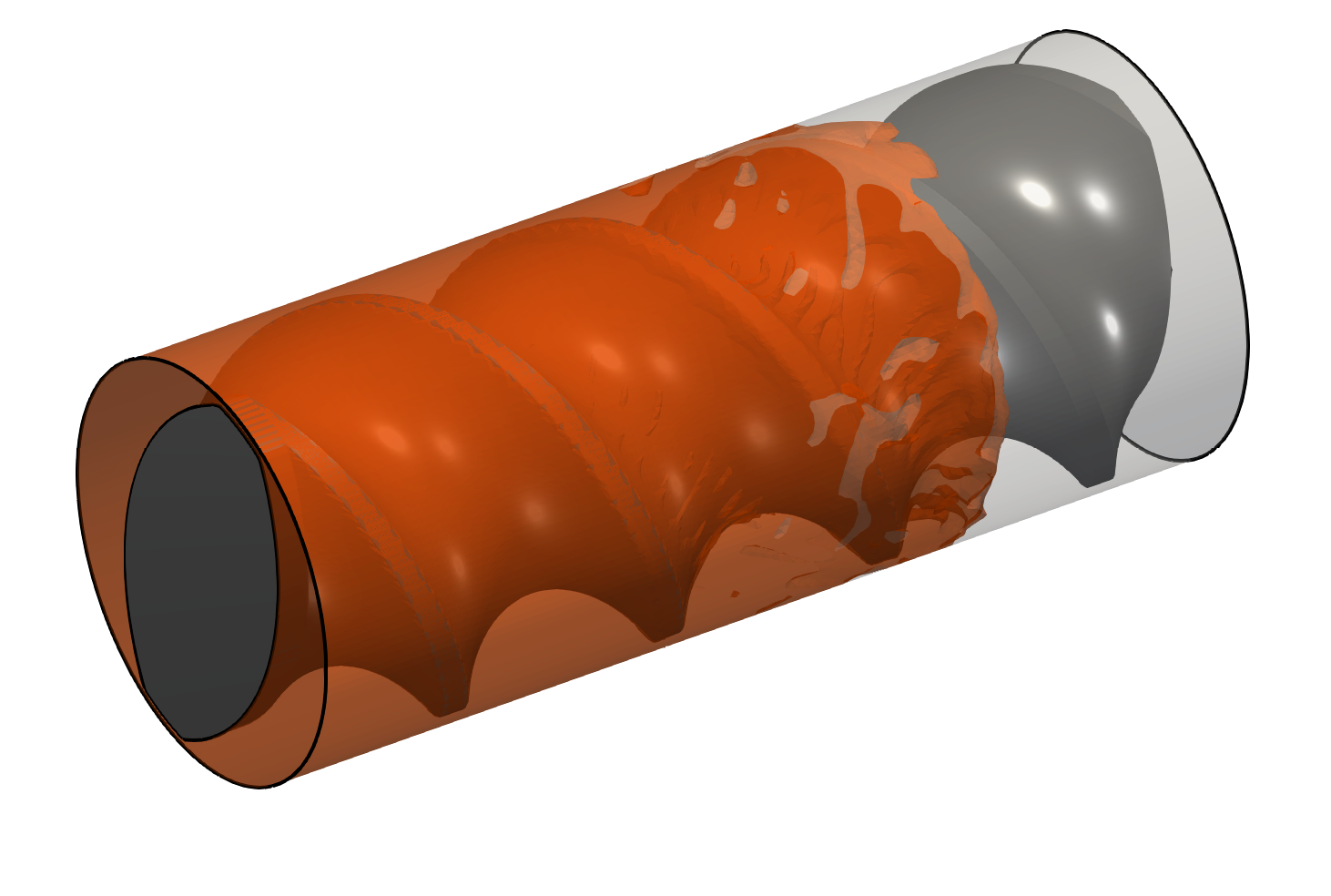}
\includegraphics[width=0.25\textwidth]{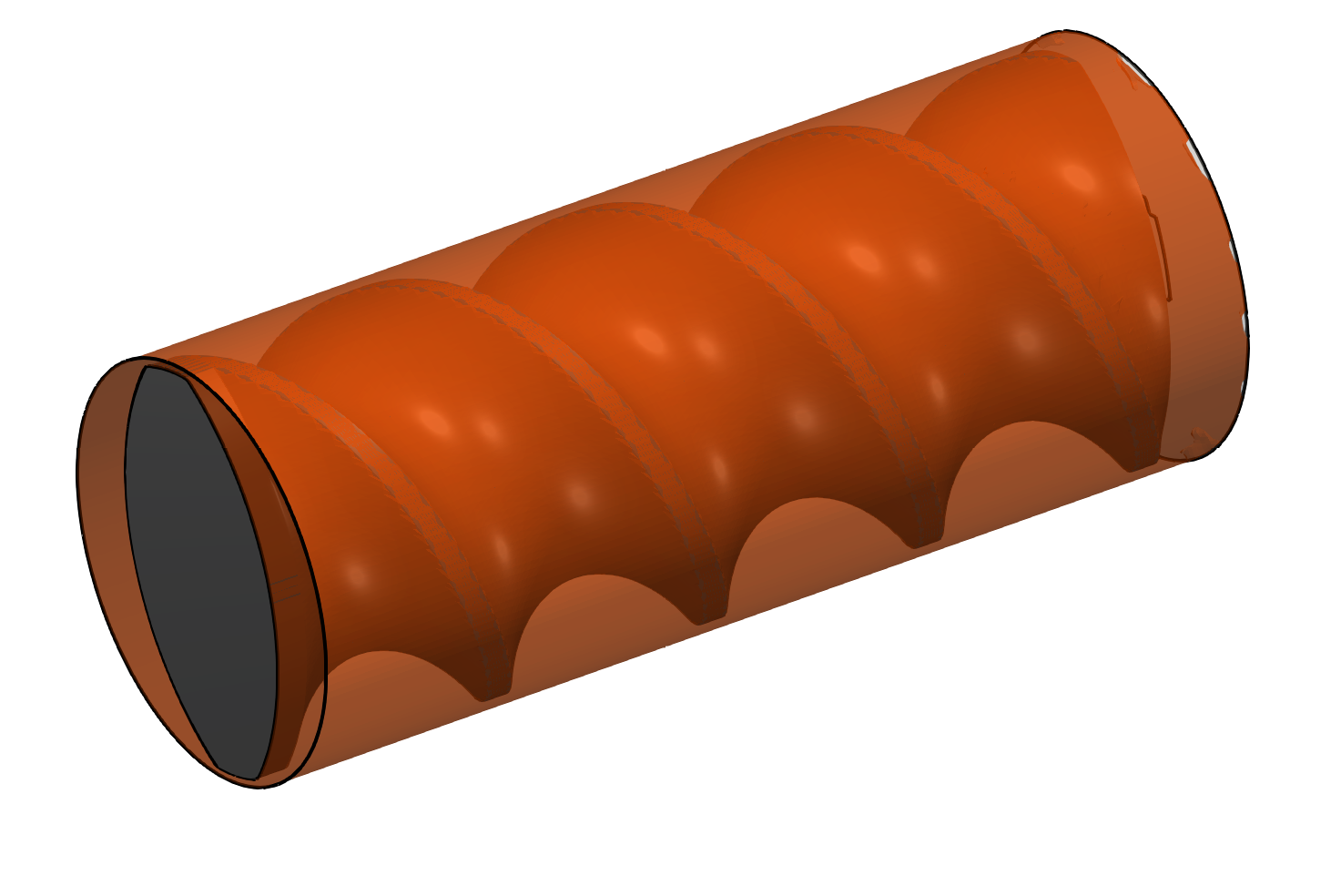}
\caption{BC-VOF solver using SRF tools.}
\label{fig:ch5-simp-SSE-2L-0.5mmGap-60RPM-2Q0-conf-high-newtonian-alpha}
\end{subfigure}
\hfill
\begin{subfigure}{\textwidth}
\centering
\includegraphics[width=0.24\textwidth]{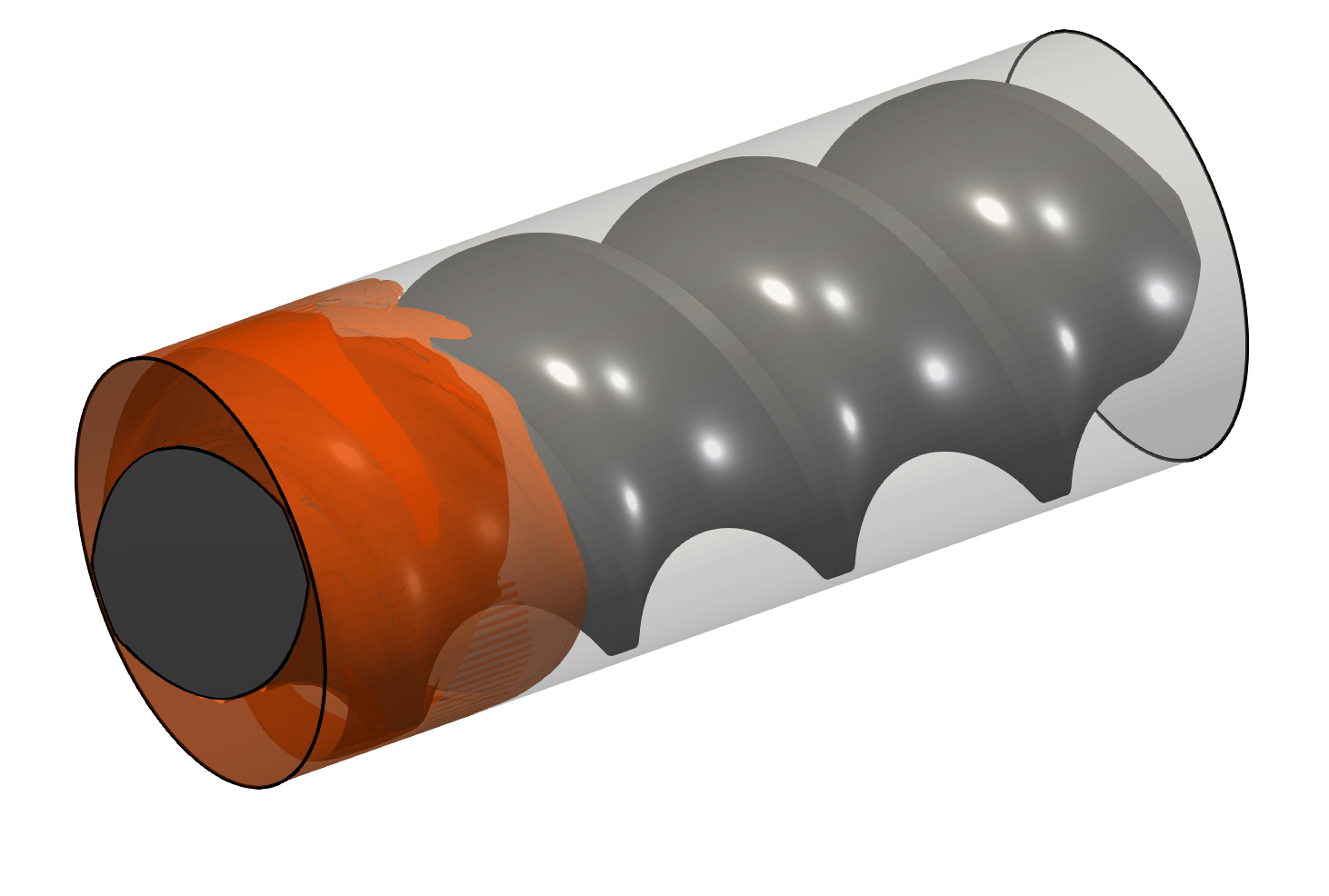}
\includegraphics[width=0.24\textwidth]{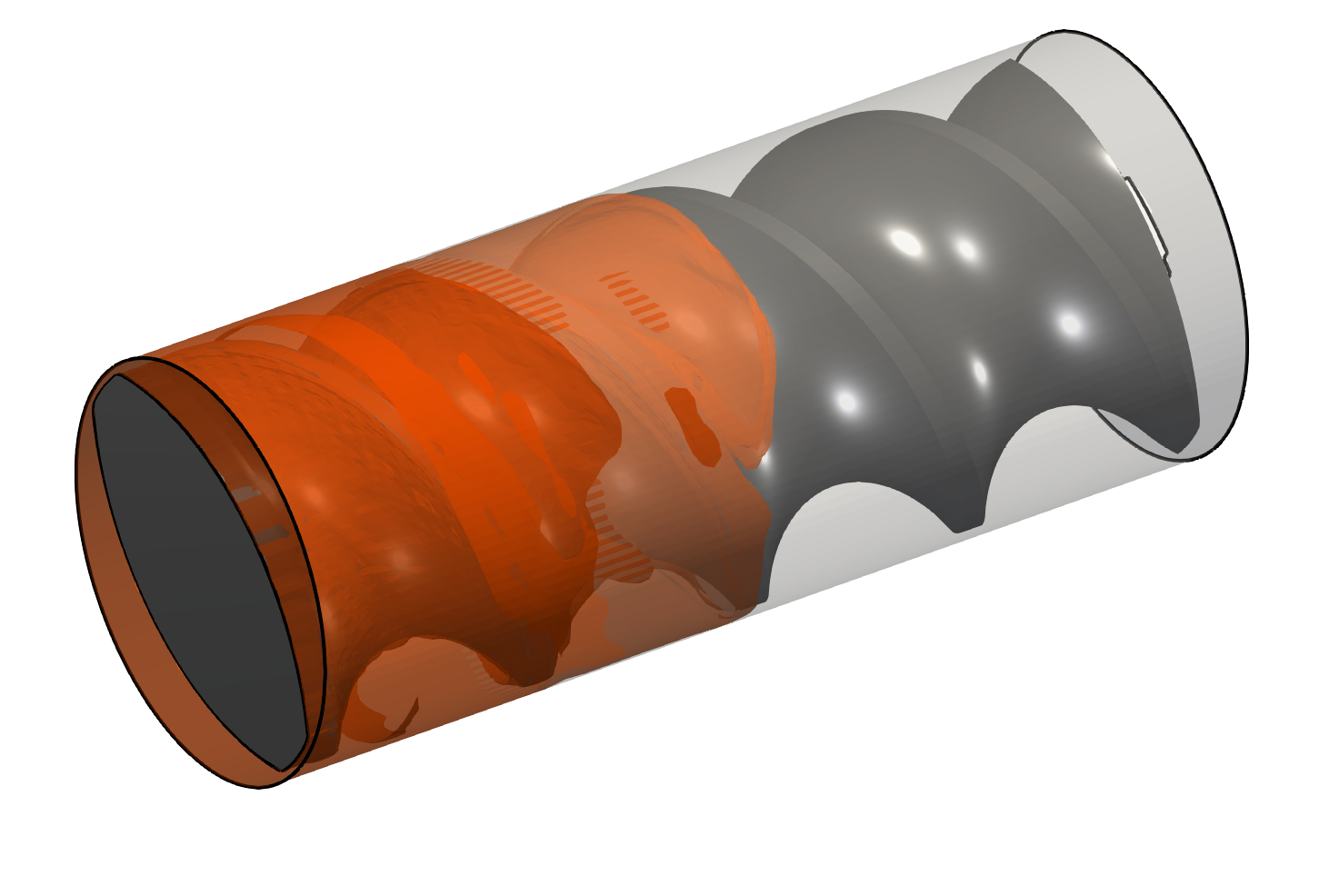}
\includegraphics[width=0.24\textwidth]{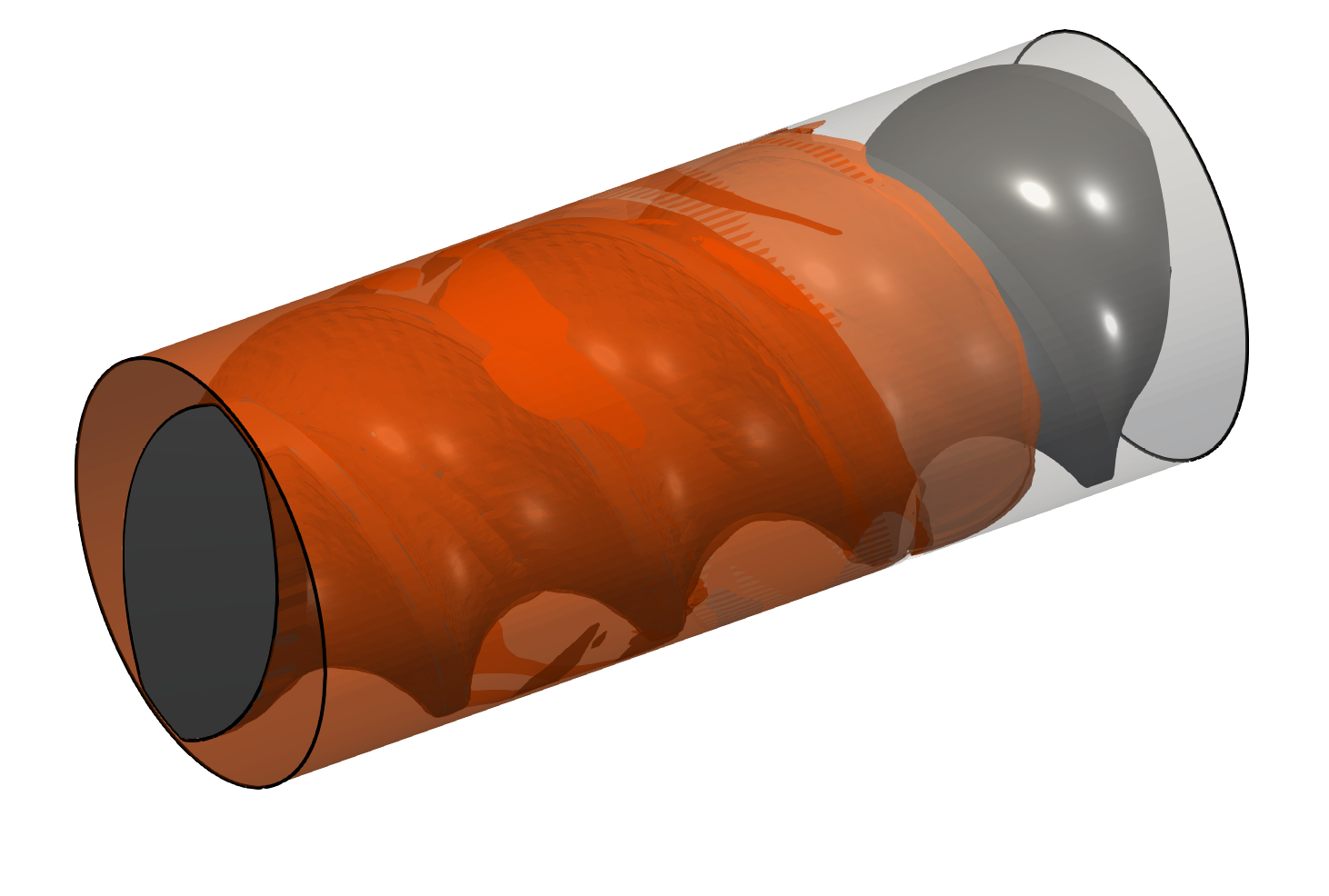}
\includegraphics[width=0.24\textwidth]{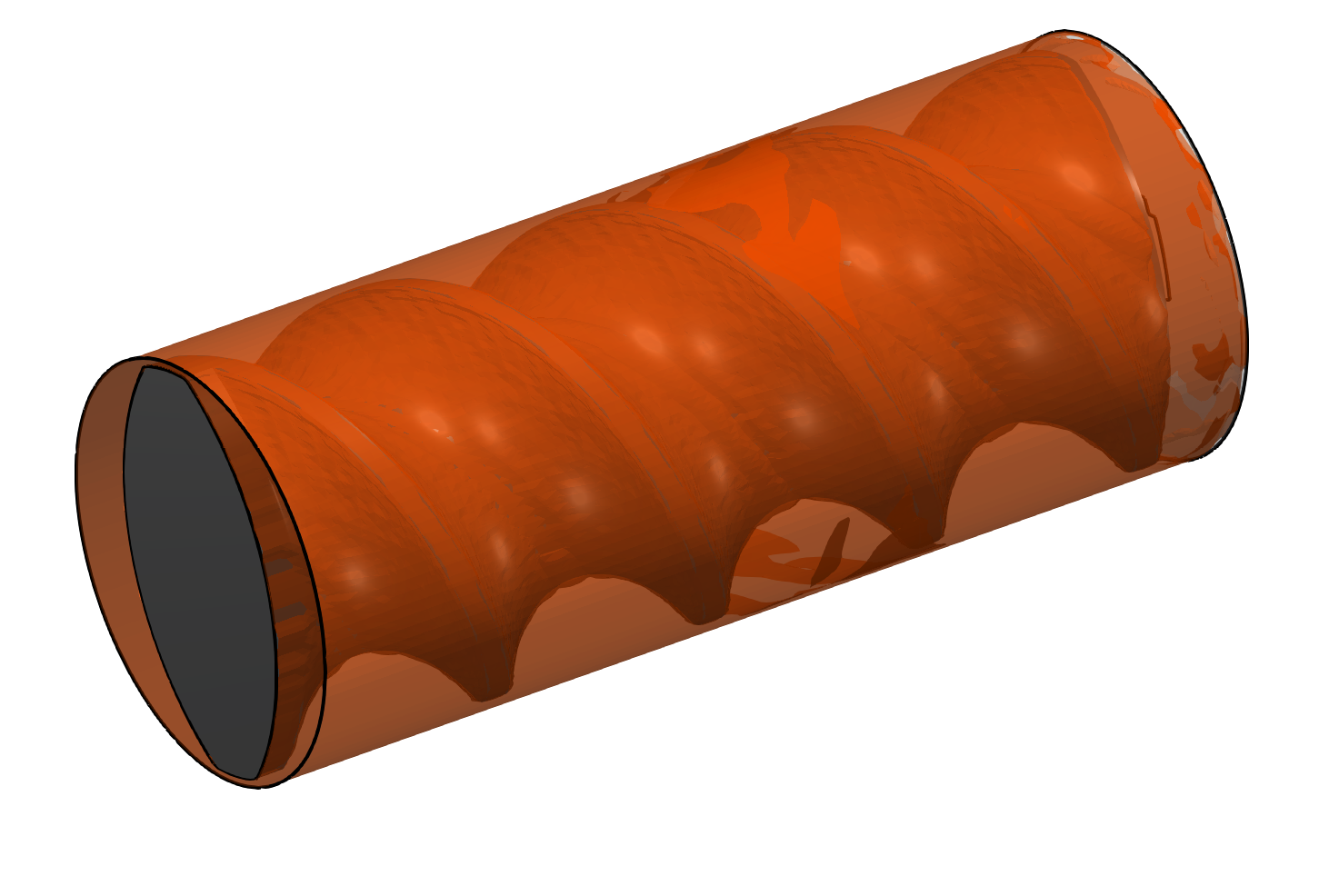}
\caption{BC-VOF-IB solver using SRF tools.}
\label{fig:ch5-simp-SSE-2L-0.5mmGap-60RPM-2Q0-IB-high-newtonian-alpha}
\end{subfigure}

\caption{Two-phase front advancement obtained with BC-VOF and BC-VOF-IB solvers at time instants $t = 0.8 \ \unit{s}, \ 1.6 \ \unit{s}, \ 2.4 \ \unit{s}, \ 5 \ \unit{s}$.}
\label{fig:ch5-simp-SSE-2L-0.5mmGap-60RPM-2Q0-high-newtonian-alpha}

\end{figure}

\begin{figure}[h!]
    \centering
    \includegraphics[width=0.8\textwidth]{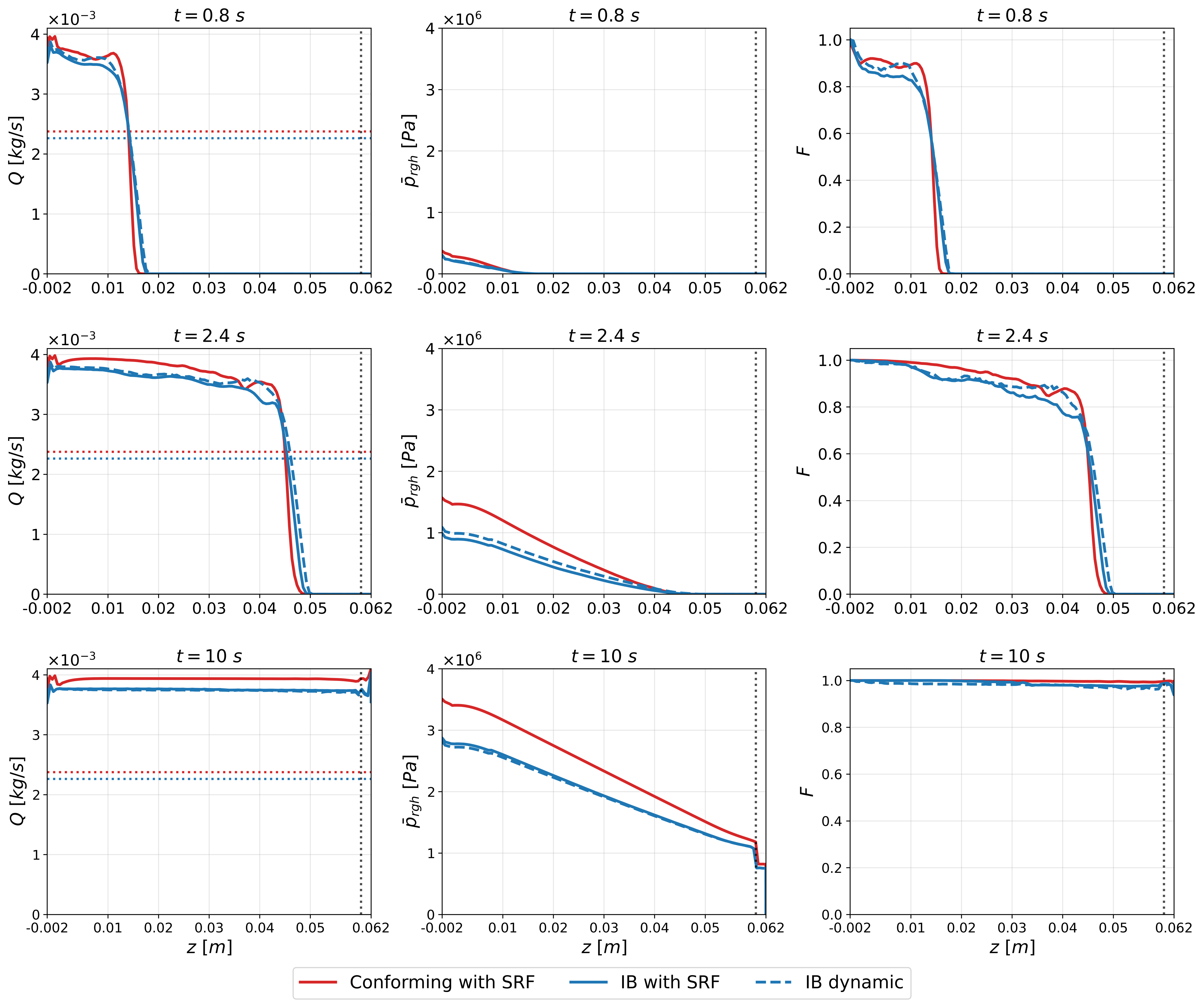}
    \caption{Plots of mass flow rate $Q$, average pressure $\bar p_{rgh}$ and filling ratio $F$ over axial sections at time-instants $t = 0.8 \ \unit{s}, 2.4 \ \unit{s}, 10 \ \unit{s}$. The dotted gray line positioned at $z = 60 \ \unit{mm}$ indicates the end of the screw. Red and blue dotted lines are $Q_{0,\text{conf}}$ and $Q_{0,\text{IB}}$, respectively.}
    \label{fig:ch5-simp-SSE-2L-0.5mmGap-60RPM-2Q0-high-newtonian-Q-p-alpha}
\end{figure}

Velocity magnitude and pressure fields obtained with BC-VOF and BC-VOF-IB are reported in Figures \ref{fig:ch5-simp-SSE-2L-0.5mmGap-60RPM-2Q0-high-newtonian-u} and \ref{fig:ch5-simp-SSE-2L-0.5mmGap-60RPM-2Q0-high-newtonian-p}, respectively, for the highly viscous Newtonian phase (fluid 1) that is progressively transported in the simplified SSE geometry, while Figure \ref{fig:ch5-simp-SSE-2L-0.5mmGap-60RPM-2Q0-high-newtonian-alpha} displays the two-phase front advancement in the whole SSE domain.
The results obtained with the two approaches present some discrepancies, especially concerning the pressure field, as one can notice also from the plots of average pressure values on axial sections in Figure \ref{fig:ch5-simp-SSE-2L-0.5mmGap-60RPM-2Q0-high-newtonian-Q-p-alpha}.

In Figure \ref{fig:ch5-simp-SSE-2L-0.5mmGap-60RPM-2Q0-high-newtonian-Q-p-alpha} we present the evolution of \textit{mass flow rate} $Q$ (left column), \textit{average pressure} $\bar p_{rgh}$ (central column) and \textit{filling ratio} $F$ (right column), computed for fluid 1 along axial sections, comparing the results obtained with the BC-VOF and BC-VOF-IB solvers. The latter is employed in two different configurations: the first one solves the equations in non-inertial reference frame (SRF tools of OpenFOAM), so the IB surface of the screw is fixed over time; secondly, dynamic immersed boundary tools are used, meaning that the momentum equation is solved in the absolute reference frame leading to a moving IB mask that follows the screw rotation.

At each axial section $\widetilde S$ we define
\begin{itemize}
    \item The \textit{mass flow rate}: $Q_\text{conf} = \int_{\widetilde S_\text{conf}} \alpha \rho_1 u_z \ dS \text{ and } Q_\text{IB} = \int_{\widetilde S_\text{IB}} \chi_\text{IB} \ \alpha \rho_1 u_z \ dS$;
    \item The \textit{average pressure}: $\bar p_{rgh,\text{conf}}= \frac{\int_{\widetilde S_\text{conf}} \alpha p_{rgh}\ dS}{\int_{\widetilde S_\text{conf}} \alpha dS} \text{ and }
        \bar p_{rgh,\text{IB}} = \frac{\int_{\widetilde S_\text{IB}} \chi_\text{IB} \ \alpha p_{rgh}\ dS}{\int_{\widetilde S_\text{IB}} \chi_\text{IB} \ \alpha \ dS}$;
    \item The \textit{filling ratio}: $F_{\text{conf}}= \frac{\int_{\widetilde S_\text{conf}} \alpha \ dS}{\int_{\widetilde S_\text{conf}} dS} \text{ and } 
        F_{\text{IB}} = \frac{\int_{\widetilde S_\text{IB}} \chi_\text{IB} \ \alpha \ dS}{\int_{\widetilde S_\text{IB}} \chi_\text{IB} \ dS}$;
\end{itemize}

Despite the discrepancies, both solvers provide physically reasonable solutions. Prescribing $Q_\text{in} > Q_0$ leads to a pressure decrease from inlet to outlet, due to the fact that the flow is dominated by the inlet mass flow rate that moves the material forward, since the screw is not rotating fast enough to contrast that pressure drop. Additionally, in the proximity of the screw ending at $z = 0.06 \ \unit{m}$ (vertical dotted line in Figure \ref{fig:ch5-simp-SSE-2L-0.5mmGap-60RPM-2Q0-high-newtonian-Q-p-alpha}) we observe a stronger pressure gradient due to the presence of the outlet restriction, that allows the material to be discharged from the die. As expected, at the final time $t = 10 \ \unit{s}$, the material eventually fills the whole domain leading to a filling ratio $F\sim1$ on all sections.

\subsubsection{The Twin Screw Extruder}
\label{sec:TSE}

The real geometry of a TSE is addressed, characterized by the presence of two screws positioned side-by-side, providing improved mixing efficiency and process control compared to the simpler SSE. Specifically, we consider a \textit{co-rotating intermeshing} TSE \cite{rauwendaal2014TSE}, where both screws rotate in the same direction and the flights of one screw interpenetrate the channels of the other, creating the so-called \textit{self-wiping effect}, where the screws constantly clean each other, thus preventing material from sticking and degrading.
The main advantage of the TSE lies in its enhanced mixing capability, thanks to the presence of \textit{kneading blocks},  i.e.\ screw elements made of discs staggered by $45^\circ$ or $90^\circ$, specifically designed to shear, stretch, and fold the polymer.

\begin{figure}[h!]
    \centering
    \begin{subfigure}{0.3\textwidth}
        \centering
        \includegraphics[width=\textwidth]{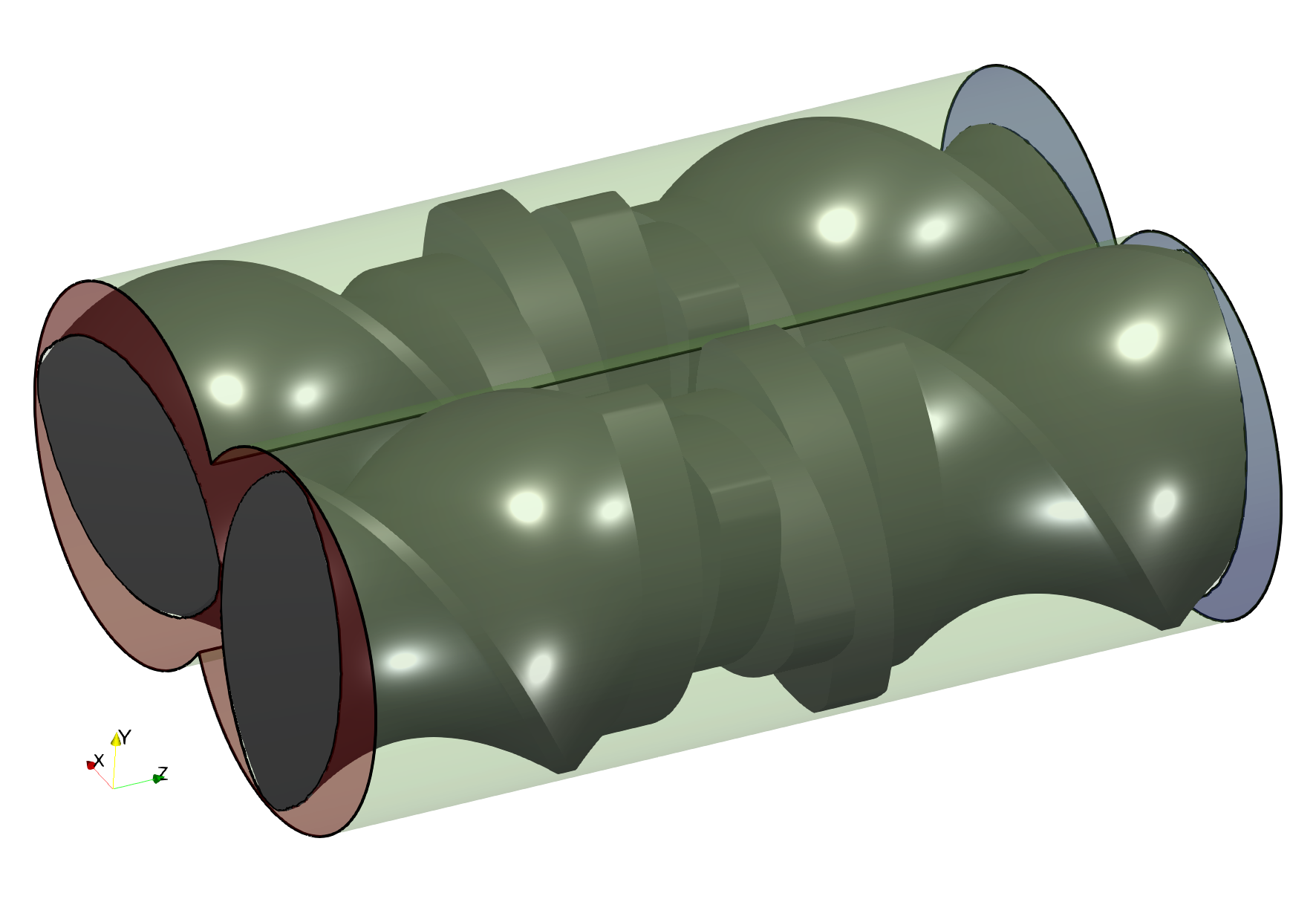}
        \caption{Computational domain.} 
        \label{fig:ch5-tse-domain}
    \end{subfigure}
    \begin{subfigure}{0.3\textwidth}
        \centering
        \includegraphics[width=\textwidth]{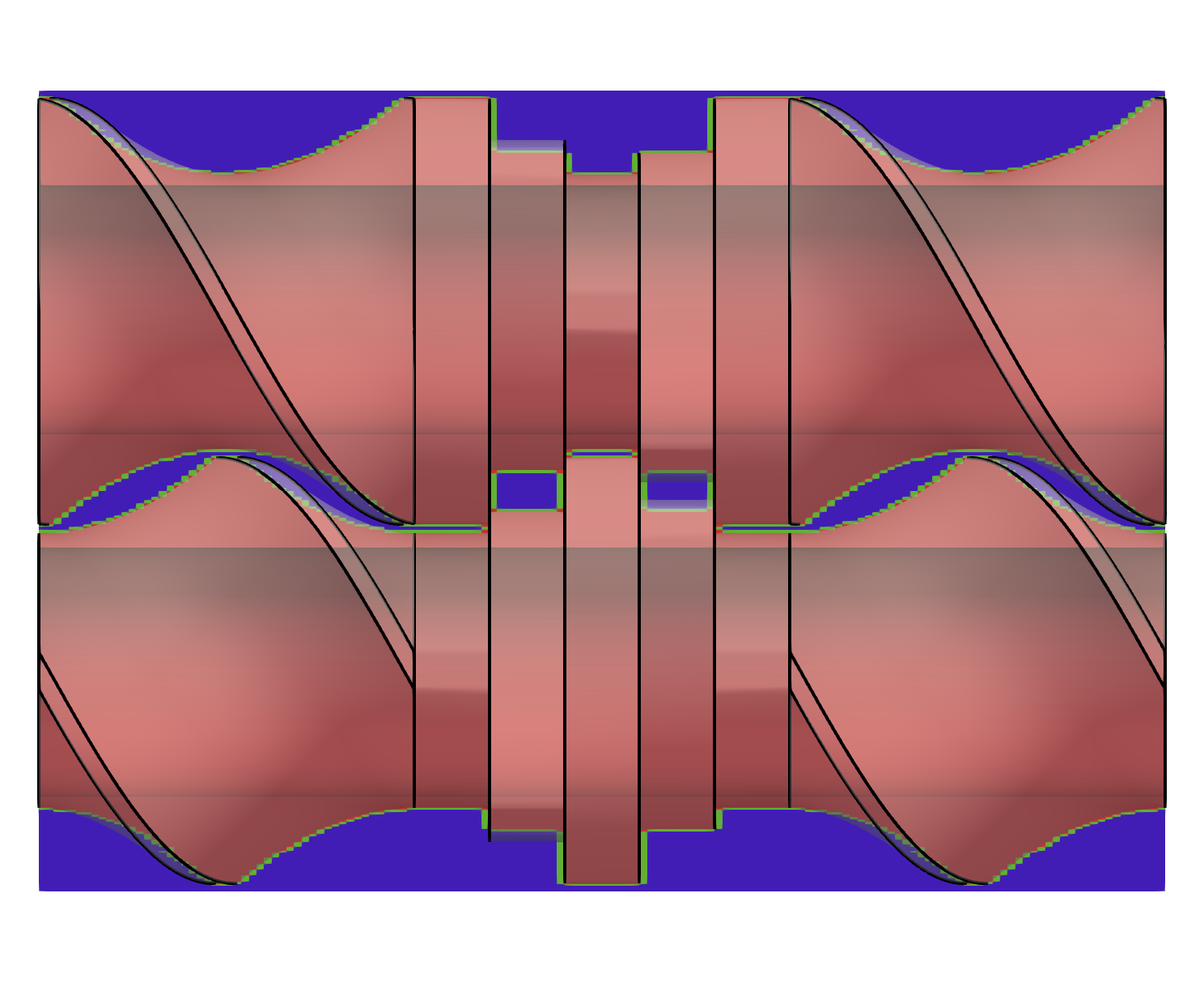}
        \caption{View from above.}
        \label{fig:ch5-tse-mesh-above}
    \end{subfigure}
    \begin{subfigure}{0.38\textwidth}
        \centering
        \includegraphics[width=\textwidth]{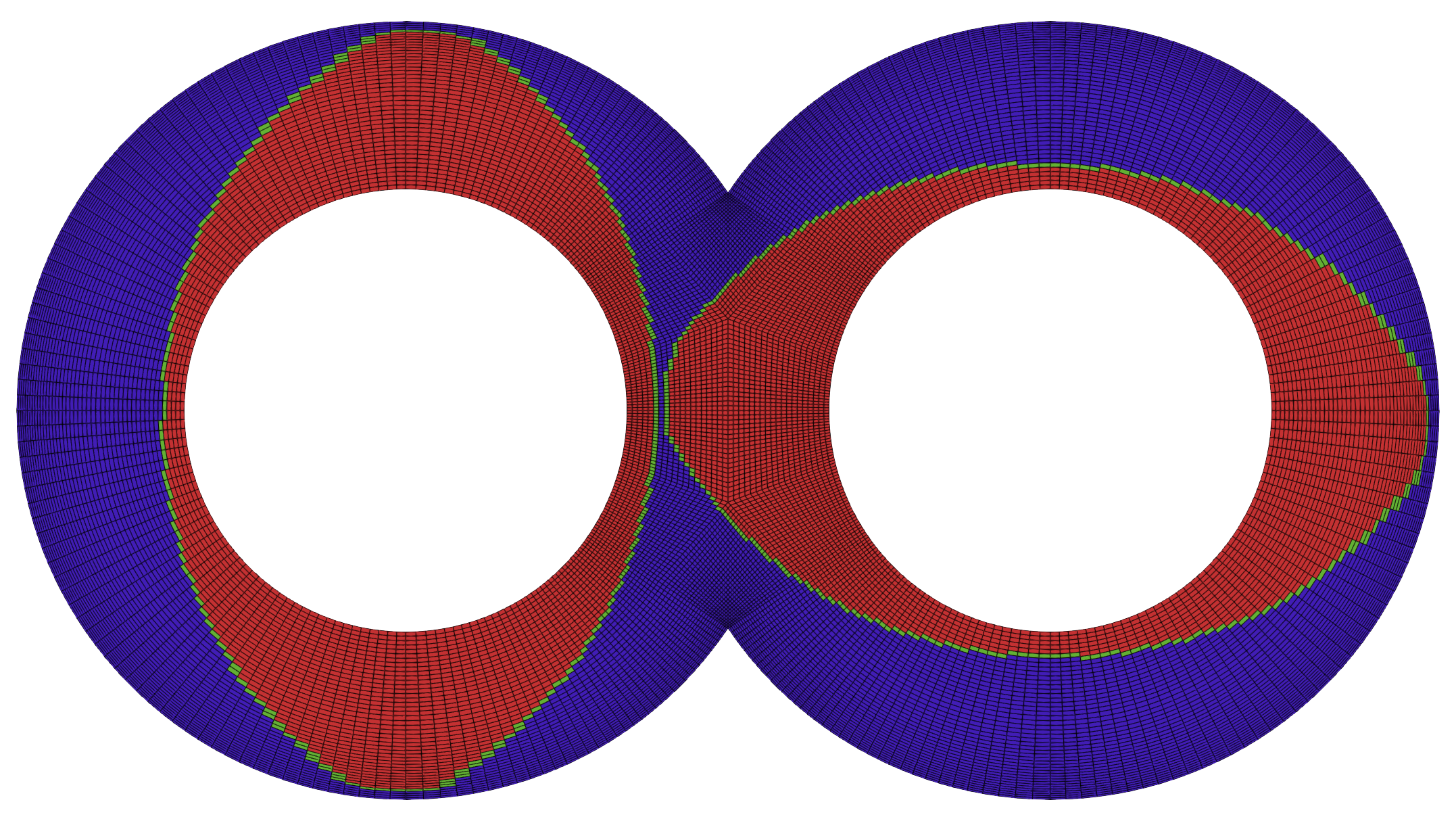}
        \caption{Front view.}
        \label{fig:ch5-tse-mesh-front}
    \end{subfigure}
    \caption{Geometry of the TSE test case. The subdivision of the non-conforming grid into IB cells (green), fluid cells (blue) and solid cells (red) is represented.} 
    \label{fig:ch5-tse-mesh}
\end{figure}

The geometry of the TSE device used for the simulations is reported in Figure \ref{fig:ch5-tse-domain}. We denote by $\ell$ the characteristic length being the horizontal distance (along $x-$axis) between the axes of the two screws. Then, the total width of the barrel is $L_{B,x} = 2 \ell$ and its axial length is $L_{B,z} = 2.8 \ell$, where $\ell$ is a few centimeters. The gap between screws and barrel is of the order of millimeters, as well as the gap between the two screws.
The screws are divided into three different portions: the first and last elements consist of \textit{transport modules} while the middle section is characterized by \textit{kneading modules} devoted to mixing rather than transport.

A non-conforming grid with respect to the screws is generated (Figure \ref{fig:ch5-tse-mesh}) to use the IB approximation \cite{negrini2025IBM,negrini2023phdthesis}, since the use of a conforming grid would be unfeasible in this case, requiring the adoption of dynamic mesh functionalities. 
The computational mesh is composed of \num{2767800} elements; the screw–barrel gap is resolved by four fluid cells and the gap
between the two screws by a single fluid cell. The latter is a coarse resolution for the WLS reconstruction, so that the flow in
the intermeshing region is captured only approximately. The most relevant mesh quality indicators are reported in Table \ref{tab:ch5-TSE-grids}.

\begin{table}[h!]
    \renewcommand{\arraystretch}{1.3}
    \setlength{\tabcolsep}{7.5pt}
    \small
    \centering
    \begin{NiceTabular}{|c|c|c|c|c|c|c|c|} 
        \noalign{\hrule height 1.2pt}
        \Block{2-1}{$\#$DOFs} & \multicolumn{2}{c}{$\#$DOFs across}  & \multicolumn{2}{c}{Non-orthogonality} & Skewness & \multicolumn{2}{c}{Cell volume $[\unit{m}^3]$} \\
        \cline{2-8}  
        & screw-barrel gap & intermeshing gap & $\max(\tau_{ij})$ & $\text{avg}(\tau_{ij})$ & $\max(\psi_{ij})$ & $\displaystyle \min_{K_i \in \mathcal{T}_h}|K_i|$ & $\displaystyle \max_{K_i \in \mathcal{T}_h}|K_i|$\\
        \noalign{\hrule height 0.8pt}
        \num{2767800} & 5 & 1 & 20.16 & 3.39 & 0.42 & $3.62 \cdot 10^{-11}$ & $6.77 \cdot 10^{-10}$ \\
        \noalign{\hrule height 1.2pt}
    \end{NiceTabular}    
    \caption{Mesh size and key quality indicators for the TSE test case.}
    \label{tab:ch5-TSE-grids}
\end{table}

Fluid 1 is a high-viscosity power-law shear-thinning polymeric material, with coefficients of the rheological law being calibrated based on experimental data provided by the industrial partner which cannot be disclosed (see the Data Availability Statement). The polymer flows inside the TSE device whose screws are rotating at 100 rpm.

As it is done for the SSE case, a preliminary single-phase isothermal simulation is first carried out to determine the drag flow rate $Q_0$ and to set two-phase simulations accordingly.

\paragraph{Free-surface Simulation of Fully Filled TSE}

The first two-phase simulation inside the TSE geometry of the highly viscous non-Newtonian polymer is obtained injecting the polymeric material from the inlet patch at the initial time with a constant inlet flow rate $Q_{\text{in}} > Q_0$, with $Q_0$ coming from the single-phase simulation. The domain is initially full of air, the screw velocity is kept at 100 rpm and no-slip conditions are imposed at barrel walls: they are adopted here as the simplest choice; the influence of wall slip, which can be significant for polymer melts, will be assessed in future work using the Navier-slip condition \eqref{eq:navierSlipBC} mentioned in Section \ref{sec:model}.
The BC-VOF-IB solver is employed for the simulation.

\begin{figure}[h!]
\centering

\begin{subfigure}{\textwidth}
\centering
\includegraphics[width=0.45\textwidth]{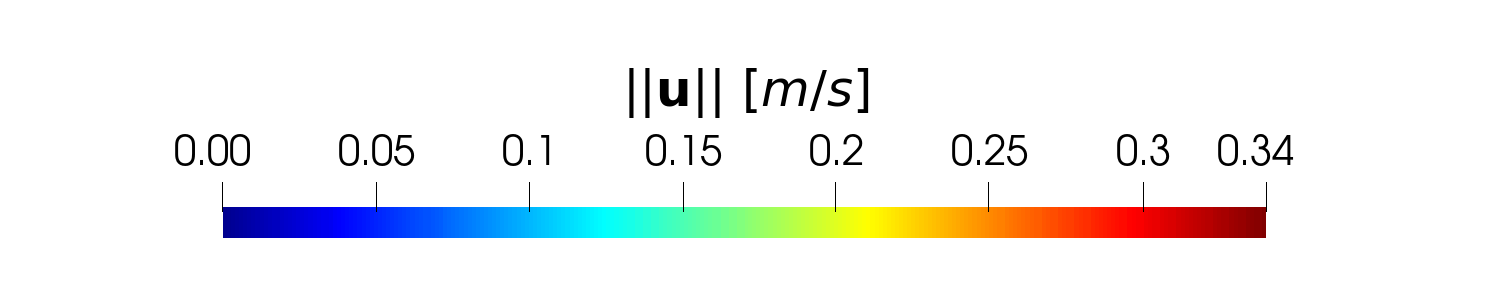} \\
\includegraphics[width=0.24\textwidth]{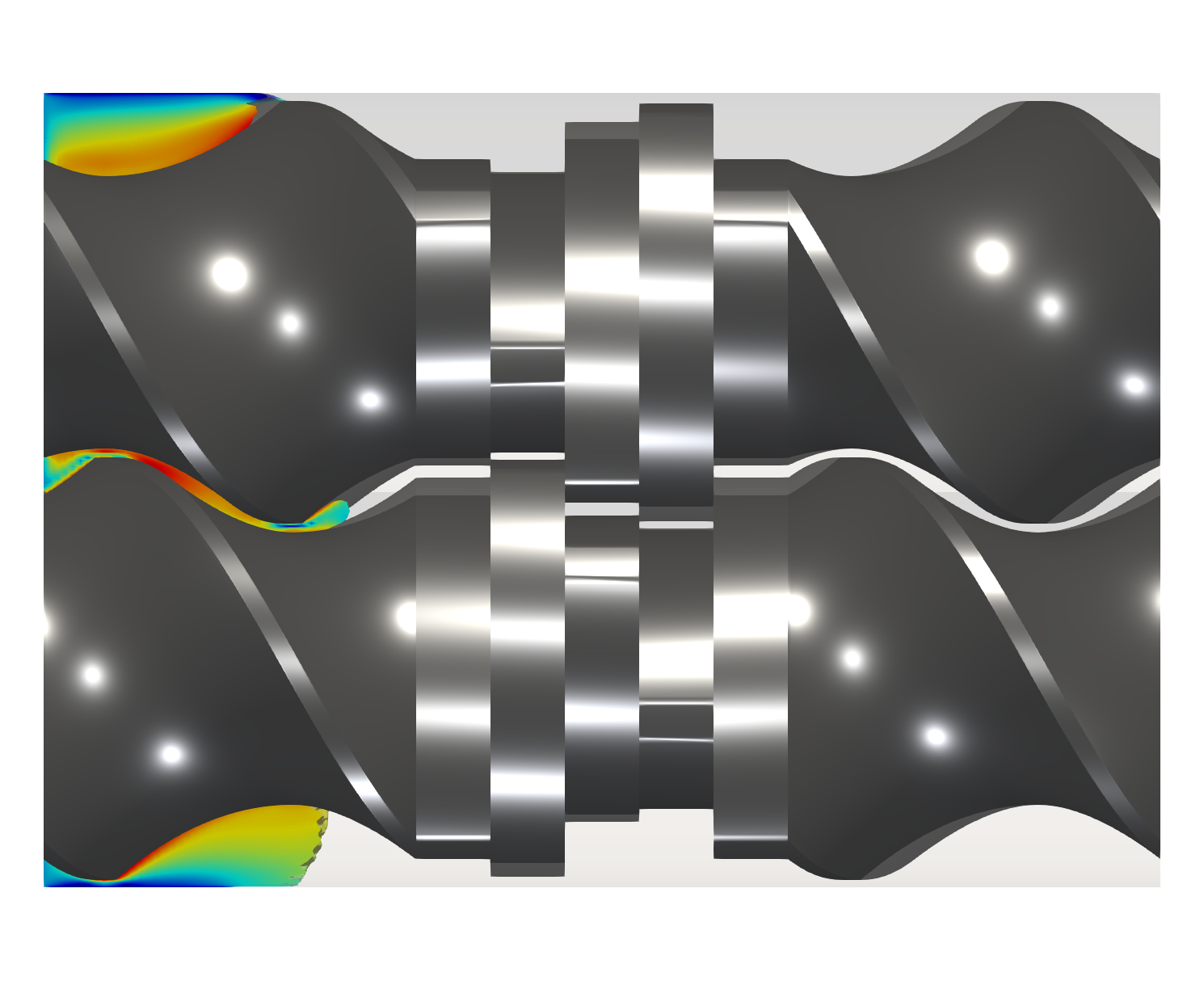}
\includegraphics[width=0.24\textwidth]{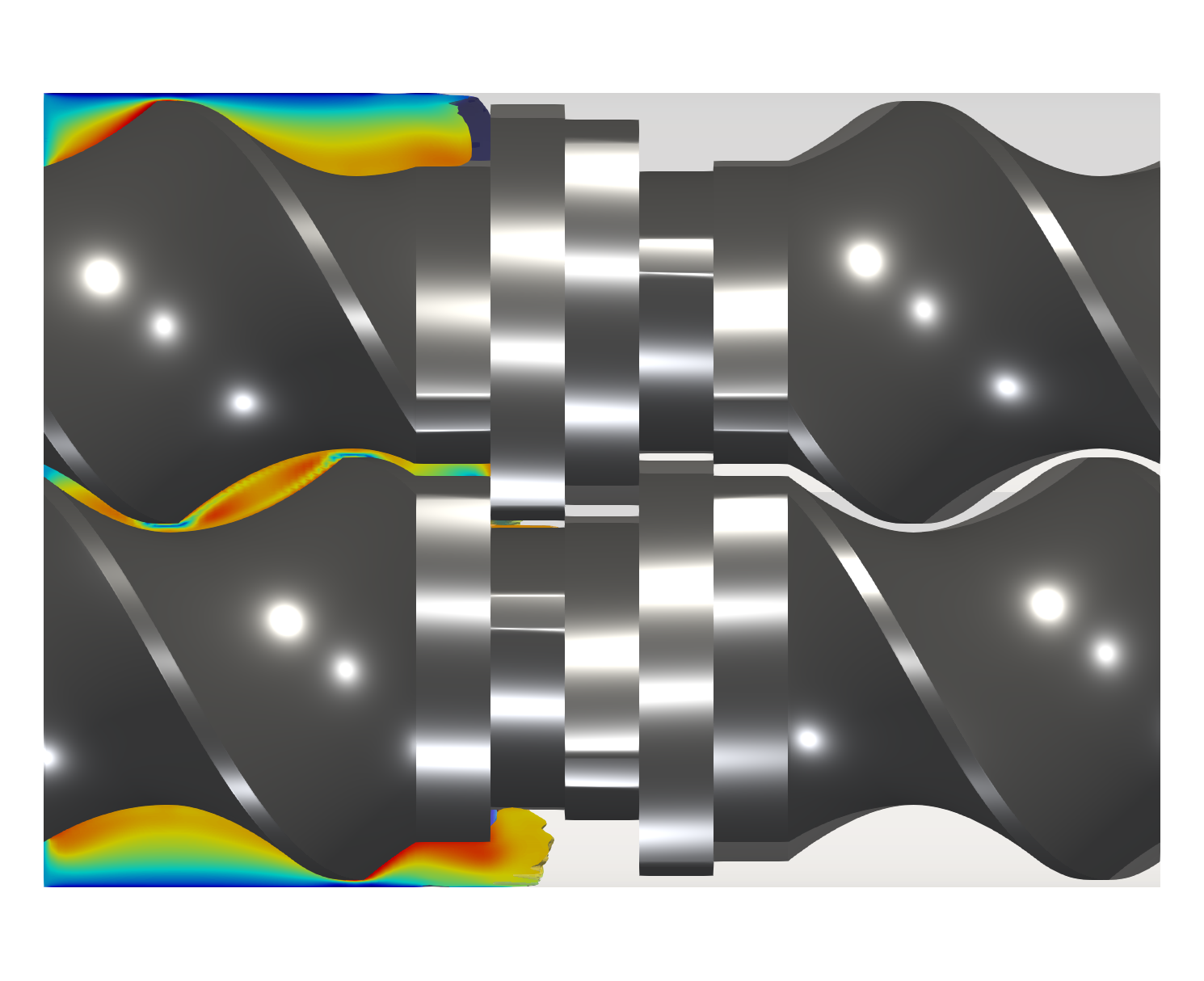}
\includegraphics[width=0.24\textwidth]{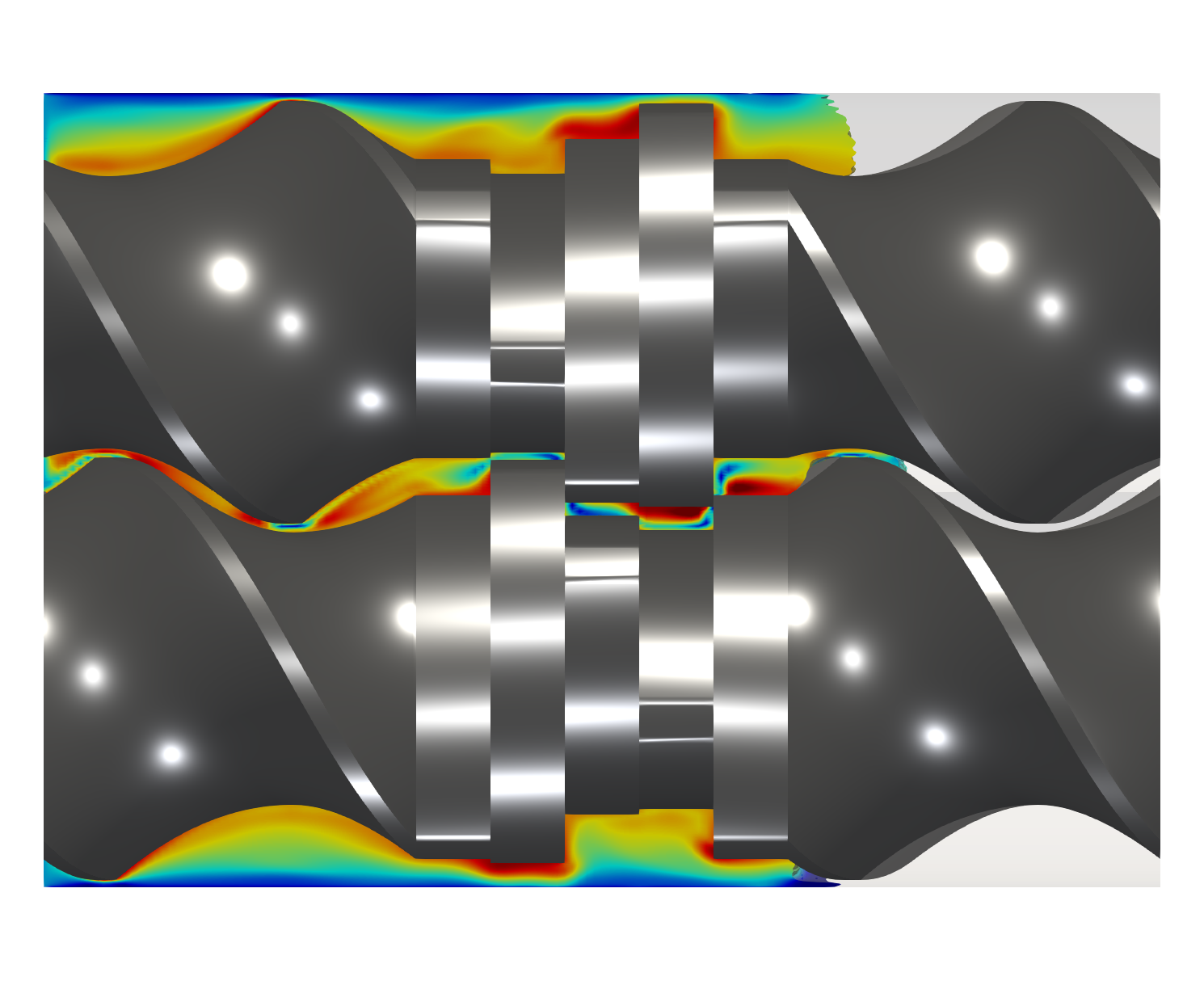}
\includegraphics[width=0.24\textwidth]{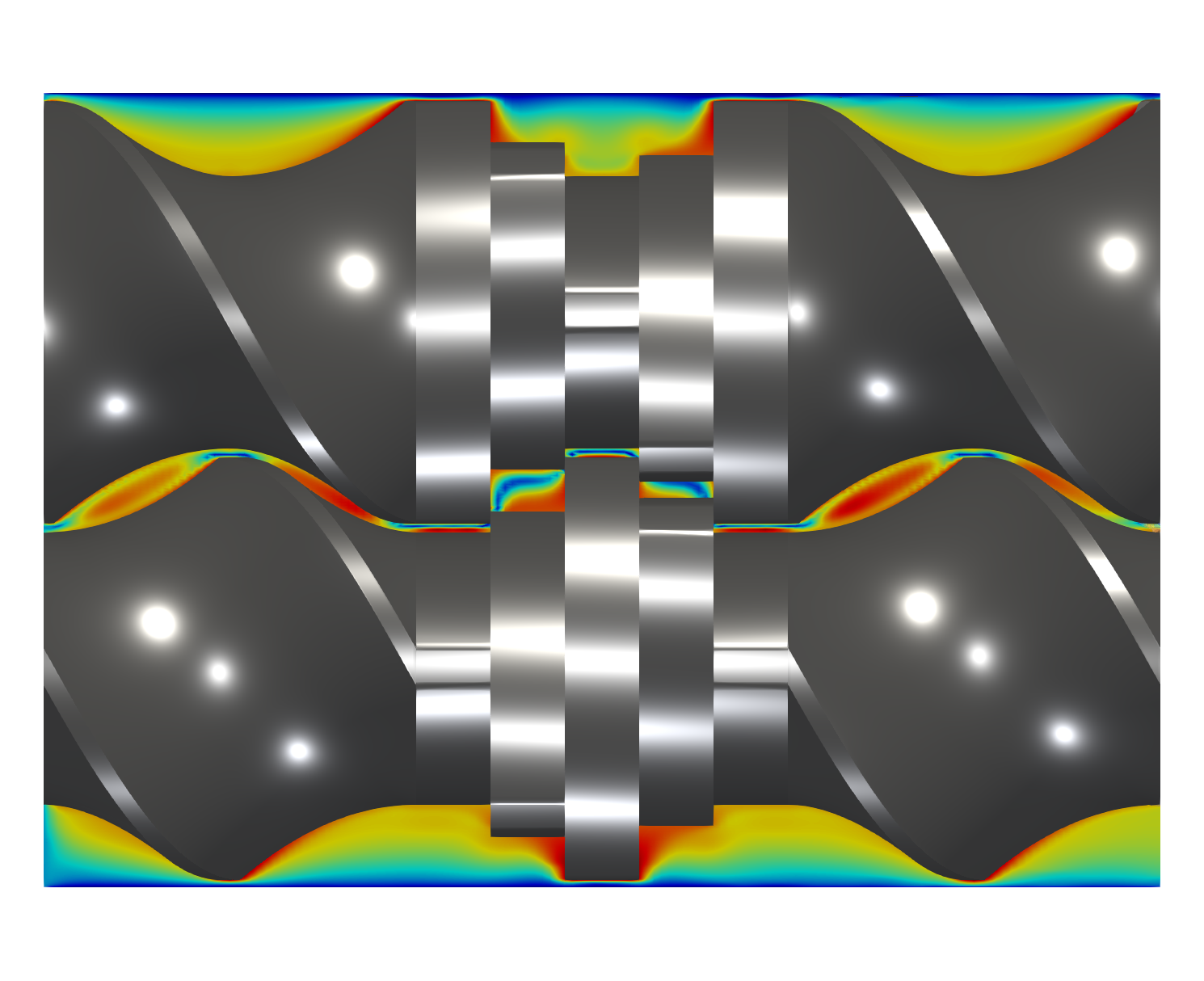}
\caption{Velocity field.}
\label{fig:ch5-tse-twophase-filling-u}
\end{subfigure}
\hfill
\begin{subfigure}{\textwidth}
\centering
\includegraphics[width=0.45\textwidth]{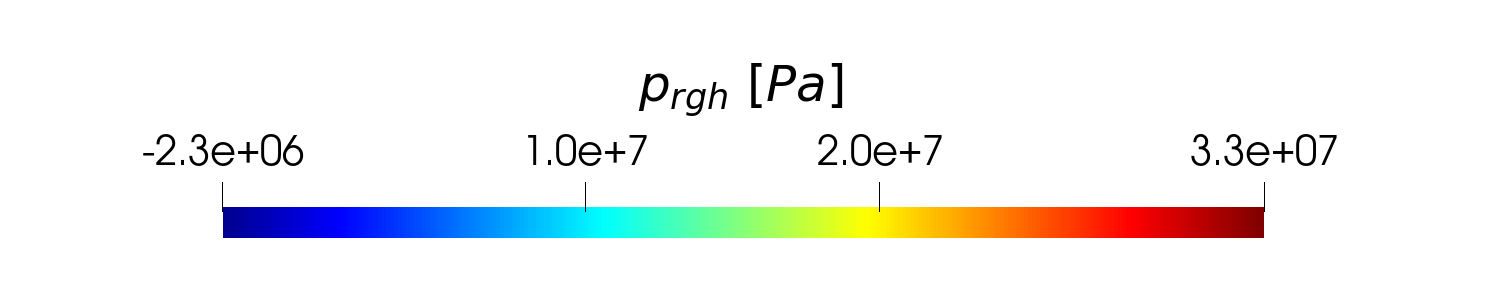}\\
\includegraphics[width=0.24\textwidth]{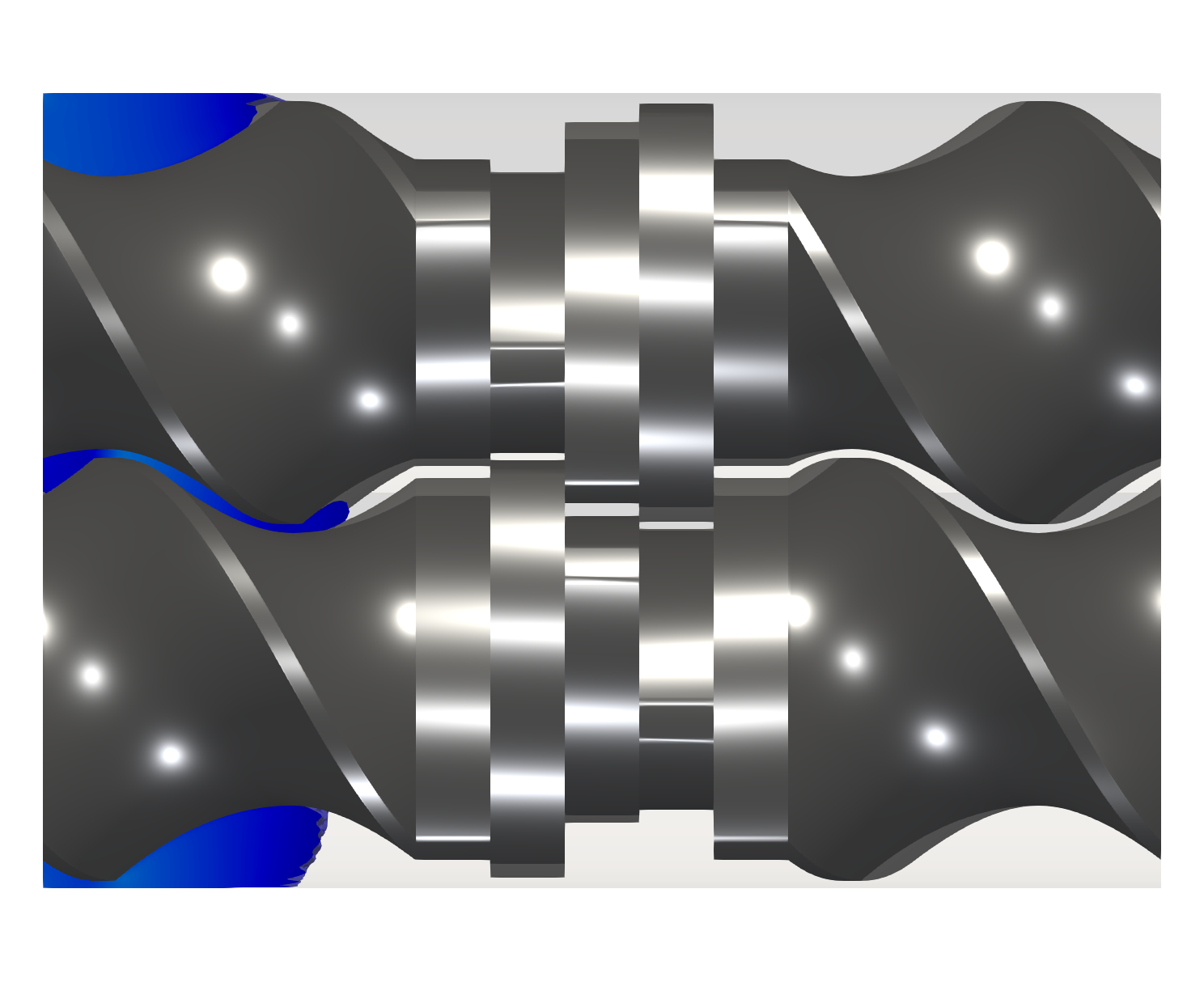}
\includegraphics[width=0.24\textwidth]{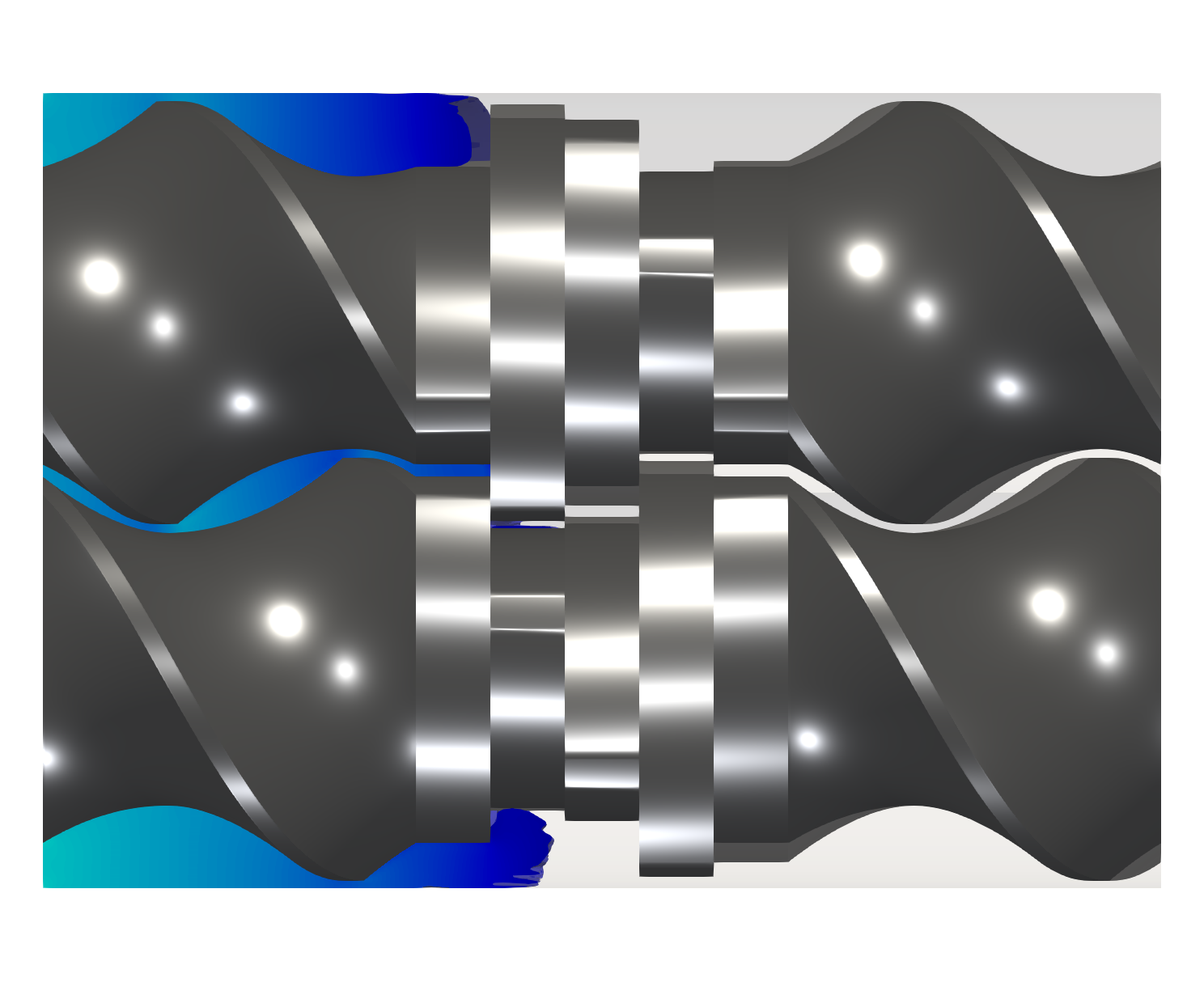}
\includegraphics[width=0.24\textwidth]{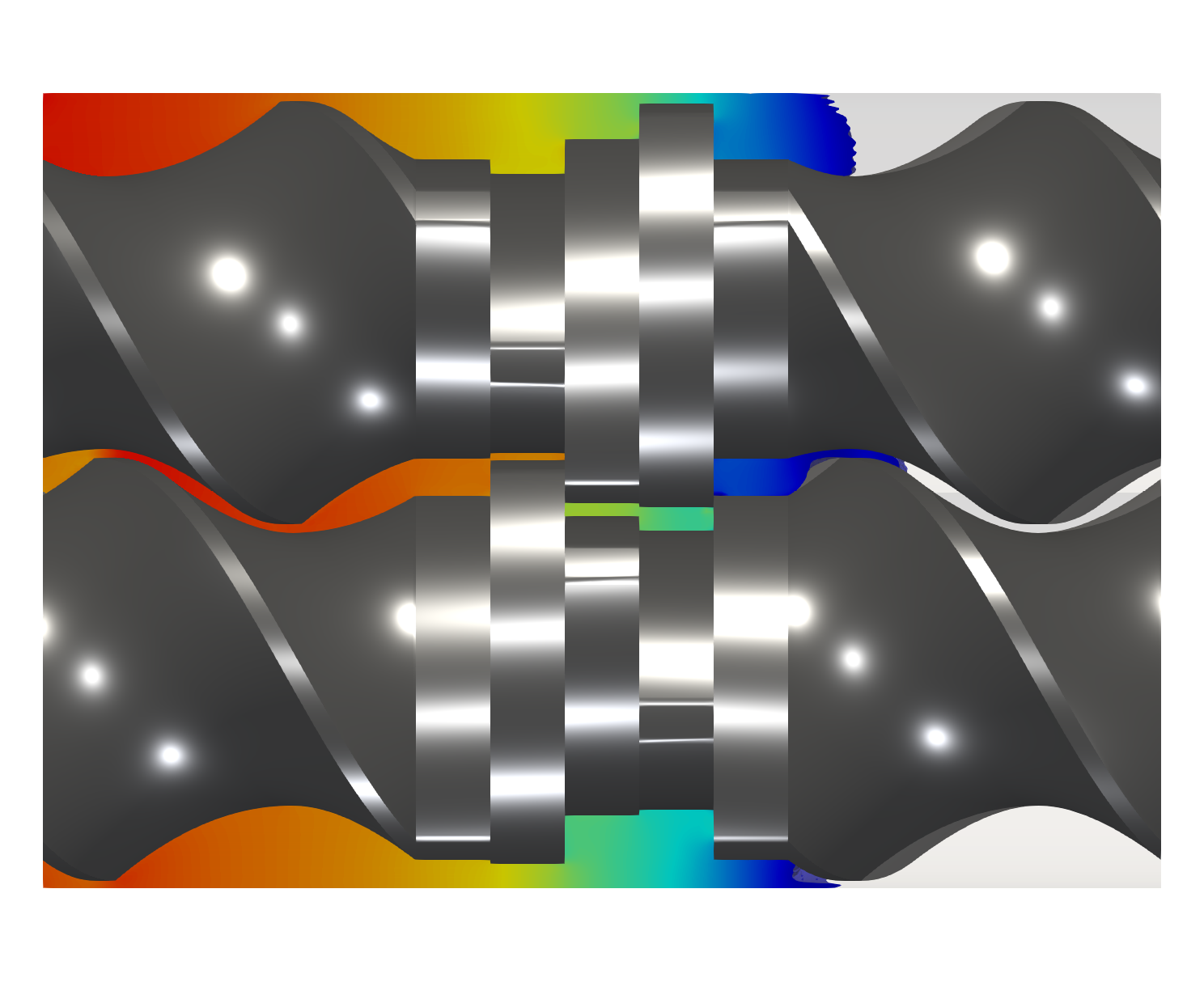}
\includegraphics[width=0.24\textwidth]{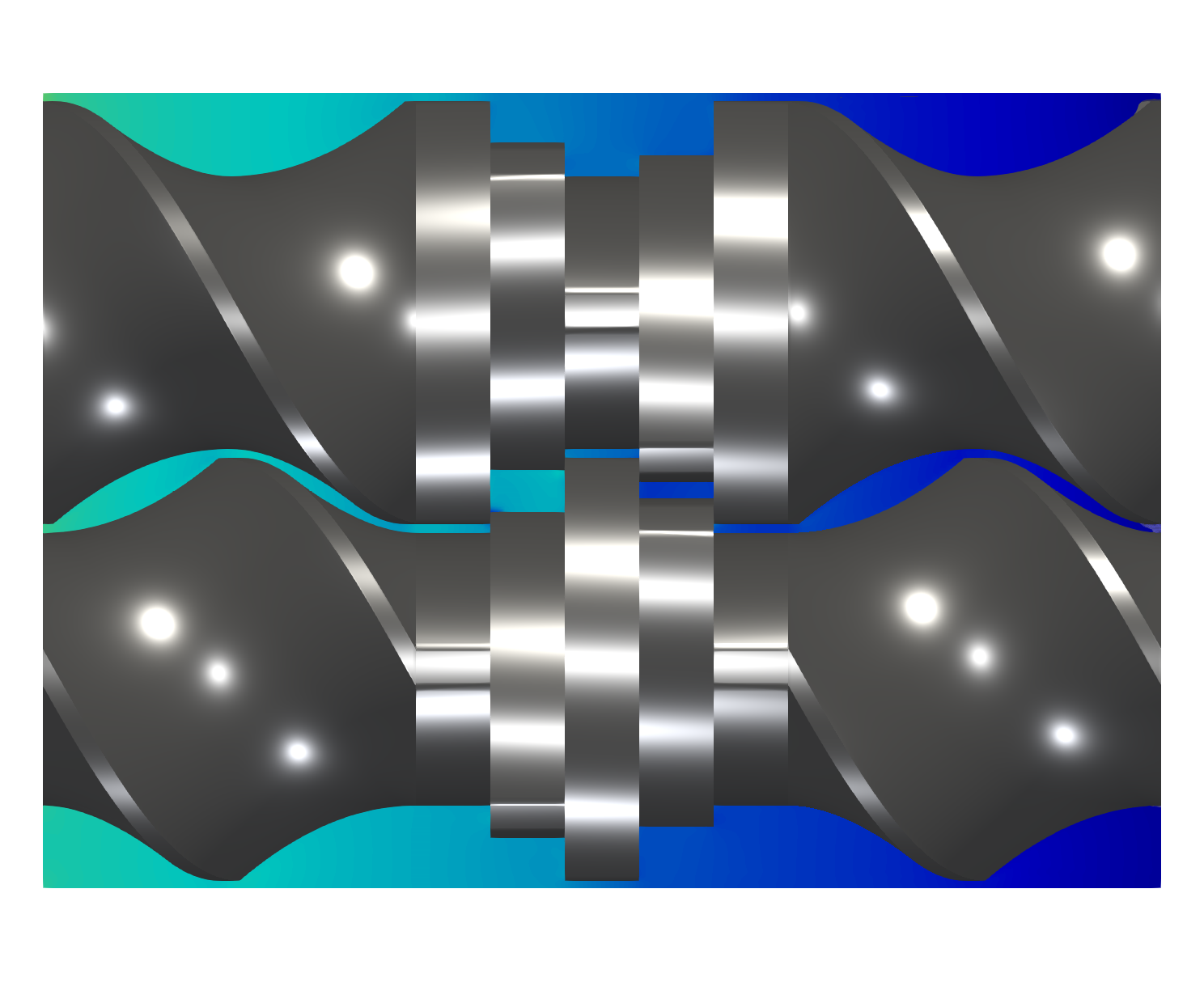}
\caption{Pressure field.}
\label{fig:ch5-tse-twophase-filling-p}
\end{subfigure}

\caption{Case of inlet flow rate $Q_\text{in} > Q_0$ simulated with the BC-VOF-IB solver. Fields are reported on a slice along the $x-z$ plane (top view) at time instants $t = 0.5 \ \unit{s}, \ 1 \ \unit{s}, \ 2 \ \unit{s}, \ 3 \ \unit{s}$.}
\label{fig:ch5-tse-twophase-filling-u-p}

\end{figure}

\begin{figure}[H]
    \begin{subfigure}{0.3\textwidth}
    \centering
    \includegraphics[width=0.7\textwidth]{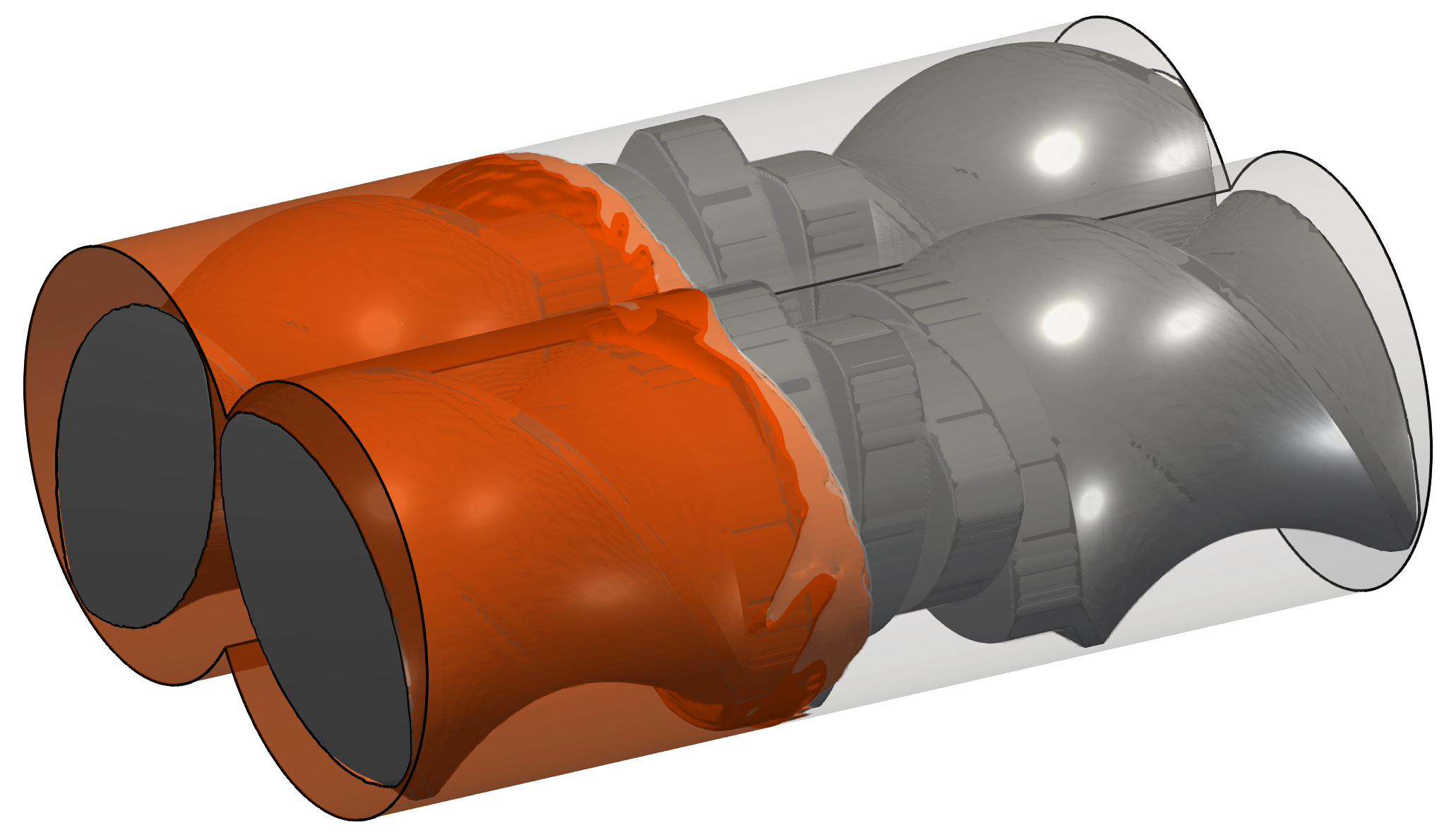}\\
    \vspace{0.5cm}
    \includegraphics[width=0.7\textwidth]{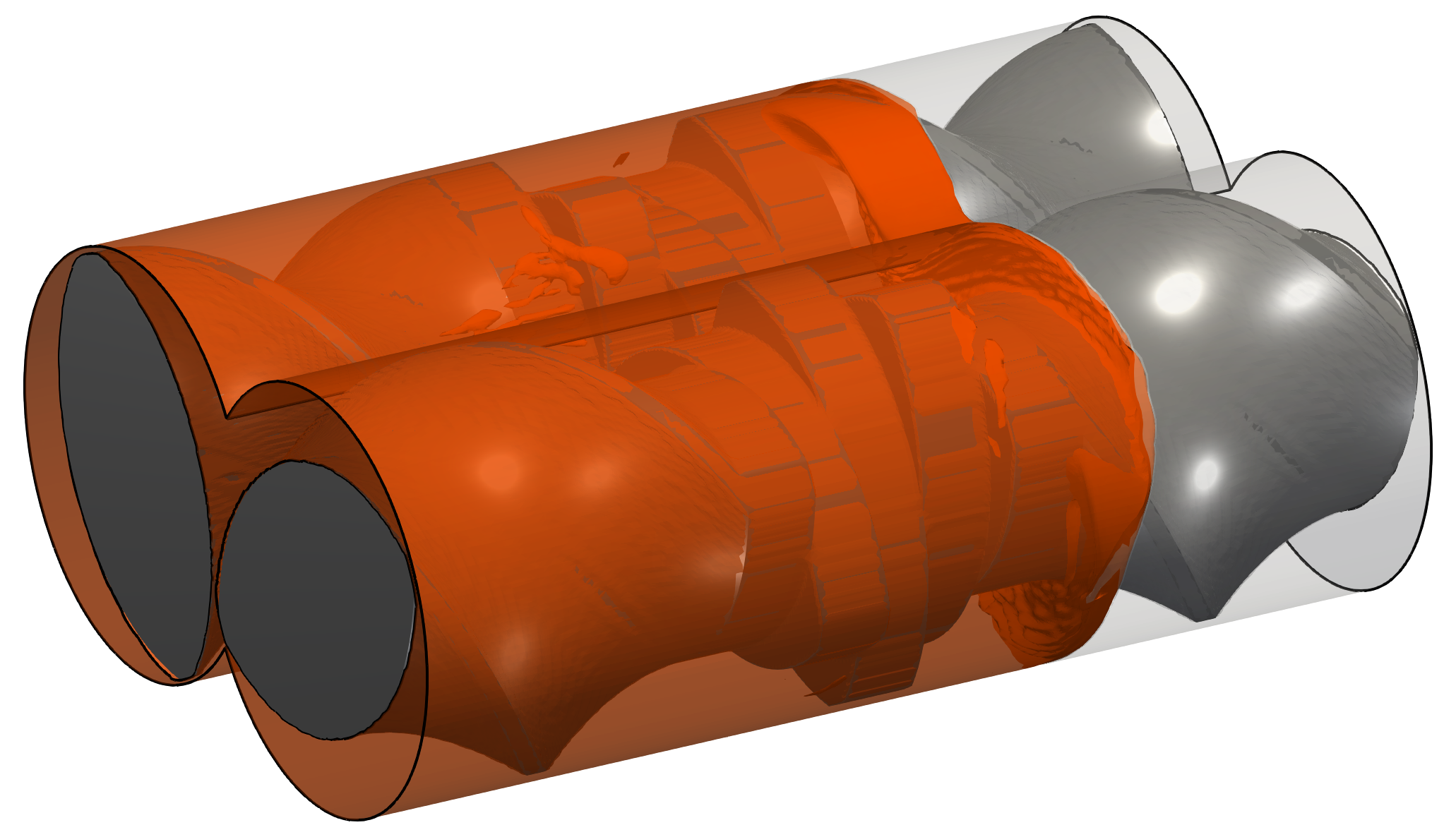}\\
    \vspace{0.5cm}
    \includegraphics[width=0.7\textwidth]{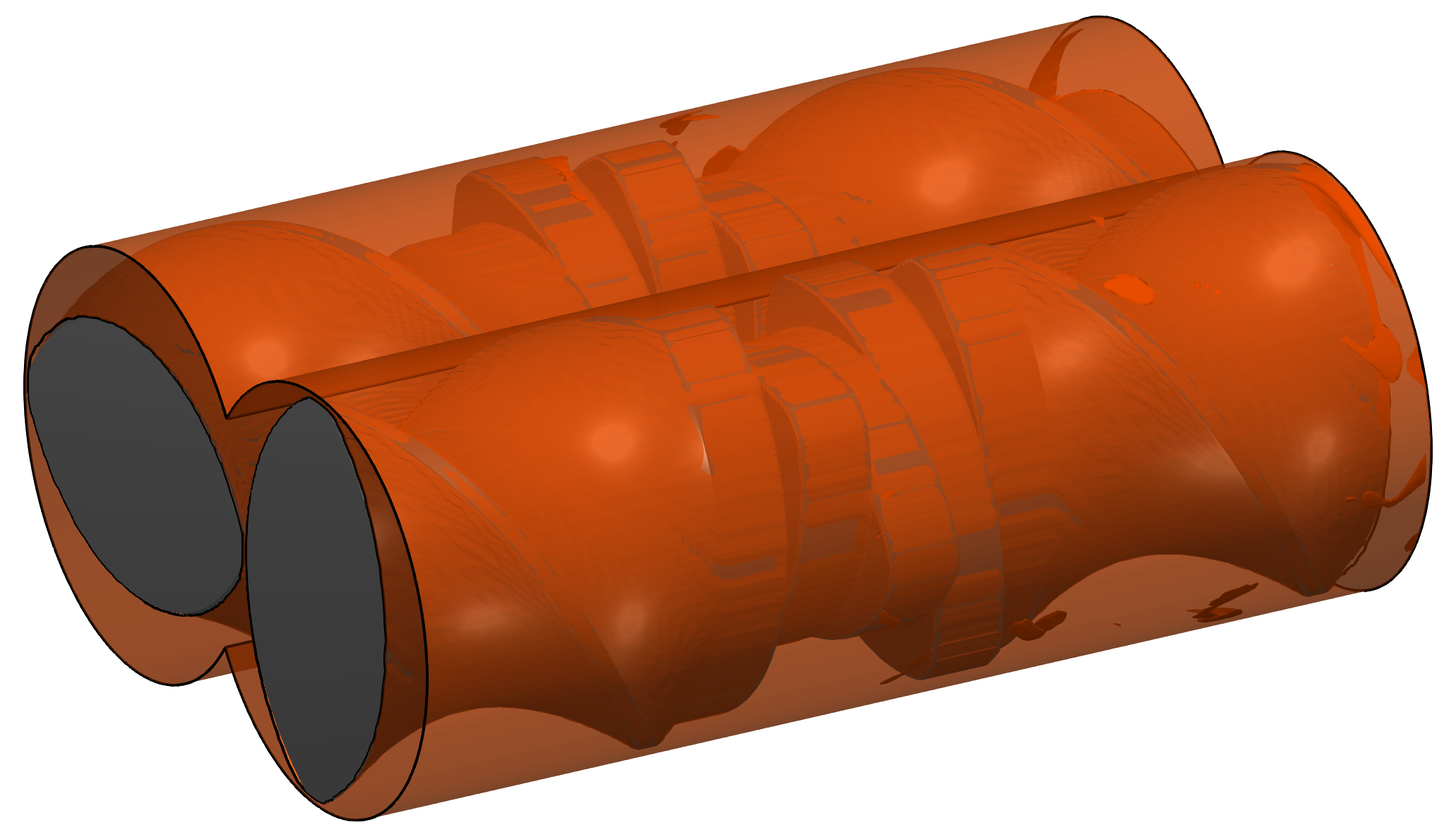}\\
    \vspace{0.7cm}
    \caption{Two-phase front evolution.}
    \label{fig:ch5-tse-twophase-filling-alpha}
    \end{subfigure}
    \begin{subfigure}{0.65\textwidth}
    \centering
    \includegraphics[width=0.8\textwidth]{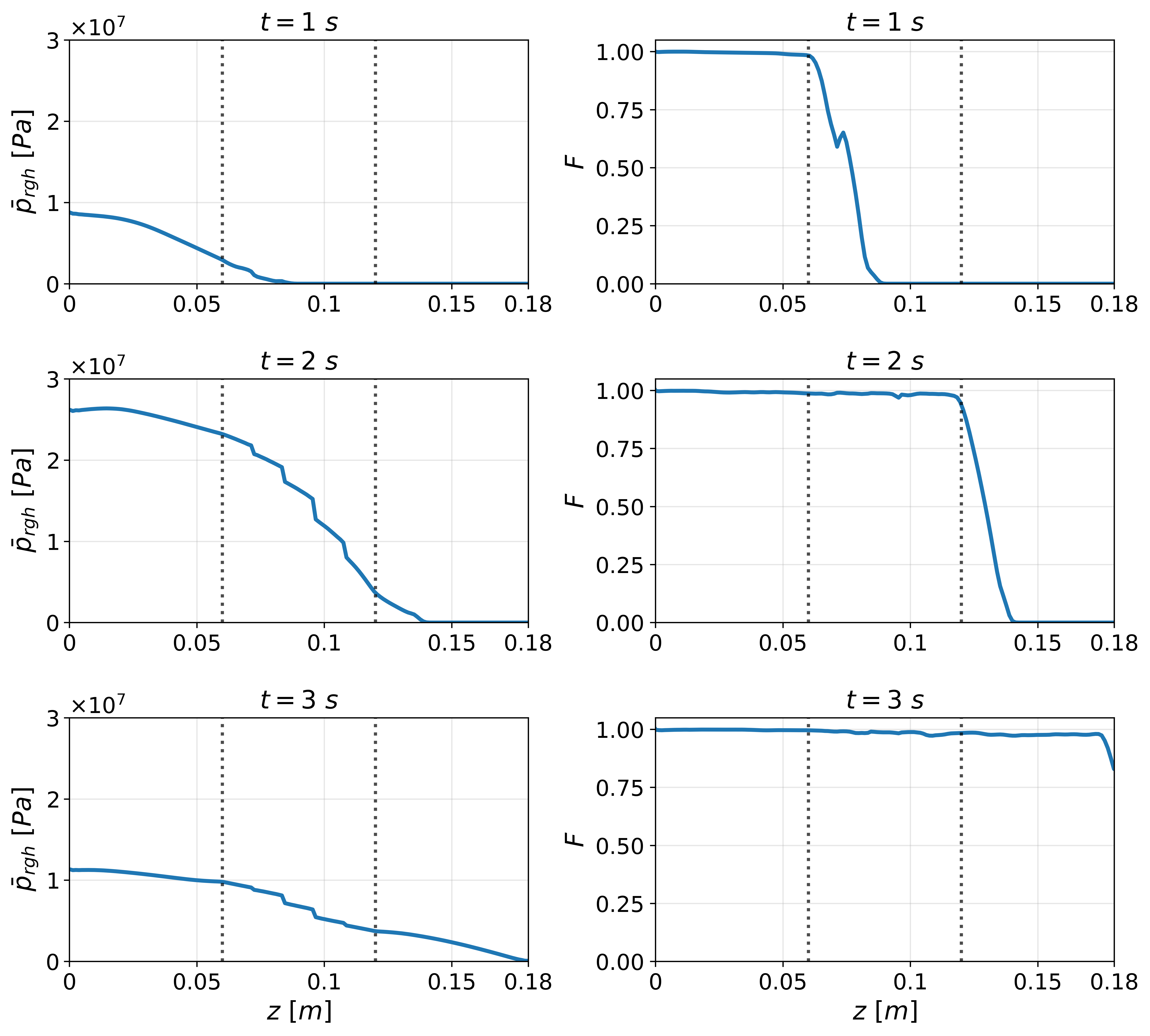}
    \caption{Average quantities. Grey dotted lines represent start and ending of kneading modules.}
    \label{fig:ch5-tse-twophase-filling-p-alpha-av}
    \end{subfigure}

    \caption{Case of inlet flow rate $Q_\text{in} > Q_0$ simulated with the BC-VOF-IB solver. Temporal evolution of the two-phase interface (left) and plots of average pressure $\bar p_{rgh}$ and filling ratio $F$ along axial sections (right) are reported for times $t = 1 \ \unit{s}, \ 2 \ \unit{s}, \ 3 \ \unit{s}$.}
    \label{fig:tse-twophase-filling-p-alpha}
\end{figure}

The results obtained at consecutive times are displayed in Figure \ref{fig:ch5-tse-twophase-filling-u-p}, where the velocity magnitude and the pressure fields are plotted on a slice of the $x-z$ plane (view from above), while the evolution of the two-phase front in the whole TSE domain is reported in Figure \ref{fig:ch5-tse-twophase-filling-alpha} together with the corresponding axial distributions of average pressure and filling ratio in Figure \ref{fig:ch5-tse-twophase-filling-p-alpha-av}.

The same behavior we observed for the simplified SSE when imposing $Q_\text{in} > Q_0$ is reproduced for the TSE. At the final simulation time the device is completely filled with polymer, indeed $F \sim 1$ on all axial sections, and a negative pressure gradient originates from inlet to outlet.

\paragraph{Free-Surface Simulation of Partially Filled TSE}

The second setup for the two-phase simulation inside the TSE device is aimed at reproducing a partially filled scenario. To this end, the polymer is injected through the inlet patch at a reduced flow rate $Q_\text{in} < Q_0$.
All the other simulation settings remain unchanged with respect to the previous case.

\begin{figure}[h!]
\centering
\begin{subfigure}{\textwidth}
\centering
\includegraphics[width=0.45\textwidth]{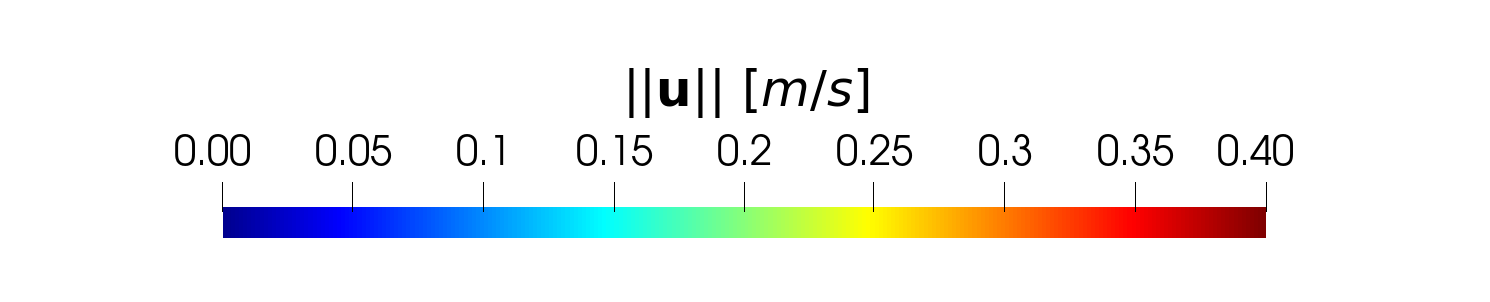} \\
\includegraphics[width=0.24\textwidth]{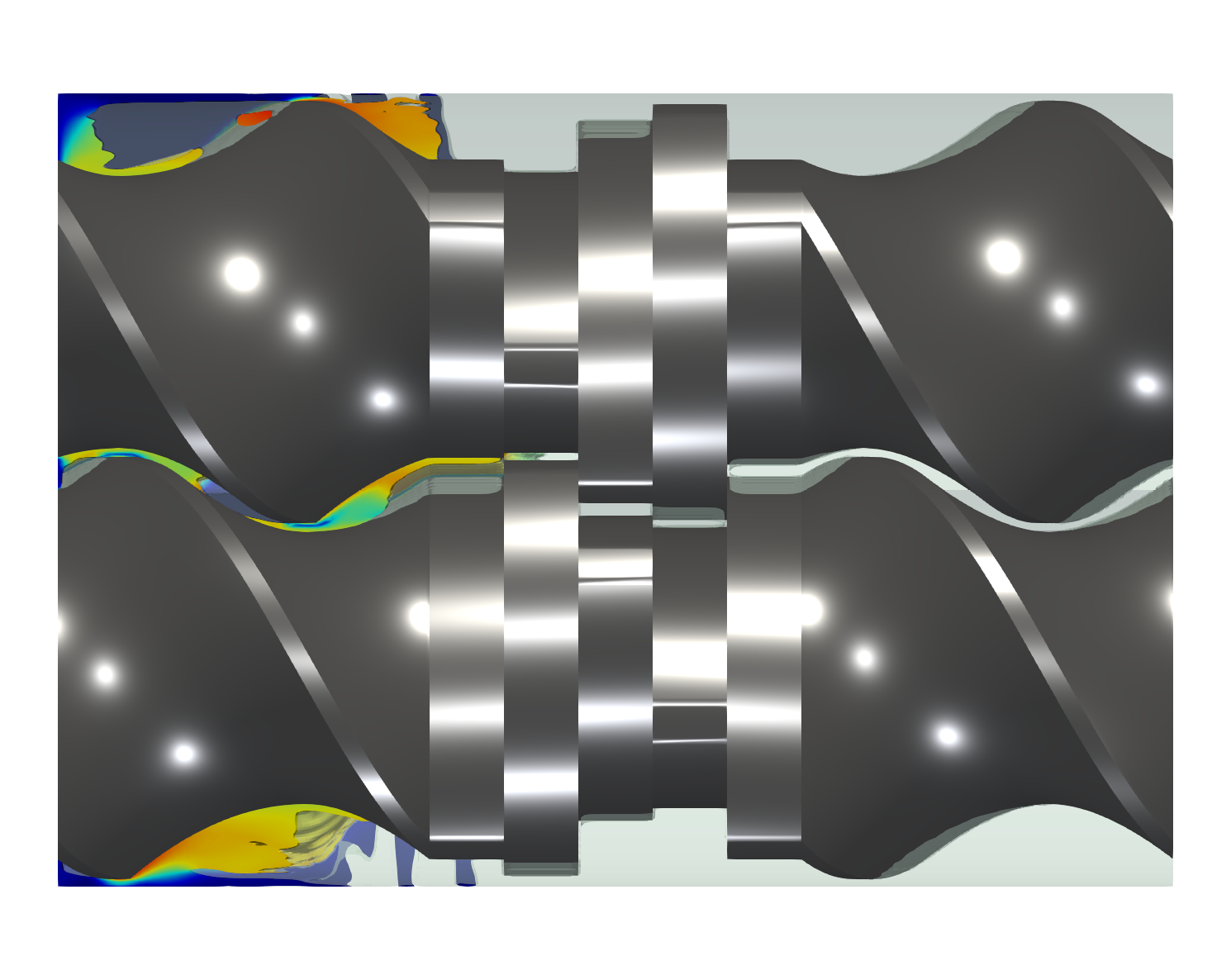}
\includegraphics[width=0.24\textwidth]{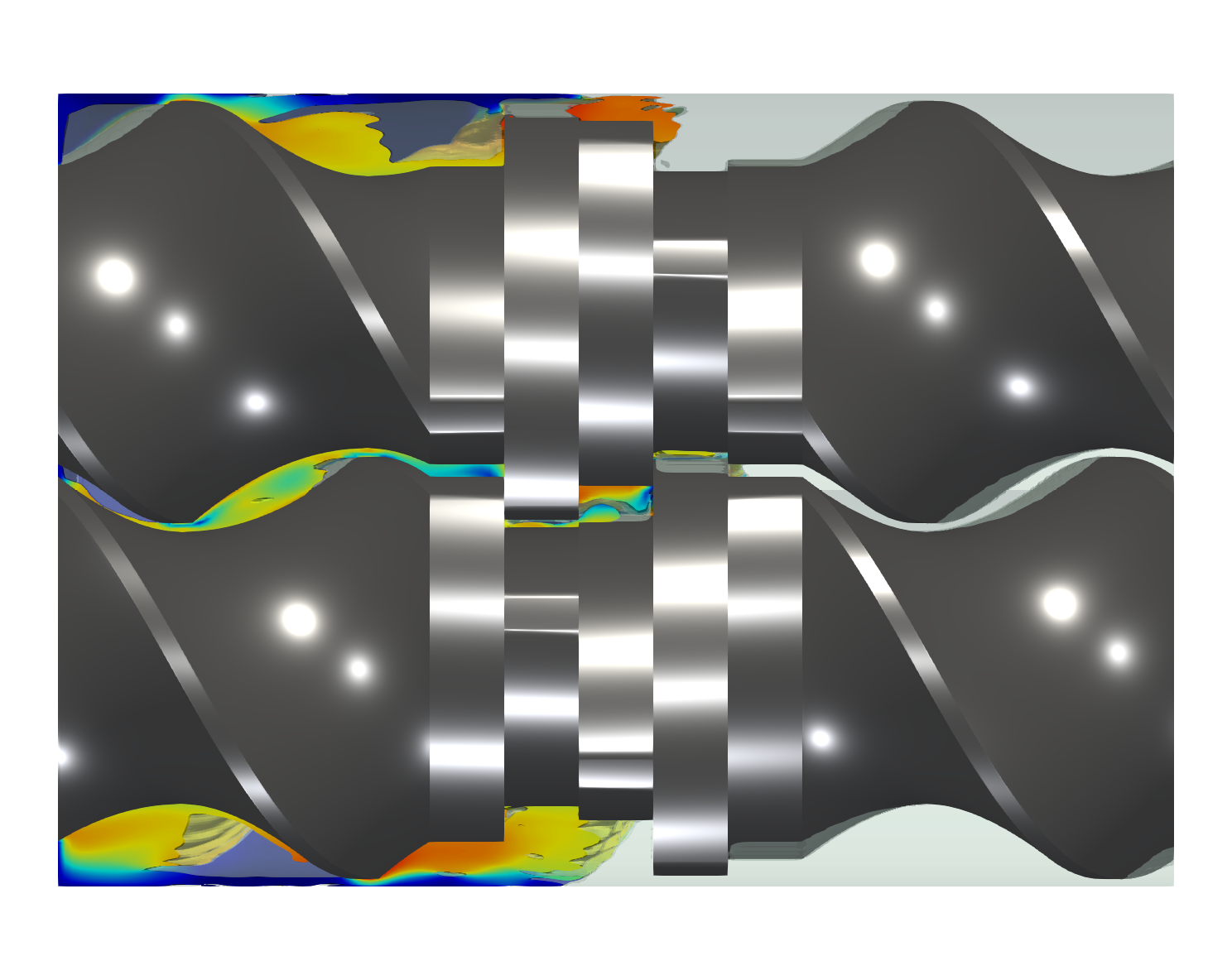}
\includegraphics[width=0.24\textwidth]{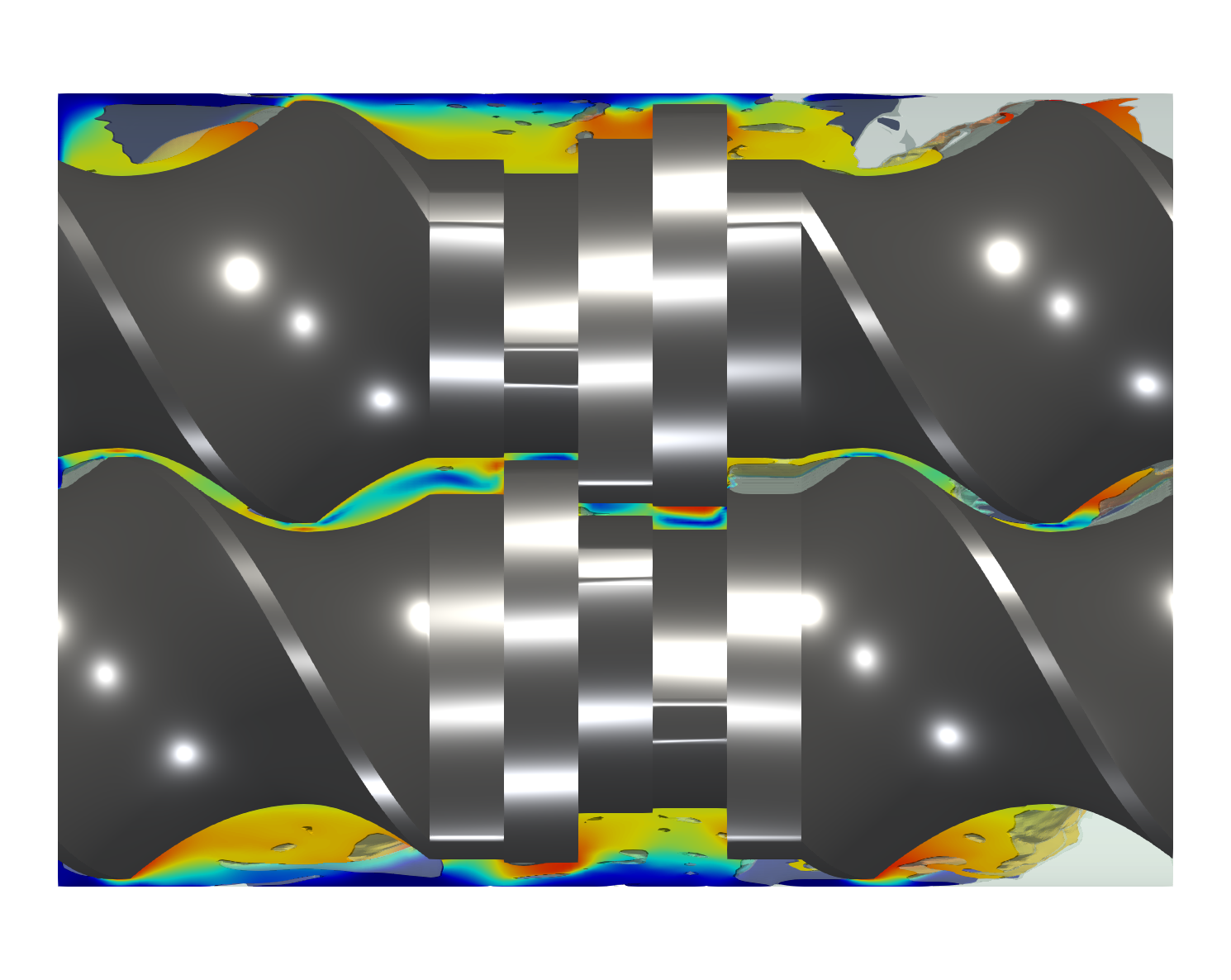}
\includegraphics[width=0.24\textwidth]{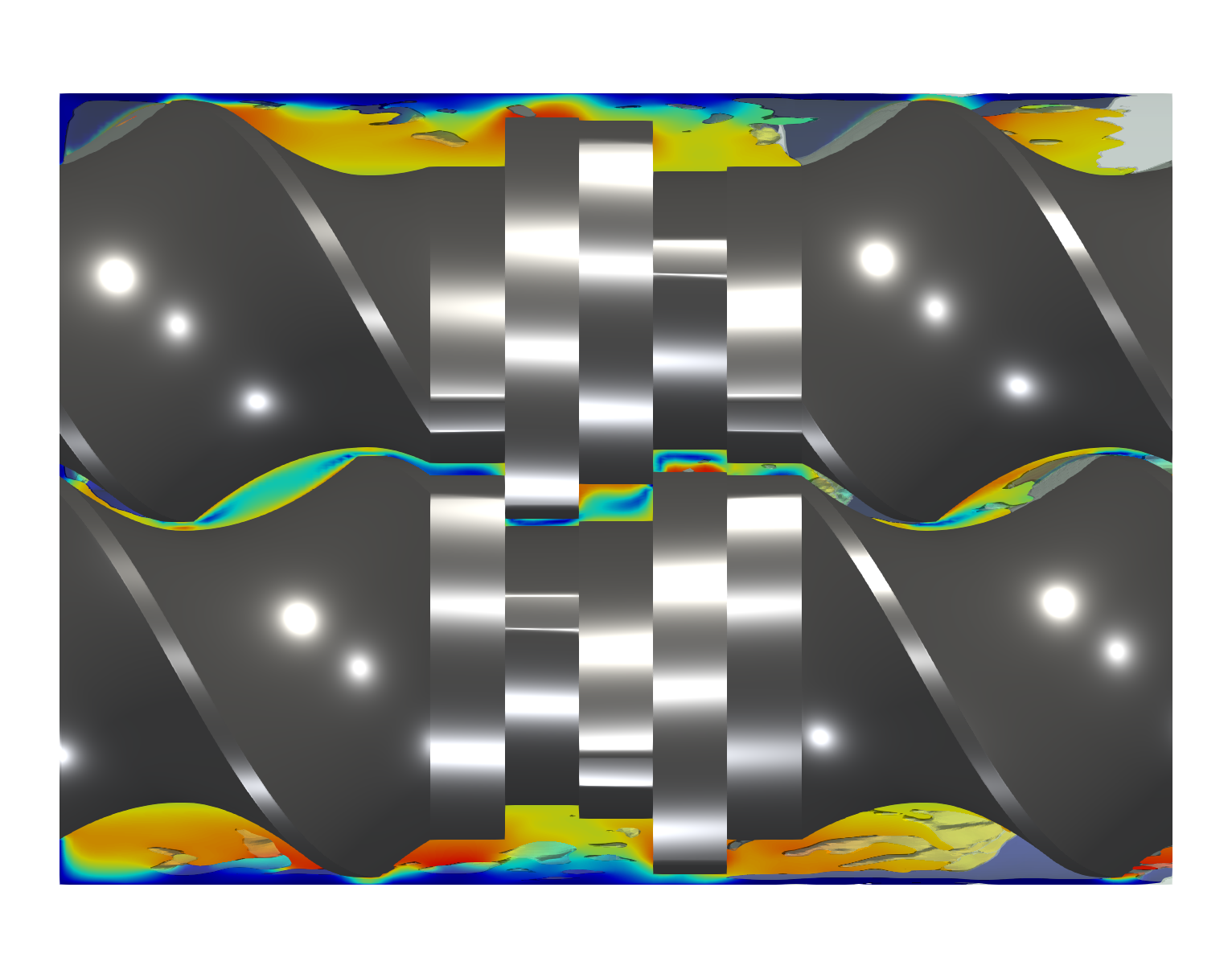}
\caption{Velocity magnitude field.}
\label{fig:ch5-tse-twophase-starving-u}
\end{subfigure}
\hfill
\begin{subfigure}{\textwidth}
\centering
\includegraphics[width=0.45\textwidth]{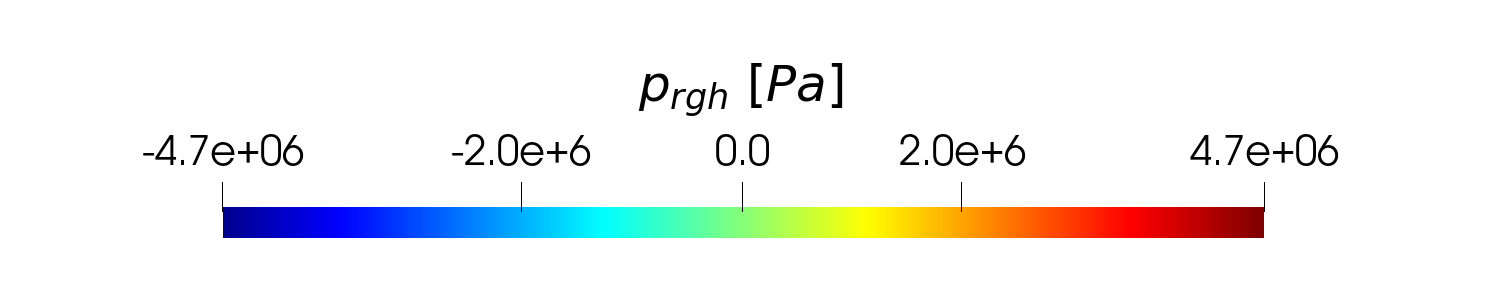} \\
\includegraphics[width=0.24\textwidth]{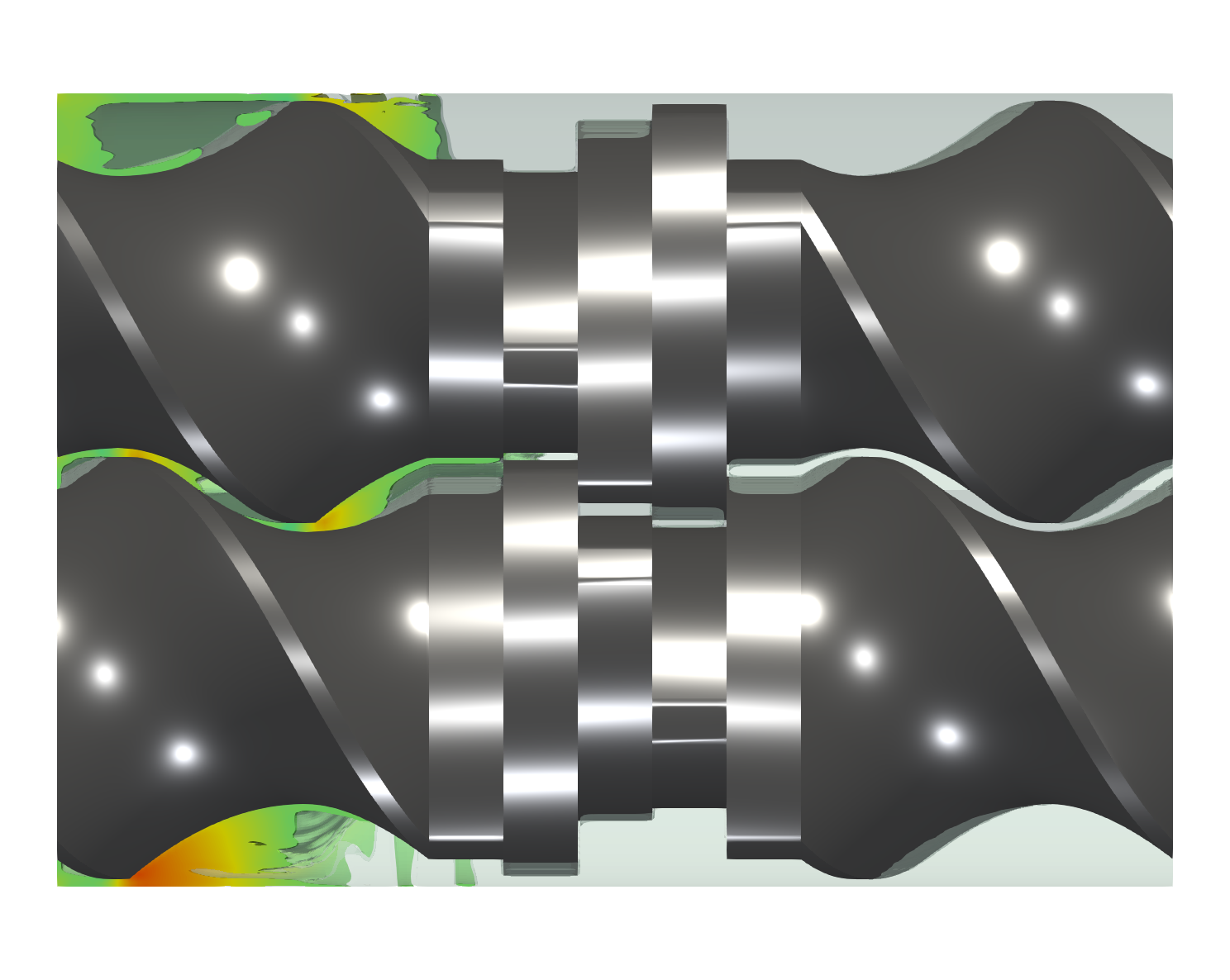}
\includegraphics[width=0.24\textwidth]{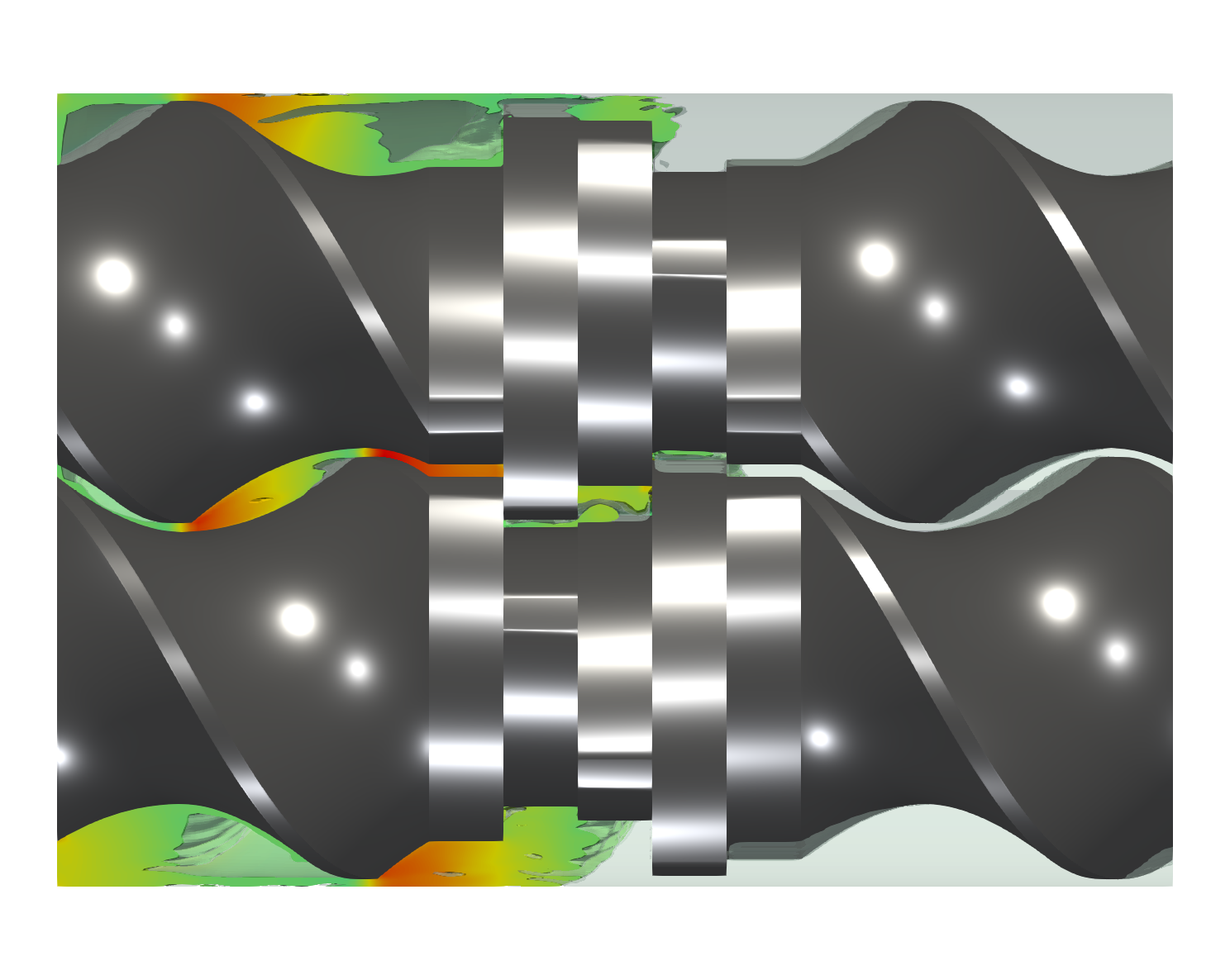}
\includegraphics[width=0.24\textwidth]{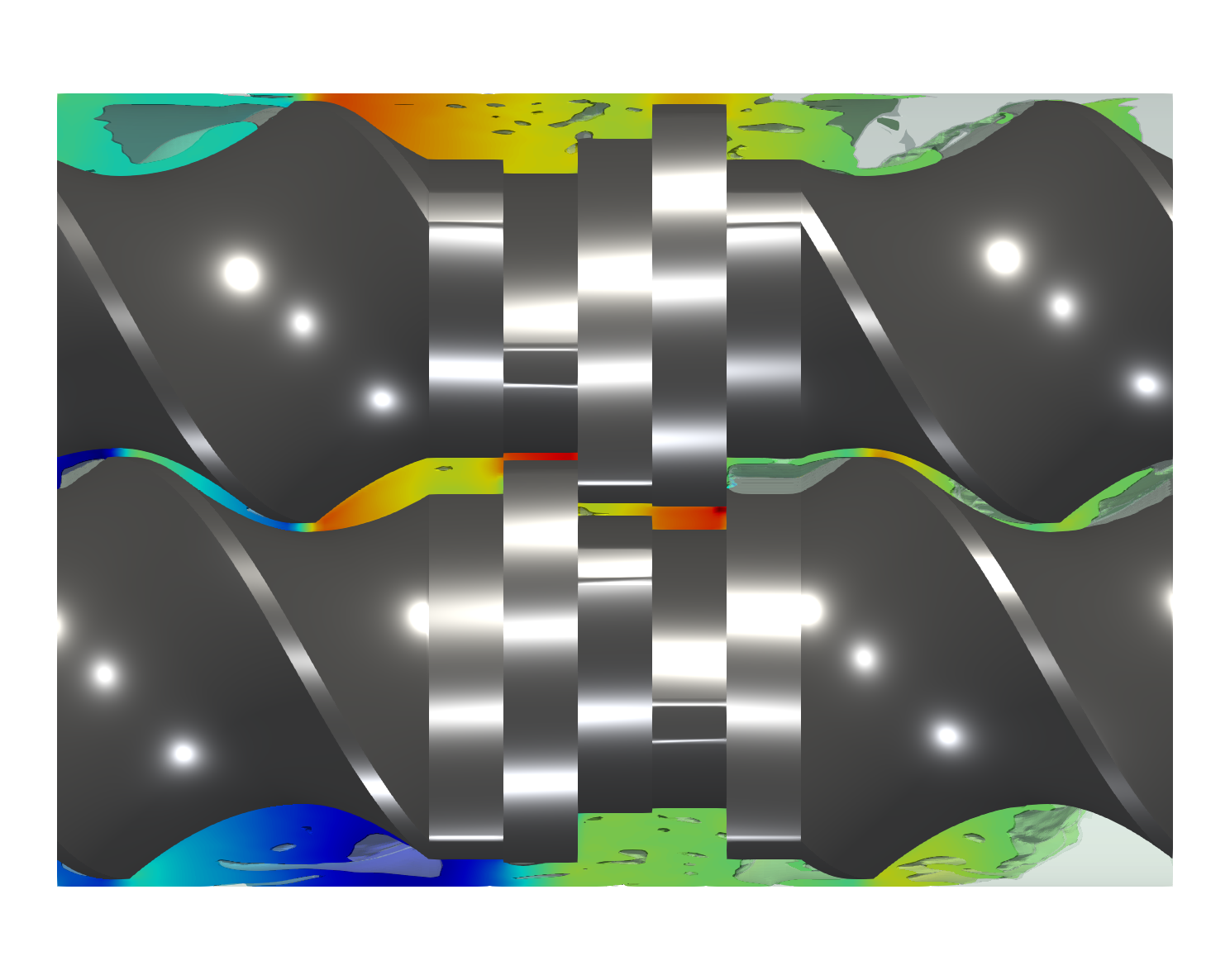}
\includegraphics[width=0.24\textwidth]{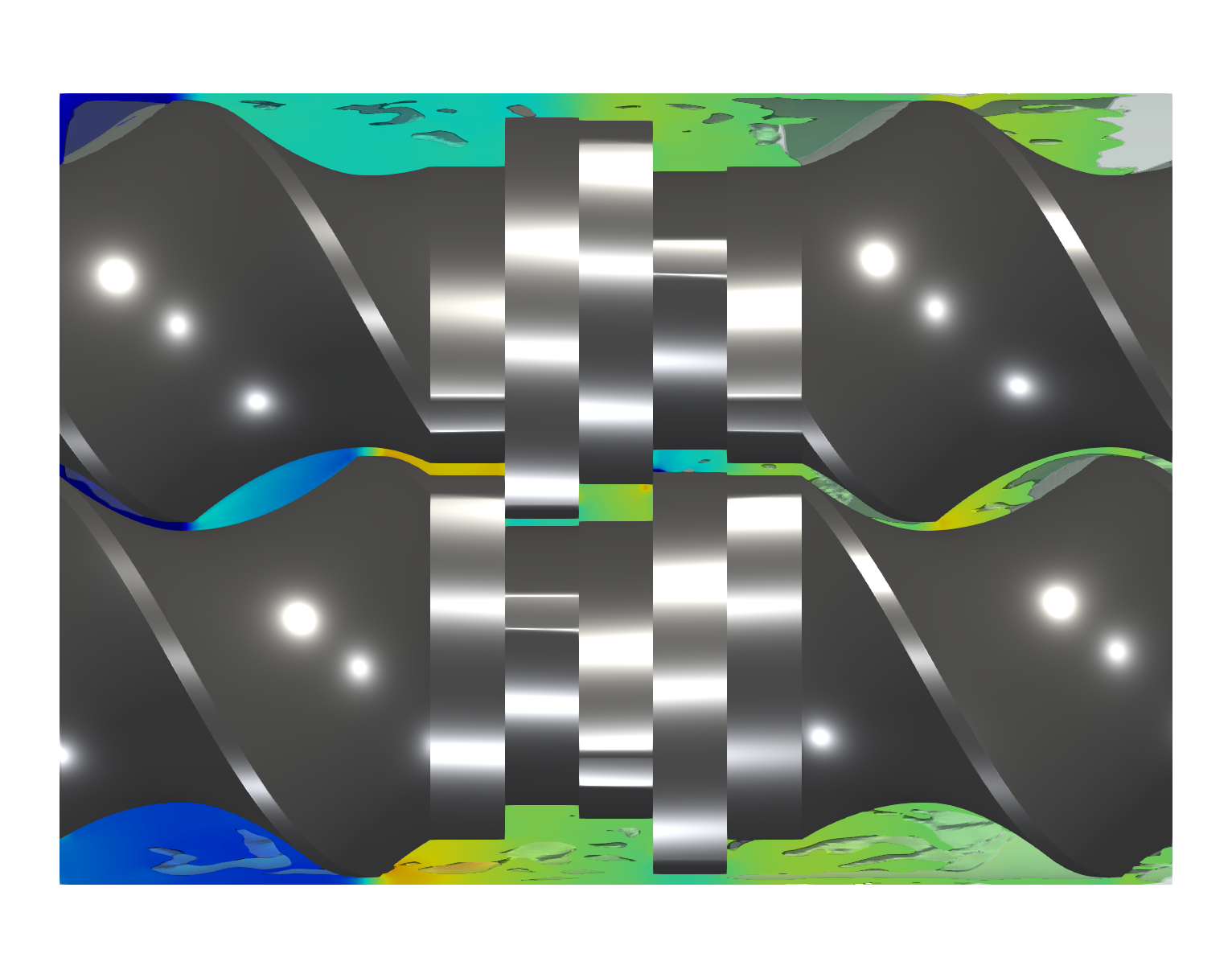}
\caption{Pressure field.}
\label{fig:ch5-tse-twophase-starving-p}
\end{subfigure}

\caption{Case of inlet flow rate $Q_\text{in} < Q_0$ simulated with the BC-VOF-IB solver. Fields are reported on a slice along the $x-z$ plane (top view) at time instants $t = 2 \ \unit{s}, \ 4 \ \unit{s}, \ 8 \ \unit{s}, \ 10 \ \unit{s}$.}
\label{fig:ch5-tse-twophase-starving-u-p}

\end{figure}

\begin{figure}[h!]
    \begin{subfigure}{0.3\textwidth}
    \centering
    \includegraphics[width=0.7\textwidth]{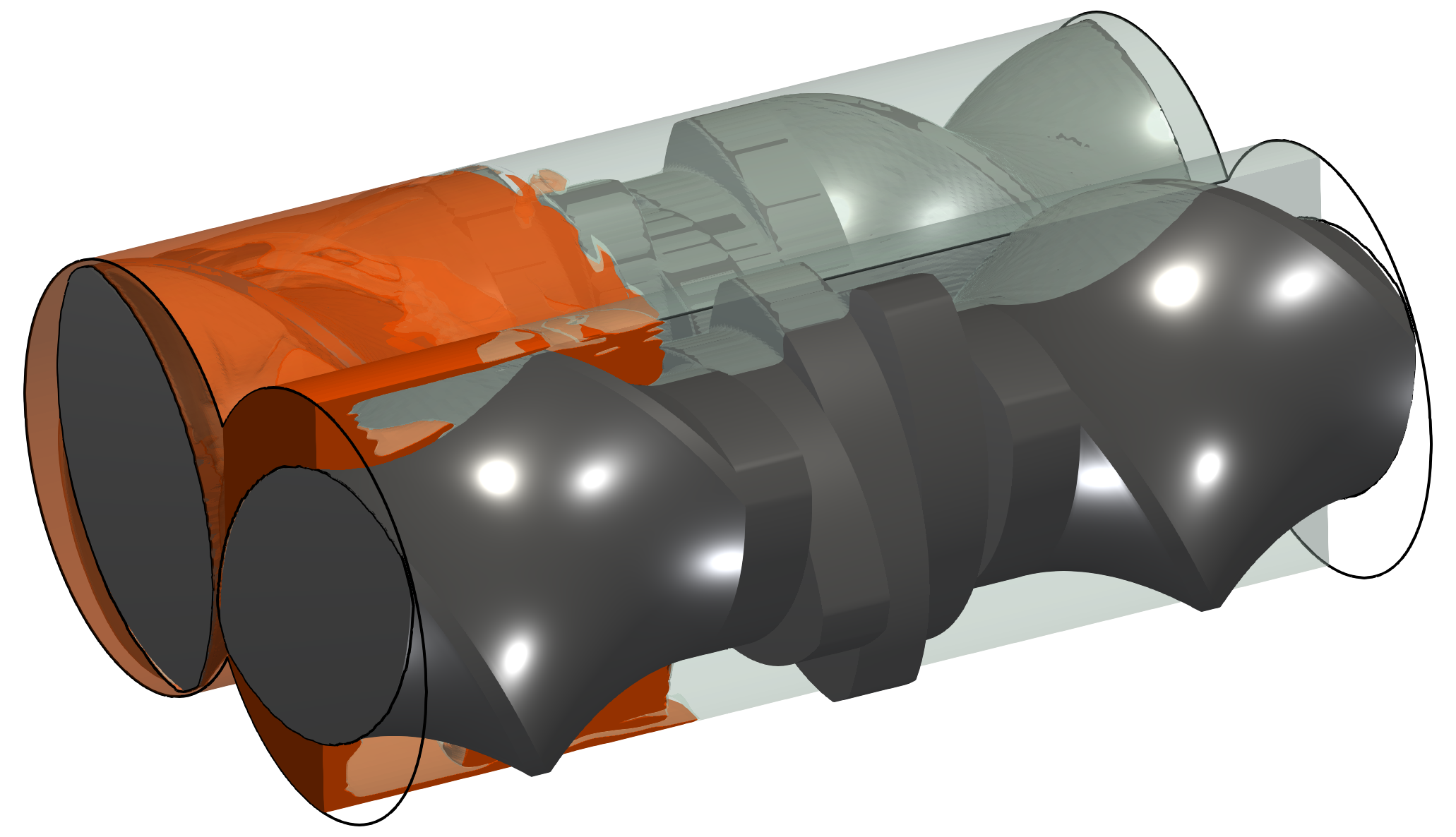}\\
    \vspace{0.5cm}
    \includegraphics[width=0.7\textwidth]{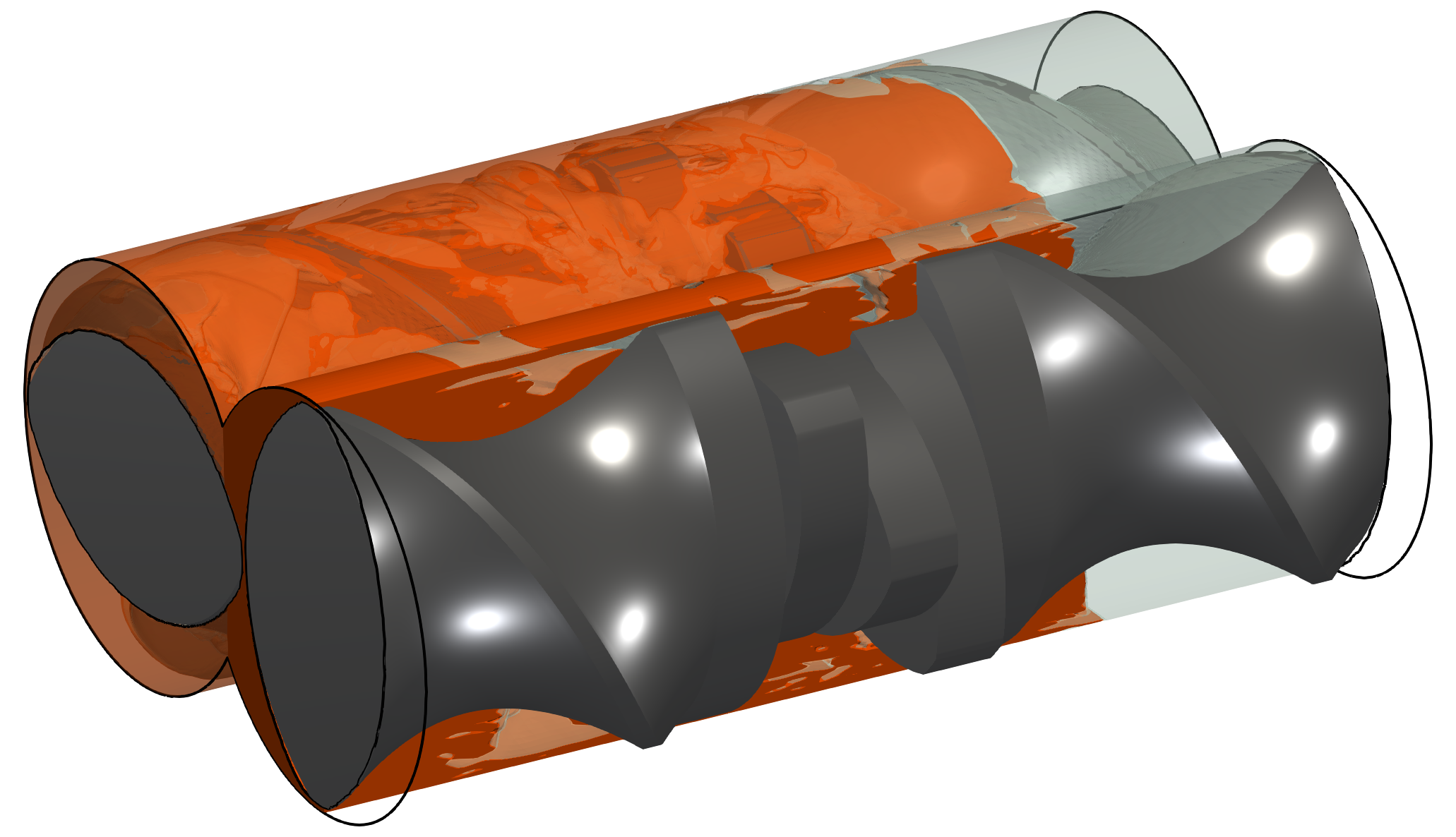}\\
    \vspace{0.5cm}
    \includegraphics[width=0.7\textwidth]{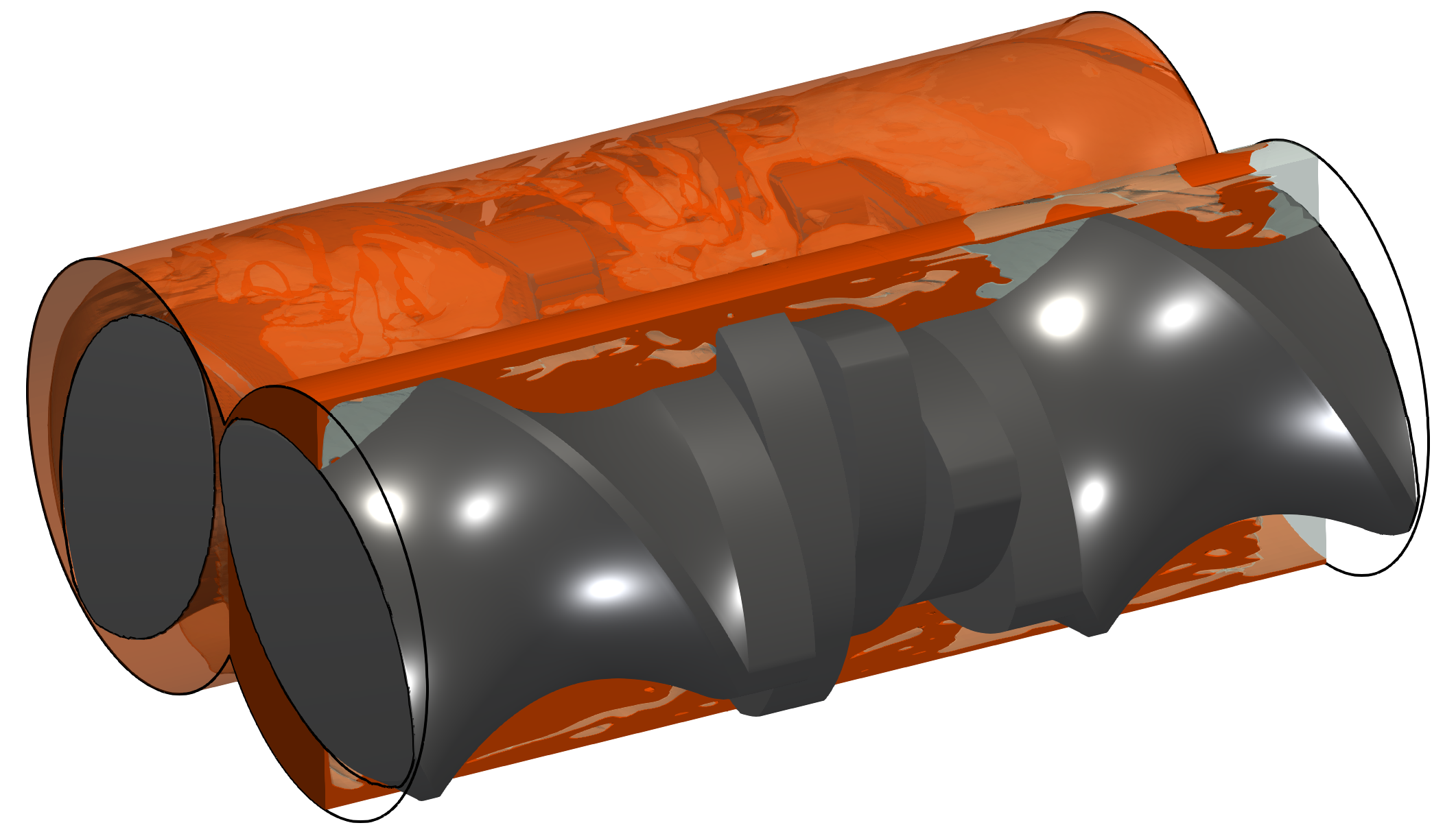}\\
    \vspace{0.7cm}
    \caption{Two-phase front evolution.}
    \label{fig:ch5-tse-twophase-starving-alpha}
    \end{subfigure}
    \hfill
    \begin{subfigure}{0.65\textwidth}
    \centering
    \includegraphics[width=0.8\textwidth]{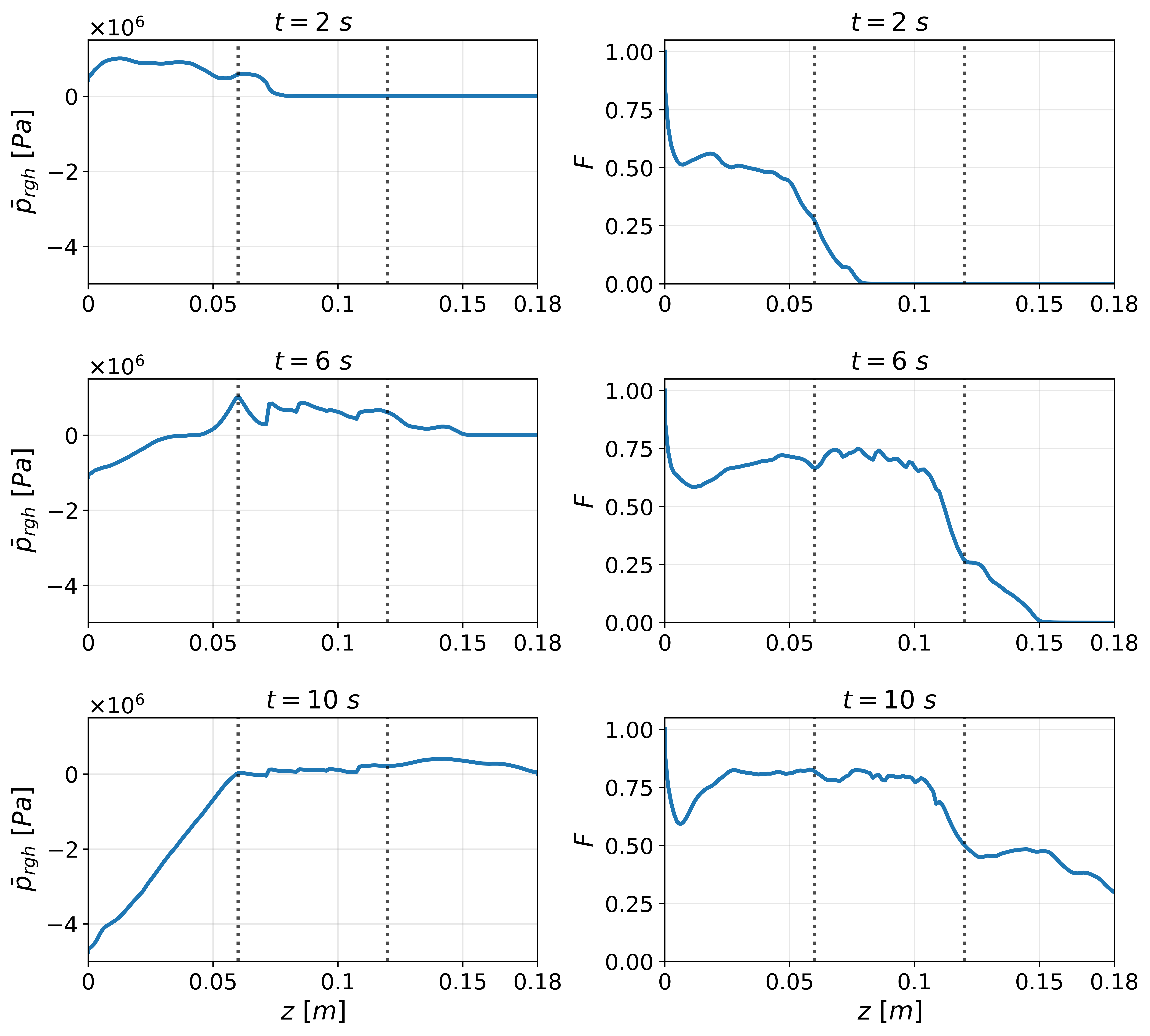}
    \caption{Average pressure and filling ratio over axial sections. Grey dotted lines represent start and ending of kneading modules.}
    \label{fig:ch5-tse-twophase-starving-p-alpha-av}
    \end{subfigure}

    \caption{Case of inlet flow rate $Q_\text{in} < Q_0$ simulated with the BC-VOF-IB solver. Temporal evolution of the two-phase interface (left) and plots of average pressure $\bar p_{rgh}$ and filling ratio $F$ along axial sections (right) is reported for times $t = 2 \ \unit{s}, \ 4 \ \unit{s}, \ 6 \ \unit{s}, \ 8 \ \unit{s}, \ 10 \ \unit{s}$.}
    \label{fig:tse-twophase-starving-p-alpha}
\end{figure}

Figure \ref{fig:ch5-tse-twophase-starving-u-p} shows the numerical results for the velocity magnitude and the pressure fields plotted on a slice of the $x-z$ plane (view from above). We represent also the two-phase front advancement in the TSE domain in Figure \ref{fig:ch5-tse-twophase-starving-alpha}, where at the last time step, after the material front has surpassed the kneading elements, more polymer accumulates in that portion of the device since the kneading modules are devoted to mixing rather than transport, as expected from experimental evidence.
Plots of average pressure and filling ratio along axial slices are reported in Figure \ref{fig:ch5-tse-twophase-starving-p-alpha-av}, where we observe an opposite behavior for pressure with respect to the previous full filling case. Here, prescribing an inlet flow rate $Q_\text{in} < Q_0$ produces a pressure increase from inlet to outlet as soon as the material first reaches the outflow boundary. 
 
For the sake of completeness, Table \ref{tab:ch5-TSE-setup} collects all simulation settings of the two TSE test cases presented in this work.
\begin{table}[h!]
    \renewcommand{\arraystretch}{1.3}
    \setlength{\tabcolsep}{5pt}  
    \small
    \centering
    \begin{NiceTabular}{|c|cc|}
        \noalign{\hrule height 1.2pt}
        Parameter & Filled & Starve-fed \\
        \noalign{\hrule height 0.8pt}
        $c_\alpha$ & 1 & 1  \\
        \hline
        \texttt{MULESCorr} & no & no \\
        \hline
        $N_\text{sc}$ & 10 & 50 \\        
        \hline
        $N_{\alpha,\text{corr}}$ & 1 & 1 \\
        \hline
        $N_\text{inner}$ & 5 & 10 \\
        \hline
        $N_\text{outer}$ & 10 &  10 \\
        \noalign{\hrule height 0.8pt}
        $p_{rgh}$ solver & \multicolumn{2}{c|}{PCG (GAMG) $tol = 10^{-6}$} \\
        \hline
        $\bm u$ solver & \multicolumn{2}{c|}{GMRES (LUSGS) $tol = 10^{-5}$} \\
        \noalign{\hrule height 0.8pt} 
        $N_T$ & \num{6929} & \num{57685} \\
        \hline
        $\text{Co}_{h,\max}$ & 20 & 50 \\
        \hline
        $\text{Co}_{\alpha,h,\max}$ & 1.5 & 2\\
        \hline
        $\Delta t^0 \ [\unit{s}]$ & $10^{-5}$ & $10^{-5}$ \\
        \hline
        $\Delta t_{\max} \ [\unit{s}]$ & 0.001 & 0.001 \\
        \hline
        $T_f \ [\unit{s}]$ & 3 s & 10 s \\
        \noalign{\hrule height 1.2pt}
    \end{NiceTabular}
    \caption{Numerical settings for the TSE test cases.}
    \label{tab:ch5-TSE-setup}
\end{table}


\section{Conclusions}
\label{sec:discussion}

The central contribution of the present work is a Volume of Fluid Immersed Boundary (VOF-IB) method, which couples a VOF interface-capturing solver with a non-conforming immersed boundary approach, enabling two-phase flow simulations in complex geometries without the need for body-fitted meshes. The IB conditions for velocity and volume fraction are imposed at the algebraic level through the WLS reconstruction, and the pressure equation is derived to be consistent with both the IB constraint and the volume-fraction transport.

The time-step restrictions and instabilities that affect standard segregated VOF solvers in the presence of large viscosity contrasts are addressed by extending the block-coupled discretization of Cardiff et al. \cite{cardiff2016block} to incompressible two-phase flows. In this discretization, the full viscous stress, including its transposed-gradient part, is treated implicitly, and the resulting block-coupled momentum system is combined with the IB constraint.

A comparison on an injection-molding test case (Section~\ref{sec:dogbone}) shows that, at the time steps allowed by the block-coupled solver, segregated solvers either diverge or produce spurious air-phase velocities, including a segregated variant with the same algorithmic structure. The block-coupled solver yields a physical solution and reduces the computing time by a factor of about 2.7 with respect to the fastest segregated run that remains physical (from 1 h 4 min to 24 min).
A systematic assessment of the stability limits of the two solvers as functions of the viscosity ratio and of the mesh size, for instance on a two-layer circular Couette flow with an analytical solution, is left for future work.

The method has then been applied to free-surface simulations of single- and twin-screw extruders under fully and partially filled conditions, capturing the filling dynamics and the resulting pressure build-up along the screw.

The present formulation has five main limitations, each of which defines a direction for future work. First, the simulations are isothermal, whereas viscous heating and temperature-dependent viscosity play a major role in polymer mixing \cite{macosko1994, tadmor2013principles, Larsen2023sigma, Ahmed2019fill}; the next step is therefore to solve the energy equation for non-isothermal free-surface flows. 
Second, the polymer melt is currently described by generalized Newtonian models that do not account for elastic effects; viscoelastic constitutive models \cite{Favero2010viscoelastic, Habla2011viscoelastic, PimentaAlves2017} would provide a more faithful rheological description.
Third, with no-slip walls the contact line moves only through numerical slip, so that its position near walls depends on the mesh, as reflected by the behavior of the Hausdorff distance in Table~\ref{tab:ch4-dogbone-non-newtonian-high-distances}, and the static contact angle is restricted to $90^\circ$; a systematic use of the available Navier-slip condition, together with a dynamic contact-angle model at walls and immersed boundaries, would address both issues.
Fourth, although the block-coupled solver reduces the computing time substantially, simulating a complete industrial process remains expensive; profiling and optimization of the implementation, together with tuned simulation settings, are needed to simulate full-length extruders over process-relevant time scales, and a study of the memory requirements and parallel scalability of the block-coupled solver, whose extended stencil increases the number of non-zero entries per row, is part of this effort.
Fifth, a formal verification of the method, based on manufactured solutions and on space and time refinement studies \cite{roache2002code,celik2008procedure}, has not yet been carried out, and, partly because of the computational cost and partly because suitable experimental data are scarce, the extruder results have not yet been validated quantitatively. Systematic comparison with literature results \cite{Dong2022, Dong2023SPH} and with industrial measurements is essential to establish the predictive capability of the method in real-world applications.
Finally, a compressible formulation of the gas phase \cite{Georgoulas2021compressibleVOF} would allow the dynamics of entrapped air bubbles to be captured under realistic processing conditions.

\vspace{1cm}

\subsection*{Authors' Contributions} 

\textbf{Emilia Capuano:} Writing - review \& editing, Writing - original draft, Software, Methodology, Formal analysis and investigation, Numerical simulations, Validation, Conceptualization. \textbf{Daniele Cerroni:} Writing - review \& editing, Conceptualization. \textbf{Holger Marschall:} Writing - review \& editing, Methodology, Formal analysis and investigation, Conceptualization. \textbf{Giorgio Negrini:} Writing - review \& editing, Software, Methodology, Formal analysis and investigation, Conceptualization. \textbf{Nicola Parolini:}  Writing - review \& editing, Methodology, Formal analysis and investigation, Conceptualization, Supervision. \textbf{Marco Verani:} Writing - review \& editing, Supervision

\subsection*{Acknowledgments} 

This research is part of the activities of Dipartimento di Eccellenza 2023-2027. EC, NP, MV are members of the INdAM group GNCS ``Gruppo Nazionale per il Calcolo Scientifico'' (National Group for Scientific Computing). The authors acknowledge financial support from the Ministerial Decree no. 352 of 9th April 2022, based on the NRRP - funded by the European Union - NextGenerationEU - Mission 4 ``Education and Research'', Component 2 ``From Research to Business'', Investment 3.3, and by the company Pirelli Tyre S.p.A. 

\subsection*{Competing Interests} 

The authors declare no competing interests.

\subsection*{Data Availability Statement}

The source code of the solvers developed in this work is part of an in-house library, and the geometry and physical parameters of the twin-screw extruder were provided by the industrial partner. For confidentiality reasons, neither is publicly available. The parameters needed to reproduce the dog-bone test case are reported in Section \ref{sec:dogbone}.

\bibliographystyle{abbrv}
\bibliography{references.bib}
\end{document}